\documentclass{article}
\usepackage{epubtk}
\usepackage{epsf}
\usepackage{graphicx}
\usepackage{amsmath}
\usepackage{amssymb}
\usepackage{color}

\showlistoftablesfalse

\DeclareMathOperator{\curl}{curl}
\DeclareMathOperator{\diag}{diag}

\newcommand{\D}{\vec\nabla}
\newcommand{\cu}{\rho_{{\mathcal E}}}
\newcommand{\cq}{q^{{\mathcal E}}}
\newcommand{\cp}{\pi^{{\mathcal E}}}
\newcommand{\bcq}{\bar{q}^{{\mathcal E}}}
\newcommand{\bcp}{\bar{\pi}^{{\mathcal E}}}
\newcommand*{\di}{\partial}
\renewcommand*{\b}{\text{b}}
\newcommand*{\KK}{\mathcal{E}}
\newcommand*{\eff}{\text{\tiny{ref}}}
\newcommand*{\fiveD}{\text{\tiny{5D}}}
\newcommand*{\prim}{\text{\tiny{inf}}}
\newcommand*{\GR}{\text{\tiny{GR}}}
\renewcommand*{\c}{\text{c}}

\newcommand{\new}[1]{\textcolor{blue}{#1}}

\begin{document}

\title{Brane-World Gravity}

\author{%
\epubtkAuthorData{Roy Maartens}
        {Institute of Cosmology \& Gravitation \\
         University of Portsmouth \\
         Portsmouth PO1~3FX, U.K.}
        {roy.maartens@port.ac.uk}
        {http://research.icg.port.ac.uk/members}
\\
\and \\
\epubtkAuthorData{Kazuya Koyama}
        {Institute of Cosmology \& Gravitation \\
         University of Portsmouth \\
         Portsmouth PO1~3FX, U.K.}
        {kazuya.koyama@port.ac.uk}
        {http://research.icg.port.ac.uk/members}
}

\date{}
\maketitle


\begin{abstract}
  The observable universe could be a 1+3-surface (the ``brane'')
  embedded in a 1+3+\textit{d}-dimensional spacetime (the ``bulk''), with
  Standard Model particles and fields trapped on the brane while gravity
  is free to access the bulk. At least one of the \textit{d} extra spatial
  dimensions could be very large relative to the Planck scale, which
  lowers the fundamental gravity scale, possibly even down to the
  electroweak ($\sim$~TeV) level. This revolutionary picture arises in
  the framework of recent developments in M~theory. The 1+10-dimensional
  M~theory encompasses the known 1+9-dimensional superstring theories,
  and is widely considered to be a promising potential route to quantum
  gravity.
  At low energies, gravity is localized at the brane and
  general relativity is recovered, but at high energies gravity
  ``leaks'' into the bulk, behaving in a truly higher-dimensional
  way. This introduces significant changes to gravitational dynamics
  and perturbations, with interesting and potentially testable
  implications for high-energy astrophysics, black holes, and
  cosmology. Brane-world models offer a phenomenological way to test
  some of the novel predictions and corrections to general
  relativity that are implied by M~theory. This review analyzes the
  geometry, dynamics and perturbations of simple brane-world models
  for cosmology and astrophysics, mainly focusing on warped
  5-dimensional brane-worlds based on the Randall--Sundrum models. We
  also cover the simplest brane-world models in which 4-dimensional
  gravity on the brane is modified at \emph{low} energies -- the 5-dimensional
  Dvali--Gabadadze--Porrati models. Then we discuss co-dimension two
  branes in 6-dimensional models.
\end{abstract}

\epubtkKeywords{Brane-world models}

\newpage

\epubtkUpdate
    [Id=A,
     ApprovedBy=subjecteditor,
     AcceptDate={21 April 2010},
     PublishDate={21 April 2010},
     Type=major]{%
Inserted subsections on numerical solutions in
    Sections~\ref{section_6} and \ref{section_7} and five
    figures. Added new Subsection~\ref{section_2-2} on the
    Randall--Sundrum model in string theory. Added new
    Section~\ref{section_9} on Dvali--Gabadadze--Porrati models and
    nine figures. Added new Section~\ref{section_10} on 6-dimensional
    models and one figure. More than 80 new references.}

\newpage


\section{Introduction}
\label{section_1}

At high enough energies, Einstein's theory of general relativity
breaks down, and will be superceded by a quantum gravity theory.
The classical singularities predicted by general relativity in
gravitational collapse and in the hot big bang will be removed by
quantum gravity. But even below the fundamental energy scale that
marks the transition to quantum gravity, significant corrections
to general relativity will arise. These corrections could have a
major impact on the behaviour of gravitational collapse, black
holes, and the early universe, and they could leave a trace -- a
``smoking gun'' -- in various observations and experiments. Thus it
is important to estimate these corrections and develop tests for
detecting them or ruling them out. In this way, quantum gravity
can begin to be subject to testing by astrophysical and
cosmological observations.

Developing a quantum theory of gravity and a unified theory of all
the forces and particles of nature are the two main goals of
current work in fundamental physics. There is as yet no generally
accepted (pre-)quantum gravity theory. Two of the main contenders
are M~theory (for reviews see, e.g., \cite{mtheory_1,
  mtheory_2, mtheory_3}) and quantum geometry (loop quantum
gravity; for reviews see, e.g., \cite{loop_1, loop_2}). It is
important to explore the astrophysical
and cosmological predictions of both these approaches. This review
considers only models that arise within the framework of M~theory.

In this review, we focus on RS brane-worlds (mainly the RS 1-brane model) and
their generalizations, with the emphasis on geometry and
gravitational dynamics (see~\cite{m2, rev_1, rev_2, rev_3, rev_4,
  rev_5, rev_6, rev_7, rev_8, rev_9, lan} for previous reviews
with a broadly similar approach). Other reviews focus on
string-theory aspects, e.g., \cite{que_1, que_2, que_3, que_4}, or on particle physics
aspects, e.g., \cite{r_1, r_2, r_3, r_4, cav}. We also discuss the 5D
DGP models, which modify general relativity at low energies, unlike
the RS models; these models have become important examples in
cosmology for achieving late-time acceleration of the universe without
dark energy. Finally, we give brief overviews of 6D models, in
which the brane has co-dimension two, introducing very different
features to the 5D case with co-dimension one branes.

\epubtkUpdateA{Extended a paragraph moved here from the end of
  Section~\ref{section_1.2}.}


\subsection{Heuristics of higher-dimensional gravity}

One of the fundamental aspects of string theory is the need for
extra spatial dimensions\epubtkFootnote{We do not consider timelike
  extra dimensions: see~\cite{varun1} for an interesting
  example.}. This revives the original higher-dimensional ideas of
Kaluza and Klein in the 1920s, but in a new context of quantum
gravity. An important consequence of extra dimensions is that the

4-dimensional Planck scale $M_\mathrm{p}\equiv M_{4}$ is no longer the
fundamental scale, which is $M_{4+d}$, where $d$ is the number of
extra dimensions. This can be seen from the modification of the
gravitational potential. For an Einstein--Hilbert gravitational action
we have
\begin{eqnarray}
  S_\mathrm{gravity} &=&
  {1\over 2\kappa_{4+d}^2}\int d^4x\, d^dy\,\sqrt{-^{(4+d)\!}g}
  \left[ {}^{(4+d)\!}R- 2\Lambda_{4+d} \right],
  \\
  {}^{(4+d)\!}G_{AB} & \equiv & \;{}^{(4+d)\!}R_{AB}-{1\over2}
  \;{}^{(4+d)\!} R \;{}^{(4+d)\!}g_{AB} = -\Lambda_{4+d}
  \;{}^{(4+d)\!}g_{AB}+ \kappa_{4+d}^2 \;{}^{(4+d)\!}T_{AB},
  \label{defe}
\end{eqnarray}%
where $X^A=(x^\mu,y^1, \dots, y^d)$, and $\kappa_{4+d}^2$ is the
gravitational coupling constant,
\begin{equation}
  \kappa_{4+d}^2=8\pi G_{4+d}={8\pi\over M_{4+d}^{2+d}}.
\end{equation}
The static weak field limit of the field equations leads to the
$4+d$-dimensional Poisson equation, whose solution is the
gravitational potential,
\begin{equation}
  V(r) \propto {\kappa_{4+d}^2\over r^{1+d}}.
  \label{v}
\end{equation}
If the length scale of the extra dimensions is $L$, then on scales
$r\lesssim L$, the potential is $4+d$-dimensional, $V\sim
r^{-(1+d)}$. By contrast, on scales large relative to $L$, where
the extra dimensions do not contribute to variations in the
potential, $V$ behaves like a 4-dimensional potential, i.e.,
$r\sim L$ in the $d$ extra dimensions, and $V \sim L^{-d}r^{-1}$.
This means that the usual Planck scale becomes an effective
coupling constant, describing gravity on scales much larger than
the extra dimensions, and related to the fundamental scale via the
volume of the extra dimensions:
\begin{equation}
  M_\mathrm{p}^2 \sim M_{4+d}^{2+d}\,L^d.
\end{equation}
If the extra-dimensional volume is Planck scale, i.e., $L\sim
M_\mathrm{p}^{-1}$, then $M_{4+d}\sim M_\mathrm{p}$. But if the extra-dimensional
volume is significantly above Planck scale, then the true
fundamental scale $M_{4+d}$ can be much less than the effective
scale $M_\mathrm{p} \sim 10^{19} \mathrm{\ GeV}$. In this case, we understand
the weakness of gravity as due to the fact that it ``spreads'' into
extra dimensions and only a part of it is felt in 4 dimensions.

A lower limit on $M_{4+d}$ is given by null results in table-top
experiments to test for deviations from Newton's law in 4
dimensions, $V\propto r^{-1}$. These experiments
currently~\cite{exp} probe sub-millimetre scales, so that
\begin{equation}
  L \lesssim 10^{-1} \mathrm{\ mm} \sim  (10^{-15} \mathrm{\ TeV})^{-1}
  \quad \Rightarrow \quad
  M_{4+d}\gtrsim 10^{(32-15d)/(d+2)} \mathrm{\ TeV}.
  \label{tt}
\end{equation}
Stronger bounds for brane-worlds with compact flat extra
dimensions can be derived from null results in particle
accelerators and in high-energy
astrophysics~\cite{cav, cheung, hanraf_1, hanraf_2}.


\subsection{Brane-worlds and M~theory}
\label{section_1.2}

String theory thus incorporates the possibility that the
fundamental scale is much less than the Planck scale felt in 4
dimensions. There are five distinct 1+9-dimensional superstring
theories, all giving quantum theories of gravity. Discoveries in
the mid-1990s of duality transformations that relate these
superstring theories and the 1+10-dimensional supergravity theory,
led to the conjecture that all of these theories arise as
different limits of a single theory, which has come to be known as
M~theory. The 11th dimension in M~theory is related to the string
coupling strength; the size of this dimension grows as the
coupling becomes strong. At low energies, M~theory can be
approximated by 1+10-dimensional supergravity.

It was also discovered that p-branes, which are extended objects
of higher dimension than strings (1-branes), play a fundamental
role in the theory. In the weak coupling limit, p-branes ($p>1$)
become infinitely heavy, so that they do not appear in the
perturbative theory. Of particular importance among p-branes are
the D-branes, on which open strings can end. Roughly speaking,
open strings, which describe the non-gravitational sector, are
attached at their endpoints to branes, while the closed strings of
the gravitational sector can move freely in the bulk. Classically,
this is realised via the localization of matter and radiation
fields on the brane, with gravity propagating in the bulk (see
Figure~\ref{figure_01}).

\epubtkImage{figure01.png}{%
 \begin{figure}[htbp]
   \def\epsfsize#1#2{0.5#1}
   \centerline{\epsfbox{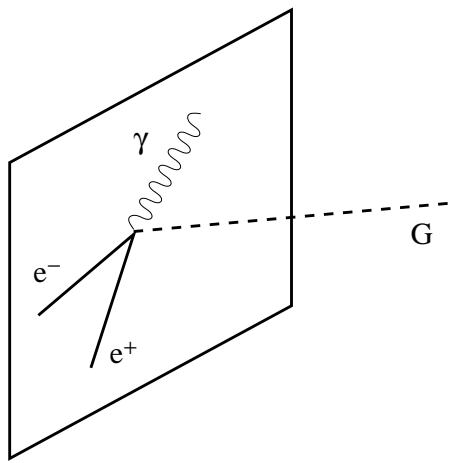}}
   \caption{Schematic of confinement of matter to the brane, while
     gravity propagates in the bulk (from~\cite{cav}).}
   \label{figure_01}
 \end{figure}}

In the Horava--Witten solution~\cite{hv}, gauge fields of the
standard model are confined on two 1+9-branes located at the end
points of an $S^1/Z_2$ orbifold, i.e., a circle folded on itself
across a diameter. The 6 extra dimensions on the branes are
compactified on a very small scale close to the fundamental
scale, and their effect on the dynamics is felt through ``moduli''
fields, i.e., 5D scalar fields. A 5D realization of the
Horava--Witten theory and the corresponding brane-world cosmology
is given in~\cite{low_1, low_2, low_3}.

\epubtkImage{figure02.png}{%
 \begin{figure}[htbp]
   \def\epsfsize#1#2{0.5#1}
   \centerline{\epsfbox{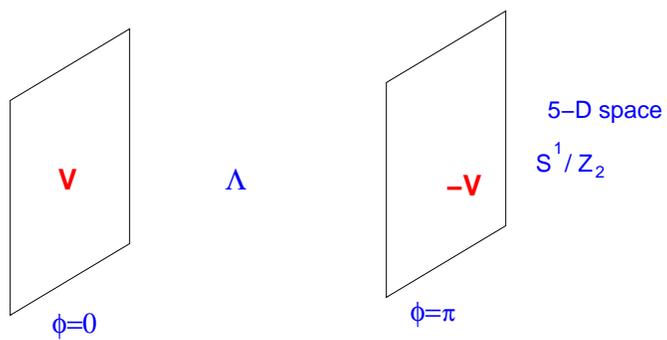}}
   \caption{The RS 2-brane model. (Figure taken from~\cite{cheung}.)}
   \label{figure_02}
 \end{figure}}

These solutions can be thought of as effectively 5-dimensional,
with an extra dimension that can be large relative to the
fundamental scale. They provide the basis for the Randall--Sundrum
(RS) 2-brane models of 5-dimensional gravity~\cite{rs1} (see
Figure~\ref{figure_02}). The single-brane Randall--Sundrum
models~\cite{rs2} with infinite extra dimension arise when the
orbifold radius tends to infinity. The RS models are not the only
phenomenological realizations of M~theory ideas. They were preceded by the
Arkani--Hamed--Dimopoulos--Dvali (ADD) brane-world models~\cite{add_1,
  add_2, add_3, add_4, add_5, add_6, add_7, add_8},
which put forward the idea that a large volume for the compact
extra dimensions would lower the fundamental Planck scale,
\begin{equation}
  M_\mathrm{ew}\sim 1 \mathrm{\ TeV} \lesssim M_{4+d} \leq M_\mathrm{p}
  \sim 10^{16} \mathrm{\ TeV},
  \label{scales}
\end{equation}
where $M_\mathrm{ew}$ is the electroweak scale. If $M_{4+d}$ is close
to the lower limit in Equation~(\ref{scales}), then this would address
the long-standing ``hierarchy'' problem, i.e., why there is such a
large gap between $M_\mathrm{ew}$ and $M_\mathrm{p}$.

In the ADD models, more than one extra dimension is required for
agreement with experiments, and there is ``democracy'' amongst the
equivalent extra dimensions, which are typically flat. By
contrast, the RS models have a ``preferred'' extra dimension, with
other extra dimensions treated as ignorable (i.e., stabilized
except at energies near the fundamental scale). Furthermore, this
extra dimension is curved or ``warped'' rather than flat: The bulk
is a portion of anti-de Sitter ($\mathrm{AdS}_5$) spacetime. As in the
Horava--Witten solutions, the RS branes are $Z_2$-symmetric (mirror
symmetry), and have a tension, which serves to counter the
influence of the negative bulk cosmological constant on the brane.
This also means that the self-gravity of the branes is
incorporated in  the RS models. The novel feature of the RS models
compared to previous higher-dimensional models is that the
observable 3 dimensions are protected from the large extra
dimension (at low energies) by curvature rather than
straightforward compactification.

The RS brane-worlds and their generalizations (to include matter
on the brane, scalar fields in the bulk, etc.) provide
phenomenological models that reflect at least some of the features
of M~theory, and that bring exciting new geometric and particle
physics ideas into play. The RS models also provide a framework
for exploring holographic ideas that have emerged in M~theory.
Roughly speaking, holography suggests that higher-dimensional
gravitational dynamics may be determined from knowledge of the
fields on a lower-dimensional boundary. The AdS/CFT correspondence
is an example, in which the classical dynamics of the
higher-dimensional gravitational field are equivalent to the
quantum dynamics of a conformal field theory (CFT) on the
boundary. The RS model with its $\mathrm{AdS}_5$ metric satisfies this
correspondence to lowest perturbative order~\cite{acft} (see
also~\cite{acftcosmo_1, acftcosmo_2, acftcosmo_8, acftcosmo_3,
  acftcosmo_4, acftcosmo_5, acftcosmo_6, acftcosmo_7} for the AdS/CFT
correspondence in a cosmological context).


Before turning to a more detailed
analysis of RS brane-worlds, We discuss the notion of Kaluza--Klein
(KK) modes of the graviton.


\subsection{Heuristics of KK modes}

The dilution of gravity via extra dimensions not only weakens
gravity on the brane, it also extends the range of graviton modes
felt on the brane beyond the massless mode of 4-dimensional
gravity. For simplicity, consider a flat brane with one flat extra
dimension, compactified through the identification
$y\leftrightarrow y+2\pi n L$, where $n=0,1,2,\dots$. The
perturbative 5D graviton amplitude can be Fourier expanded as
\begin{equation}
  f(x^a,y)=\sum_n e^{iny/L}\,f_n(x^a),
\end{equation}
where $f_n$ are the amplitudes of the KK modes, i.e., the effective
4D modes of the 5D graviton. To see that these KK modes are
massive from the brane viewpoint, we start from the 5D wave
equation that the massless 5D field $f$ satisfies (in a suitable
gauge):
\begin{equation}
  {}^{(5)\!}\Box f=0
  \quad \Rightarrow \quad
  \Box f+\partial_y^2 f=0.
\end{equation}
It follows that the KK modes satisfy a 4D Klein--Gordon equation
with an effective 4D mass $m_n$,
\begin{equation}
  \Box f_n=m_n^2\,f_n,
  \qquad
  m_n={n\over L}.
\end{equation}
The massless mode $f_0$ is the usual 4D graviton mode. But there
is a tower of massive modes, $L^{-1},2L^{-1},\dots$, which
imprint the effect of the 5D gravitational field on the 4D brane.
Compactness of the extra dimension leads to discreteness of the
spectrum. For an infinite extra dimension, $L\to\infty$, the
separation between the modes disappears and the tower forms a
continuous spectrum. In this case, the coupling of the KK modes to
matter must be very weak in order to avoid exciting the lightest
massive modes with $m\gtrsim 0$.

From a geometric viewpoint, the KK modes can also be understood
via the fact that the projection of the null graviton 5-momentum
${}^{(5)\!}p_A$ onto the brane is timelike. If the unit normal to the
brane is $n_A$, then the induced metric on the brane is
\begin{equation}
  g_{AB}= {}^{(5)\!}g_{AB}-n_An_B,
  \qquad
  {}^{(5)\!}g_{AB}n^An^B=1,
  \qquad
  g_{AB}n^B=0,
\end{equation}
and the 5-momentum may be decomposed as
\begin{equation}
  {}^{(5)\!}p_A=mn_A+ p_A,
  \qquad
  p_An^A=0,
  \qquad
  m= {}^{(5)\!}p_A\, n^A,
\end{equation}
where $p_A=g_{AB} \;{}^{(5)\!}p^B$ is the projection along the brane,
depending on the orientation of the 5-momentum relative to the
brane. The effective 4-momentum of the 5D graviton is thus $p_A$.
Expanding ${}^{(5)\!}g_{AB}{}{} \;{}^{(5)\!}p^A \;{}^{(5)\!}p^B=0$, we find that
\begin{equation}
  g_{AB}p^Ap^B=-m^2.
\end{equation}
It follows that the 5D graviton has an effective mass $m$ on the
brane. The usual 4D graviton corresponds to the zero mode, $m=0$,
when ${}^{(5)\!}p_A$ is tangent to the brane.

The extra dimensions lead to new scalar and vector degrees of
freedom on the brane. In 5D, the spin-2 graviton is represented by
a metric perturbation ${}^{(5)\!}h_{AB}$ that is transverse traceless:
\begin{equation}
  {}^{(5)\!}h^A{}_{A}=0=\partial_B \;{}^{(5)\!}h_{A}{}^{B}.
\end{equation}
In a suitable gauge, ${}^{(5)\!}h_{AB}$ contains a 3D transverse
traceless perturbation $h_{ij}$, a 3D transverse vector
perturbation $\Sigma_i$, and a scalar perturbation $\beta$, which
each satisfy the 5D wave equation~\cite{durkoc}:
\begin{eqnarray}
  & {}^{(5)\!}h_{AB}
  \quad \longrightarrow \quad
  h_{ij}, \Sigma_i, \beta, &
  \label{5dg} \\
  & h^i{}_i = 0 = \partial_j h^{ij}, &
  \\
  & \partial_i \Sigma^i = 0, &
  \\
  & (\Box + \partial_y^2)
  \left(
    \begin{array}{c}
      \beta \\
      \Sigma_i \\
      h_{ij}
    \end{array}
  \right) = 0. &
\end{eqnarray}%
The other components of ${}^{(5)\!}h_{AB}$ are determined via
constraints once these wave equations are solved. The 5 degrees of
freedom (polarizations) in the 5D graviton are thus split into 2
($h_{ij}$) + 2 ($\Sigma_i$) +1 ($\beta$) degrees of freedom in 4D.
On the brane, the 5D graviton field is felt as
\begin{itemize}
\item a 4D spin-2 graviton $h_{ij}$ (2 polarizations),
\item a 4D spin-1 gravi-vector (gravi-photon) $\Sigma_i$ (2
  polarizations), and
\item a 4D spin-0 gravi-scalar $\beta$.
\end{itemize}
The massive modes of the 5D graviton are represented via massive
modes in all three of these fields on the brane. The standard 4D
graviton corresponds to the massless zero-mode of $h_{ij}$.

In the general case of $d$ extra dimensions, the number of degrees
of freedom in the graviton follows from the irreducible tensor
representations of the isometry group as ${1\over2}(d+1)(d+4)$.

\newpage


\section{Randall--Sundrum Brane-Worlds}
\label{section_2}

RS brane-worlds do not rely on compactification to localize
gravity at the brane, but on the curvature of the bulk (sometimes
called ``warped compactification''). What prevents gravity from
`leaking' into the extra dimension at low energies is a negative
bulk cosmological constant,
\begin{equation}
  \Lambda_5=-{6\over \ell^2}=-6\mu^2,
\end{equation}
where $\ell$ is the curvature radius of $\mathrm{AdS}_5$ and $\mu$ is the
corresponding energy scale. The curvature radius determines the
magnitude of the Riemann tensor:
\begin{equation}
  {}^{(5)\!}R_{ABCD}=-{1\over \ell^2}
  \left[ {}^{(5)\!}g_{AC} \;{}^{(5)\!}g_{BD} - {}^{(5)\!}g_{AD} \;{}^{(5)\!}g_{BC} \right].
\end{equation}
The bulk cosmological constant acts to ``squeeze'' the
gravitational field closer to the brane. We can see this clearly
in Gaussian normal coordinates $X^A=(x^\mu,y)$ based on the brane
at $y=0$, for which the $\mathrm{AdS}_5$ metric takes the form
\begin{equation}
  {}^{(5)\!}ds^2=e^{-2|y|/\ell} \eta_{\mu\nu}dx^\mu dx^\nu + dy^2,
\end{equation}
with $\eta_{\mu\nu}$ being the Minkowski metric. The exponential warp
factor reflects the confining role of the bulk cosmological
constant. The $Z_2$-symmetry about the brane at $y=0$ is
incorporated via the $|y|$ term. In the bulk, this metric is a
solution of the 5D Einstein equations,
\begin{equation}
  {}^{(5)\!}G_{AB}=- \Lambda_5 \;{}^{(5)\!}g_{AB},
  \label{rsefe}
\end{equation}
i.e., ${}^{(5)}T_{AB}=0$ in Equation~(\ref{defe}). The brane is a flat
Minkowski spacetime,
$g_{AB}(x^\alpha,0)=\eta_{\mu\nu}\delta^\mu{}_A \delta^\nu{}_B$,
with self-gravity in the form of brane tension. One can also use
Poincare coordinates, which bring the metric into manifestly
conformally flat form,
\begin{equation}
  {}^{(5)\!}ds^2={\ell^2\over z^2}
  \left[ \eta_{\mu\nu}dx^\mu dx^\nu + dz^2 \right],
  \label{poinc}
\end{equation}
where $z=\ell e^{y/\ell}$. \\

\noindent
The two RS models are distinguished as follows:
\begin{description}
\item[RS 2-brane:] There are two branes in this model~\cite{rs1}, at
  $y=0$ and $y=L$, with $Z_2$-symmetry identifications
  \begin{equation}
    y \leftrightarrow -y,
    \qquad
    y+L \leftrightarrow L-y.
  \end{equation}
  The branes have equal and opposite tensions $\pm\lambda$, where
  \begin{equation}
    \lambda={3M_\mathrm{p}^2 \over 4\pi \ell^2}.
    \label{rst}
  \end{equation}
  The positive-tension brane has fundamental scale $M_5$ and is
  ``hidden''. Standard model fields are confined on the negative
  tension (or ``visible'') brane. Because of the exponential warping
  factor, the effective Planck scale on the visible brane at $y=L$ is given by
  \begin{equation}
    M_\mathrm{p}^2=M_5^3\, \ell\left[e^{2L/\ell}-1\right].
  \end{equation}
  So the RS 2-brane model gives a new approach to the hierarchy
  problem: even if $M_5 \sim \ell^{-1} \sim \mathrm{TeV}$, we can
  recover $M_\mathrm{p}\sim 10^{16} \mathrm{\ TeV}$ by choosing
  $L/\ell$ large enough. Because of the finite separation between the
  branes, the KK spectrum is discrete. Furthermore, at low energies
  gravity on the branes becomes Brans--Dicke-like, with the sign of
  the Brans--Dicke parameter equal to the sign of the brane
  tension~\cite{gt}. In order to recover 4D general relativity at low
  energies, a mechanism is required to stabilize the inter-brane
  distance, which corresponds to a scalar field degree of freedom
  known as the radion~\cite{goldwise,2b_1,2b_2,2b_3}.
\item[RS 1-brane:] In this model~\cite{rs2}, there is only one,
  positive tension, brane. It may be thought of as arising from
  sending the negative tension brane off to infinity,
  $L\to\infty$. Then the energy scales are related via
  \begin{equation}
    M_5^3={M_\mathrm{p}^2 \over \ell}.
    \label{tt2}
  \end{equation}
  The infinite extra dimension makes a finite contribution to the 5D
  volume because of the warp factor:
  \begin{equation}
    \int d^5X\sqrt{-{}^{(5)\!}g}=2\int d^4x \int_0^\infty \!\!\! dy
    e^{-4y/\ell}= {\ell \over 2} \int d^4x.
  \end{equation}
  Thus the effective size of the extra dimension probed by the 5D
  graviton is $\ell$.
\end{description}

\noindent
We will concentrate mainly on RS 1-brane from now on, referring to
RS 2-brane occasionally. The RS 1-brane models are in some sense
the most simple and geometrically appealing form of a brane-world
model, while at the same time providing a framework for the
AdS/CFT correspondence~\cite{acft, acftcosmo_1, acftcosmo_2,
  acftcosmo_8, acftcosmo_3, acftcosmo_4, acftcosmo_5, acftcosmo_6,
  acftcosmo_7}. The RS 2-brane introduce the added complication of
radion stabilization, as well as possible complications arising from
negative tension. However, they remain important and will occasionally
be discussed.


\subsection{KK modes in RS 1-brane}

In RS 1-brane, the negative $\Lambda_5$ is offset by the
positive brane tension $\lambda$. The fine-tuning in
Equation~(\ref{rst}) ensures that there is a zero effective cosmological
constant on the brane, so that the brane has the induced geometry
of Minkowski spacetime. To see how gravity is localized at low
energies, we consider the 5D graviton perturbations of the
metric~\cite{rs2, gt, morepert_1, morepert_2},
\begin{equation}
  {}^{(5)\!}g_{AB} \to {}^{(5)\!}g_{AB} + e^{-2|y|/\ell} \;{}^{(5)\!}h_{AB},
  \qquad
  {}^{(5)\!}h_{Ay}=0= {}^{(5)\!}h^\mu{}_\mu = {}^{(5)\!}h^{\mu\nu}{}_{,\nu}
\end{equation}
(see Figure~\ref{figure_03}). This is the RS gauge, which is different from the
gauge used in Equation~(\ref{5dg}), but which also has no remaining
gauge freedom. The 5 polarizations of the 5D graviton are
contained in the 5 independent components of $^{(5)\!}h_{\mu\nu}$ in the
RS gauge.

\epubtkImage{figure03.png}{%
 \begin{figure}[htbp]
   \def\epsfsize#1#2{0.7#1}
   \centerline{\epsfbox{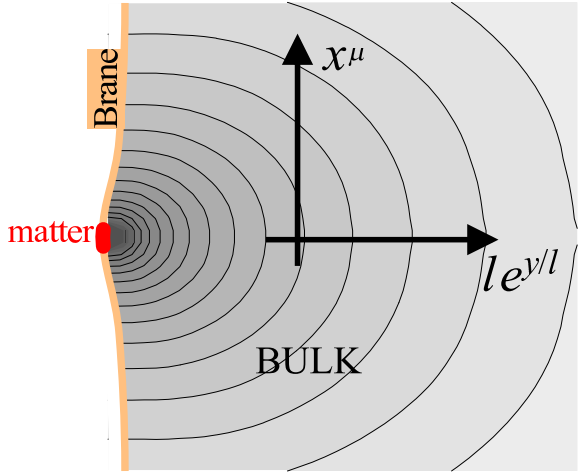}}
   \caption{Gravitational field of a small point particle on the
     brane in RS gauge. (Figure taken from~\cite{gt}.)}
   \label{figure_03}
 \end{figure}}

We split the amplitude $f$ of ${}^{(5)\!}h_{AB}$ into 3D Fourier modes,
and the linearized 5D Einstein equations lead to the wave equation
($y>0$)
\begin{equation}
  \ddot f+k^2f=e^{-2y/\ell}\left[f''- {4\over \ell}\, f'\right].
\end{equation}
Separability means that we can write
\begin{equation}
  f(t,y)=\sum_m \varphi_m(t)\,f_m(y),
\end{equation}
and the wave equation reduces to
\begin{eqnarray}
  \ddot{\varphi}_m+(m^2+k^2)\varphi_m &=& 0,
  \\
  f_m''- {4\over \ell}\, f_m'+e^{2y/\ell}m^2f_m&=& 0.
\end{eqnarray}%
The zero mode solution is
\begin{eqnarray}
  \varphi_0(t)&=& A_{0+}e^{+ikt}+ A_{0-}e^{-ikt},
  \\
  f_0(y)&=& B_0+C_0e^{4 y/\ell},
\end{eqnarray}%
and the $m>0$ solutions are
\begin{eqnarray}
  \varphi_m(t)&=& A_{m+}\exp\left(+i\sqrt{m^2+k^2}\,t\right)+
  A_{m-}\exp \left(-i\sqrt{m^2+k^2}\,t \right),
  \\
  f_m(y) &=& e^{2y/\ell} \left[ B_mJ_2\left(m\ell e^{y/\ell}\right) +
  C_m Y_2\left(m\ell e^{y/\ell} \right)\right].
\end{eqnarray}%

The boundary condition for the perturbations arises from the
junction conditions, Equation~(\ref{ext}), discussed below, and leads
to $f'(t,0)=0$, since the transverse traceless part of the
perturbed energy-momentum tensor on the brane vanishes. This
implies
\begin{equation}
  C_0=0,
  \qquad
  C_m=-{J_1(m\ell) \over Y_1(m\ell)}\,B_m.
  \label{rsbc}
\end{equation}
The zero mode is normalizable, since
\begin{equation}
  \left| \int_0^\infty \!\!\! B_0 e^{-2y/\ell} dy \right|<\infty.
\end{equation}
Its contribution to the gravitational potential $V={1\over2}
\;{}^{(5)\!}h_{00}$ gives the 4D result, $V \propto r^{-1}$. The contribution
of the massive KK modes sums to a correction of the 4D potential.
For $r\ll \ell$, one obtains
\begin{equation}
  V(r)\approx {GM\ell \over r^2},
  \label{newt2}
\end{equation}
which simply reflects the fact that the potential becomes truly 5D
on small scales. For $r\gg\ell$,
\begin{equation}
  V(r)\approx {GM \over r}\left(1+{2\ell^2 \over 3r^2}\right),
  \label{newt}
\end{equation}
which gives the small correction to 4D gravity at low energies
from extra-dimensional effects. These effects serve to slightly
strengthen the gravitational field, as expected.

Table-top tests of Newton's laws currently find no deviations down
to ${\cal O}(10^{-1} \mathrm{\ mm})$, so that
$\ell\lesssim 0.1 \mathrm{\ mm}$ in
Equation~(\ref{newt}). Then by Equations~(\ref{rst}) and~(\ref{tt2}), this
leads to lower limits on the brane tension and the fundamental
scale of the  RS 1-brane model:
\begin{equation}
  \lambda > (1 \mathrm{\ TeV})^4,
  \qquad
  M_5 > 10^5 \mathrm{\ TeV}.
  \label{rslimit}
\end{equation}
These limits do not apply to the 2-brane case.

For the 1-brane model, the boundary condition, Equation~(\ref{rsbc}),
admits a continuous spectrum $m>0$ of KK modes. In the 2-brane
model, $f'(t,L)=0$ must hold in addition to Equation~(\ref{rsbc}). This
leads to conditions on $m$, so that the KK spectrum is discrete:
\begin{equation}
  m_n= {x_n \over \ell}\,e^{-L/\ell}
  \qquad \mbox{where } J_1(x_n)=\frac{J_1(m\ell)}{Y_1(m\ell)}\, Y_1(x_n).
\end{equation}
The limit Equation~(\ref{rslimit}) indicates that there are no observable
collider, i.e., ${\cal O}(\mathrm{TeV})$, signatures for the RS 1-brane
model. The 2-brane model by contrast, for suitable choice of $L$
and $\ell$ so that $m_1={\cal O}(\mathrm{TeV})$, does predict collider
signatures that are distinct from those of the ADD
models~\cite{hanraf_1, hanraf_2}.

\subsection{\new{RS model in string theory}}
\label{section_2-2}

\epubtkUpdateA{Added Subsection~\ref{section_2-2} ``RS model in string theory''.}

There have been interesting developments in realizing the
Randall--Sundrum brane-world model in string theory. A concrete
example was found in type IIB supergravity~\cite{Giddings:2001yu}. The
bosonic part of the 10-dimensional action is given by (see for
example~\cite{Polchinski:1998rr})
\begin{eqnarray}
S_{\mathrm{IIB}} &=& {1\over 2\kappa_{10}^2} \int d^{10} x \sqrt{-g_{\mathrm{s}}}\Biggl\{
e^{-2\phi}
\left[{\cal R}_{\mathrm{s}} + 4(\nabla \phi)^2 \right] - {F_{(1)}^2\over 2} -
{1\over 2 \cdot 3!} G_{(3)} \cdot \bar{G}_{(3)} -
{\tilde{F}_{(5)}^2 \over 4\cdot5!} \Biggr\} \nonumber\\
&& + {1 \over 8 i \kappa_{10}^2}
\int e^{\phi}{C_{(4)}\wedge
G_{(3)}\wedge\bar{G}_{(3)}}\ +\ S_{\mathrm{loc}} \ .
\label{IIBS}
\end{eqnarray}
Here $g_{\mathrm{s}}$ denotes the string metric.  We have also defined the
combined three-flux, $G_{(3)}= F_{(3)} - \tau H_{(3)}$, where $\tau=
C_{(0)} + i e^{-\phi}$, and
\begin{equation}
{\tilde F}_{(5)} = F_{(5)} - {1\over 2} C_{(2)}\wedge H_{(3)} + {1\over 2}
B_{(2)}\wedge F_{(3)}\ .
\end{equation}
The term $S_{\mathrm{loc}}$ is the action of localized
objects, such as branes.

There is a 10-dimensional solution similar to the one in the RS model
\begin{equation}
ds_{10}^2 = h(y)^{-\frac{1}{2}} \eta_{\mu\nu} dx^\mu dx^\nu
+ h(y)^{\frac{1}{2}} \tilde{g}_{mn} dy^m dy^n,
\label{mansatz}
\end{equation}
where $x^\mu$ are four-dimensional coordinates and
$y^m$ are coordinates on the 6-dimensional compact manifold ${\cal M}_6$.
If there are N coincident D3-branes on the manifold, the 10-dimensional spacetime is
described by $AdS_5 \times X_5$ where $X_5$ is a five-dimensional Einstein manifold.
The warp factor is given by~\cite{Verlinde:1999fy}
\begin{equation}
h(r)  \sim \left( \frac{L}{r} \right)^4,
\end{equation}
where $r$ is the distance from the D3 branes in the $\tilde{g}_{mn}$
metric and $L$ is given by
\begin{equation}
L=4 \pi a g_s N \alpha'^2.
\end{equation}
Here $g_s$ is the string coupling constant, $\alpha'$ is related to
the string mass scale $m_s$ as $\alpha'=1/m_s^2$ and $a$ depends on $X_5$.

Near the D3 branes, the geometry is $AdS_5 \times S^5$. At large $r$,
the product structure of the 10-dimensional metric breaks down and
$r$ ceases to be a good coordinate, and the 6D extra-dimensional space
${\cal M}_6$ is not a product of a five sphere and one-dimensional
spacetime. This gives a minimum distance $r_0$. This cut-off acts as a
TeV brane in the RS model. On the other hand, for $r >r_{\mathrm{max}}$,
the AdS geometry smoothly glues into a Calabi--Yau
compactification. Thus, the compact manifold plays the role of the
Planck brane in the RS model, acting as a cut-off of the
spacetime. The minimum warp factor $h(r_0)^{-1/2}$ is determined by
the flux associated with $F_3$ and $H_3$ and thanks to the warping, it
is possible to realize a large hierarchy~\cite{Giddings:2001yu}
\begin{equation}\label{}
{h(r_0) \over h(r_{\mathrm{max}})} \gg 1\,.
\end{equation}

\newpage


\section{Covariant Approach to Brane-World Geometry and Dynamics}
\label{section_3}

The RS models and the subsequent generalization from a Minkowski
brane to a Friedmann--Robertson--Walker (FRW) brane~\cite{bdel,
  morers2_1, morers2_2, morers2_3, morers2_4, morers2_5, morers2_6,
  morers2_7, gs} were derived as solutions in
particular coordinates of the 5D Einstein equations, together with
the junction conditions at the $Z_2$-symmetric
brane\epubtkFootnote{Brane-worlds without $Z_2$ symmetry have also
  been considered; see e.g., \cite{gergasym} for recent work.}. A
broader perspective, with useful insights into the inter-play between 4D
and 5D effects, can be obtained via the covariant
Shiromizu--Maeda--Sasaki approach~\cite{sms}, in which the brane and
bulk metrics remain general. The basic idea is to use the
Gauss--Codazzi equations to project the 5D curvature along the
brane. (The general formalism for relating the geometries of a
spacetime and of hypersurfaces within that spacetime is given
in~\cite{wald}.)

The 5D field equations determine the 5D curvature tensor; in the
bulk, they are
\begin{equation}
  {}^{(5)}\!G_{AB}=-\Lambda_5 \;{}^{(5)\!}g_{AB}+ \kappa_5^2 \;{}^{(5)}T_{AB},
  \label{5efe}
\end{equation}
where $^{(5)}T_{AB}$ represents any 5D energy-momentum of the
gravitational sector (e.g., dilaton and moduli scalar fields, form
fields).

Let $y$ be a Gaussian normal coordinate orthogonal to the brane
(which is at $y=0$ without loss of generality), so that
$n_AdX^A=dy$, with $n^A$ being the unit normal. The 5D metric in terms
of the induced metric on $\{y=\mathrm{const.}\}$ surfaces is locally
given by
\begin{equation}
  {}^{(5)}\!g_{AB}=g_{AB}+ n_An_B,
  \qquad
  {}^{(5)\!}ds^2=g_{\mu\nu}(x^\alpha,y)dx^\mu dx^\nu +dy^2.
  \label{gn}
\end{equation}
The extrinsic curvature of $\{y=\mathrm{const.}\}$ surfaces
describes the embedding of these surfaces. It can be defined via
the Lie derivative or via the covariant derivative:
\begin{equation}
  K_{AB}={1\over2} \mbox{\bf \pounds}_{\bf n}\,g_{AB}=g_A{}^C
  \;{}^{(5)}\nabla_C n_B,
  \label{exc}
\end{equation}
so that
\begin{equation}
  K_{[AB]}=0=K_{AB}n^B,
\end{equation}
where square brackets denote anti-symmetrization. The Gauss
equation gives the 4D curvature tensor in terms of the projection
of the 5D curvature, with extrinsic curvature corrections:
\begin{equation}
  R_{ABCD}= {}^{(5)\!}{R}_{EFGH} g_A{}^E g_B{}^F g_C{}^G g_D{}^H +
  2 K_{A[C}K_{D]B},
  \label{gauss}
\end{equation}
and the Codazzi equation determines the change of $K_{AB}$ along
$\{y=\mathrm{const.}\}$ via
\begin{equation}
  \nabla_BK^B{}_A-\nabla_AK= {}^{(5)\!}{R}_{BC} \, g_A{}^B n^C,
  \label{cod}
\end{equation}
where $K=K^A{}_A$.

Some other useful projections of the 5D curvature are:
\begin{eqnarray}
  {}^{(5)\!}{R}_{EFGH} \, g_A{}^E g_B{}^F g_C{}^G n^H &=&
  2\nabla_{[A}K_{B]C},
  \\
  {}^{(5)\!}{R}_{EFGH} \, g_A{}^E n^F g_B{}^G n^H &=&
  - \mbox{\bf \pounds}_\mathbf{n} K_{AB}+K_{AC}K^C{}_B,
  \\
  {}^{(5)\!}{R}_{CD} \, g_A{}^C g_B{}^D &=&
  R_{AB} - \mbox{\bf \pounds}_\mathbf{n} K_{AB}-KK_{AB}+2K_{AC}K^C{}_B.
\end{eqnarray}%
The 5D curvature tensor has Weyl (tracefree) and Ricci parts:
\begin{equation}
  {}^{(5)\!}{R}_{ABCD} =
  {}^{(5)\!}C_{ACBD}+{2\over3}
  \left( {}^{(5)\!}g_{A[C} \;{}^{(5)\!}R_{D]B} -
  {}^{(5)\!}g_{B[C} \;{}^{(5)\!}R_{D]A} \right) -
  {1\over6} {}^{(5)\!}g_{A[C} \;{}^{(5)\!}g_{D]B} \;{}^{(5)\!}R.
\end{equation}


\subsection{Field equations on the brane}

Using Equations~(\ref{5efe}) and~(\ref{gauss}), it follows that
\begin{eqnarray}
  {G}_{\mu\nu}&=&-{1\over2}{\Lambda}_5 g_{\mu\nu}+{2\over3}
  \kappa_5^2 \left[{}^{(5)}T_{AB}g_\mu{}^A g_\nu{}^B +
  \left( {}^{(5)}T_{AB}n^An^B-{1\over 4} \;{}^{(5)}T \right)
  g_{\mu\nu} \right]
  \nonumber \\
  && + K K_{\mu\nu}-K_\mu{}^\alpha K_{\alpha\nu} +
  {1\over2}\left[K^{\alpha\beta}K_{\alpha\beta}-K^2 \right]g_{\mu\nu} -
  {\cal E}_{\mu\nu},
  \label{ein}
\end{eqnarray}%
where ${}^{(5)}T = {}^{(5)}T^A{}_{A}$, and where
\begin{equation}
  {\cal E}_{\mu\nu} =
  {}^{(5)\!}C_{ACBD} \, n^Cn^D g_\mu{}^A g_\nu{}^B,
\end{equation}
is the projection of the bulk Weyl tensor orthogonal to $n^A$.
This tensor satisfies
\begin{equation}
  {\cal E}_{AB}n^B= 0 ={\cal E}_{[AB]}={\cal E}_{A}{}^A,
\end{equation}
by virtue of the Weyl tensor symmetries. Evaluating
Equation~(\ref{ein}) on the brane (strictly, as $y\to\pm 0$, since
${\cal E}_{AB}$ is not defined on the brane~\cite{sms}) will give
the field equations on the brane.

First, we need to determine $K_{\mu\nu}$ at the brane from the
junction conditions. The total energy-momentum tensor on the brane
is
\begin{equation}
  T_{\mu\nu}^\mathrm{brane} =T_{\mu\nu}-\lambda g_{\mu\nu},
\end{equation}
where $T_{\mu\nu}$ is the energy-momentum tensor of particles and
fields confined to the brane (so that $T_{AB}n^B=0$). The 5D field
equations, including explicitly the contribution of the brane, are
then
\begin{equation}
  ^{(5)}\!G_{AB}=-\Lambda_5 \, {}^{(5)}\!g_{AB}+
  \kappa_5^2\left[{}^{(5)}T_{AB}+ T_{AB}^\mathrm{brane}\delta(y)\right].
  \label{feb}
\end{equation}
Here the delta function enforces in the classical theory the
string theory idea that Standard Model fields are confined to the
brane. This is not a gravitational confinement, since there is in
general a nonzero acceleration of particles normal to the
brane~\cite{m1}.

Integrating Equation~(\ref{feb}) along the extra dimension from
$y=-\epsilon$ to $y=+\epsilon$, and taking the limit $\epsilon\to
0$, leads to the Israel--Darmois junction conditions at the brane,
\begin{eqnarray}
  g^+_{\mu\nu}-g^-_{\mu\nu}&=&0, \\
  K_{\mu\nu}^{+}-K_{\mu\nu}^{-}&=&
  -\kappa_5^2 \left[T_{\mu\nu}^\mathrm{brane}-
  {1\over3} T^\mathrm{brane}g_{\mu\nu}\right],
  \label{jun}
\end{eqnarray}%
where $T^\mathrm{brane}=g^{\mu\nu}T_{\mu\nu}^\mathrm{brane}$. The $Z_2$
symmetry means that when you approach the brane from one side and
go through it, you emerge into a bulk that looks the same, but
with the normal reversed, $n^A \to -n^A$. Then Equation~(\ref{exc})
implies that
\begin{equation}
  K_{\mu\nu}^{-}=-K_{\mu\nu}^{+},
  \label{z2}
\end{equation}
so that we can use the junction condition Equation~(\ref{jun}) to
determine the extrinsic curvature on the brane:
\begin{equation}
  K_{\mu\nu}=-{1\over2}\kappa_5^2 \left[T_{\mu\nu}+
  {1\over3} \left(\lambda-T\right)g_{\mu\nu} \right],
  \label{ext}
\end{equation}
where $T=T^\mu{}_\mu $, where we have dropped the $(+)$, and where we
evaluate quantities on the brane by taking the limit $y\to+0$.

Finally we arrive at the induced field equations on the brane, by
substituting Equation~(\ref{ext}) into Equation~(\ref{ein}):
\begin{equation}
  G_{\mu\nu} = - \Lambda g_{\mu\nu} + \kappa^2 T_{\mu\nu} +
  6\frac{\kappa^2}{\lambda} {\cal S}_{\mu\nu} - {\cal E}_{\mu\nu}+
  4\frac{\kappa^2}{\lambda}{\cal F}_{\mu\nu}.
  \label{e:einstein1}
\end{equation}
The 4D gravitational constant is an effective coupling constant
inherited from the fundamental coupling constant, and the 4D
cosmological constant is nonzero when the RS balance between the
bulk cosmological constant and the brane tension is broken:
\begin{eqnarray}
  \kappa^2 &\equiv & \kappa^2_{4}={1\over6}\lambda\kappa^4_5,
  \\
  \Lambda &=& {1\over 2}\left[ \Lambda_5+\kappa^2\lambda
\right].
\end{eqnarray}%

The first correction term relative to Einstein's theory is
quadratic in the energy-momentum tensor, arising from the
extrinsic curvature terms in the projected Einstein tensor:
\begin{equation}
  {\cal S}_{\mu\nu}= {{1\over12}}T T_{\mu\nu}
  -{{1\over4}}T_{\mu\alpha}T^\alpha{}_\nu + {{1\over24}}g_{\mu\nu}
  \left[3 T_{\alpha\beta} T^{\alpha\beta}-T^2 \right].
\end{equation}
The second correction term is the projected Weyl term. The last
correction term on the right of Equation~(\ref{e:einstein1}), which
generalizes the field equations in~\cite{sms}, is
\begin{equation}
  {\cal F}_{\mu\nu}= {}^{(5)}T_{AB} g_\mu{}^A g_\nu{}^B +
  \left[{}^{(5)}T_{AB}n^An^B-{1\over 4} \;{}^{(5)}T \right]
  g_{\mu\nu},
\end{equation}
where ${}^{(5)}T_{AB}$ describes any stresses in the bulk apart
from the cosmological constant (see~\cite{mw} for the case of a
scalar field).

What about the conservation equations? Using Equations~(\ref{5efe}),
(\ref{cod}) and~(\ref{ext}), one obtains
\begin{equation}
  \nabla^\nu T_{\mu\nu}=-2 \;{}^{(5)}T_{AB} n^A g^B{}_\mu.
  \label{cong}
\end{equation}
Thus in general there is exchange of energy-momentum between the
bulk and the brane. From now on, we will assume that
\begin{equation}
  {}^{(5)}T_{AB}=0
  \quad \Rightarrow \quad
  {\cal F}_{\mu\nu}=0,
\end{equation}
so that
\begin{eqnarray}
  {}^{(5)}\!G_{AB} & = & -\Lambda_5 \;{}^{(5)}\!g_{AB}
  \qquad \qquad \qquad \qquad \qquad \quad \, \mbox{in the bulk},
  \\
  G_{\mu\nu} & = & - \Lambda g_{\mu\nu} + \kappa^2 T_{\mu\nu} +
  6\frac{\kappa^2}{\lambda} {\cal S}_{\mu\nu} - {\cal E}_{\mu\nu}
  \qquad \mbox{on the brane}.
  \label{bife}
\end{eqnarray}%
One then recovers from Equation~(\ref{cong}) the standard 4D
conservation equations,
\begin{equation}
  \nabla^\nu T_{\mu\nu}=0.
  \label{lc}
\end{equation}
This means that there is no exchange of energy-momentum between
the bulk and the brane; their interaction is purely gravitational.
Then the 4D contracted Bianchi identities ($\nabla^\nu
G_{\mu\nu}=0$), applied to Equation~(\ref{e:einstein1}), lead to
\begin{equation}
  \nabla^\mu{\cal E}_{\mu\nu}=
  {6\kappa^2\over\lambda}\,\nabla^\mu{\cal S}_{\mu\nu},
  \label{nlc}
\end{equation}
which shows qualitatively how $1+3$ spacetime variations in the
matter-radiation on the brane can source KK modes.

The induced field equations~(\ref{bife}) show two key
modifications to the standard 4D Einstein field equations arising
from extra-dimensional effects:
\begin{itemize}
\item ${\cal S}_{\mu\nu}\sim (T_{\mu\nu})^2$ is the high-energy
  correction term, which is negligible for $\rho\ll\lambda$, but
  dominant for $\rho\gg\lambda$:
  \begin{equation}
    {|\kappa^2{\cal S}_{\mu\nu}/\lambda | \over |\kappa^2 T_{\mu\nu}|}
    \sim {|T_{\mu\nu}| \over \lambda} \sim {\rho\over\lambda}.
  \end{equation}
\item ${\cal E}_{\mu\nu}$ is the projection of the bulk Weyl tensor
  on the brane, and encodes corrections from 5D graviton effects (the
  KK modes in the linearized case). From the brane-observer viewpoint,
  the energy-momentum corrections in ${\cal S}_{\mu\nu}$ are local,
  whereas the KK corrections in ${\cal E}_{\mu\nu}$ are nonlocal,
  since they incorporate 5D gravity wave modes. These nonlocal
  corrections cannot be determined purely from data on the brane. In
  the perturbative analysis of RS 1-brane which leads to the
  corrections in the gravitational potential, Equation~(\ref{newt}),
  the KK modes that generate this correction are responsible for a
  nonzero ${\cal E}_{\mu\nu}$; this term is what carries the
  modification to the weak-field field equations. The 9 independent
  components in the tracefree ${\cal E}_{\mu\nu}$ are reduced to 5
  degrees of freedom by Equation~(\ref{nlc}); these arise from the 5
  polarizations of the 5D graviton.
  Note that the covariant formalism applies also to the two-brane
  case. In that case, the gravitational influence of the second brane
  is felt via its contribution to ${\cal E}_{\mu\nu}$.
\end{itemize}


\subsection{5-dimensional equations and the initial-value problem}

The effective field equations are not a closed system. One needs
to supplement them by 5D equations governing ${\cal E}_{\mu\nu}$,
which are obtained from the 5D Einstein and Bianchi equations.
This leads to the coupled system~\cite{ssm}
\begin{eqnarray}
  \mbox {\bf \pounds}_{\bf n} K_{\mu\nu} &=&
  K_{\mu\alpha}K^\alpha {}_\nu -
  {\cal E}_{\mu\nu}-{1\over6}\Lambda_5 g_{\mu\nu},
  \\
  \mbox {\bf \pounds}_{\bf n} {\cal E}_{\mu\nu} &=&
  \nabla^\alpha {\cal B}_{\alpha(\mu\nu)} +
  \frac{1}{6} \Lambda_5\left(K_{\mu\nu} - g_{\mu\nu}K\right) +
  K^{\alpha\beta}R_{\mu\alpha\nu\beta} +
  3K^\alpha{}_{(\mu}{\cal E}_{\nu)\alpha} -
  K{\cal E}_{\mu\nu} \nonumber
  \\
  & & {} + \left( K_{\mu\alpha}K_{\nu\beta} -
  K_{\alpha\beta}K_{\mu\nu} \right) K^{\alpha\beta},
  \label{eq:EEE} \\
  \mbox {\bf \pounds}_{\bf n} {\cal B}_{\mu\nu\alpha} &=&
  - 2\nabla_{[\mu}{\cal E}_{\nu]\alpha} +
  K_\alpha{}^\beta {\cal B}_{\mu\nu\beta} -
  2{\cal B}_{\alpha\beta [\mu }K_{\nu]}{}^\beta,
  \label{eq:BBB} \\
  \mbox {\bf \pounds}_{\bf n} R_{\mu\nu\alpha\beta} &=&
  -2R_{\mu\nu\gamma [\alpha}K_{\beta]}{}^\gamma -
  \nabla_{\mu}{\cal B}_{\alpha\beta\nu} +
  \nabla_{\mu}{\cal B}_{\beta\alpha\nu},
  \label{eq:bianchi3}
\end{eqnarray}%
where the ``magnetic'' part of the bulk Weyl tensor, counterpart to
the ``electric'' part ${\cal E}_{\mu\nu}$, is
\begin{equation}
  {\cal B}_{\mu\nu\alpha}= g_\mu{}^A g_\nu{}^B
  g_\alpha{}^C \;{}^{(5)\!}C_{ABCD} n^D.
\end{equation}
These equations are to be solved subject to the boundary
conditions at the brane,
\begin{eqnarray}
  \nabla^\mu {\cal E}_{\mu\nu} &\doteq&
  \kappa_5^4 \nabla^\mu {\cal S}_{\mu\nu},
  \\
  {\cal B}_{\mu\nu\alpha} &\doteq &
  2\nabla_{[\mu}K_{\nu]\alpha} \doteq
  -\kappa_5^2\nabla_{[\mu}
  \left( T_{\nu] \alpha}-\frac{1}{3}g_{\nu] \alpha}T \right),
  \label{bcwall}
\end{eqnarray}%
where $A\doteq B$ denotes $A|_\mathrm{brane}=B|_\mathrm{brane}$.

The above equations have been used to develop a covariant analysis
of the weak field~\cite{ssm}. They can also be used to develop a
Taylor expansion of the metric about the brane. In Gaussian normal
coordinates, Equation~(\ref{gn}), we have $\mbox {\bf \pounds}_{\bf
n}=\partial/\partial y$. Then we find
\begin{eqnarray}
  g_{\mu\nu}(x,y) &=& g_{\mu\nu}(x,0) -
  \kappa_5^2\left[ T_{\mu\nu}+{1\over 3}(\lambda-T)g_{\mu\nu}
  \right]_{y=0+} \!\!\! |y| \nonumber\\
  && + \left[ -{\cal E}_{\mu\nu} + {1\over4}\kappa_5^4 \left(
  T_{\mu\alpha} T^\alpha{}_\nu +{2\over3} (\lambda-T)T_{\mu\nu} \right)
  + {1\over6} \left( {1\over6} \kappa_5^4(\lambda-T)^2 - \Lambda_5
  \right) g_{\mu\nu}\right]_{y=0+} \!\!\! y^2 + \dots
  \nonumber \\
  \label{tay}
\end{eqnarray}%

In a non-covariant approach based on a specific form of the bulk
metric in particular coordinates, the 5D Bianchi identities would
be avoided and the equivalent problem would be one of solving the
5D field equations, subject to suitable 5D initial conditions and
to the boundary conditions Equation~(\ref{ext}) on the metric. The
advantage of the covariant splitting of the field equations and
Bianchi identities along and normal to the brane is the clear
insight that it gives into the interplay between the 4D and 5D
gravitational fields. The disadvantage is that the splitting is
not well suited to dynamical evolution of the equations. Evolution
off the timelike brane in the spacelike normal direction does not
in general constitute a well-defined initial value
problem~\cite{antav}. One needs to specify initial data on a 4D
spacelike (or null) surface, with boundary conditions at the
brane(s) ensuring a consistent evolution~\cite{ichnak_1, ichnak_2}. Clearly
the evolution of the observed universe is dependent upon initial
conditions which are inaccessible to brane-bound observers; this
is simply another aspect of the fact that the brane dynamics is
not determined by 4D but by 5D equations. The initial conditions
on a 4D surface could arise from models for creation of the 5D
universe~\cite{gs, ksb_1, ksb_2, ksb_3, ksb_4}, from dynamical attractor
behaviour~\cite{mukcol} or from suitable conditions (such as no
incoming gravitational radiation) at the past Cauchy horizon if
the bulk is asymptotically AdS.


\subsection{The brane viewpoint: A 1\,+\,3-covariant analysis}

Following~\cite{m1}, a systematic analysis can be developed from the
viewpoint of a brane-bound observer. (See also~\cite{gergnew}.) The effects of bulk
gravity are conveyed, from a brane observer viewpoint, via the
local (${\cal S}_{\mu\nu}$) and nonlocal (${\cal E}_{\mu\nu}$)
corrections to Einstein's equations. (In the more general case,
bulk effects on the brane are also carried by ${\cal F}_{\mu\nu}$,
which describes any 5D fields.) The ${\cal E}_{\mu\nu}$ term
cannot in general be determined from data on the brane, and the 5D
equations above (or their equivalent) need to be solved in order
to find ${\cal E}_{\mu\nu}$.

The general form of the brane energy-momentum tensor for any
matter fields (scalar fields, perfect fluids, kinetic gases,
dissipative fluids, etc.), including a combination of different
fields, can be covariantly given in terms of a chosen 4-velocity
$u^\mu$ as
\begin{equation}
  T_{\mu\nu}=\rho u_\mu u_\nu +ph_{\mu\nu}+
  \pi_{\mu\nu}+q_{\mu}u_{\nu}+q_\nu u_\mu.
 \label{3_prime_prime}
\end{equation}
Here $\rho$ and $p$ are the energy density and isotropic pressure,
respectively, and
\begin{equation}
  h_{\mu\nu}=g_{\mu\nu}+u_\mu  u_\nu =
  \;{}^{(5)\!}g_{\mu\nu}-n_\mu n_\nu +u_\mu  u_\nu
\end{equation}
projects into the comoving rest space orthogonal to $u^\mu$ on the
brane. The momentum density and anisotropic stress obey
\begin{equation}
  q_{\mu}=q_{\langle \mu \rangle},
  \qquad
  \pi_{\mu\nu}=\pi_{\langle \mu\nu \rangle},
\end{equation}
where angled brackets denote the spatially projected, symmetric,
and tracefree part:
\begin{equation}
  V_{\langle \mu \rangle}=h_\mu \; {}^\nu V_\nu,
  \qquad
  W_{\langle \mu\nu \rangle}=
  \left[h_{(\mu}{}^\alpha h_{\nu)}{}^\beta-
  {1\over3}h^{\alpha\beta}h_{\mu\nu}\right]W_{\alpha\beta}.
\end{equation}
In an inertial frame at any point on the brane, we have
\begin{equation}
  u^\mu=(1,\vec 0),
  \qquad
  h_{\mu\nu}=\diag(0,1,1,1),
  \qquad
  V_\mu=(0,V_i),
  \qquad
  W_{\mu 0} =0= \sum W_{ii}=W_{ij}- W_{ji},
\end{equation}
where $i,j = 1,2,3$.

The tensor ${\cal S}_{\mu\nu}$, which carries local bulk effects
onto the brane, may then be irreducibly decomposed as
\begin{eqnarray}
  {\cal S}_{\mu\nu}&=&{{1\over24}}
  \left[2\rho^2-3\pi_{\alpha\beta} \pi^{\alpha\beta}\right]u_\mu u_\nu +
  {1\over24}\left[2\rho^2+4\rho p+\pi_{\alpha\beta} \pi^{\alpha\beta}-
  4q_\alpha q^\alpha\right]h_{\mu\nu}
  \nonumber \\
  && - {1\over12}(\rho+3p)\pi_{\mu\nu}-{1\over4}\pi_{ \alpha \langle \mu}
  \pi_{\nu \rangle}{}^\alpha +{1\over4} q_{\langle \mu}q_{\nu \rangle}+
  {1\over3}\rho q_{(\mu}u_{\nu)}-
  {1\over2} q^\alpha \pi_{\alpha(\mu}u_{\nu)}.
  \label{3_prime_prime_prime}
\end{eqnarray}%
This simplifies for a perfect fluid or minimally-coupled scalar
field to
\begin{equation}
  {\cal S}_{\mu\nu}={{1\over12}}\rho
  \left[\rho u_\mu u_\nu +\left(\rho+2 p\right)h_{\mu\nu}\right].
\end{equation}

The tracefree ${\cal E}_{\mu\nu}$ carries nonlocal bulk effects
onto the brane, and contributes an effective ``dark'' radiative
energy-momentum on the brane, with energy density $\rho_{\cal E}$,
pressure $\rho_{\cal E}/3$, momentum density $q^{\cal E}_\mu $, and
anisotropic stress $\pi^{\cal E}_{\mu\nu}$:
\begin{equation}
  -{1\over\kappa^2} {\cal E}_{\mu\nu} =
  \cu\left(u_\mu  u_\nu +{1\over3} h_{\mu\nu}\right)+
  {\cq_\mu } u_{\nu} + {\cq_\nu} u_{\mu}+\cp_{\mu\nu}.
\end{equation}
We can think of this as a KK or Weyl ``fluid''. The brane ``feels''
the bulk gravitational field through this effective fluid. More
specifically:
\begin{itemize}
\item The KK (or Weyl) anisotropic stress $\cp_{\mu\nu}$ incorporates
  the scalar or spin-0 (``Coulomb''), the vector (transverse) or
  spin-1 (gravimagnetic), and the tensor (transverse traceless) or
  spin-2 (gravitational wave) 4D modes of the spin-2 5D graviton.
\item The KK momentum density $\cq_\mu$ incorporates spin-0 and spin-1
  modes, and defines a velocity $v^{\cal E}_\mu $ of the Weyl fluid
  relative to $u^\mu$ via $\cq_\mu =\cu v^{\cal E}_\mu $.
\item The KK energy density $\cu $, often called the ``dark
  radiation'', incorporates the spin-0 mode.
\end{itemize}

In special cases, symmetry will impose simplifications on this
tensor. For example, it must vanish for a conformally flat bulk,
including $\mathrm{AdS}_5$,
\begin{equation}
  {}^{(5)\!}g_{AB}~\mbox{conformally flat}
  \quad \Rightarrow \quad
  {\cal E}_{\mu\nu}=0.
\end{equation}
The RS models have a Minkowski brane in an $\mathrm{AdS}_5$ bulk. This bulk
is also compatible with an FRW brane. However, the most general
vacuum bulk with a Friedmann brane is Schwarzschild-anti-de Sitter
spacetime~\cite{birk_1, birk_2}. Then it follows from the FRW symmetries
that
\begin{equation}
  \mbox{Schwarzschild AdS}_5 \mbox{bulk, FRW brane:}
  \qquad
  \cq_\mu = 0 = \cp_{\mu\nu},
\end{equation}
where $\cu=0$ only if the mass of the black hole in the bulk is
zero. The presence of the bulk black hole generates via Coulomb
effects the dark radiation on the brane.

For a static spherically symmetric brane (e.g., the exterior of a
static star or black hole)~\cite{dmpr},
\begin{equation}
  \mbox{static spherical brane:}
  \qquad \cq_\mu = 0.
\end{equation}
This condition also holds for a Bianchi~I brane~\cite{mss}. In
these cases, $\cp_{\mu\nu}$ is not determined by the symmetries,
but by the 5D field equations. By contrast, the symmetries of a
G\"odel brane fix $\cp_{\mu\nu}$~\cite{tsab}.

The brane-world corrections can conveniently be consolidated into
an effective total energy density, pressure, momentum density, and
anisotropic stress:
\begin{eqnarray}
  \rho_\mathrm{tot} &=& \rho+{1\over 4\lambda}
  \left(2\rho^2 - 3 \pi_{\mu\nu} \pi^{\mu\nu}\right) + \cu,
  \label{a} \\
  p_\mathrm{tot} &=& p+ {1\over 4\lambda}
  \left(2\rho^2+4\rho p + \pi_{\mu\nu}\pi^{\mu\nu}-4q_\mu q^\mu\right) +
  {\cu\over 3},
  \label{b}\\
  q^\mathrm{tot}_\mu &=& q_\mu +{1\over 2\lambda}
  \left(2\rho q_\mu-3\pi_{\mu\nu}q^\nu\right)+ \cq_\mu,
  \label{d} \\
  \pi^\mathrm{tot}_{\mu\nu} &=& \pi_{\mu\nu}+{1\over 2\lambda}
  \left[-(\rho+3p)\pi_{\mu\nu}-
  3\pi_{\alpha\langle \mu}\pi_{\nu \rangle}{}^\alpha+
  3q_{\langle \mu}q_ { \nu \rangle}\right] + \cp_{\mu\nu}.
  \label{c}
\end{eqnarray}%
These general expressions simplify in the case of a perfect fluid
(or minimally coupled scalar field, or isotropic one-particle
distribution function), i.e., for $q_\mu=0=\pi_{\mu\nu}$, to
\begin{eqnarray}
  \rho_\mathrm{tot} &=&
  \rho\left(1 +\frac{\rho}{2\lambda} + \frac{\rho_{\cal E}}{\rho} \right),
  \label{rtot} \\
  p_\mathrm{tot} &=&
  p + \frac{\rho}{2\lambda} (2p+\rho)+\frac{\rho_{\cal E}}{3},
  \label{ptot} \\
  q^\mathrm{tot}_\mu  &=& q^{\cal E}_\mu,
  \\
  \pi^\mathrm{tot}_{\mu\nu} &=& \pi^{\cal E}_{\mu\nu}.
  \label{e:pressure2}
\end{eqnarray}%
Note that nonlocal bulk effects can contribute to effective
imperfect fluid terms even when the matter on the brane has
perfect fluid form: There is in general an effective momentum
density and anisotropic stress induced on the brane by massive KK
modes of the 5D graviton.

The effective total equation of state and sound speed follow from
Equations~(\ref{rtot}) and~(\ref{ptot}) as
\begin{eqnarray}
  w_\mathrm{tot} & \equiv & {p_\mathrm{tot}\over\rho_\mathrm{tot}} =
  {w+(1+2w)\rho/2\lambda+ \cu/3\rho \over 1+\rho/2\lambda +\cu/\rho},
  \label{vh1} \\
  c_\mathrm{tot}^2 & \equiv &
  {\dot{p}_\mathrm{tot}\over\dot{\rho}_\mathrm{tot}} =
  \left[c_\mathrm{s}^2+{\rho+p \over \rho+\lambda} +
  {4\cu\over 9(\rho+p)(1+\rho/\lambda)}\right]
  \left[1+ {4\cu\over 3(\rho+p)(1+\rho/\lambda)}\right]^{-1}\!\!\!\!\!\!\!\!,
  \label{vh2}
\end{eqnarray}%
where $w=p/\rho$ and $c_\mathrm{s}^2=\dot p/\dot\rho$. At very high
energies, i.e., $\rho\gg\lambda$, we can generally neglect $\cu$
(e.g., in an inflating cosmology), and the effective equation of
state and sound speed are stiffened:
\begin{equation}
  w_\mathrm{tot}\approx 2w+1,
  \qquad
  c_\mathrm{tot}^2 \approx c_\mathrm{s}^2+w+1.
\end{equation}
This can have important consequences in the early universe and
during gravitational collapse. For example, in a very high-energy
radiation era, $w=1/3$, the effective cosmological equation
of state is ultra-stiff: $w_\mathrm{tot}\approx 5/3$. In
late-stage gravitational collapse of pressureless matter, $w=0$,
the effective equation of state is stiff, $w_\mathrm{tot}\approx 1$,
and the effective pressure is nonzero and dynamically important.


\subsection{Conservation equations}
\label{coneq}

Conservation of $T_{\mu\nu}$ gives the standard general relativity
energy and momentum conservation equations, in the general,
nonlinear case:
\begin{eqnarray}
  \dot{\rho}+\Theta(\rho+p)+\D^\mu q_\mu+2A^\mu q_\mu +
  \sigma^{\mu\nu }\pi_{\mu\nu}&=&0,
  \label{c1} \\
  \dot{q}_{\langle \mu\rangle}+{{4\over3}}\Theta q_\mu+
  \D_\mu p+(\rho+p)A_\mu+\D^\nu \pi_{\mu\nu}+
  A^\nu\pi_{\mu\nu}+\sigma_{\mu\nu}q^\nu-
  \varepsilon_{\mu\nu\alpha}\omega^\nu q^\alpha&=&0.
  \label{c2}
\end{eqnarray}%
In these equations, an overdot denotes $u^\nu\nabla_\nu$,
$\Theta=\nabla^\mu u_\mu$ is the volume expansion rate of the
$u^\mu$ worldlines, $A_\mu=\dot{u}_\mu=A_{\langle \mu\rangle}$ is
their 4-acceleration,
$\sigma_{ \mu\nu}=\D_{\langle \mu}u_{ \nu\rangle}$ is their shear
rate, and
$\omega_\mu =-{1\over2}\curl u_\mu =\omega_{\langle \mu\rangle}$
is their vorticity rate.

On a Friedmann brane, we get
\begin{equation}
  A_\mu =\omega_\mu =\sigma_{\mu\nu}=0,
  \qquad
  \Theta=3H,
\end{equation}
where $H=\dot a/a$ is the Hubble rate. The covariant spatial curl
is given by
\begin{equation}
  \curl V_\mu =\varepsilon_{\mu\alpha\beta}\D^\alpha V^\beta,
  \qquad
  \curl W_{\mu\nu}=\varepsilon_{\alpha\beta(\mu}\D^\alpha W^\beta{}_{\nu)},
\end{equation}
where $\varepsilon_{\mu\alpha\beta}$ is the projection orthogonal to $u^
\mu$ of the 4D brane alternating tensor, and $\D_\mu $ is the
projected part of the brane covariant derivative, defined by
\begin{equation}
  \D_\mu  F^{\alpha\dots}{}{}_{\dots \beta}=
  \left(\nabla_\mu  F^{\alpha\dots}{}{}_{\dots \beta}\right)_{\perp u}=
  h_\mu{}^\nu h^\alpha{}_\gamma \dots
  h_\beta{}^\delta \nabla_\nu F^{\gamma \dots}{}_{\dots \delta}.
\end{equation}
In a local inertial frame at a point on the brane, with $u^
\mu=\delta^ \mu{}_0$, we have: $0=A_0=\omega_0=\sigma_{0\mu}=
\varepsilon_{0\alpha\beta}= \curl V_0 =\curl W_{0\mu}$, and
\begin{equation}
  \D_\mu  F^{\alpha \dots}{}_{\dots \beta} =
  \delta_\mu{}^i \delta^\alpha{}_j \dots
  \delta_\beta{}^k \nabla_i F^{j \dots}{}_{\dots k}
  \qquad \mbox{(local inertial frame)},
\end{equation}
where $i,j,k=1,2,3$.

The absence of bulk source terms in the conservation equations is
a consequence of having $\Lambda_5$ as the only 5D source in the
bulk. For example, if there is a bulk scalar field, then there is
energy-momentum exchange between the brane and bulk (in addition
to the gravitational interaction)~\cite{mw, sca_1, sca_2, sca_3, sca_4, sca_5, sca_6}.

Equation~(\ref{nlc}) may be called the ``nonlocal conservation
equation''. Projecting along $u^\mu$ gives the nonlocal energy
conservation equation, which is a propagation equation for $\cu$.
In the general, nonlinear case, this gives
\begin{eqnarray}
  && \dot{\rho}_{\cal E}+{{4\over3}}\Theta{\cu}+\D^\mu \cq_\mu+
  2A^\mu \cq_\mu +\sigma^{\mu\nu}\cp_{\mu\nu} =
  \nonumber \\
  && \qquad \qquad {1\over 2\lambda}
  \Bigl[3\pi^{\mu\nu}\dot{\pi}_{\mu\nu}+
  3(\rho+p)\sigma^{\mu\nu} \pi_{\mu\nu}+
  \Theta \left(2q^\mu  q_\mu + \pi^{\mu\nu}\pi_{\mu\nu}\right) +
  6A^\mu q^\nu \pi_{\mu\nu}
  \nonumber \\
  && \qquad \qquad \qquad -2q^\mu \D_\mu \rho+3q^\mu \D^\nu \pi_{\mu\nu} +
  3\pi^{\mu\nu}\D_\mu  q_\nu +3
  \sigma^{\mu\nu}\pi_{\alpha\mu}\pi_\nu{}^\alpha-
  3\sigma^{\mu\nu}q_\mu q_\nu \Bigr].
  \label{c1_prime}
\end{eqnarray}%
Projecting into the comoving rest space gives the nonlocal
momentum conservation equation, which is a propagation equation
for $\cq_\mu $:
\begin{eqnarray}
  && \dot{q}^{\cal E}_{\langle \mu\rangle}+
  {4\over3}\Theta\cq_\mu+{1\over3}\D_\mu {\cu}+
  {4\over3}{\cu}A_\mu+\D^\nu\cp_{\mu\nu}+
  A^\nu \cp_{\mu\nu}+\sigma_{\mu}{}^\nu\cq_\nu-
  \varepsilon_{\mu}{}^{\nu\alpha}\omega_\nu \cq_\alpha =
  \nonumber \\
  && \qquad \qquad {1\over 4 \lambda}
  \biggl[ -4(\rho+p)\D_\mu \rho +6(\rho+p)\D^\nu \pi_{\mu\nu} +
  6q^\nu\dot{\pi}_{\langle \mu\nu \rangle}+2\pi_\mu{}^\nu \D_\nu \rho
  \nonumber \\
  && \qquad \qquad \qquad -6\pi^{\alpha\beta}
  \left(\D_\mu \pi_{\alpha\beta}-\D_\alpha \pi_{\beta\mu}\right)-
  3\pi_{\mu\alpha}\D_\beta \pi^{\alpha\beta}+
  12q^\nu \D_\mu q_\nu-6q^\nu\D_\nu q_\mu-2 q\mu\D^\nu q_\nu
  \nonumber \\
  && \qquad \qquad \qquad
  -6\pi_{\mu\alpha}\sigma^{\alpha\beta}q_\beta
  +6\sigma_{\mu\alpha} \pi^{\alpha\beta}q_\beta+
  6\pi_{\mu\nu}\varepsilon^{\nu\alpha\beta}\omega_\alpha q_\beta-
  6\varepsilon_{\mu\alpha\beta}\omega^\alpha \pi^{\beta\nu}q_\nu
  \nonumber \\
  && \qquad \qquad \qquad
  +  4(\rho+p)\Theta q_\mu + 2q_\mu A^\nu  q_\nu +
  6A_\mu  q^\nu  q_\nu +
  4q_\mu \sigma^{\alpha\beta} \pi_{\alpha\beta}\biggr].
  \label{c2_prime}
\end{eqnarray}%
The $1+3$-covariant decomposition shows two key features:
\begin{itemize}
\item Inhomogeneous and anisotropic effects from the 4D
  matter-radiation distribution on the brane are a source for the 5D
  Weyl tensor, which nonlocally ``backreacts'' on the brane via its
  projection ${\cal E}_{\mu\nu}$.
\item There are evolution equations for the dark radiative (nonlocal,
  Weyl) energy ($\cu$) and momentum ($\cq_\mu $) densities (carrying
  scalar and vector modes from bulk gravitons), but there is no
  evolution equation for the dark radiative anisotropic stress
  ($\cp_{\mu\nu}$) (carrying tensor, as well as scalar and vector,
  modes), which arises in both evolution equations.
\end{itemize}

In particular cases, the Weyl anisotropic stress $\cp_{\mu\nu}$
may drop out of the nonlocal conservation equations, i.e., when we
can neglect $\sigma^{\mu\nu}\cp_{\mu\nu}$, $\D^\nu \cp_{\mu\nu}$,
and $A^\nu \cp_{\mu\nu}$. This is the case when we consider
linearized perturbations about an FRW background (which remove the
first and last of these terms) and further when we can neglect
gradient terms on large scales (which removes the second term).
This case is discussed in Section~\ref{section_6}. But in general, and especially
in astrophysical contexts, the $\cp_{\mu\nu}$ terms cannot be
neglected. Even when we can neglect these terms, $\cp_{\mu\nu}$
arises in the field equations on the brane.

All of the matter source terms on the right of these two
equations, except for the first term on the right of
Equation~(\ref{c2_prime}), are imperfect fluid terms, and most of these
terms are quadratic in the imperfect quantities $q_\mu $ and
$\pi_{\mu\nu}$. For a single perfect fluid or scalar field, only
the $\D_\mu \rho$ term on the right of Equation~(\ref{c2_prime}) survives,
but in realistic cosmological and astrophysical models, further
terms will survive. For example, terms linear in $\pi_{\mu\nu}$
will carry the photon quadrupole in cosmology or the shear viscous
stress in stellar models. If there are two fluids (even if both
fluids are perfect), then there will be a relative velocity $v_\mu
$ generating a momentum density $q_\mu =\rho v_\mu $, which will
serve to source nonlocal effects.

In general, the 4 independent equations in Equations~(\ref{c1_prime})
and~(\ref{c2_prime}) constrain 4 of the 9 independent components of
${\cal E}_{\mu\nu}$ on the brane. What is missing is an evolution
equation for $\cp_{\mu\nu}$, which has up to 5 independent
components. These 5 degrees of freedom correspond to the 5
polarizations of the 5D graviton. Thus in general, the projection
of the 5-dimensional field equations onto the brane does not lead
to a closed system, as expected, since there are bulk degrees of
freedom whose impact on the brane cannot be predicted by brane
observers. The KK anisotropic stress $\cp_{\mu\nu}$ encodes the
nonlocality.

In special cases the missing equation does not matter. For
example, if $\cp_{\mu\nu}=0$ by symmetry, as in the case of an FRW
brane, then the evolution of ${\cal E}_{\mu\nu}$ is determined by
Equations~(\ref{c1_prime}) and~(\ref{c2_prime}). If the brane is stationary (with
Killing vector parallel to $u^\mu $), then evolution equations are
not needed for ${\cal E}_{\mu\nu}$, although in general
$\cp_{\mu\nu}$ will still be undetermined. However, small
perturbations of these special cases will immediately restore the
problem of missing information.

If the matter on the brane has a perfect-fluid or scalar-field
energy-momentum tensor, the local conservation
equations~(\ref{c1}) and~(\ref{c2}) reduce to
\begin{eqnarray}
  \dot{\rho}+\Theta(\rho+p)&=&0,
  \label{pc1} \\
  \D_\mu p+(\rho+p)A_\mu&=&0,
  \label{pc2}
\end{eqnarray}%
while the nonlocal conservation equations~(\ref{c1_prime})
and~(\ref{c2_prime}) reduce to
\begin{eqnarray}
  \dot{\cu}+{{4\over3}}\Theta{\cu}+\D^\mu \cq_\mu +2A^\mu \cq_\mu
  +\sigma^{\mu\nu}\cp_{\mu\nu}&=&0,
  \label{pc1_prime} \\
  \dot{q}^{\cal E}_{\langle \mu\rangle}+{{4\over3}}\Theta\cq_\mu
  +{{1\over3}}\D_\mu {\cu}+{{4\over3}}{\cu}A_\mu  +\D^\nu
  \cp_{\mu\nu}+A^\nu \cp_{\mu\nu}+\sigma_{\mu}{}^\nu\cq_\nu
  -\varepsilon_{\mu}{}^{\nu\alpha}\omega_\nu \cq_\alpha
  &=&-{(\rho+p)\over\lambda} \D_\mu \rho. \qquad
  \label{pc2_prime}
\end{eqnarray}%

Equation~(\ref{pc2_prime}) shows that~\cite{sms}
\begin{itemize}
\item if ${\cal E}_{\mu\nu}=0$ and the brane energy-momentum tensor
  has perfect fluid form, then the density $\rho$ must be homogeneous,
  $\D_\mu \rho=0$;
\item the converse does not hold, i.e., homogeneous density does
  \emph{not} in general imply vanishing ${\cal E}_{\mu\nu}$.
\end{itemize}

A simple example of the latter point is the FRW case:
Equation~(\ref{pc2_prime}) is trivially satisfied, while Equation~(\ref{pc1_prime})
becomes
\begin{equation}
  \dot{\rho}_{\cal E}+4H{\cu}=0.
\end{equation}
This equation has the dark radiation solution
\begin{equation}
  {\cu}=\rho_{{\cal E}\,0}\left({a_0\over a}\right)^4\!\!\!.
  \label{dr}
\end{equation}

If ${\cal E}_{\mu\nu}=0$, then the field equations on the brane
form a closed system. Thus for perfect fluid branes with
homogeneous density and ${\cal E}_{\mu\nu}=0$, the brane field
equations form a consistent closed system. However, this is
unstable to perturbations, and there is also no guarantee that the
resulting brane metric can be embedded in a regular bulk.

It also follows as a corollary that inhomogeneous density requires
nonzero ${\cal E}_{\mu\nu}$:
\begin{equation}
  \D_\mu \rho\neq 0
  \quad \Rightarrow \quad
  {\cal E}_{\mu\nu}\neq0.
\end{equation}
For example, stellar solutions on the brane necessarily have
${\cal E}_{\mu\nu}\neq0$ in the stellar interior if it is
non-uniform. Perturbed FRW models on the brane also must have
${\cal E}_{\mu\nu}\neq0$. Thus a nonzero ${\cal E}_{\mu\nu}$, and
in particular a nonzero $\cp_{\mu\nu}$, is inevitable in realistic
astrophysical and cosmological models.


\subsection{Propagation and constraint equations on the brane}

The propagation equations for the local and nonlocal energy
density and momentum density are supplemented by further
$1+3$-covariant propagation and constraint equations for the
kinematic quantities $\Theta$, $A_\mu $, $\omega_\mu $,
$\sigma_{\mu\nu}$, and for the free gravitational field on the
brane. The kinematic quantities govern the relative motion of
neighbouring fundamental world-lines. The free gravitational field
on the brane is given by the brane Weyl tensor $C_{\mu\nu
\alpha\beta}$. This splits into the gravito-electric and
gravito-magnetic fields on the brane:
\begin{equation}
  E_{\mu\nu}=C_{\mu\alpha\nu\beta}u^\alpha u^\beta=E_{\langle \mu\nu\rangle},
  \qquad
  H_{\mu\nu}={{1\over2}}\varepsilon_{\mu\alpha\beta}
  C^{\alpha\beta}{}{}_{\nu\gamma}u^\gamma=H_{\langle \mu\nu\rangle},
\end{equation}
where $E_{ \mu\nu}$ is not to be confused with ${\cal
E}_{\mu\nu}$. The Ricci identity for $u^\mu$
\begin{equation}
  \nabla_{[\mu}\nabla_{\nu]}u_\alpha={1\over2}
  R_{\alpha\nu\mu\beta}u^\beta,
\end{equation}
and the Bianchi identities
\begin{equation}
  \nabla^\beta C_{\mu\nu \alpha\beta} =
  \nabla_{[\mu}\left(-R_{\nu]\alpha} + {1\over6}Rg_{ \nu] \alpha}\right),
\end{equation}
produce the fundamental evolution and constraint equations
governing the above covariant quantities. The field equations are
incorporated via the algebraic replacement of the Ricci tensor
$R_{ \mu\nu}$ by the effective total energy-momentum tensor,
according to Equation~(\ref{e:einstein1}). The brane equations are
derived directly from the standard general relativity versions by
simply replacing the energy-momentum tensor terms $\rho,\dots$ by
$\rho_\mathrm{tot},\dots$. For a general fluid source, the equations
are given in~\cite{m1, gergnew}. In the case of a single perfect fluid or
minimally-coupled scalar field, the
equations reduce to the following nonlinear equations:
\begin{itemize}
\item Generalized Raychaudhuri equation (expansion propagation):
  \begin{equation}
    \dot{\Theta}+{1\over3}\Theta^2+\sigma_{ \mu\nu} \sigma^{
      \mu\nu} -2\omega_  \mu\omega^\mu  -\D^\mu  A_\mu -A_\mu A^\mu
    +{\kappa^2\over2}(\rho + 3p) -\Lambda =
    -{\kappa^2\over2}(2\rho+3p){\rho\over\lambda}- \kappa^2\cu.
    \label{pr}
  \end{equation}
\item
  Vorticity propagation:
  \begin{equation}
    \dot{\omega}_{\langle \mu\rangle} +{2\over3}\Theta\omega_\mu
    +{1\over2}\curl A_\mu -\sigma_{\mu\nu}\omega^\nu =0.
    \label{pe4}
  \end{equation}
\item Shear propagation:
  \begin{equation}
    \dot{\sigma}_{\langle \mu\nu\rangle}+{2\over3}\Theta\sigma_{\mu\nu}+
    E_{\mu\nu}-\D_{\langle\mu}A_{\nu\rangle} +
    \sigma_{\alpha\langle\mu}\sigma_{\nu\rangle}{}^\alpha +
    \omega_{\langle \mu}\omega_{\nu\rangle} -
    A_{\langle\mu}A_{\nu\rangle} ={\kappa^2\over 2}\cp_{\mu\nu}.
    \label{pe5}
  \end{equation}
\item Gravito-electric propagation (Maxwell--Weyl E-dot equation):
  \begin{eqnarray}
    & & \dot{E}_{\langle \mu\nu\rangle} \!+\!\Theta E_{\mu\nu}
    \!-\!\curl H_{\mu\nu}\!+\!{\kappa^2\over2}(\rho+p)\sigma_{\mu\nu}
    \!-\!2A^\alpha \varepsilon_{\alpha\beta(\mu}H_{\nu)}{}^\beta \!-\!
    3\sigma_{\alpha\langle \mu}E_{\nu\rangle}{}^\alpha \!+\!
    \omega^\alpha \varepsilon_{\alpha\beta(\mu}E_{\nu)}{}^\beta \!=
    \nonumber \\
    && \quad -{{\kappa^2\over2}} (\rho+p){\rho\over\lambda}\sigma_{\mu\nu}
    \nonumber \\
    && \quad -{\kappa^2\over6}\!\left[4\cu \sigma_{\mu\nu}+
    3\dot{\pi}^{\cal E}_{\langle \mu\nu\rangle} +\Theta \cp_{\mu\nu}
    +3\D_{\langle\mu}\cq_{\nu\rangle} +6A_{\langle \mu}\cq_{\nu\rangle}+
    3\sigma^\alpha {}_{\langle\mu} \cp_{ \nu\rangle \alpha}+
    3 \omega^\alpha \varepsilon_{\alpha\beta(\mu} \cp_{\nu)}{}^\beta \right]\!.
    \qquad
    \label{pe6}
  \end{eqnarray}%
\item Gravito-magnetic propagation (Maxwell--Weyl H-dot equation):
  \begin{eqnarray}
    &&\dot{H}_{\langle \mu\nu\rangle} +\Theta H_{\mu\nu} +\curl E_{\mu\nu}-
    3\sigma_{\alpha\langle \mu}H_{\nu\rangle}{}^\alpha +
    \omega^\alpha \varepsilon_{\alpha\beta(\mu}H_{\nu)}{}^\beta +
    2A^\alpha \varepsilon_{\alpha\beta(\mu}E_{\nu)}{}^\beta =
    \nonumber \\
    && \qquad \qquad {\kappa^2\over2}\left[ \curl\cp_{\mu\nu}-
    3\omega_{\langle\mu} \cq_{ \nu\rangle} +
    \sigma_{\alpha( \mu}\varepsilon_{\nu)}{}^{\alpha\beta}\cq_\beta\right].
    \label{pe7}
  \end{eqnarray}%
\item Vorticity constraint:
  \begin{equation}
    \D^\mu \omega_\mu -A^\mu \omega_\mu =0.
    \label{pcc1}
  \end{equation}
\item Shear constraint:
  \begin{equation}
    \D^\nu \sigma_{\mu\nu}-\curl\omega_\mu -{2\over3}\D_\mu \Theta
    +2\varepsilon_{\mu\nu\alpha}\omega^\nu A^\alpha = -\kappa^2\cq_\mu.
    \label{pcc2}
  \end{equation}
\item Gravito-magnetic constraint:
  \begin{equation}
    \curl\sigma_{ \mu\nu}+\D_{\langle \mu}\omega_{\nu\rangle} -
    H_{\mu\nu}+2A_{\langle \mu} \omega_{\nu\rangle}=0.
    \label{pcc3}
  \end{equation}
\item Gravito-electric divergence (Maxwell--Weyl div-E equation):
  \begin{eqnarray}
    && \D^\nu E_{\mu\nu}-{{\kappa^2\over3}}\D_\mu \rho -
    \varepsilon_{\mu\nu\alpha}\sigma^\nu {}_\beta H^{\alpha\beta} +
    3H_{\mu\nu}\omega^\nu =
    \nonumber \\
    && \qquad \qquad {\kappa^2\over3} {\rho\over \lambda} \D_\mu \rho +
    {\kappa^2\over6}\left(2\D_\mu \cu-2\Theta \cq_\mu -
    3\D^\nu \cp_{ \mu\nu} +3\sigma_{\mu}{}^\nu \cq_\nu -
    9\varepsilon_{\mu}{}^{\nu\alpha}\omega_\nu \cq_\alpha \right).
    \label{pcc4}
  \end{eqnarray}%
\item Gravito-magnetic divergence (Maxwell--Weyl div-H equation):
  \begin{eqnarray}
    && \D^\nu H_{\mu\nu}-\kappa^2(\rho+p)\omega_\mu +
    \varepsilon_{\mu\nu\alpha}\sigma^\nu {}_\beta E^{\alpha\beta} -
    3E_{\mu\nu}\omega^\nu =
    \nonumber \\
    && \qquad \qquad \kappa^2(\rho+ p){\rho\over\lambda} \omega_\mu +
    {\kappa^2\over6}\left(8 \cu \omega_\mu -3\curl\cq_\mu -
    3\varepsilon_\mu{}^{\nu\alpha}\sigma_\nu {}^\beta\cp_{\alpha\beta}-
    3\cp_{\mu\nu}\omega^\nu \right).
    \label{pcc5}
  \end{eqnarray}%
\item Gauss--Codazzi equations on the brane (with $\omega_\mu =0$):
  \begin{eqnarray}
    R^\perp_{\langle \mu\nu\rangle}+
    \dot{\sigma}_{\langle \mu\nu\rangle }+
    \Theta\sigma_{\mu\nu} -\D_{\langle \mu}A_{\nu\rangle} -
    A_{\langle \mu}A_{\nu\rangle}&=&\kappa^2\cp_{\mu\nu},
    \label{gc1} \\
    R^\perp+ {{2\over3}}\Theta^2-\sigma_{ \mu\nu} \sigma^{\mu\nu}-
    2\kappa^2\rho -2\Lambda &=& {\kappa^2}{\rho^2\over\lambda}+
    2\kappa^2\cu,
    \label{gc2}
  \end{eqnarray}%
  where $R^\perp_{\mu\nu}$ is the Ricci tensor for 3-surfaces
  orthogonal to $u^\mu $ on the brane, and
  $R^\perp=h^{\mu\nu}R^\perp_{\mu\nu}$.
\end{itemize}

\epubtkImage{figure04.png}{%
 \begin{figure}[htbp]
   \def\epsfsize#1#2{0.9#1}
   \centerline{\epsfbox{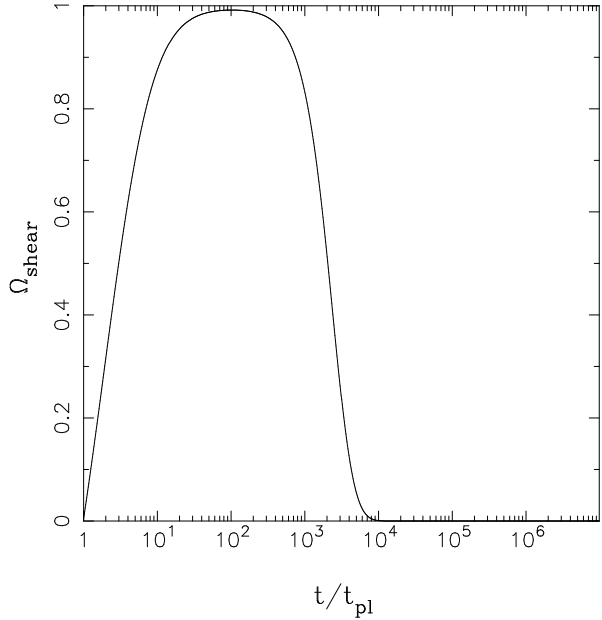}}
   \caption{The evolution of the dimensionless shear parameter
     $\Omega_\mathrm{shear} = \sigma^2/6H^2$ on a Bianchi~I brane, for a
     $V={1\over2}m^2\phi^2$ model. The early and late-time expansion
     of the universe is isotropic, but the shear dominates during an
     intermediate anisotropic stage. (Figure taken from~\cite{mss}.)}
   \label{figure_04}
 \end{figure}}

The standard 4D general relativity results are regained when
$\lambda^{-1}\to0$ and ${\cal E}_{\mu\nu}=0$, which sets all right
hand sides to zero in Equations~(\ref{pr}, \ref{pe4}, \ref{pe5},
\ref{pe6}, \ref{pe7}, \ref{pcc1}, \ref{pcc2}, \ref{pcc3}, \ref{pcc4},
\ref{pcc5}, \ref{gc1}, \ref{gc2}). Together with Equations~(\ref{pc1},
\ref{pc2}, \ref{pc1_prime}, \ref{pc2_prime}), these equations govern
the dynamics of the matter and gravitational fields on the brane,
incorporating both the local, high-energy (quadratic
energy-momentum) and nonlocal, KK (projected 5D Weyl) effects from
the bulk. High-energy terms are proportional to $\rho/\lambda$,
and are significant only when $\rho>\lambda$. The KK terms contain
$\cu$, $\cq_\mu $, and $\cp_{\mu\nu }$, with the latter two
quantities introducing imperfect fluid effects, even when the
matter has perfect fluid form.

Bulk effects give rise to important new driving and source terms
in the propagation and constraint equations. The vorticity
propagation and constraint, and the gravito-magnetic constraint
have no direct bulk effects, but all other equations do.
High-energy and KK energy density terms are driving terms in the
propagation of the expansion $\Theta$. The spatial gradients of
these terms provide sources for the gravito-electric field
$E_{\mu\nu}$. The KK anisotropic stress is a driving term in the
propagation of shear $\sigma_{\mu\nu}$ and the
gravito-electric/gravito-magnetic fields, $E$ and $H_{\mu\nu}$
respectively, and the KK momentum density is a
source for shear and the gravito-magnetic field. The 4D
Maxwell--Weyl equations show in detail the contribution to the 4D
gravito-electromagnetic field on the brane, i.e., $(E_{\mu\nu},H_{
\mu\nu})$, from the 5D Weyl field in the bulk.

An interesting example of how high-energy effects can modify
general relativistic dynamics arises in the analysis of
isotropization of Bianchi spacetimes. For a Binachi type~I brane,
Equation~(\ref{gc2}) becomes~\cite{mss}
\begin{equation}
  H^2={\kappa^2\over3}\rho\left(1+{\rho\over2\lambda}\right)+
  {\Sigma^2\over a^6},
\end{equation}
if we neglect the dark radiation, where $a$ and $H$ are the
average scale factor and expansion rate, and $\Sigma$ is the shear
constant. In general relativity, the shear term dominates as
$a\to0$, but in the brane-world, the high-energy $\rho^2$ term
will dominate if $w>0$, so that the matter-dominated early
universe is isotropic~\cite{mss, cs_1, cs_2, b1_1, b1_2, b1_3,
  b1_4}. This is illustrated in Figure~\ref{figure_04}.

Note that this conclusion is sensitive to the assumption that
$\cu\approx0$, which by Equation~(\ref{pc1_prime}) implies the
restriction
\begin{equation}
  \sigma^{\mu\nu}\cp_{\mu\nu} \approx 0.
\end{equation}
Relaxing this assumption can lead to non-isotropizing
solutions~\cite{ruth_1, ruth_2, ruth_3}.

The system of propagation and constraint equations, i.e.,
Equations~(\ref{pc1}, \ref{pc2}, \ref{pc1_prime}, \ref{pc2_prime})
and~(\ref{pr}, \ref{pe4}, \ref{pe5}, \ref{pe6}, \ref{pe7}, \ref{pcc1},
\ref{pcc2}, \ref{pcc3}, \ref{pcc4}, \ref{pcc5}, \ref{gc1},
\ref{gc2}), is exact and nonlinear, applicable to both cosmological
and astrophysical modelling, including strong-gravity effects. In
general the system of equations is not closed: There is no evolution
equation for the KK anisotropic stress $\cp_{\mu\nu}$.

\newpage


\section{Gravitational Collapse and Black Holes on the Brane}
\label{section_4}

The physics of brane-world compact objects and gravitational
collapse is complicated by a number of factors, especially the
confinement of matter to the brane, while the gravitational field
can access the extra dimension, and the nonlocal (from the brane
viewpoint) gravitational interaction between the brane and the
bulk. Extra-dimensional effects mean that the 4D matching
conditions on the brane, i.e., continuity of the induced metric
and extrinsic curvature across the 2-surface boundary, are much
more complicated to implement~\cite{germ_1, germ_2, germ_3, bgm}. High-energy
corrections increase the effective density and pressure of stellar
and collapsing matter. In particular this means that the effective
pressure does not in general vanish at the boundary 2-surface,
changing the nature of the 4D matching conditions on the brane.
The nonlocal KK effects further complicate the matching problem on
the brane, since they in general contribute to the effective
radial pressure at the boundary 2-surface. Gravitational collapse
inevitably produces energies high enough, i.e., $\rho\gg\lambda$,
to make these corrections significant.

We expect that extra-dimensional effects will be negligible
outside the high-energy, short-range regime. The corrections to
the weak-field potential, Equation~(\ref{newt}), are at the second
post-Newtonian (2PN) level~\cite{gr1, inta}. However, modifications
to Hawking radiation may bring significant corrections even for
solar-sized black holes, as discussed below.

A vacuum on the brane, outside a star or black hole, satisfies the
brane field equations
\begin{equation}
  R_{\mu\nu}=-{\cal E}_{\mu\nu},
  \qquad
  R^\mu {}_\mu =0={\cal E}^\mu {}_\mu,
  \qquad
  \nabla^\nu {\cal E}_{\mu\nu}=0.
  \label{vac}
\end{equation}
The Weyl term ${\cal E}_{\mu\nu}$ will carry an imprint of
high-energy effects that source KK modes (as discussed above).
This means that high-energy stars and the process of gravitational
collapse will in general lead to deviations from the 4D general
relativity problem. The weak-field limit for a static spherical
source, Equation~(\ref{newt}), shows that ${\cal E}_{\mu\nu}$ must be
nonzero, since this is the term responsible for the corrections to
the Newtonian potential.


\subsection{The black string}

The projected Weyl term vanishes in the simplest candidate for a
black hole solution. This is obtained by assuming the exact
Schwarzschild form for the induced brane metric and ``stacking'' it
into the extra dimension~\cite{chr},
\begin{eqnarray}
  {}^{(5)\!}ds^2 &=& e^{-2|y|/\ell}\tilde{g}_{\mu\nu}dx^\mu dx^\nu + dy^2,
  \label{bs1} \\
  \tilde{g}_{\mu\nu} &=&e^{2|y|/\ell}{g}_{\mu\nu}=
  -(1-{2GM/ r})dt^2+ {dr^2\over 1-2GM/ r}+r^2d \Omega^2.
  \label{bs2}
\end{eqnarray}%
(Note that Equation~(\ref{bs1}) is in fact a solution of the 5D field
equations~(\ref{rsefe}) if $\tilde{g}_{\mu\nu}$ is any 4D Einstein
vacuum solution, i.e., if $\tilde{R}_{\mu\nu}=0$, and this can be
generalized to the case $\tilde{R}_{\mu\nu}=-\tilde\Lambda
\tilde{g}_{\mu\nu}\,$~\cite{anlid, bmsv}.)

Each $\{y=\mathrm{const.}\}$ surface is a 4D Schwarzschild
spacetime, and there is a line singularity along $r=0$ for all
$y$. This solution is known as the Schwarzschild black string,
which is clearly not localized on the brane $y=0$. Although
${}^{(5)\!}C_{ABCD}\neq0$, the projection of the bulk Weyl tensor
along the brane is zero, since there is no correction to the 4D
gravitational potential:
\begin{equation}
  V(r)={GM \over r}
  \quad \Rightarrow \quad
  {\cal E}_{\mu\nu}=0.
\end{equation}
The violation of the perturbative corrections to the potential
signals some kind of non-$\mathrm{AdS}_5$ pathology in the bulk. Indeed,
the 5D curvature is unbounded at the Cauchy horizon, as
$y\to\infty$~\cite{chr}:
\begin{equation}
  {}^{(5)\!}R_{ABCD} \;{}^{(5)\!}R^{ABCD} = {40\over \ell^4}+
  {48G^2M^2 \over r^6} \, e^{4|y|/\ell}.
\end{equation}
Furthermore, the black string is unstable to large-scale
perturbations~\cite{g}.

Thus the ``obvious'' approach to finding a brane black hole fails.
An alternative approach is to seek solutions of the brane field
equations with nonzero ${\cal E}_{\mu\nu}$~\cite{dmpr}. Brane
solutions of static black hole exteriors with 5D corrections to
the Schwarzschild metric have been found~\cite{dmpr, germ_1, germ_2, germ_3, bhsol_1, bhsol_2, bhsol_3},
but the bulk metric for these solutions has not been found.
Numerical integration into the bulk, starting from static black
hole solutions on the brane, is plagued with
difficulties~\cite{num_1, num_2}.


\subsection{Taylor expansion into the bulk}

One can use a Taylor expansion, as in Equation~(\ref{tay}), in order to
probe properties of a static black hole on the brane~\cite{dms}.
(An alternative expansion scheme is discussed in~\cite{cas}.) For
a vacuum brane metric,
\begin{eqnarray}
  \tilde{g}_{\mu\nu}(x,y) & = &
  \tilde {g}_{\mu\nu}(x,0)-{\cal E}_{\mu\nu}(x,0+)y^2-
  \frac{2}{\ell} {\cal E}_{\mu\nu}(x,0+)|y|^3
  \nonumber \\
  & & + \frac{1}{12}\left[ \Box {\cal E}_{\mu\nu}-
  \frac{32}{\ell^2} {\cal E}_{\mu\nu} +
  2R_{\mu\alpha\nu\beta}{\cal E}^{\alpha\beta} +
  6{\cal E}_\mu{}^\alpha {\cal E}_{\alpha\nu} \right]_{y=0+}
  \!\!\!\!\! y^4 + \dots
  \label{tayv}
\end{eqnarray}%
This shows in particular that the propagating effect of 5D gravity
arises only at the fourth order of the expansion. For a static
spherical metric on the brane,
\begin{equation}
  \tilde{g}_{\mu\nu}dx^\mu dx^\nu =
  -F(r)dt^2+{dr^2\over H(r)}+r^2d\Omega^2,
  \label{sss}
\end{equation}
the projected Weyl term on the brane is given by
\begin{eqnarray}
  {\cal E}_{00} & = & \frac{F}{r}\left[ H'-\frac{1-H}{r} \right],
  \label{e00} \\
  {\cal E}_{rr} & = & -\frac{1}{rH}\left[{F'\over F}-\frac{1-H}{r} \right],
  \\
  {\cal E}_{\theta\theta}&=&-1+H +\frac{r}{2}H\left(\frac{F'}{F}+
  \frac{H'}{H} \right).
\end{eqnarray}%
These components allow one to evaluate the metric coefficients in
Equation~(\ref{tayv}). For example, the area of the 5D horizon is
determined by $\tilde{g}_{\theta\theta}$; defining $\psi(r)$ as
the deviation from a Schwarzschild form for $H$, i.e.,
\begin{equation}
  H(r)=1-{2m\over r} + \psi(r),
\end{equation}
where $m$ is constant, we find
\begin{equation}
  \tilde{g}_{\theta\theta}(r,y)=
  r^2-\psi'\left(1+{2\over\ell}|y|\right)y^2 +
  {1\over 6r^2}\left[\psi'+{1\over2}(1+\psi')(r\psi'-\psi)'\right]y^4
  + \dots
\end{equation}
This shows how $\psi$ and its $r$-derivatives determine the change
in area of the horizon along the extra dimension. For the black
string $\psi=0$, and we have $\tilde{g}_{\theta\theta}(r,y)=r^2$.
For a large black hole, with horizon scale $\gg \ell$, we have
from Equation~(\ref{newt}) that
\begin{equation}
  \psi \approx -{4m\ell^2 \over 3r^3}.
\end{equation}
This implies that $\tilde{g}_{\theta\theta}$ is decreasing as we
move off the brane, consistent with a pancake-like shape of the
horizon. However, note that the horizon shape is tubular in
Gaussian normal coordinates~\cite{giaren}.


\subsection{The ``tidal charge'' black hole}

The equations~(\ref{vac}) form a system of constraints on the
brane in the stationary case, including the static spherical case,
for which
\begin{equation}
  \Theta=0=\omega_\mu =\sigma_{\mu\nu},
  \qquad
  \dot{\rho}_{\cal E}=0=\cq_\mu=\dot{\pi}^{\cal E}_{\mu\nu}.
\end{equation}
The nonlocal conservation equations $\nabla^\nu {\cal
E}_{\mu\nu}=0$ reduce to
\begin{equation}
  {1\over3}{\D}_{\mu}\cu+{4\over3}\cu A_\mu  + \D^\nu
  \cp_{\mu\nu}+A^\nu \cp_{\mu\nu}=0,
\end{equation}
where, by symmetry,
\begin{equation}
  \cp_{\mu\nu}=
  \Pi_{\cal E}\left({1\over3}h_{\mu\nu}-r_\mu r_\nu\right),
  \label{ass}
\end{equation}
for some $\Pi_{\cal E}(r)$, with $r_\mu $ being the unit radial vector.
The solution of the brane field equations requires the input of
${\cal E}_{\mu\nu}$ from the 5D solution. In the absence of a 5D
solution, one can make an assumption about ${\cal E}_{\mu\nu}$ or
$g_{\mu\nu}$ to close the 4D equations.

If we assume a metric on the brane of Schwarzschild-like form,
i.e., $H=F$ in Equation~(\ref{sss}), then the general solution of the
brane field equations is~\cite{dmpr}
\begin{eqnarray}
  F&=&1-{2GM\over  r} +{2G\ell Q\over r^2},
  \label{bh} \\
  {\cal E}_{\mu\nu }&=&-{2G\ell Q\over r^4}
  \left[u_\mu u_\nu -2r_\mu r_\nu +h_{\mu\nu } \right],
\end{eqnarray}%
where $Q$ is a constant. It follows that the KK energy density and
anisotropic stress scalar (defined via Equation~(\ref{ass})) are given
by
\begin{equation}
  \cu = {\ell Q \over 4\pi\, r^4} = {1\over 2} \Pi_{\cal E}.
\end{equation}

The solution~(\ref{bh}) has the form of the general relativity
Reissner--Nordstr\"om solution, but there is \emph{no} electric
field on the brane. Instead, the nonlocal Coulomb effects
imprinted by the bulk Weyl tensor have induced a ``tidal'' charge
parameter $Q$, where $Q=Q(M)$, since $M$ is the source of the bulk
Weyl field. We can think of the gravitational field of $M$ being
``reflected back'' on the brane by the negative bulk cosmological
constant~\cite{dad}. If we impose the small-scale perturbative
limit ($r\ll\ell$) in Equation~(\ref{newt2}), we find that
\begin{equation}
  Q=-2M.
\end{equation}
Negative $Q$ is in accord with the intuitive idea that the tidal
charge strengthens the gravitational field, since it arises from
the source mass $M$ on the brane. By contrast, in the
Reissner--Nordstr\"om solution of general relativity, $Q\propto
+q^2$, where $q$ is the electric charge, and this weakens the
gravitational field. Negative tidal charge also preserves the
spacelike nature of the singularity, and it means that there is
only one horizon on the brane, outside the Schwarzschild horizon:
\begin{equation}
  r_\mathrm{h}=GM\left[1+\sqrt {1 -{2\ell Q \over GM^2}}\right]=
  GM\left[1+\sqrt {1 +{4\ell \over GM}}\right].
\end{equation}

The tidal-charge black hole metric does not satisfy the far-field
$r^{-3}$ correction to the gravitational potential, as in
Equation~(\ref{newt}), and therefore cannot describe the end-state of
collapse. However, Equation~(\ref{bh}) shows the correct 5D behaviour
of the potential ($\propto r^{-2}$) at short distances, so that
the tidal-charge metric could be a good approximation in the
strong-field regime for small black holes.


\subsection{Realistic black holes}

Thus a simple brane-based approach, while giving useful insights,
does not lead to a realistic black hole solution. There is no
known solution representing a realistic black hole localized on
the brane, which is stable and without naked singularity. This
remains a key open question of nonlinear brane-world gravity.
(Note that an exact solution is known for a black hole on a
$1+2$-brane in a 4D bulk~\cite{ehm}, but this is a very special
case.) Given the nonlocal nature of ${\cal E}_{\mu\nu}$, it is
possible that the process of gravitational collapse itself leaves
a signature in the black hole end-state, in contrast with general
relativity and its no-hair theorems. There are contradictory
indications about the nature of the realistic black hole solution
on the brane:
\begin{itemize}
\item Numerical simulations of highly relativistic static stars on
  the brane~\cite{w} indicate that general relativity remains a good
  approximation.
\item Exact analysis of Oppenheimer--Snyder collapse on the brane
  shows that the exterior is non-static~\cite{bgm}, and this is
  extended to general collapse by arguments based on a generalized
  AdS/CFT correspondence~\cite{t, efk}.
\end{itemize}
The first result suggests that static black holes could exist as
limits of increasingly compact static stars, but the second result
and conjecture suggest otherwise. This remains an open question.
More recent numerical evidence is also not conclusive, and it
introduces further possible subtleties to do with the size of the
black hole~\cite{ktn, Yoshino:2008rx}.

On very small scales relative to the $\mathrm{AdS}_5$ curvature scale,
$r\ll\ell$, the gravitational potential becomes 5D, as shown in
Equation~(\ref{newt2}),
\begin{equation}
  V(r) \approx {G\ell M \over r^2}={G_5 M \over r^2}.
\end{equation}
In this regime, the black hole is so small that it does not ``see''
the brane, so that it is approximately a 5D Schwarzschild (static)
solution. However, this is always an approximation because of the
self-gravity of the brane (the situation is different in ADD-type
brane-worlds where there is no brane tension). As the black hole
size increases, the approximation breaks down. Nevertheless, one
might expect that static solutions exist on sufficiently small
scales. Numerical investigations appear to confirm
this~\cite{ktn, Yoshino:2008rx}: Static metrics satisfying the asymptotic $\mathrm{AdS}_5$
boundary conditions are found if the horizon is small compared to
$\ell$, but no numerical convergence can be achieved close to
$\ell$. The numerical instability that sets in may mask the fact
that even the very small black holes are not strictly static. Or
it may be that there is a transition from static to non-static
behaviour. Or it may be that static black holes do exist on all
scales.

The 4D Schwarzschild metric cannot describe the final state of
collapse, since it cannot incorporate the 5D behaviour of the
gravitational potential in the strong-field regime (the metric is
incompatible with massive KK modes). A non-perturbative exterior
solution should have nonzero ${\cal E}_{\mu\nu}$ in order to be
compatible with massive KK modes in the strong-field regime. In
the end-state of collapse, we expect an ${\cal E}_{\mu\nu}$ which
goes to zero at large distances, recovering the Schwarzschild
weak-field limit, but which grows at short range. Furthermore,
${\cal E}_{\mu\nu}$ may carry a Weyl ``fossil record'' of the
collapse process.


\subsection{Oppenheimer--Snyder collapse gives a non-static black hole}

The simplest scenario in which to analyze gravitational collapse
is the Oppenheimer--Snyder model, i.e., collapsing homogeneous and
isotropic dust~\cite{bgm}. The collapse region on the brane has an
FRW metric, while the exterior vacuum has an unknown metric. In 4D
general relativity, the exterior is a Schwarzschild spacetime; the
dynamics of collapse leaves \emph{no} imprint on the exterior.

The collapse region has the metric
\begin{equation}
  ds^2=-d\tau^2+
  {a(\tau)^2\left[dr^2+r^2d\Omega^2\right] \over (1+ kr^2/4)^{2}},
  \label{os1}
\end{equation}
where the scale factor satisfies the modified Friedmann equation
(see below),
\begin{equation}
  {\dot{a}^2 \over a^2}= {8\pi G\over 3}
  \rho\left(1 + {\rho \over 2\lambda}+ {\cu \over \rho} \right).
  \label{evol}
\end{equation}
The dust matter and the dark radiation evolve as
\begin{equation}
  \rho=\rho_0\left({a_0\over\mu}\right)^3,
  \qquad
  \cu=\rho_{{\cal E}\,0}\left({a_0\over\mu}\right)^4,
\end{equation}
where $a_0$ is the epoch when the cloud started to collapse. The
proper radius from the centre of the cloud is
$R(\tau)=r a(\tau)/(1+ { {1\over4}} kr^2)$. The collapsing boundary
surface $\Sigma$ is given in the interior comoving coordinates as a
free-fall surface, i.e.\ $r=r_0= \mathrm{const.}$, so that
$R_\Sigma(\tau)= r_0 a(\tau)/(1+ {1\over4} kr_0^2)$.

We can rewrite the modified Friedmann equation on the interior
side of $\Sigma$ as
\begin{equation}
  \dot{R}^2= {2GM\over R}+ {3GM^2\over 4\pi\lambda R^4}+ {Q \over R^2}+ E,
  \label{geo1}
\end{equation}
where the ``physical mass'' $M$ (total energy per proper star
volume), the total ``tidal charge'' $Q$, and the ``energy'' per unit
mass $E$ are given by
\begin{eqnarray}
  M &=&{4\pi a_0^3 r_0^3\rho_0 \over 3(1+ {{1\over4}}kr_0^2)^3},
  \\
  Q &=& {\rho_{{\cal E}\,0}a_0^4r_0^4 \over(1+ { {1\over4}}kr_0^2)^4},
  \\
  E&=&-{kr_0^2 \over(1+ { {1\over4}}kr_0^2)^2}>-1.
\end{eqnarray}%

Now we \emph{assume} that the exterior is static, and satisfies the
standard 4D junction conditions. Then we check whether this
exterior is physical by imposing the modified Einstein
equations~(\ref{vac}). We will find a contradiction.

The standard 4D Darmois--Israel matching conditions, which we
assume hold on the brane, require that the metric and the
extrinsic curvature of $\Sigma$ be continuous (there are no
intrinsic stresses on $\Sigma$). The extrinsic curvature is
continuous if the metric is continuous and if $\dot R$ is
continuous. We therefore need to match the metrics and $\dot R$
across $\Sigma$.

The most general static spherical metric that could match the
interior metric on $\Sigma$ is
\begin{equation}
  ds^2 = -F(R)^2\left[1-{2Gm(R)\over R}\right]dt^2 +
  {dR^2 \over 1-{2Gm(R)/ R}} +R^2d\Omega^2.
  \label{s1}
\end{equation}
We need two conditions to determine the functions $F(R)$ and
$m(R)$. Now $\Sigma$ is a freely falling surface in both metrics,
and the radial geodesic equation for the exterior metric gives
$\dot{R}^2=-1+2Gm(R)/R+{\tilde{E}/ F(R)^2},$ where $\tilde{E}$
is a constant and the dot denotes a proper time derivative, as
above. Comparing this with Equation~(\ref{geo1}) gives one condition.
The second condition is easier to derive if we change to null
coordinates. The exterior static metric, with
\begin{displaymath}
  dv=dt+\frac{dR}{(1-2Gm/R)F},
\end{displaymath}
\begin{equation}
  ds^2 = -F^2 \left( 1-\frac{2Gm}{R} \right) dv^2 +2F\,dv\,dR+R^2\,d\Omega^2.
  \label{s1_prime}
\end{equation}
The interior Robertson--Walker metric takes the form~\cite{bgm}
\begin{equation}
  ds^2= -{\tau_{,v}^2\left[ 1- (k+\dot{a}^2)R^2/a^2 \right] dv^2
  \over (1-kR^2/a^2)}+ {2 \tau_{,v} dvdR \over \sqrt{1-kR^2/a^2}}
  +R^2d\Omega^2,
  \label{s1_prime_prime}
\end{equation}
where
\begin{displaymath}
  d\tau=\tau_{,v}dv+\left(1+{1\over4}kr^2\right)
  \frac{dR}{r\dot a-1+{1\over4}kr^2}.
\end{displaymath}
Comparing Equations~(\ref{s1_prime}) and~(\ref{s1_prime_prime})
on $\Sigma$ gives the second condition. The two conditions
together imply that $F$ is a constant, which we can take as
$F(R)=1$ without loss of generality (choosing $\tilde E=E+1$), and
that
\begin{equation}
  m(R)= M+{3M^2 \over 8\pi\lambda R^3}+ {Q\over 2G R}.
  \label{s3}
\end{equation}
In the limit $\lambda^{-1}=0=Q$, we recover the 4D Schwarzschild
solution. In the general brane-world case, Equations~(\ref{s1})
and~(\ref{s3}) imply that the brane Ricci scalar is
\begin{equation}
  R^\mu{}_\mu ={9GM^2 \over 2\pi\lambda R^6}.
  \label{m1}
\end{equation}
However, this contradicts the field equations~(\ref{vac}), which
require
\begin{equation}
  R^\mu{}_\mu =0.
  \label{vac2}
\end{equation}
It follows that a static exterior is only possible if
$M/\lambda=0,$ which is the 4D general relativity limit. In the
brane-world, collapsing homogeneous and isotropic dust leads to a
\emph{non-static} exterior. Note that this no-go result does not
require any assumptions on the nature of the bulk spacetime, which
remains to be determined.

Although the exterior metric is not determined (see~\cite{govdad}
for a toy model), we know that its non-static nature arises from
\begin{itemize}
\item 5D bulk graviton stresses, which transmit effects nonlocally
  from the interior to the exterior, and
\item the non-vanishing of the effective pressure at the boundary,
  which means that dynamical information from the interior can be
  conveyed outside via the 4D matching conditions.
\end{itemize}

The result suggests that gravitational collapse on the brane may
leave a signature in the exterior, dependent upon the dynamics of
collapse, so that astrophysical black holes on the brane may in
principle have KK ``hair''. It is possible that the non-static
exterior will be transient, and will tend to a static geometry at
late times, close to Schwarzschild at large distances.


\subsection{AdS/CFT and black holes on 1-brane RS-type models}

Oppenheimer--Snyder collapse is very special; in particular, it is
homogeneous. One could argue that the non-static exterior arises
because of the special nature of this model. However, the
underlying reasons for non-static behaviour are not special to
this model; on the contrary, the role of high-energy corrections
and KK stresses will if anything be enhanced in a general,
inhomogeneous collapse. There is in fact independent heuristic
support for this possibility, arising from the AdS/CFT
correspondence.

The basic idea of the correspondence is that the classical
dynamics of the $\mathrm{AdS}_5$ gravitational field correspond to the
quantum dynamics of a 4D conformal field theory on the brane. This
correspondence holds at linear perturbative order~\cite{acft}, so
that the RS 1-brane infinite $\mathrm{AdS}_5$ brane-world (without matter
fields on the brane) is equivalently described by 4D general
relativity coupled to conformal fields,
\begin{equation}
  G_{\mu\nu}=8\pi G T^\mathrm{(cft)}_{\mu\nu}.
  \label{cft}
\end{equation}
According to a conjecture~\cite{t}, the correspondence holds also
in the case where there is strong gravity on the brane, so that
the classical dynamics of the bulk gravitational field of the
brane black hole are equivalent to the dynamics of a
quantum-corrected 4D black hole (in the dual CFT-plus-gravity
description). In other words~\cite{t, efk}:
\begin{itemize}
\item Quantum backreaction due to Hawking radiation in the 4D picture
  is described as classical dynamics in the 5D picture.
\item The black hole evaporates as a classical process in the 5D
  picture, and there is thus no stationary black hole solution in RS
  1-brane.
\end{itemize}

A further remarkable consequence of this conjecture is that
Hawking evaporation is dramatically enhanced, due to the very
large number of CFT modes of order $(\ell/\ell_\mathrm{p})^2$. The
energy loss rate due to evaporation is
\begin{equation}
  {\dot M \over M} \sim N \left({1 \over G^2M^3}\right),
\end{equation}
where $N$ is the number of light degrees of freedom. Using $N\sim
\ell^2/G$, this gives an evaporation timescale~\cite{t}
\begin{equation}
  t_\mathrm{evap} \sim \left({M\over M_\odot}\right)^3
  \left({1 \mathrm{\ mm} \over \ell}\right)^2 \mathrm{\ yr}.
\end{equation}
A more detailed analysis~\cite{egk} shows that this expression
should be multiplied by a factor $\approx 100$. Then the existence
of stellar-mass black holes on long time scales places limits on
the $\mathrm{AdS}_5$ curvature scale that are more stringent than the
table-top limit, Equation~(\ref{tt}). The existence of black hole X-ray
binaries implies
\begin{equation}
  \ell \lesssim  10^{-2} \mathrm{\ mm},
\end{equation}
already an order of magnitude improvement on the table-top limit.

One can also relate the Oppenheimer--Snyder result to these
considerations. In the AdS/CFT picture, the non-vanishing of the
Ricci scalar, Equation~(\ref{m1}), arises from the trace of the Hawking
CFT energy-momentum tensor, as in Equation~(\ref{cft}). If we evaluate
the Ricci scalar at the black hole horizon, $R\sim 2GM$, using
$\lambda=6M_5^6/M_\mathrm{p}^2$, we find
\begin{equation}
  R^\mu{}_\mu  \sim {M_5{}^{12} \, \ell^6 \over M^4}.
\end{equation}
The CFT trace on the other hand is given by
$T^\mathrm{(cft)}\sim N T_\mathrm{h}^4/M_\mathrm{p}^2$, so that
\begin{equation}
  8\pi GT^\mathrm{(cft)}\sim {M_5{}^{12} \, \ell^6 \over M^4}.
\end{equation}
Thus the Oppenheimer--Snyder result is qualitatively consistent
with the AdS/CFT picture.

Clearly the black hole solution, and the collapse process that
leads to it, have a far richer structure in the brane-world than
in general relativity, and deserve further attention. In
particular, two further topics are of interest:
\begin{itemize}
\item Primordial black holes in 1-brane RS-type cosmology have been
  investigated in~\cite{inta, Guedens:2002km, Guedens:2002sd, clan_3, Clancy:2003zd,
  Sendouda:2003dc}. High-energy effects in the early universe (see the
  next Section~\ref{section_5}) can significantly modify the
  evaporation and accretion processes, leading to a prolonged survival
  of these black holes. Such black holes evade the enhanced Hawking
  evaporation described above when they are formed, because they are
  much smaller than $\ell$.
\item Black holes will also be produced in particle collisions at
  energies $\gtrsim M_5$, possibly well below the Planck scale. In ADD
  brane-worlds, where $M_{4+d} = {\cal O}(\mathrm{TeV})$ is not ruled
  out by current observations if $d>1$, this raises the exciting
  prospect of observing black hole production signatures in the
  next-generation colliders and cosmic ray detectors (see~\cite{cav,
    gid_1, gid_2}).
\end{itemize}

\newpage


\section{Brane-World Cosmology: Dynamics}
\label{section_5}

A $1+4$-dimensional spacetime with spatial 4-isotropy (4D
spherical/plane/hyperbolic symmetry) has a natural foliation into the
symmetry group orbits, which are $1+3$-dimensional surfaces with
3-isotropy and 3-homogeneity, i.e., FRW surfaces. In particular,
the $\mathrm{AdS}_5$ bulk of the RS brane-world, which admits a foliation
into Minkowski surfaces, also admits an FRW foliation since it is
4-isotropic. Indeed this feature of 1-brane RS-type cosmological
brane-worlds underlies the importance of the AdS/CFT correspondence in
brane-world cosmology~\cite{acftcosmo_1, acftcosmo_2, acftcosmo_8,
  acftcosmo_3, acftcosmo_4, acftcosmo_5, acftcosmo_6, acftcosmo_7}.

The generalization of $\mathrm{AdS}_5$ that preserves 4-isotropy and solves
the vacuum 5D Einstein equation~(\ref{rsefe}) is
Schwarzschild--$\mathrm{AdS}_5$, and this bulk therefore admits an FRW
foliation. It follows that an FRW brane-world, the cosmological
generalization of the RS brane-world, is a part of
Schwarzschild--$\mathrm{AdS}_5$, with the $Z_2$-symmetric FRW brane at the
boundary. (Note that FRW branes can also be embedded in non-vacuum
generalizations, e.g., in Reissner--Nordstr\"om--$\mathrm{AdS}_5$ and
Vaidya--$\mathrm{AdS}_5$.)

In natural static coordinates, the bulk metric is
\begin{eqnarray}
  {}^{(5)\!}ds^2 &=& - F(R)dT^2 + {dR^2 \over F(R)} +
  R^2 \left( {dr^2 \over 1 - K r^2} + r^2 d \Omega^2 \right),
  \label{sads} \\
  F(R) &=& K + {R^2 \over \ell^2} - {m \over R^2},
  \label{sads2}
\end{eqnarray}%
where $K=0,\pm1$ is the FRW curvature index, and $m$ is the mass
parameter of the black hole at $R=0$ (recall that the 5D
gravitational potential has $R^{-2}$ behaviour). The bulk black
hole gives rise to dark radiation on the brane via its Coulomb
effect. The FRW brane moves radially along the 5th dimension, with
$R=a(T)$, where $a$ is the FRW scale factor, and the junction
conditions determine the velocity via the Friedmann equation for
$a$~\cite{birk_1, birk_2}. Thus one can interpret the expansion of the
universe as motion of the brane through the static bulk. In the
special case $m=0$ and $da/dT=0$, the brane is fixed and has
Minkowski geometry, i.e., the original RS 1-brane brane-world is
recovered in different coordinates.

The velocity of the brane is coordinate-dependent, and can be set
to zero. We can use Gaussian normal coordinates, in which the
brane is fixed but the bulk metric is not manifestly
static~\cite{bdel}:
\begin{equation}
  {}^{(5)\!}ds^2 = -N^2(t,y)dt^2+A^2(t,y)
  \left[{dr^2 \over 1 - Kr^2}+r^2d\Omega^2\right]+dy^2.
  \label{gnm}
\end{equation}
Here $a(t)=A(t,0)$ is the scale factor on the FRW brane at $y=0$,
and $t$ may be chosen as proper time on the brane, so that
$N(t,0)=1$. In the case where there is no bulk black hole ($m=0$),
the metric functions are
\begin{eqnarray}
  N &=& {\dot {A}(t,y) \over \dot {a}(t)},
  \label{gnm1} \\
  A &=& a(t)\left[ \cosh \left({y \over \ell}\right) -
  \left\{1+{\rho(t)\over\lambda}\right\}
  \sinh \left({|y| \over \ell}\right)\right].
  \label{gnm2}
\end{eqnarray}%
Again, the junction conditions determine the Friedmann equation.
The extrinsic curvature at the brane is
\begin{equation}
  K^\mu{}_\nu =
  \diag \left({N'\over N}, {A'\over A}, {A'\over A},
  {A'\over A}\right)_\mathrm{brane}\!\!\!\!\!\!\!\!\!\!\!\!.
\end{equation}
Then, by Equation~(\ref{ext}),
\begin{eqnarray}
  {N'\over N}\Big|_\mathrm{brane}&=&
  {\kappa_5^2 \over 6}(2\rho+3p-\lambda),
  \\
  {A'\over A}\Big|_\mathrm{brane}&=&
  -{\kappa_5^2 \over 6}(\rho+\lambda).
  \label{biga}
\end{eqnarray}%
The field equations yield the first integral~\cite{bdel}
\begin{equation}
  (AA')^2-{A^2\over N^2}\dot{A}^2+{\Lambda_5 \over6}A^4+m=0,
\end{equation}
where $m$ is constant. Evaluating this at the brane, using
Equation~(\ref{biga}), gives the modified Friedmann
equation~(\ref{mf}).

The dark radiation carries the imprint on the brane of the bulk
gravitational field. Thus we expect that ${\cal E}_{\mu\nu}$ for
the Friedmann brane contains bulk metric terms evaluated at the
brane. In Gaussian normal coordinates (using the field equations
to simplify),
\begin{equation}
  {\cal E}^0{}_0=3{A''\over A}\Big|_\mathrm{brane}+{\Lambda_5 \over 2},
  \qquad
  {\cal E}^i{}_j=-\left({1\over 3} {\cal E}^0{}_0\right)\delta^i{}_j.
\end{equation}

Either form of the cosmological metric, Equation~(\ref{sads})
or~(\ref{gnm}), may be used to show that 5D gravitational wave
signals can take ``short-cuts'' through the bulk in travelling
between points A and B on the brane~\cite{speed_1, speed_2, speed_3}. The travel time
for such a graviton signal is less than the time taken for a
photon signal (which is stuck to the brane) from A to B.

Instead of using the junction conditions, we can use the covariant
3D Gauss--Codazzi equation~(\ref{gc2}) to find the modified
Friedmann equation:
\begin{equation}
  H^2 = \frac{\kappa^2}{3} \rho\left(1+{\rho\over 2\lambda}\right) +
  {m\over a^4}+ \frac{1}{3} \Lambda - \frac{K}{a^2},
  \label{mf}
\end{equation}
on using Equation~(\ref{dr}), where
\begin{equation}
  m= \frac{\kappa^2}{3}\rho_{{\cal E}\,0}a_0^4.
\end{equation}
The covariant Raychauhuri equation~(\ref{pr}) yields
\begin{equation}
  \dot H= - {\kappa^2\over 2}(\rho+p)
  \left(1+ {\rho\over \lambda}\right)-2{m\over a^4}+{K\over a^2},
\end{equation}
which also follows from differentiating Equation~(\ref{mf}) and using
the energy conservation equation.

When the bulk black hole mass vanishes, the bulk geometry reduces
to $\mathrm{AdS}_5$, and $\cu=0$. In order to avoid a naked singularity, we
assume that the black hole mass is non-negative, so that
$\rho_{{\cal E}\,0}\geq0$. (By Equation~(\ref{sads2}), it is possible
to avoid a naked singularity with negative $m$ when $K=-1$,
provided $|m| \leq \ell^2/4$.) This additional effective
relativistic degree of freedom is constrained by nucleosynthesis
and CMB observations to be no more than $\sim 5\%$ of the radiation
energy density~\cite{lmsw, bm, dr_1, dr_2}:
\begin{equation}
  \left. {\cu \over \rho_\mathrm{rad}}\right|_\mathrm{nuc}
  \!\!\! \lesssim 0.05.
\end{equation}
The other modification to the Hubble rate is via the high-energy
correction $\rho/\lambda$. In order to recover the observational
successes of general relativity, the high-energy regime where
significant deviations occur must take place before
nucleosynthesis, i.e., cosmological observations impose the lower
limit
\begin{equation}
  \lambda > (1 \mathrm{\ MeV})^4
  \quad \Rightarrow \quad
  M_5> 10^4 \mathrm{\ GeV}.
\end{equation}
This is much weaker than the limit imposed by table-top
experiments, Equation~(\ref{rslimit}). Since $\rho^2/\lambda$ decays as
$a^{-8}$ during the radiation era, it will rapidly become
negligible after the end of the high-energy regime,
$\rho=\lambda$.

If $\cu=0$ and $K=0=\Lambda$, then the exact solution of the
Friedmann equations for $w=p/\rho=\mathrm{const.}$ is~\cite{bdel}
\begin{equation}
  a=\mathrm{const.}\times[t(t+t_{\lambda})]^{1/3(w+1)},
  \qquad
  t_{\lambda}={M_\mathrm{p}\over\sqrt{3\pi\lambda}(1+w)}
  \lesssim (1+w)^{-1}10^{-9} \mathrm{\ s},
  \label{ex1}
\end{equation}
where $w> -1$. If $\cu\neq0$ (but $K=0=\Lambda$), then the
solution for the radiation era ($w={1\over3}$) is~\cite{bm}
\begin{equation}
  a=\mathrm{const.}\times[t(t+t_{\lambda})]^{1/4},
  \qquad
  t_{\lambda}={\sqrt{3}\,M_\mathrm{p}\over 4\sqrt{\pi\lambda}\,(1+\cu/\rho)}.
  \label{ex2}
\end{equation}
For $t\gg t_\lambda$ we recover from Equations~(\ref{ex1})
and~(\ref{ex2}) the standard behaviour, $a\propto t^{2/3(w+1)}$,
whereas for $t\ll t_\lambda$, we have the very different behaviour
of the high-energy regime,
\begin{equation}
  \rho\gg\lambda
  \quad \Rightarrow \quad
  a \propto t^{1/3(w+1)}.
\end{equation}

When $w=-1$ we have $\rho=\rho_0$ from the conservation equation.
If $K=0=\Lambda$, we recover the de Sitter solution for $\cu=0$
and an asymptotically de Sitter solution for $\cu>0$:
\begin{eqnarray}
  a & = & a_0\exp[H_0(t-t_0)],
  \qquad
  H_0=\kappa\sqrt{{\rho_0\over 3} \left(1+ {\rho_0\over 2\lambda}\right)},
  \qquad \mbox{for } \cu=0,
  \\
  a^2 & = & \sqrt{m \over H_0} \sinh[2H_0(t-t_0)]
  \qquad \qquad \qquad \qquad \qquad \qquad \:\: \mbox{for } \cu>0.
\end{eqnarray}%

A qualitative analysis of the Friedmann equations is given
in~\cite{cs_1, cs_2}.


\subsection{Brane-world inflation}

In 1-brane RS-type brane-worlds, where the bulk has only a vacuum
energy, inflation on the brane must be driven by a 4D scalar field
trapped on the brane. In more general brane-worlds, where the bulk
contains a 5D scalar field, it is possible that the 5D field
induces inflation on the brane via its effective
projection~\cite{hs_1, hs_2, hs_3, hs_4, hs_5, hs_6, hs_7, hs_8, hs_9,
  hs_10, hs_11, hs_12, hs_13, hs_14, hs_15, hs_16, hs_17}.

More exotic possibilities arise from the interaction between two
branes, including possible collision, which is mediated by a 5D
scalar field and which can induce either inflation~\cite{kss_1, kss_2} or a
hot big-bang radiation era, as in the ``ekpyrotic'' or cyclic
scenario~\cite{ek_1, ek_2, ek_3, ek_4, ek_5, ek_6, ek_7}, or in
colliding bubble scenarios~\cite{bub_1, bub_2, bub_3}. (See
also~\cite{ek5_1, ek5_2, ek5_3} for colliding branes in an M~theory
approach.) Here we discuss the simplest case of a 4D scalar field
$\phi$ with potential $V(\phi)$ (see~\cite{lidrev} for a review).

High-energy brane-world modifications to the dynamics of inflation
on the brane have been investigated~\cite{mwbh, morers2_2, inf_2, inf_3,
  inf_4, inf_5, inf_6, inf_7, inf_8, inf_9, inf_10, inf_11, inf_12,
  inf_13, inf_14, mss}. Essentially, the high-energy corrections
provide increased Hubble damping, since $\rho\gg\lambda$ implies that
$H$ is larger for a given energy than in 4D general relativity. This
makes slow-roll inflation possible even for potentials that would be
too steep in standard cosmology~\cite{mwbh, steep_1, steep_2, steep_3,
  steep_4, steep_5, hulid1}.

The field satisfies the Klein--Gordon equation
\begin{equation}
  \ddot{\phi}+3H\dot{\phi}+V'(\phi)=0.
  \label{5}
\end{equation}
In 4D general relativity, the condition for inflation,
$\ddot{a}>0$, is $\dot{\phi}^2<V(\phi)$, i.e., $p<-{1\over3}\rho$,
where $\rho={1\over2}\dot{\phi}^2+V$ and
$p={1\over2}\dot{\phi}^2-V$. The modified Friedmann equation leads
to a stronger condition for inflation: Using Equation~(\ref{mf}), with
$m=0=\Lambda=K$, and Equation~(\ref{5}), we find that
\begin{equation}
  \ddot{a}>0
  \quad \Rightarrow \quad
  w<-{1\over3}\left[{1+2\rho/\lambda \over 1+ \rho/\lambda}\right],
  \label{6}
\end{equation}
where the square brackets enclose the brane correction to the
general relativity result. As $\rho/\lambda\rightarrow 0$, the 4D
result $w<-{1\over3}$ is recovered, but for $\rho>\lambda$, $w$
must be more negative for inflation. In the very high-energy limit
$\rho/\lambda\rightarrow\infty$, we have $w<-{2\over3}$. When the
only matter in the universe is a self-interacting scalar field,
the condition for inflation becomes
\begin{equation}
  \dot\phi^2 - V + \left[{{1\over2}\dot\phi^2 + V \over \lambda}
  \left({5\over4}\dot\phi^2-{1\over2}V\right)\right] < 0,
  \label{endinf}
\end{equation}
which reduces to $\dot{\phi}^2<V$ when $\rho_\phi =
{1\over2}\dot\phi^2+V \ll\lambda$.

In the slow-roll approximation, we get
\begin{eqnarray}
  H^2 &\approx& {\kappa^2\over3} V\left[ 1+{V\over2\lambda} \right],
  \label{7} \\
  \dot\phi &\approx& -{V'\over 3H}.
  \label{8}
\end{eqnarray}%
The brane-world correction term $V/\lambda$ in Equation~(\ref{7})
serves to enhance the Hubble rate for a given potential energy,
relative to general relativity. Thus there is enhanced Hubble
`friction' in Equation~(\ref{8}), and brane-world effects will
reinforce slow-roll at the same potential energy. We can see this
by defining slow-roll parameters that reduce to the standard
parameters in the low-energy limit:
\begin{eqnarray}
  \epsilon &\equiv&
  -{\dot H \over H^2}={M_\mathrm{p}^2 \over 16\pi}
  \left( {V' \over V} \right)^2
  \left[ {1+V/\lambda \over(1+V/2\lambda)^2} \right],
  \label{epsilon}
  \\
  \eta &\equiv &
  - {\ddot\phi \over H \dot\phi}={M_\mathrm{p}^2 \over 8\pi}
  \left( {V'' \over V} \right) \left[ {1 \over 1+V/2\lambda} \right].
  \label{eta}
\end{eqnarray}%
Self-consistency of the slow-roll approximation then requires
$\epsilon,|\eta|\ll 1$. At low energies, $V\ll\lambda$, the
slow-roll parameters reduce to the standard form. However at high
energies, $V\gg\lambda$, the extra contribution to the Hubble
expansion helps damp the rolling of the scalar field, and the new
factors in square brackets become $\approx\lambda/V$:
\begin{equation}
  \epsilon\approx\epsilon_\mathrm{gr}\left[{ {4\lambda\over V}}\right],
  \qquad
  \eta\approx\eta_\mathrm{gr}\left[{ {2\lambda\over V}}\right],
\end{equation}
where $\epsilon_\mathrm{gr},\eta_\mathrm{gr}$ are the standard general
relativity slow-roll parameters. In particular, this means that
steep potentials which do not give inflation in general
relativity, can inflate the brane-world at high energy and then
naturally stop inflating when $V$ drops below $\lambda$. These
models can be constrained because they typically end inflation in
a kinetic-dominated regime and thus generate a blue spectrum of
gravitational waves, which can disturb
nucleosynthesis~\cite{steep_1, steep_2, steep_3, steep_4, steep_5}. They also allow for the novel
possibility that the inflaton could act as dark matter or
quintessence at low energies~\cite{steep_1, steep_2, steep_3, steep_4, steep_5, dq_1, dq_2, dq_3, dq_4, dq_5}.

The number of e-folds during inflation, $N = \int Hdt$, is, in the
slow-roll approximation,
\begin{equation}
  N \approx - {8\pi \over M_\mathrm{p}^2}
  \int_{\phi_\mathrm{i}}^{\phi_\mathrm{f}}{V\over V'}
  \left[ 1+{V \over 2\lambda} \right] d\phi.
  \label{efold}
\end{equation}
Brane-world effects at high energies increase the Hubble rate by a
factor $V/2\lambda$, yielding more inflation between any two
values of $\phi$ for a given potential. Thus we can obtain a given
number of e-folds for a smaller initial inflaton value
$\phi_\mathrm{i}$. For $V\gg\lambda$, Equation~(\ref{efold}) becomes
\begin{equation}
  N \approx - {128\pi^3\over 3 M_5^6}
  \int_{\phi_\mathrm{i}}^{\phi_\mathrm{f}}{V^2\over V'} d\phi.
\end{equation}

The key test of any modified gravity theory during inflation will
be the spectrum of perturbations produced due to quantum
fluctuations of the fields about their homogeneous background
values. We will discuss brane-world cosmological perturbations in
the next Section~\ref{section_6}. In general, perturbations on the
brane are coupled to bulk metric perturbations, and the problem is very
complicated. However, on large scales on the brane, the density
perturbations decouple from the bulk metric
perturbations~\cite{m1, lmsw, gm, fk}.
For 1-brane RS-type models, there is no scalar zero-mode of the
bulk graviton, and in the extreme
slow-roll (de Sitter) limit, the massive scalar modes are heavy
and stay in their vacuum state during inflation~\cite{fk}. Thus it
seems a reasonable approximation in slow-roll to neglect the KK
effects carried by ${\cal E}_{\mu\nu}$ when
computing the density perturbations.

\epubtkImage{figure05.png}{%
 \begin{figure}[htbp]
   \def\epsfsize#1#2{0.5#1}
   \centerline{\epsfbox{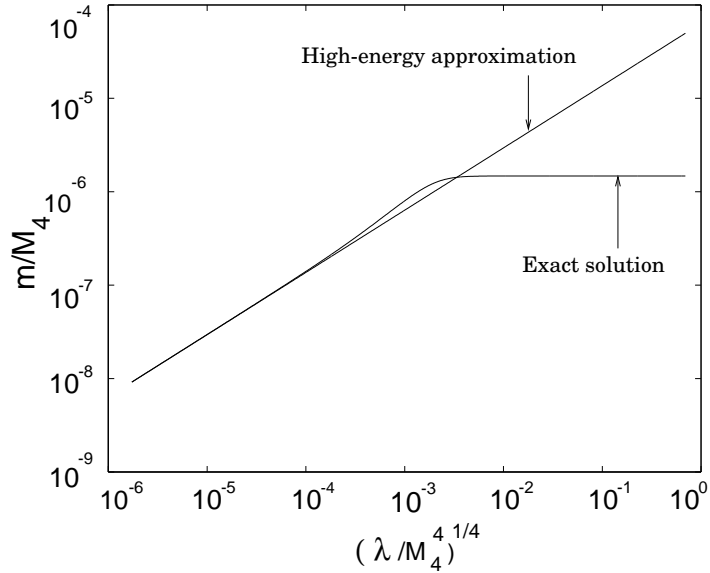}}
   \caption{The relation between the inflaton mass $m/M_4$
     ($M_4\equiv M_\mathrm{p}$) and the brane tension
     $(\lambda/M_4^4)^{1/4}$ necessary to satisfy the COBE
     constraints. The straight line is the approximation used in
     Equation~(\ref{phi55}), which at high energies is in excellent
     agreement with the exact solution, evaluated numerically in
     slow-roll. (Figure taken from~\cite{mwbh}.)}
   \label{figure_05}
 \end{figure}}

To quantify the amplitude of scalar (density) perturbations we
evaluate the usual gauge-invariant quantity
\begin{equation}
  \zeta \equiv {\cal R}-{H\over\dot\rho}\delta\rho,
  \label{defzeta}
\end{equation}
which reduces to the curvature perturbation ${\cal R}$ on
uniform density hypersurfaces ($\delta\rho=0$). This is conserved
on large scales for purely adiabatic perturbations as a
consequence of energy conservation (independently of the field
equations)~\cite{wmll}. The curvature perturbation on uniform
density hypersurfaces is given in terms of the scalar field
fluctuations on spatially flat hypersurfaces $\delta\phi$ by
\begin{equation}
  \zeta = H {\delta\phi\over\dot\phi}.
  \label{9}
\end{equation}
%
\new{
If one makes the assumption that backreaction due to metric perturbations
in the bulk can be neglected,
}
the field fluctuations at Hubble crossing ($k=aH$) in the
slow-roll limit are given by
$\langle\delta\phi^2\rangle\approx\left({H/2\pi} \right)^2$, a
result for a massless field in de Sitter space that is also
independent of the gravity theory~\cite{wmll}. For a single scalar
field the perturbations are adiabatic and hence the curvature
perturbation $\zeta$ can be related to the density perturbations
when modes re-enter the Hubble scale during the matter dominated
era which is given by $A_\mathrm{s}^2 = 4\langle \zeta^2 \rangle/25$.
Using the slow-roll equations and Equation~(\ref{9}), this gives
\begin{equation}
  A_\mathrm{s}^2 \approx \left. \left({512\pi\over75 M_\mathrm{p}^6}\,
  {V^3 \over V^{\prime2}}\right)\left[ {2\lambda + V \over 2\lambda}
  \right]^3 \right|_{k=aH}\!\!\!\!\!\!\!\!\!\!\!\!\!.
  \label{AS}
\end{equation}
Thus the amplitude of scalar perturbations is \emph{increased}
relative to the standard result at a fixed value of $\phi$ for a
given potential.

\epubtkUpdateA{Added two paragraphs discussing the neglection of
  backreaction due to metric perturbations in the bulk.}
\new{
A crucial assumption is
that backreaction due to metric perturbations in the bulk can be
neglected.  In the extreme slow-roll limit this is necessarily
correct because the coupling between inflaton fluctuations and
metric perturbations vanishes; however, this is not necessarily the
case when slow-roll corrections are included in the calculation.
Previous work~\cite{Koyama:2004ap, Koyama:2005gh, Koyama:2005ek}
has shown that such bulk effects
can be subtle and interesting (see also~\cite{deRham:2004yt, deRham:2007db}
for other approaches). In particular, subhorizon inflaton fluctuations on a
brane excite an infinite ladder of Kaluza--Klein modes of the bulk
metric perturbations at first order in slow-roll parameters, and a
naive slow-roll expansion breaks down in the high-energy regime once
one takes into account the backreaction of the bulk metric
perturbations, as confirmed by direct numerical
simulations~\cite{Hiramatsu:2006cv}. However, an order-one correction to the
behaviour of inflaton fluctuations on subhorizon scales does not
necessarily imply that the amplitude of the inflaton perturbations
receives corrections of order one on large scales; one must
consistently quantise the coupled brane inflaton fluctuations and
bulk metric perturbations.
This requires a detailed analysis of the coupled brane-bulk
system~\cite{Cardoso:2006nh, Koyama:2007as}.}

\new{
It was shown that the
coupling to bulk metric perturbations cannot be ignored in the
equations of motion.  Indeed, there are
order-unity differences between the classical solutions without
coupling and with slow-roll induced coupling.  However, the change
in the amplitude of quantum-generated perturbations is at
next-to-leading order~\cite{Koyama:2007as} because there is still no
mixing at leading order between positive and negative frequencies
when scales observable today crossed the horizon, so the Bogoliubov
coefficients receive no corrections at leading order.  The amplitude
of perturbations generated is also subject to the usual slow-roll
corrections on super-horizon scales. The next-order slow-roll
corrections from bulk gravitational perturbations are calculated
in~\cite{Koyama:2007su} and they are the same order as the usual
Stewart--Lyth correction~\cite{Stewart:1993bc}. These results also
show that the ratio of tensor-to-scalar perturbation amplitudes are
not influenced by brane-bulk interactions at leading order in
slow-roll. It is remarkable that the predictions from inflation
theories should be so robust that this result holds in spite of the
leading-order change to the solutions of the classical equations of
motion.
}

The scale-dependence of the perturbations is described by the
spectral tilt
\begin{equation}
  n_\mathrm{s}-1\equiv {d\ln A_\mathrm{s}^2 \over d\ln k} \approx
  -4\epsilon + 2\eta,
  \label{15}
\end{equation}
where the slow-roll parameters are given in Equations~(\ref{epsilon})
and~(\ref{eta}). Because these slow-roll parameters are both
suppressed by an extra factor $\lambda/V$ at high energies, we see
that the spectral index is driven towards the Harrison--Zel'dovich
spectrum, $n_\mathrm{s}\to1$, as $V/\lambda\to\infty$; however, as
explained below, this does not necessarily mean that the
brane-world case is closer to scale-invariance than the general
relativity case.

As an example, consider the simplest chaotic inflation model
$V={1\over2}m^2\phi^2$. Equation~(\ref{efold}) gives the
integrated expansion from $\phi_\mathrm{i}$ to $\phi_\mathrm{f}$ as
\begin{equation}
  N\approx {2\pi\over M_\mathrm{p}^2}
  \left(\phi_\mathrm{i}^2-\phi_\mathrm{f}^2\right)+
  {\pi^2m^2\over3 M_5^6}\left(\phi_\mathrm{i}^4-\phi_\mathrm{f}^4\right).
  \label{20}
\end{equation}
The new high-energy term on the right leads to more inflation for
a given initial inflaton value $\phi_\mathrm{i}$.

The standard chaotic inflation scenario requires an inflaton mass
$m\sim 10^{13} \mathrm{\ GeV}$ to match the observed level of anisotropies in
the cosmic microwave background (see below). This corresponds to
an energy scale $\sim 10^{16} \mathrm{\ GeV}$ when the relevant scales left
the Hubble scale during inflation, and also to an inflaton field
value of order $3M_\mathrm{p}$. Chaotic inflation has been criticised
for requiring super-Planckian field values, since these can lead
to nonlinear quantum corrections in the potential.

If the brane tension $\lambda$ is much below $10^{16} \mathrm{\ GeV}$,
corresponding to $M_5<10^{17} \mathrm{\ GeV}$, then the terms quadratic in
the energy density dominate the modified Friedmann equation. In
particular the condition for the end of inflation given in
Equation~(\ref{endinf}) becomes $\dot\phi^2<{2\over5}V$. In the
slow-roll approximation (using Equations~(\ref{7}) and~(\ref{8}))
$\dot\phi\approx-M_5^3/2\pi\phi$, and this yields
\begin{equation}
  \phi_\mathrm{end}^4 \approx
  {5 \over 4\pi^2}\left({M_5\over m}\right)^2M_5^4.
\end{equation}
In order to estimate the value of $\phi$ when scales corresponding
to large-angle anisotropies on the microwave background sky left
the Hubble scale during inflation, we take $N_\mathrm{cobe}\approx55$
in Equation~(\ref{20}) and $\phi_\mathrm{f}=\phi_\mathrm{end}$. The second
term on the right of Equation~(\ref{20}) dominates, and we obtain
\begin{equation}
  \phi_\mathrm{cobe}^4 \approx {165\over\pi^2}
  \left({M_5 \over m}\right)^2M_5^4.
  \label{phi55}
\end{equation}
Imposing the COBE normalization on the curvature perturbations
given by Equation~(\ref{AS}) requires
\begin{equation}
  A_\mathrm{s}\approx
  \left({8\pi^2\over45}\right){m^4\phi_\mathrm{cobe}^5\over M_5^6}
  \approx 2\times10^{-5}.
  \label{22}
\end{equation}
Substituting in the value of $\phi_\mathrm{cobe}$ given by
Equation~(\ref{phi55}) shows that in the limit of strong brane
corrections, observations require
\begin{equation}
  m \approx 5\times 10^{-5}\, M_5,
  \qquad
  \phi_\mathrm{cobe}\approx 3\times 10^2\,M_5.
  \label{23}
\end{equation}
Thus for $M_5<10^{17} \mathrm{\ GeV}$, chaotic inflation can occur for
field values below the 4D Planck scale, $\phi_\mathrm{cobe}<M_\mathrm{p}$, although still above the 5D scale $M_5$. The relation
determined by COBE constraints for arbitrary brane tension is
shown in Figure~\ref{figure_05}, together with the high-energy approximation used
above, which provides an excellent fit at low brane tension
relative to $M_4$.

It must be emphasized that in comparing the high-energy
brane-world case to the standard 4D case, we implicitly require
the same potential energy. However, precisely because of the
high-energy effects, large-scale perturbations will be generated
at different values of $V$ than in the standard case, specifically
at lower values of $V$, closer to the reheating minimum. Thus
there are two competing effects, and it turns out that the shape
of the potential determines which is the dominant
effect~\cite{lidsmi}. For the quadratic potential, the lower
location on $V$ dominates, and the spectral tilt is slightly
further from scale invariance than in the standard case. The same
holds for the quartic potential. Data from WMAP and 2dF can be
used to constrain inflationary models via their deviation from
scale invariance, and the high-energy brane-world versions of the
quadratic and quartic potentials are thus under more pressure from
data than their standard counterparts~\cite{lidsmi}, as shown in
Figure~\ref{figure_06}.

\epubtkImage{figure06.png}{%
 \begin{figure}[htbp]
   \def\epsfsize#1#2{0.8#1}
   \centerline{\epsfbox{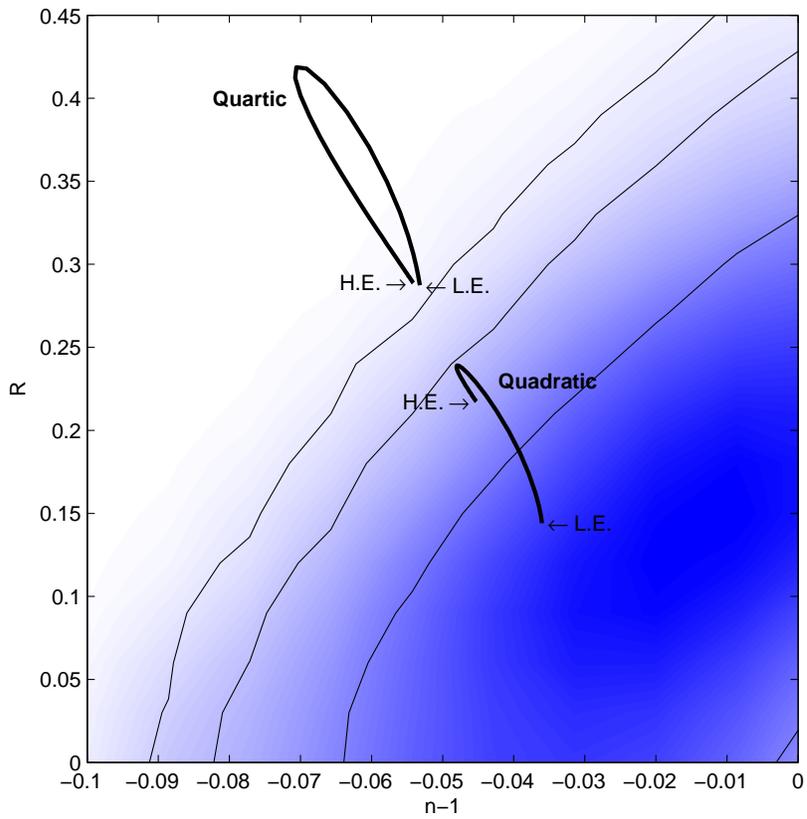}}
   \caption{Constraints from WMAP data on inflation models with
     quadratic and quartic potentials, where \textit{R} is the ratio of
     tensor to scalar amplitudes and \textit{n} is the scalar spectral
     index. The high energy (H.E.) and low energy (L.E.) limits are
     shown, with intermediate energies in between, and the 1-$\sigma$
     and 2-$\sigma$ contours are also shown. (Figure taken
     from~\cite{lidsmi}.)}
   \label{figure_06}
 \end{figure}}

Other perturbation modes have also been investigated:
\begin{itemize}
\item High-energy inflation on the brane also generates a zero-mode
  (4D graviton mode) of tensor perturbations, and stretches it to
  super-Hubble scales, as will be discussed below. This zero-mode has
  the same qualitative features as in general relativity, remaining
  frozen at constant amplitude while beyond the Hubble horizon. Its
  amplitude is enhanced at high energies, although the enhancement is
  much less than for scalar perturbations~\cite{lmw}:
  \begin{eqnarray}
    A_\mathrm{t}^2 &\approx&
    \left({32V\over 75 M_\mathrm{p}^2}\right)
    \left[ {3V^2 \over 4\lambda^2}\right],
    \label{higw} \\
    {A_\mathrm{t}^2\over A_\mathrm{s}^2} &\approx&
    \left({M_\mathrm{p}^2\over 16\pi} {V'^2\over V^2}\right)
    \left[ {6\lambda\over V} \right].
    \label{ten}
  \end{eqnarray}%
  Equation~(\ref{ten}) means that brane-world effects suppress the
  large-scale tensor contribution to CMB anisotropies. The tensor
  spectral index at high energy has a smaller magnitude than in
  general relativity,
  \begin{equation}
    n_\mathrm{t}=-3\epsilon,
  \end{equation}
  but remarkably the same consistency relation as in general
  relativity holds~\cite{hulid1}:
  \begin{equation}
    n_\mathrm{t} = -2{A_\mathrm{t}^2\over A_\mathrm{s}^2}.
  \end{equation}
  This consistency relation persists when $Z_2$ symmetry is
  dropped~\cite{hulid2} (and in a two-brane model with stabilized
  radion~\cite{gklr}). It holds only to lowest order in slow-roll,
  as in general relativity, but the reason for this~\cite{seetay}
  and the nature of the corrections~\cite{cal} are not settled.

  The massive KK modes of tensor perturbations remain in the vacuum
  state during slow-roll inflation~\cite{lmw, grs}. The evolution of
  the super-Hubble zero mode is the same as in general relativity,
  so that high-energy brane-world effects in the early universe
  serve only to rescale the amplitude. However, when the zero mode
  re-enters the Hubble horizon, massive KK modes can be excited.
\item Vector perturbations in the bulk metric can support vector
  metric perturbations on the brane, even in the absence of matter
  perturbations (see the next Section~\ref{section_6}). However, there
  is no normalizable zero mode, and the massive KK modes stay in the
  vacuum state during brane-world inflation~\cite{bmwv}. Therefore,
  as in general relativity, we can neglect vector perturbations in
  inflationary cosmology.
\end{itemize}

Brane-world effects on large-scale isocurvature perturbations in
2-field inflation have also been considered~\cite{abd}.
Brane-world (p)reheating after inflation is discussed
in~\cite{preheat_1, preheat_2, preheat_3, preheat_4, preheat_5}.


\subsection{Brane-world instanton}

The creation of an inflating brane-world can be modelled as a de
Sitter instanton in a way that closely follows the 4D instanton,
as shown in~\cite{gs}. The instanton consists of two identical
patches of $\mathrm{AdS}_5$ joined together along a de Sitter brane
($\mathrm{dS}_4$) with
compact spatial sections. The instanton describes the ``birth'' of
both the inflating brane and the bulk spacetime, which are
together ``created from nothing'', i.e., the point at the south pole
of the de Sitter 4-sphere. The Euclidean $\mathrm{AdS}_5$ metric is
\begin{equation}
  {}^{(5)\!}ds^2_\mathrm{euclid} = dr^2+\ell^2\sinh^2(r/\ell)
  \left[ d\chi^2+\sin^2 \chi \, d\Omega^2_{(3)} \right],
\end{equation}
where $d\Omega^2_{(3)}$ is a 3-sphere, and $r\leq r_0$. The
Euclidean instanton interpolates between $r=0$ (``nothing'') and
$r=r_0$ (the created universe), which is a spherical brane of
radius
\begin{equation}
H_0^{-1} \equiv \ell \sinh (r_0/\ell).
\end{equation}
After creation, the brane-world evolves according to the
Lorentzian continuation, $\chi \to iH_0t+\pi/2$,
\begin{equation}
  {}^{(5)\!}ds^2 = dr^2+(\ell H_0)^2\sinh^2(r/\ell)
  \left[ -dt^2+H_0^{-2}\cosh^2(H_0t)\, d\Omega^2_{(3)} \right]
\end{equation}
(see Figure~\ref{figure_07}).

\epubtkImage{figure07.png}{%
 \begin{figure}[htbp]
   \def\epsfsize#1#2{0.6#1}
   \centerline{\epsfbox{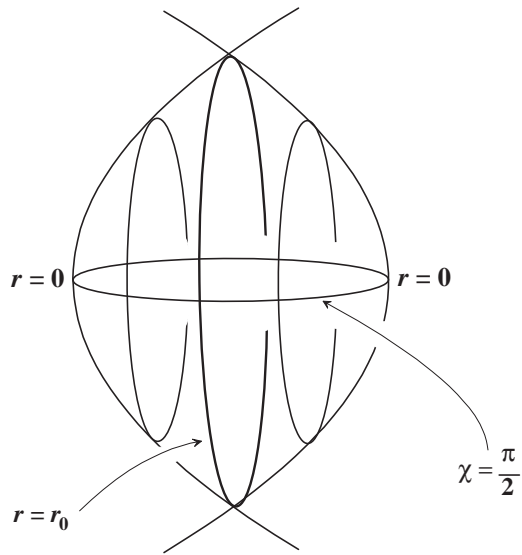}}
   \caption{Brane-world instanton. (Figure taken from~\cite{gs}.)}
   \label{figure_07}
 \end{figure}}


\subsection{Models with non-empty bulk}

The single-brane cosmological model can be generalized to include
stresses other than $\Lambda_5$ in the bulk:
\begin{itemize}
\item The simplest example arises from considering a charged bulk
  black hole, leading to the Reissner--Nordstr\"om $\mathrm{AdS}_5$
  bulk metric~\cite{bv}. This has the form of Equation~(\ref{sads}),
  with
  \begin{equation}
    F(R) = K+ {R^2\over \ell^2} -{m \over R^2}+{q^2\over R^4},
    \label{rnads}
  \end{equation}
  where $q$ is the ``electric'' charge parameter of the bulk black
  hole. The metric is a solution of the 5D Einstein--Maxwell
  equations, so that ${}^{(5)}T_{AB}$ in Equation~(\ref{5efe}) is the
  energy-momentum tensor of a radial static 5D ``electric'' field. In
  order for the field lines to terminate on the boundary brane, the
  brane should carry a charge $-q$. Since the RN$\mathrm{AdS}_5$
  metric is 4-isotropic, it is still possible to embed a FRW brane in
  it, which is moving in the coordinates of Equation~(\ref{sads}).
  The effect of the black hole charge on the brane arises via the
  junction conditions and leads to the modified Friedmann
  equation~\cite{bv},
  \begin{equation}
    H^2 = \frac{\kappa^2}{3} \rho\left(1+{\rho\over 2\lambda}\right)+
    {m\over a^4}-{q^2\over a^6}+ \frac{1}{3} \Lambda - \frac{K}{a^2}.
    \label{mf2}
  \end{equation}
  The field lines that terminate on the brane imprint on the brane
  an effective negative energy density $-3q^2/(\kappa^2a^6)$, which
  redshifts like stiff matter ($w=1$). The negativity of this term
  introduces the possibility that at high energies it can bring the
  expansion rate to zero and cause a turn-around or bounce (but
  see~\cite{hovmy} for problems with such bounces).

  Apart from negativity, the key difference between this ``dark
  stiff matter'' and the dark radiation term $m/a^4$ is that the
  latter arises from the bulk Weyl curvature via the ${\cal
    E}_{\mu\nu}$ tensor, while the former arises from non-vacuum
  stresses in the bulk via the ${\cal F}_{\mu\nu}$ tensor in
  Equation~(\ref{e:einstein1}). The dark stiff matter does not arise
  from massive KK modes of the graviton.
\item Another example is provided by the Vaidya--$\mathrm{AdS}_5$
  metric, which can be written after transforming to a new coordinate
  $v=T+\int dR/F$ in Equation~(\ref{sads}), so that
  $v=\mathrm{const.}$ are null surfaces, and
  \begin{eqnarray}
    {}^{(5)\!}ds^2 &=& -F(R,v)dv^2+2dv\,dR+
    R^2\left({dr^2 \over 1 - Kr^2}+r^2d\Omega^2\right),
    \\
    F(R,v) &=& K+ {R^2\over \ell^2}-{m(v) \over R^2}.
  \end{eqnarray}%
  This model has a moving FRW brane in a 4-isotropic bulk (which is
  not static), with either a radiating bulk black hole ($dm/dv<0$),
  or a radiating brane ($dm/dv>0$)~\cite{ckn_1, ckn_2, ckn_3, ckn_4}. The metric satisfies
  the 5D field equations~(\ref{5efe}) with a null-radiation
  energy-momentum tensor,
  \begin{equation}
    {}^{(5)\!}T_{AB} = \psi k_Ak_B,
    \qquad
    k_Ak^A=0,
    \qquad
    k_Au^A=1,
    \label{vem}
  \end{equation}
  where $\psi\propto dm/dv$. It follows that
  \begin{equation}
    {\cal F}_{\mu\nu}= \kappa_5^{-2} \psi h_{\mu\nu}.
    \label{vf}
  \end{equation}
  In this case, the same effect, i.e., a varying mass parameter $m$,
  contributes to both ${\cal E}_{\mu\nu}$ and ${\cal F}_{\mu\nu}$ in
  the brane field equations. The modified Friedmann equation has the
  standard 1-brane RS-type form, but with a dark radiation term that
  no longer behaves strictly like radiation:
  \begin{equation}
    H^2 = \frac{\kappa^2}{3} \rho\left(1+{\rho\over 2\lambda}\right)+
    {m(t)\over a^4}+ \frac{1}{3} \Lambda - \frac{K}{a^2}.
    \label{mf3}
  \end{equation}

  By Equations~(\ref{cong}) and~(\ref{vem}), we arrive at the matter
  conservation equations,
  \begin{equation}
    \nabla^\nu T_{\mu\nu}=-2\psi u_\mu.
  \end{equation}
  This shows how the brane loses ($\psi>0$) or gains ($\psi<0$)
  energy in exchange with the bulk black hole. For an FRW brane,
  this equation reduces to
  \begin{equation}
    \dot\rho+3H(\rho+p)=-2\psi.
  \end{equation}
  The evolution of $m$ is governed by the 4D contracted Bianchi
  identity, using Equation~(\ref{vf}):
  \begin{equation}
    \nabla^\mu{\cal E}_{\mu\nu}={6\kappa^2\over\lambda}
    \nabla^\mu{\cal S}_{\mu\nu}+{2\over3}
    \left[ \kappa_5^2\left(\dot\psi+\Theta\psi\right)-
    3\kappa^2 \psi\right]u_\mu +{2\over3} \kappa_5^2\D_\mu \psi.
    \label{nlcv}
  \end{equation}
  For an FRW brane, this yields
  \begin{equation}
    \dot{\rho}_{\cal E}+4H\cu=2\psi-{2\over3}
    {\kappa_5^2\over\kappa^2}\left(\dot\psi+3H\psi\right),
  \end{equation}
  where $\cu=3m(t)/(\kappa^2a^4)$.
\item A more complicated bulk metric arises when there is a
  self-interacting scalar field $\Phi$ in the bulk~\cite{mw, sca_1,
    sca_2, sca_3, sca_4, sca_5, sca_6}. In the simplest case, when
  there is no coupling between the bulk field and brane matter, this
  gives
  \begin{equation}
    {}^{(5)}T_{AB}=\Phi_{,A}\Phi_{,B}-{}^{(5)\!}g_{AB}
    \left[ V(\Phi)+{1\over 2} {}^{(5)\!}g^{CD}\Phi_{,C}\Phi_{,D} \right],
    \label{bsf}
  \end{equation}
  where $\Phi(x,y)$ satisfies the 5D Klein--Gordon equation,
  \begin{equation}
    {}^{(5)}\Box\Phi-V'(\Phi)=0.
  \end{equation}
  The junction conditions on the field imply that
  \begin{equation}
    \partial_y\Phi(x,0)=0.
  \end{equation}
  Then Equations~(\ref{cong}) and~(\ref{bsf}) show that matter
  conservation continues to hold on the brane in this simple case:
  \begin{equation}
    \nabla^\nu T_{\mu\nu}=0.
  \end{equation}
  From Equation~(\ref{bsf}) one finds that
  \begin{equation}
    {\cal F}_{\mu\nu}={1\over 4\kappa_5^2}
    \left[4\phi_{,\mu}\phi_{,\nu}-
    g_{\mu\nu}\left(3V(\phi)+ {5\over 2}
    g^{\alpha\beta}\phi_{,\alpha}\phi_{,\beta}\right) \right],
  \end{equation}
  where
  \begin{equation}
    \phi(x)=\Phi(x,0),
  \end{equation}
  so that the modified Friedmann equation becomes
  \begin{equation}
    H^2 = \frac{\kappa^2}{3} \rho\left(1+{\rho\over 2\lambda}\right)+
    {m\over a^4}+ {\kappa_5^2\over6}\left[{1\over2}\dot\phi^2+
    V(\phi) \right]+ \frac{1}{3} \Lambda - \frac{K}{a^2}.
    \label{mf4}
  \end{equation}

  When there is coupling between brane matter and the bulk scalar
  field, then the Friedmann and conservation equations are more
  complicated~\cite{mw, sca_1, sca_2, sca_3, sca_4, sca_5, sca_6}.
\end{itemize}

\newpage


\section{Brane-World Cosmology: Perturbations}
\label{section_6}

The background dynamics of brane-world cosmology are simple
because the FRW symmetries simplify the bulk and rule out nonlocal
effects. But perturbations on the brane immediately release the
nonlocal KK modes. Then the 5D bulk perturbation equations must be
solved in order to solve for perturbations on the brane. These 5D
equations are partial differential equations for the 3D Fourier
modes, with both initial and boundary conditions needed.

The theory of gauge-invariant perturbations in brane-world
cosmology has been extensively investigated and
developed~\cite{m1, lmsw, bm, mwbh, steep_1, steep_2, steep_3,
  steep_4, steep_5, gm, lmw, grs, mu_1, mu_2, pert_1, pert_2, pert_3,
  pert_4, pert_5, pert_6, pert_7, hs_1, pert_9, pert_10, pert_11,
  pert_12, pert_13, pert_14, pert_15, pert_16, pert_17, hs_4,
  inf_7, pert_20, pert_21, pert_22, pert_23, pert_24, pert_25,
  bmw_1, bmw_2, l1, l2}
and is qualitatively well understood. The key task is
integration of the coupled brane-bulk perturbation equations.
Special cases have been solved, where these equations effectively
decouple~\cite{lmsw, bm, l1, l2}, and approximation schemes have
been developed~\cite{sod_1, sod_2, sod_3, sod_4, sod_5, koy,
  rbbd_1, rbbd_2, kkt, elmw, mbb_1, mbb_2, mbb_3} for the
more general cases where the coupled system must be solved. Below we will also present the results of full numerical integration of the 5D perturbation equations in the RS case.

From
the brane viewpoint, the bulk effects, i.e., the high-energy
corrections and the KK modes, act as source terms for the brane
perturbation equations. At the same time, perturbations of matter
on the brane can generate KK modes (i.e., emit 5D gravitons into
the bulk) which propagate in the bulk and can subsequently
interact with the brane. This nonlocal interaction amongst the
perturbations is at the core of the complexity of the problem. It
can be elegantly expressed via integro-differential
equations~\cite{mu_1, mu_2}, which take the form (assuming no incoming 5D
gravitational waves)
\begin{equation}
  A_k(t)=\int dt'\,{\cal G}(t,t') B_k(t'),
  \label{ide}
\end{equation}
where ${\cal G}$ is the bulk retarded Green's function evaluated
on the brane, and $A_k, B_k$ are made up of brane metric and
matter perturbations and their (brane) derivatives, and include
high-energy corrections to the background dynamics. Solving for
the bulk Green's function, which then determines ${\cal G}$, is
the core of the 5D problem.

We can isolate the KK anisotropic stress $\cp_{\mu\nu}$ as the
term that must be determined from 5D equations. Once
$\cp_{\mu\nu}$ is determined in this way, the perturbation
equations on the brane form a closed system. The solution will be
of the form (expressed in Fourier modes):
\begin{equation}
  \pi^{\cal E}_k(t) \propto \int d t'\,{\cal G}(t,t') F_k(t'),
  \label{e:soln}
\end{equation}
where the functional $F_k$ will be determined by the covariant
brane perturbation quantities and their derivatives. It is known
in the case of a Minkowski background~\cite{ssm}, but not in the
cosmological case.

The KK terms act as source terms modifying the standard general
relativity perturbation equations, together with the high-energy
corrections. For example, the linearization of the shear
propagation equation~(\ref{pe5}) yields
\begin{equation}
  \dot{\sigma}_{\mu\nu}+2H\sigma_{\mu\nu}+
  E_{\mu\nu}-{\kappa^2\over2} \pi_{\mu\nu}-
  \D_{\langle \mu}A_{\nu\rangle} =
  {\kappa^2\over 2}\cp_{\mu\nu}- {\kappa^2\over 4}(1+3w)
  {\rho\over\lambda} \pi_{\mu\nu}.
\end{equation}
In 4D general relativity, the right hand side is zero. In the
brane-world, the first source term on the right is the KK term,
and the second term is the high-energy modification. The other
modification is a straightforward high-energy correction of the
background quantities $H$ and $\rho$ via the modified Friedmann
equations.

As in 4D general relativity, there are various different, but
essentially equivalent, ways to formulate linear cosmological
perturbation theory. First we describe the covariant brane-based
approach.


\subsection{1\,+\,3-covariant perturbation equations on the brane}

In the $1+3$-covariant approach~\cite{m1, l1, mjap}, perturbative
quantities are projected vectors, $V_\mu =V_{\langle \mu\rangle}$,
and projected symmetric tracefree tensors, $W_{\mu\nu}=W_{\langle
\mu \nu\rangle}$, which are gauge-invariant since they vanish in
the background. These are decomposed into (3D) scalar, vector, and
tensor modes as
\begin{eqnarray}
  V_\mu  &=& \D_\mu  V+\bar{V}_\mu, \\
  W_{\mu\nu} &=& \D_{\langle\mu}\D_{\nu\rangle}{W}+
  \D_{\langle\mu}\bar{W}_{\nu\rangle}+\bar{W}_{\mu\nu},
\end{eqnarray}%
where $\bar{W}_{\mu\nu}=\bar{W}_{\langle \mu\nu\rangle}$ and an
overbar denotes a (3D) transverse quantity,
\begin{equation}
  \D^\mu\bar{V}_\mu =0= \D^\nu \bar{W}_{\mu\nu}.
\end{equation}
In a local inertial frame comoving with $u^\mu$, i.e.,
$u^\mu=(1,\vec 0)$, all time components may be set to zero:
$V_\mu=(0, V_i)$, $W_{0\mu}=0$,
$\vec\nabla_\mu=(0,\vec\nabla_i)$.

Purely scalar perturbations are characterized by the fact that
vectors and tensors are derived from scalar potentials, i.e.,
\begin{equation}
  \bar{V}_\mu =\bar{W}_\mu =\bar{W}_{\mu\nu}=0.
\end{equation}
Scalar perturbative quantities are formed from the potentials via
the (3D) Laplacian, e.g., ${\cal V}=\D^\mu\D_\mu  V\equiv \D^2 V$.
Purely vector perturbations are characterized by
\begin{equation}
  V_\mu =\bar{V}_\mu,
  \qquad
  W_{\mu\nu}=\D_{\langle\mu}\bar{W}_{\nu\rangle},
  \qquad
  \curl\D_\mu  f=-2\dot{f}\omega_\mu,
  \label{vec}
\end{equation}
where $\omega_\mu $ is the vorticity, and purely tensor by
\begin{equation}
  \D_\mu f=0=V_\mu,
  \qquad
  W_{\mu\nu}=\bar{W}_{\mu\nu}.
\end{equation}

The KK energy density produces a scalar mode $\D_\mu {\cu}$ (which
is present even if $\cu=0$ in the background). The KK momentum
density carries scalar and vector modes, and the KK anisotropic
stress carries scalar, vector, and tensor modes:
\begin{eqnarray}
  {\cq_\mu }&=&\D_\mu \cq+{\bcq_\mu},
  \label{q_star} \\
  {\cp_{\mu\nu}}&=&\D_{\langle\mu}\D_{\nu\rangle}\cp +
  \D_{\langle \mu}{\bcp_{\nu\rangle}}+{\bcp_{\mu\nu}}.
  \label{p_star}
\end{eqnarray}%

Linearizing the conservation equations for a single adiabatic
fluid, and the nonlocal conservation equations, we obtain
\begin{eqnarray}
  \dot\rho+\Theta(\rho+p)&=&0,
  \label{ecl} \\
  c_\mathrm{s}^2\D_\mu +(\rho+p)A_\mu&=&0,
  \label{momcl} \\
  \dot{\rho}_{\cal E}+{4\over3}\Theta{\cu}+\D^\mu{\cq_\mu }&=&0,
  \label{nlc1} \\
  \dot{q}^{\cal E}_\mu +4H{\cq_\mu}+{1\over3}\D_\mu {\cu}+
  {4\over3}{\cu}A_\mu +\D^\nu {\cp_{\mu\nu}}&=&
  -{(\rho+p)\over\lambda}\D_\mu \rho.
  \label{nlc2}
\end{eqnarray}%

Linearizing the remaining propagation and constraint equations
leads to
\begin{eqnarray}
  \dot{\Theta}+{1\over3}\Theta^2 -\D^\mu  A_\mu
  +{{1\over2}}\kappa^2(\rho + 3p) -\Lambda &=&
  -{{\kappa^2\over2}}(2\rho+3p){\rho\over\lambda}- \kappa^2\cu,
  \label{prl} \\
  \dot{\omega}_{\mu} +2H\omega_\mu
  +{{1\over2}}\curl A_\mu &=&0,
  \label{pe4l} \\
  \dot{\sigma}_{\mu\nu} +2H\sigma_{\mu\nu} +E_{\mu\nu }-
  \D_{\langle \mu}A_{ \nu\rangle } &=&{\kappa^2\over 2}\cp_{\mu\nu},
  \label{pe5l} \\
  \dot{E}_{\mu\nu} +3H E_{\mu\nu} -\curl H_{\mu\nu} +
  {\kappa^2\over2}(\rho+p)\sigma_{\mu\nu} &=&
  -{\kappa^2\over2}(\rho+p){\rho\over\lambda} \sigma_{\mu\nu}
  \nonumber \\
  &&-{\kappa^2\over6}\left[4\cu\sigma_{\mu\nu}+
  3\dot{\pi}^{\cal E}_{\mu\nu} +3H\cp_{\mu\nu}+
  3\D_{\langle\mu}\cq_{ \nu\rangle} \right],
  \label{pe6l} \\
  \dot{H}_{\mu\nu} +3H H_{\mu\nu} +\curl E_{\mu\nu}&=&
  {\kappa^2\over 2} \curl \cp_{\mu\nu},
  \label{pe7l} \\
  \D^\mu \omega_\mu &=&0,
  \label{pcc1l} \\
  \D^\nu \sigma_{\mu\nu}-\curl\omega_\mu -
  {2\over3}\D_\mu \Theta &=& -\cq_\mu,
  \label{pcc2l} \\
  \curl\sigma_{\mu\nu}+\D_{\langle \mu}\omega_{\nu\rangle}-
  H_{\mu\nu}&=&0,
  \label{pcc3l} \\
  \D^\nu E_{\mu\nu} -{{\kappa^2\over3}}\D_\mu \rho&=&
  {\kappa^2\over3}{\rho\over \lambda} \D_\mu \rho + {\kappa^2\over 6}
  \left[2\D_\mu \cu-4H \cq_\mu -3\D^\nu \cp_{\mu\nu}\right], \qquad
  \label{pcc4l} \\
  \D^\nu H_{\mu\nu}-\kappa^2(\rho+p)\omega_\mu &=&
  \kappa^2(\rho+ p){\rho\over\lambda} \omega_\mu +
  {\kappa^2\over 6}\left[8 \cu \omega_\mu-3\curl\cq_\mu \right].
  \label{pcc5l}
\end{eqnarray}%
Equations~(\ref{ecl}), (\ref{nlc1}), and~(\ref{prl}) do not provide
gauge-invariant equations for perturbed quantities, but their
spatial gradients do.

These equations are the basis for a $1+3$-covariant analysis of
cosmological perturbations from the brane observer's viewpoint,
following the approach developed in 4D general relativity (for a
review, see~\cite{covp}). The equations contain scalar, vector, and
tensor modes, which can be separated out if desired. They are not
a closed system of equations until $\cp_{\mu\nu}$ is determined by
a 5D analysis of the bulk perturbations. An extension of the
$1+3$-covariant perturbation formalism to $1+4$ dimensions would
require a decomposition of the 5D geometric quantities along a
timelike extension $u^A$ into the bulk of the brane 4-velocity
field $u^\mu$, and this remains to be done. The $1+3$-covariant
perturbation formalism is incomplete until such a 5D extension is
performed. The metric-based approach does not have this drawback.


\subsection{Metric-based perturbations}

An alternative approach to brane-world cosmological perturbations
is an extension of the 4D metric-based gauge-invariant
theory~\cite{metp_1, metp_2}. A review of this approach is given
in~\cite{bmw_1, bmw_2}. In an arbitrary gauge, and for a flat FRW
background, the perturbed metric has the form
\begin{equation}
  \renewcommand{\arraystretch}{1.5}
  \delta \:{}^{(5)\!}g_{AB} = \left[
    \begin{array}{@{\quad}c@{\quad}c@{\quad}|@{\quad}c@{\quad}}
      -2N^2\psi &
      A^2(\partial_i{\cal B}-S_i) &
      N\alpha
      \\
      A^2(\partial_j{\cal B}-S_j) &
      A^2\left\{2{\cal R} \delta_{ij} +
      2\partial_i \partial_j{\cal C} +
      2\partial_{(i}F_{j)}+f_{ij} \right\} &
      A^2(\partial_i\beta-\chi_i) \\ [0.5 em]
      \hline
      N\alpha &
      A^2(\partial_j\beta-\chi_j) &
      2\nu
    \end{array}
  \right],
  \label{pertmetric}
\end{equation}
where the background metric functions $A,N$ are given by
Equations~(\ref{gnm1}, \ref{gnm2}). The scalars $\psi,{\cal
R},{\cal C}, \alpha, \beta, \nu$ represent scalar perturbations.
The vectors $S_i$, $F_i$, and $\chi_i$ are transverse, so that they
represent 3D vector perturbations, and the tensor $f_{ij}$ is
transverse traceless, representing 3D tensor perturbations.

In the Gaussian normal gauge, the brane coordinate-position
remains fixed under perturbation,
\begin{equation}
  {}^{(5)\!}ds^2=\left[ {g}^{(0)}_{\mu\nu}(x,y)+
  \delta g_{\mu\nu}(x,y)\right]dx^\mu dx^\nu +dy^2,
\end{equation}
where ${g}^{(0)}_{\mu\nu}$ is the background metric,
Equation~(\ref{gnm}). In this gauge, we have
\begin{equation}
  \alpha= \beta=\nu=\chi_i=0.
\end{equation}

In the 5D longitudinal gauge, one gets
\begin{equation}
  -{\cal B}+\dot{\cal C}=0= -\beta+{\cal C}'.
\end{equation}
In this gauge, and for an $\mathrm{AdS}_5$ background, the metric
perturbation quantities can all be expressed in terms of a
``master variable'' $\Omega$ which obeys a wave equation~\cite{mu_1, mu_2}.
In the case of scalar perturbations, we have for example
\begin{equation}
  {\cal R}={1\over 6A}\left(\Omega''-{1\over N^2}\,
  \ddot{\Omega}-{\Lambda_5 \over 3}\,\Omega \right),
  \label{Rmaster}
\end{equation}
with similar expressions for the other quantities. All of the
metric perturbation quantities are determined once a solution is
found for the wave equation
\begin{equation}
  \left({1\over NA^3}\,\dot\Omega\right)^{\!\!\displaystyle\cdot} +
  \left({\Lambda_5 \over 6}+{k^2\over A^2} \right)
  {N \over A^3}\,\Omega = \left({N \over A^3} \Omega'\right)'\!\!.
  \label{mastergeneral}
\end{equation}

The junction conditions~(\ref{ext}) relate the off-brane
derivatives of metric perturbations to the matter perturbations:
\begin{equation}
  \partial_y\,\delta g_{\mu\nu}=
  -\kappa^2_5\left[\delta T_{\mu\nu} +
  {1\over 3}\left(\lambda- T^{(0)}\right) \delta g_{\mu\nu}-
  {1\over3} {g}^{(0)}_{\mu\nu}\delta T\right],
  \label{metjun}
\end{equation}
where
\begin{eqnarray}
  \delta T^0{}_0 &=& -\delta\rho,
  \\
  \delta T^0{}_i &=& a^2 q_i,
  \\
  \delta T^i{}_j &=& \delta p\, \delta^i{}_j+\delta\pi^i{}_j.
\end{eqnarray}%
For scalar perturbations in the Gaussian normal gauge, this gives
\begin{eqnarray}
  \partial_y \psi(x,0) &=&
  {\kappa^2_5\over6}(2\delta\rho+3\delta p),
  \\
  \partial_y {\cal B}(x,0) &=& \kappa^2_5\delta p,
  \\
  \partial_y {\cal C}(x,0) &=& -{\kappa^2_5\over2}\delta\pi,
  \\
  \partial_y {\cal R} (x,0) &=&- {\kappa^2_5\over6}\delta\rho-
  \partial_i\partial^i\,{\cal C}(x,0),
\end{eqnarray}%
where $\delta\pi$ is the scalar potential for the matter
anisotropic stress,
\begin{equation}
  \delta\pi_{ij}=\partial_i\partial_j \delta\pi-{1\over3}\delta_{ij}\,
  \partial_k\partial^k \delta\pi.
\end{equation}
The perturbed KK energy-momentum tensor on the brane is given by
\begin{eqnarray}
  \delta {\cal E}^0{}_0 &=& \kappa^2\delta\cu,
  \\
  \delta {\cal E}^0{}_i &=&-\kappa^2 a^2 \cq_i,
  \\
  \delta {\cal E}^i{}_j &=&
  -{\kappa^2 \over3}\delta\cu\, \delta^i{}_j -
  \delta\pi^{{\cal E}i}{}_j.
\end{eqnarray}%

The evolution of the bulk metric perturbations is determined by
the perturbed 5D field equations in the vacuum bulk,
\begin{equation}
  \delta \:{}^{(5)\!}G^A{}_B=0.
\end{equation}
Then the matter perturbations on the brane enter via the perturbed
junction conditions~(\ref{metjun}).

For example, for scalar perturbations in Gaussian normal gauge, we
have
\begin{equation}
  \delta \:{}^{(5)\!}G^y{}_i =
  \partial_i\left\{-\psi'+\left({A'\over A}-{N'\over N}\right)
  \psi-2{\cal R}'-{A^2\over 2N^2}\left[\dot{\cal B}'+
  \left( 5{\dot A\over A}-{\dot N \over N}\right){\cal B}'\right]\right\}.
\end{equation}
For tensor perturbations (in any gauge), the only nonzero
components of the perturbed Einstein tensor are
\begin{equation}
  \delta \:{}^{(5)\!}G^i{}_j =
  -{1\over2} \left\{- {1\over N^2} \ddot{f}^i{}_j+f''^i{}_j-
  {k^2 \over A^2} f^i{}_j+{1\over N^2}\left({\dot N \over N}-
  3{\dot A \over A}\right)\dot{f}^i{}_j+ \left({N' \over N}+
  3{A' \over A}\right)f'^i{}_j \right\}.
  \label{tensor}
\end{equation}

In the following, we will discuss various perturbation problems,
using either a $1+3$-covariant or a metric-based approach.


\subsection{Density perturbations on large scales}

In the covariant approach, we define matter density and expansion
(velocity) perturbation scalars, as in 4D general relativity,
\begin{equation}
  \Delta={a^2\over\rho}\D^2\rho,
  \qquad
  Z=a^2\D^2\Theta.
\end{equation}
Then we can define dimensionless KK perturbation
scalars~\cite{m1},
\begin{equation}
  U={a^2\over\rho}\D^2{\cu},
  \qquad
  Q={a\over\rho} \D^2 \cq,
  \qquad
  \Pi={1\over \rho}\D^2\cp,
  \label{kkpert}
\end{equation}
where the scalar potentials $\cq$ and $\cp$ are defined by
Equations~(\ref{q_star}, \ref{p_star}). The KK energy density (dark
radiation) produces a scalar fluctuation $U$ which is present even
if $\cu=0$ in the background, and which leads to a non-adiabatic
(or isocurvature) mode, even when the matter perturbations are
assumed adiabatic~\cite{gm}. We define the total effective
dimensionless entropy $S_\mathrm{tot}$ via
\begin{equation}
  p_\mathrm{tot}\,S_\mathrm{tot}=a^2\D^2 p_\mathrm{tot}-
  c_\mathrm{tot}^2a^2 \D^2\rho_\mathrm{tot},
\end{equation}
where $c_\mathrm{tot}^2=\dot{p}_\mathrm{tot}/\dot{\rho}_\mathrm{tot}$ is
given in Equation~(\ref{vh2}). Then
\begin{equation}
  S_\mathrm{tot}={9\left[c_\mathrm{s}^2 - {1\over 3} +
  \left({2\over 3}+w+c_\mathrm{s}^2\right){ {\rho/\lambda}}\right]\over
  [3(1+w)(1+\rho/\lambda)+4\cu/\rho]
  [3w+3(1+2w)\rho/ 2\lambda+\cu/\rho]}\,
  \left[ {4\over 3}{\cu\over \rho}\,\Delta -(1+w)U \right].
  \label{ent}
\end{equation}

If $\cu=0$ in the background, then $U$ is an isocurvature mode:
$S_\mathrm{tot}\propto (1+w)U$. This isocurvature mode is suppressed
during slow-roll inflation, when $1+w\approx 0$.

If $\cu\neq0$ in the background, then the weighted difference
between $U$ and $\Delta$ determines the isocurvature mode:
$S_\mathrm{tot}\propto (4\cu/ 3\rho)\Delta -(1+w)U$. At very high
energies, $\rho\gg\lambda$, the entropy is suppressed by the factor
$\lambda/\rho$.

The density perturbation equations on the brane are derived by
taking the spatial gradients of Equations~(\ref{ecl}), (\ref{nlc1}),
and~(\ref{prl}), and using Equations~(\ref{momcl})
and~(\ref{nlc2}). This leads to~\cite{gm}
\begin{eqnarray}
  \dot{\Delta} &=&3wH\Delta-(1+w)Z,
  \\
  \dot{Z} &=&-2HZ-\left({c_\mathrm{s}^2\over 1+w}\right)
  \D^2\Delta-\kappa^2\rho U-{{1\over2}}\kappa^2
  \rho\left[1+(4+3w){ {\rho\over\lambda}}-
  \left({4c_\mathrm{s}^2\over 1+w}\right){\cu\over\rho}\right] \Delta,
  \\
  \dot{U} &=& (3w-1)H U +
  \left({4c_\mathrm{s}^2\over 1+w}\right)
  \left({{\cu }\over\rho}\right) H\Delta -
  \left({4{\cu }\over3\rho}\right) Z-a\D^2 Q,
  \\
  \dot{Q} &=&(3w-1)H Q-{1\over3a}{U}-{2\over3} a \D^2\Pi+
  {1\over3\mu} \left[ \left({4c_\mathrm{s}^2\over 1+w}\right)
  {{\cu}\over\rho}-3(1+w) {\rho\over\lambda} \right] \Delta.
\end{eqnarray}%
The KK anisotropic stress term $\Pi$ occurs only via its
Laplacian, ${ \D^2\Pi}$. If we can neglect this term on large
scales, then the system of density perturbation equations closes
on super-Hubble scales~\cite{m1}. An equivalent statement applies
to the large-scale curvature perturbations~\cite{lmsw}. KK effects
then introduce two new isocurvature modes on large scales
(associated with $U$ and $Q$), and they modify the evolution
of the adiabatic modes as well~\cite{gm, l1}.

Thus on large scales the system of brane equations is closed, and
we can determine the density perturbations without solving for the
bulk metric perturbations.

\epubtkImage{figure08.png}{%
 \begin{figure}[htbp]
   \def\epsfsize#1#2{1.2#1}
   \centerline{\epsfbox{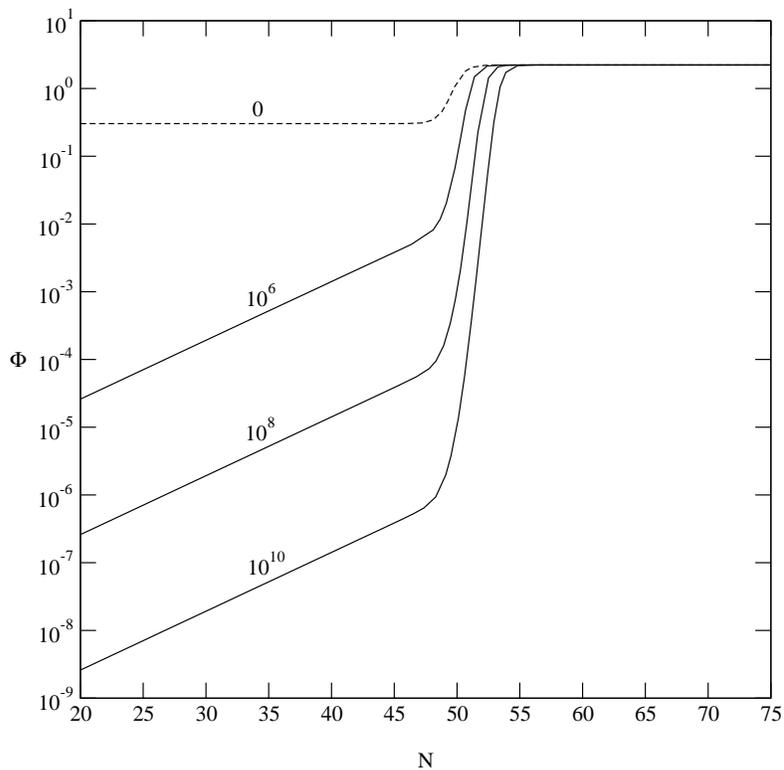}}
   \caption{The evolution of the covariant variable $\Phi$,
     defined in Equation~(\ref{cphi}) (and not to be confused with the
     Bardeen potential), along a fundamental world-line. This is a
     mode that is well beyond the Hubble horizon at $N =0$, about 50
     e-folds before inflation ends, and remains super-Hubble through
     the radiation era. A smooth transition from inflation to
     radiation is modelled by
     $w={1\over3}[(2-{3\over2}\epsilon)\tanh(N-50)-(1-{3\over2}\epsilon)]$,
     where $\epsilon$ is a small positive parameter (chosen as
     $\epsilon=0.1$ in the plot). Labels on the curves indicate the
     value of $\rho_0/\lambda$, so that the general relativistic
     solution is the dashed curve
     ($\rho_0/\lambda=0$). (Figure taken from~\cite{gm}.)}
   \label{figure_08}
 \end{figure}}

We can simplify the system as follows. The 3-Ricci tensor defined
in Equation~(\ref{gc2}) leads to a scalar covariant curvature
perturbation variable,
\begin{equation}
  C\equiv a^4\D^2R^\perp = -4a^2HZ+2\kappa^2a^2\rho
  \left( 1+ {\rho \over 2\lambda} \right)\Delta+ 2\kappa^2a^2 \rho U.
\end{equation}
It follows that $C$ is locally conserved (along $u^\mu$ flow
lines):
\begin{equation}
  C=C_0,
  \qquad
  \dot{C}_0=0.
\end{equation}
We can further simplify the system of equations via the variable
\begin{equation}
  \Phi=\kappa^2a^2\rho \Delta.
  \label{cphi}
\end{equation}
This should not be confused with the Bardeen metric perturbation
variable $\Phi_H$, although it is the covariant analogue of
$\Phi_H$ in the general relativity limit. In the brane-world,
high-energy and KK effects mean that $\Phi_H$ is a complicated
generalization of this expression~\cite{l1} involving $\Pi$, but
the simple $\Phi$ above is still useful to simplify the system of
equations. Using these new variables, we find the closed system
for large-scale perturbations:
\begin{eqnarray}
  \dot{\Phi}&=& -H\left[1+(1+w){\kappa^2\rho\over 2H^2}
  \left(1+{\rho\over \lambda}\right)\right]\Phi -
  \left[(1+w){a^2\kappa^4\rho^2\over 2 H}\right]U +
  \left[(1+w){\kappa^2 \rho\over 4H}\right]C_0,
  \label{p1_prime} \\
  \dot{U} &=& -H\left[1-3w+{2\kappa^2{\cu}\over 3H^2}\right]U -
  {2 {\cu}\over 3 a^2 H\rho}\left[1+{\rho\over\lambda} -
  {6 c_\mathrm{s}^2H^2\over (1+w)\kappa^2\rho}\right]\Phi+
  \left[{{\cu}\over 3a^2H\rho}\right] C_0.
  \label{p3}
\end{eqnarray}%

If there is no dark radiation in the background, $\cu=0$, then
\begin{equation}
  U=U_0\exp\left(-\int(1-3w)dN\right),
\end{equation}
and the above system reduces to a single equation for $\Phi$. At
low energies, and for constant $w$, the non-decaying attractor is
the general relativity solution,
\begin{equation}
  \Phi_\mathrm{low} \approx {3(1+w) \over 2(5+3w)} C_0.
  \label{low}
\end{equation}
At very high energies, for $w\geq -{1\over3}$, we get
\begin{equation}
  \Phi_\mathrm{high} \to {3\over 2}{\lambda \over \rho_0}(1+w)
  \left[ {C_0 \over 7+6w}-{2\tilde{U}_0 \over 5+6w} \right],
\end{equation}
where $\tilde{U}_0=\kappa^2a_0^2\rho_0U_0$, so that the
isocurvature mode has an influence on $\Phi$. Initially, $\Phi$ is
suppressed by the factor $\lambda/\rho_0$, but then it grows,
eventually reaching the attractor value in Equation~(\ref{low}). For
slow-roll inflation, when $1+w\sim\epsilon$, with $0<\epsilon \ll
1$ and $H^{-1}|\dot\epsilon|=|\epsilon'|\ll 1$, we get
\begin{equation}
  \Phi_\mathrm{high} \sim
  {3\over 2}\epsilon {\lambda \over \rho_0}C_0 e^{3\epsilon N},
\end{equation}
where $N=\ln(a/a_0)$, so that $\Phi$ has a growing-mode in the
early universe. This is different from general relativity, where
$\Phi$ is constant during slow-roll inflation. Thus more
amplification of $\Phi$ can be achieved than in general
relativity, as discussed above. This is illustrated for a toy
model of inflation-to-radiation in Figure~\ref{figure_08}. The early (growing) and
late time (constant) attractor solutions are seen explicitly in
the plots.

\epubtkImage{figure09.png}{%
 \begin{figure}[htbp]
   \def\epsfsize#1#2{0.8#1}
   \centerline{\epsfbox{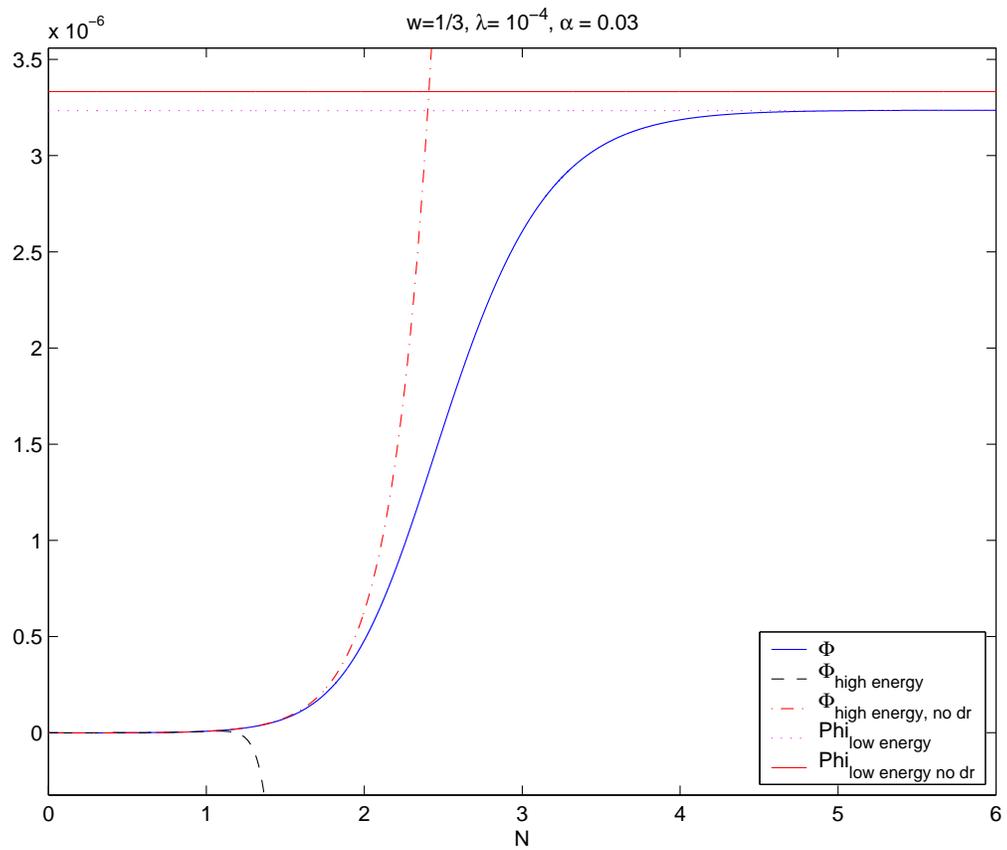}}
   \caption{The evolution of $\Phi$ in the radiation era, with
     dark radiation present in the background. (Figure taken
     from~\cite{ggm}.)}
   \label{figure_09}
 \end{figure}}

The presence of dark radiation in the background introduces new
features. In the radiation era ($w={1\over3}$), the non-decaying
low-energy attractor becomes~\cite{ggm}
\begin{eqnarray}
  \Phi_\mathrm{low} & \approx & {C_0\over 3}(1-\alpha),
  \\
  \alpha &=& {\cu\over \rho} \lesssim 0.05.
\end{eqnarray}%
The dark radiation serves to reduce the final value of $\Phi$,
leaving an imprint on $\Phi$, unlike the $\cu=0$ case,
Equation~(\ref{low}). In the very high energy limit,
\begin{equation}
  \Phi_\mathrm{high} \to {\lambda\over \rho_0}
  \left[{2\over9} C_0-{4\over7}\tilde{U}_0\right]+
  16\alpha \left({\lambda\over\rho_0}\right)^2
  \left[{C_0\over273}-{4\tilde{U}_0 \over 539}\right].
\end{equation}
Thus $\Phi$ is initially suppressed, then begins to grow, as in
the no-dark-radiation case, eventually reaching an attractor which
is less than the no-dark-radiation attractor. This is confirmed by
the numerical integration shown in Figure~\ref{figure_09}.


\subsection{Curvature perturbations and the Sachs--Wolfe effect}

The curvature perturbation ${\cal R}$ on uniform density surfaces
is defined in Equation~(\ref{pertmetric}). The associated
gauge-invariant quantity
\begin{equation}
  \zeta={\cal R}+ {\delta\rho \over 3(\rho+p)}
\end{equation}
may be defined for matter on the brane. Similarly, for the Weyl
``fluid'' if $\cu\neq0$ in the background, the curvature
perturbation on hypersurfaces of uniform dark energy density is
\begin{equation}
  \zeta_{\cal E}={\cal R}+ {\delta\rho_{\cal E} \over 4\cu}.
\end{equation}
On large scales, the perturbed dark energy conservation equation
is~\cite{lmsw}
\begin{equation}
  (\delta\cu)^{\displaystyle{\cdot}}+
  4H\delta\cu+ 4\cu \dot{\cal R}=0,
\end{equation}
which leads to
\begin{equation}
  \dot{\zeta}_{\cal E}=0.
\end{equation}
For adiabatic matter perturbations, by the perturbed matter energy
conservation equation,
\begin{equation}
  (\delta\rho)^{\displaystyle{\cdot}}+3H(\delta\rho+\delta p)+
  3(\rho +p) \dot{\cal R}=0,
\end{equation}
we find
\begin{equation}
  \dot\zeta=0.
\end{equation}
This is independent of brane-world modifications to the field
equations, since it depends on energy conservation only. For the
total, effective fluid, the curvature perturbation is defined as
follows~\cite{lmsw}: If $\cu\neq0$ in the background, we have
\begin{equation}
  {\zeta}_\mathrm{tot} =
  {\zeta}+\left[{4\cu \over 3(\rho+ p)(1+\rho/\lambda)+4\cu}\right]
  \left(\zeta_{\cal E}-\zeta\right),
\end{equation}
and if $\cu=0$ in the background, we get
\begin{eqnarray}
  {\zeta}_\mathrm{tot} &=&
  {\zeta}+ {\delta\cu \over 3(\rho+p)(1+\rho/\lambda)}
  \\
  \delta\cu &=& {\delta C_{\cal E} \over a^4},
\end{eqnarray}%
where $\delta C_{\cal E}$ is a constant. It follows that the
curvature perturbations on large scales, like the density
perturbations, can be found on the brane without solving for the
bulk metric perturbations.

Note that $\dot{\zeta}_\mathrm{\,tot}\neq 0$ even for adiabatic
matter perturbations; for example, if $\cu=0$ in the background, then
\begin{equation}
  \dot{\zeta}_\mathrm{\,tot}=
  H\left(c_\mathrm{tot}^2 -{1\over 3}\right)
  {\delta\cu \over (\rho+p)(1+\rho/\lambda)}.
\end{equation}
The KK effects on the brane contribute a non-adiabatic mode,
although $\dot{\zeta}_\mathrm{\,tot}\to 0$ at low energies.

Although the density and curvature perturbations can be found on
super-Hubble scales, the Sachs--Wolfe effect requires
$\cp_{\mu\nu}$ in order to translate from density/curvature to
metric perturbations. In the 4D longitudinal gauge of the metric
perturbation formalism, the gauge-invariant curvature and metric
perturbations on large scales are related by
\begin{eqnarray}
  \zeta_\mathrm{tot} &=&
  {\cal R}-{H \over \dot H}\left( {\dot{\cal R} \over H}-\psi\right),
  \label{curv} \\
  {\cal R}+\psi &=& -\kappa^2a^2\delta \pi_{\cal E},
  \label{metcurv}
\end{eqnarray}%
where the radiation anisotropic stress on large scales is
neglected, as in general relativity, and $\delta \pi_{\cal E}$ is
the scalar potential for $\cp_{\mu\nu}$, equivalent to the
covariant quantity $\Pi$ defined in Equation~(\ref{kkpert}). In 4D
general relativity, the right hand side of Equation~(\ref{metcurv}) is
zero. The (non-integrated) Sachs--Wolfe formula has the same form
as in general relativity:
\begin{equation}
  {\delta T\over T}\Big|_\mathrm{now}\!\!\!=
  (\zeta_\mathrm{rad}+\psi-{\cal R})|_\mathrm{dec}.
  \label{swd}
\end{equation}
The brane-world corrections to the general relativistic
Sachs--Wolfe effect are then given by~\cite{lmsw}
\begin{equation}
  {\delta T\over T} =
  \left({\delta T\over T}\right)_\mathrm{gr}\!\!\!-{8\over 3}
  \left({\rho_\mathrm{rad}\over \rho_\mathrm{cdm}}\right)
  S_{\cal E}-\kappa^2a^2\delta \pi_{\cal E}+
  {2\kappa^2\over a^{5/2}}\int da \, a^{7/2}\,\delta\pi_{\cal E},
  \label{sachsw}
\end{equation}
where $S_{\cal E}$ is the KK entropy perturbation (determined by
$\delta\cu$). The KK term $\delta\pi_{\cal E}$ cannot be
determined by the 4D brane equations, so that $\delta T/T$ cannot
be evaluated on large scales without solving the 5D equations.
(Equation~(\ref{sachsw}) has been generalized to a 2-brane model,
in which the radion makes a contribution to the Sachs--Wolfe
effect~\cite{ksw}.)

The presence of the KK (Weyl, dark) component has essentially two
possible effects:
\begin{itemize}
\item A contribution from the KK entropy perturbation $S_{\cal E}$
  that is similar to an extra isocurvature contribution.
\item The KK anisotropic stress $\delta\pi_{\cal E}$ also contributes
  to the CMB anisotropies. In the absence of anisotropic stresses,
  the curvature perturbation $\zeta_\mathrm{tot}$ would be sufficient
  to determine the metric perturbation ${\cal R}$ and hence the
  large-angle CMB anisotropies via Equations~(\ref{curv},
  \ref{metcurv}, \ref{swd}). However, bulk gravitons generate
  anisotropic stresses which, although they do not affect the
  large-scale curvature perturbation $\zeta_\mathrm{tot}$, can affect
  the relation between $\zeta_\mathrm{tot}$, ${\cal R}$, and $\psi$,
  and hence can affect the CMB anisotropies at large angles.
\end{itemize}

A simple phenomenological approximation to $\delta\pi_{\cal E}$ on
large scales is discussed in~\cite{bm}, and the Sachs--Wolfe effect
is estimated as
\begin{equation}
  {\delta T\over T} \sim
  \left({\delta\pi_{\cal E}\over\rho}\right)_\mathrm{in}
  \left({t_\mathrm{eq}\over t_\mathrm{dec}}\right)^{2/3}
  \left[{\ln(t_\mathrm{in}/t_4)\over \ln (t_\mathrm{eq}/t_4)} \right],
\end{equation}
where $t_4$ is the 4D Planck time, and $t_\mathrm{in}$ is the time
when the KK anisotropic stress is induced on the brane, which is
expected to be of the order of the 5D Planck time.

A self-consistent approximation is developed in~\cite{koy}, using the
low-energy 2-brane approximation~\cite{sod_1, sod_2, sod_3, sod_4,
  sod_5} to find an effective 4D form for ${\cal E}_{\mu\nu}$ and
hence for $\delta\pi_{\cal E}$. This is discussed below.
\new{
In a single
brane model in the AdS bulk, full numerical simulations were done to
find the behaviour of $\delta\pi_{\cal E}$~\cite{Cardoso:2007zh}, as will be
discussed in the next subsection.
}

\subsection{\new{Full numerical solutions}}
\label{section_6.5}

\epubtkUpdateA{Added Subsection~\ref{section_6.5} ``Full numerical
  solutions''.}

In order to study scalar perturbations fully, we need to numerically
solve the coupled bulk and brane equations for the master variable
$\Omega$. A thorough analysis was done in~\cite{Cardoso:2007zh}. For
this purpose, it is convenient to use the static coordinate where the
bulk equation is simple and consider a moving FRW brane;
\begin{equation}
{}^{(5)} ds^2 = \frac{\ell^2}{z^2} [\eta_{\mu \nu} dx^{\mu} dx^{\nu} + dz^2 ].
\end{equation}
The bulk master variable satisfies the following wave equation (see
Equation~(\ref{mastergeneral}))
\begin{equation}
  \label{eq:bulk wave equation}
    0 = -\frac{\di^2\Omega}{\di \tau^2} + \frac{\di^2\Omega}{\di z^2}
    + \frac{3}{z} \frac{\di\Omega}{\di z} + \left( \frac{1}{z^2} -
    k^2 \right) \Omega.
\end{equation}
From the junction condition, $\Omega$ satisfies a boundary condition
on the brane
\begin{equation}
  \label{eq:boundary condition}
    \left[ \di_n \Omega + \frac{1}{\ell} \left(1 +
    \frac{\rho}{\lambda} \right) \Omega + \frac{6\rho a^3}{\lambda
    k^2} \Delta \right]_\b = 0.
\end{equation}
where $\di_n$ is the derivative orthogonal to the brane
\begin{equation}
\di_n = \frac{1}{a} \left( -H\ell \frac{\di}{\di \tau} +
    \sqrt{1+H^2\ell^2} \frac{\di}{\di z} \right),
\end{equation}
and $\Delta$ is the density perturbation in the comoving gauge. The subscript b implies that the quantities are evaluated on the brane.
On a brane, $\Delta$ satisfies a wave equation
\begin{subequations}\label{eq:brane wave equation}
\begin{gather}
    \frac{d^2 \Delta}{d\eta^2} + (1+3c_s^2-6w) Ha \frac{d\Delta}{d\eta} +
    \left[c_s^2 k^2 + \frac{3\rho a^2}{\lambda\ell^2} A +
    \frac{3\rho^2 a^2}{\lambda^2\ell^2} B \right] \Delta =
    \frac{k^4(1+w)\Omega_\b}{3\ell a^3}, \\
    A = 6c_s^2 -1 -8w+3w^2, \qquad
    B = 3c_s^2-9w-4,
\end{gather}
\end{subequations}
where we consider a perfect fluid on a brane with an equation state
$w$ and $c_s$ is a sound speed for perturbations. The above ordinal
differential equation, the bulk wave equation~(\ref{eq:bulk wave
  equation}) and the boundary condition~(\ref{eq:boundary condition})
comprise a closed set of equations for $\Delta$ and $\Omega_c$.

On a brane, we take the longitudinal gauge
\begin{equation}
ds_b^2 = - (1 + 2 \psi) dt^2 + (1 +2 {\cal R}) \delta_{ij} dx^i dx^j.
\end{equation}
Using the expressions for metric perturbations in terms of the master
variable $\Omega$ (see Eq.~(\ref{Rmaster})) and the junction condition
Eq.~(\ref{eq:boundary condition}), $\psi$ and ${\cal R}$ are written in terms of
$\Delta$ and $\Omega$ as
\begin{subequations}
\label{eq:explicit gauge invariants}
\begin{align}
    \label{eq:explicit Phi} {\cal R} & = \frac{3a^2\rho(\rho+\lambda)}{k^2 \ell^2 \lambda^2}
    \Delta + \left( \frac{3H^2 a^2+k^2}{6\ell a^3} \right) \Omega_\b
    - \frac{H}{2\ell a^2} \frac{d\Omega_\b}{d\eta}, \\ \label{eq:explicit Psi}  \psi
    & =
    -\frac{3\rho a^2 (3w\rho+4\rho+\lambda)}{k^2\ell^2\lambda^2} \Delta -
    \left[ \frac{(3w+4)\rho^2}{2\ell^3 a\lambda^2} + \frac{(5+3w)\rho}{2\ell^3
    a\lambda} +
    \frac{k^2}{3\ell a^3} \right] \Omega_\b + \frac{3H}{2\ell a^2} \frac{d\Omega_\b}{d\eta}
    - \frac{1}{2\ell a^3} \frac{d^2\Omega_\b}{d\eta^2}.
\end{align}
\end{subequations}
Other quantities of interest are the curvature perturbation on
uniform density slices,
\begin{equation}
  \label{eq:explicit zeta}
    \zeta =  {\cal R} - \frac{Ha V}{k} + \frac{\Delta}{3(1+w)} = \left[ \frac{1}{3}
    - \frac{3\rho a^2(w\lambda - \lambda -\rho)}{k^2\ell^2\lambda^2} \right]
    \frac{\Delta}{1+w} + \frac{Ha}{k^2(1+w)} \frac{d\Delta}{d\eta}
    + \frac{k^2}{6\ell a^3} \Omega_\b,
\end{equation}
where the velocity perturbation $V$ is also written by $\Delta$ and $\Omega_\b$.
There are two independent numerical codes that can be used to solve for
$\Delta$ and $\Omega_\b$.  The first is the pseudo-spectral (PS)
method used in~\cite{Hiramatsu:2006cv} and the second is the
characteristic integration (CI) algorithm developed in
\cite{Cardoso:2006nh}.

Figure~\ref{figurev2_01} shows the output of the PS and CI codes
for a typical simulation of a mode with $\rho/\lambda =
50$ at the horizon re-enter. As expected we have excellent agreement between the two
codes, despite the fact that they use different initial conditions.
Note that for all simulations,
we recover that $\Delta$ and $\zeta$ are phase-locked plane waves,
\begin{equation}
    \Delta(\eta) \propto \cos \frac{k\eta}{\sqrt{3}}, \quad \Delta(\eta)
    \approx 4 \zeta(\eta),
\end{equation}
at sufficiently late times $k\eta \gg 1$, which is actually the same
behaviour as seen in GR.
Figure~\ref{figurev2_02} illustrates how the ordinary
superhorizon behaviour of perturbations in GR is recovered for
modes entering the Hubble horizon in the low energy era.  We see
how $\Delta$, $\psi$ and ${\cal R}$ smoothly interpolate between the
non-standard high-energy behaviour to the usual expectations in GR.
Also shown in this plot is the behaviour of the KK anisotropic stress,
which steadily decays throughout the simulation. These results confirm
that at low energies, we recover GR solutions smoothly.

\epubtkImage{figurev201.png}{%
  \begin{figure}[htbp]
    \centerline{\includegraphics[width=\textwidth]{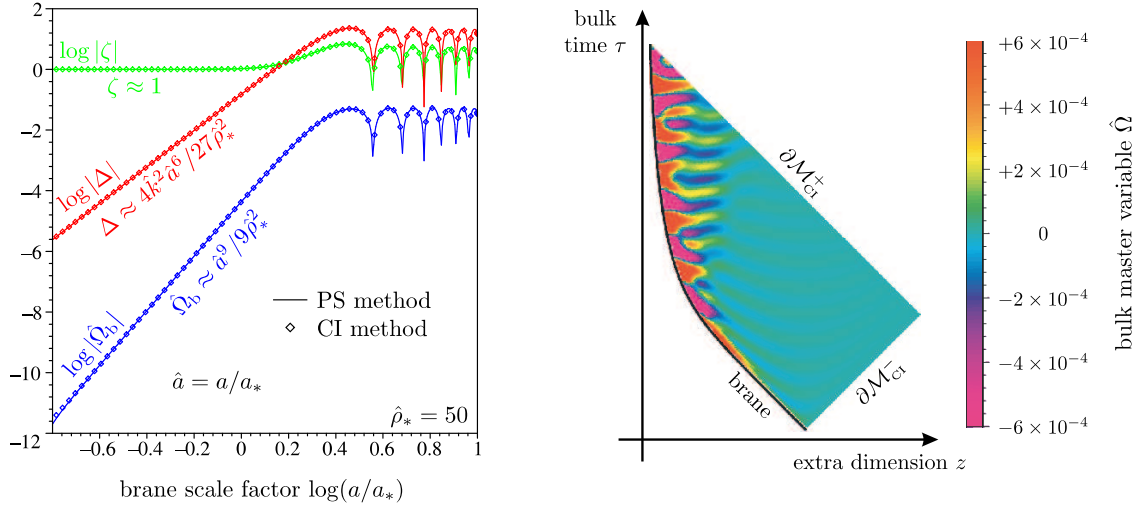}}
    \caption{Comparison between typical results of the PS and CI codes
    for various brane quantities (\emph{left}); and the typical
    behaviour of the bulk master variable (\emph{right}) as calculated
    by the CI method. Very good agreement between the two different
    numerical schemes is seen in the left panel, despite the fact that
    they use different initial conditions. Also note that on
    subhorizon scales, $\Delta$ and $\zeta$ undergo simple
    harmonic oscillations, which is consistent with the behaviour in
    GR. The bulk profile demonstrates our choice of initial
    conditions: We see that the bulk master variable $\Omega$ is
    essentially zero during the early stages of the simulation, and
    only becomes ``large'' when the mode crosses the horizon. Figure
    taken from~\cite{Cardoso:2007zh}.}
    \label{figurev2_01}
\end{figure}}

\epubtkImage{figurev202.png}{%
  \begin{figure}[htbp]
    \centerline{\includegraphics[width=\textwidth]{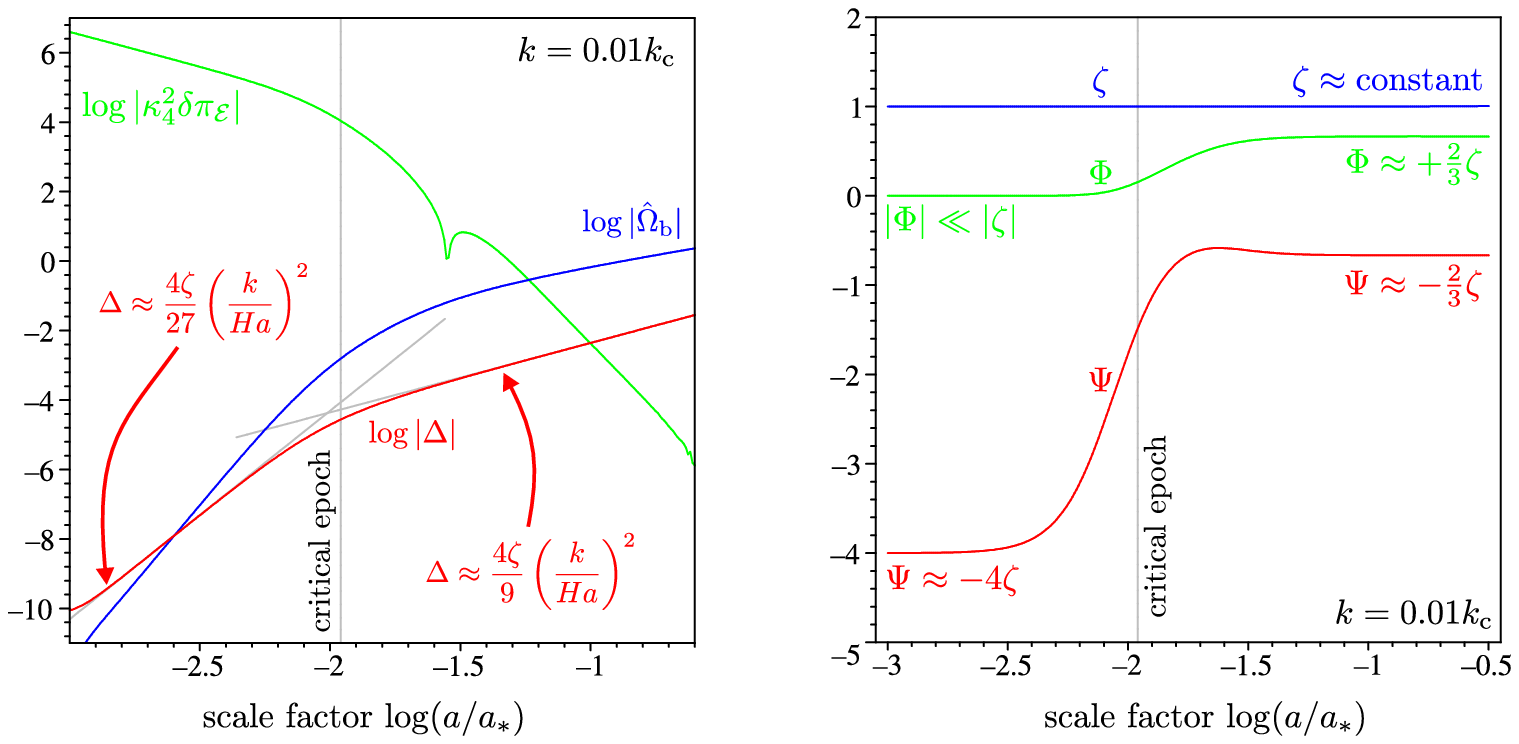}}
    \caption{The simulated behaviour of a mode on superhorizon
    scales. On the left we show how the $\Delta$ gauge
    invariant switches from the high-energy behaviour predicted to the
    familiar GR result as the universe expands through the critical
    epoch.  We also show how the KK anisotropic stress $\kappa_4^2
    \delta\pi_\KK$ steadily decays throughout the simulation, which is
    typical of all the cases we have investigated. On the right, we
    show the metric perturbations $\psi (=\Psi)$ and ${\cal R}
    (=\Phi)$ as well as the curvature perturbation $\zeta$.  Again,
    note how the GR result $\Phi \approx -\Psi \approx -2\zeta/3$ is
    recovered at low energy. Figure taken from~\cite{Cardoso:2007zh}.}
    \label{figurev2_02}
\end{figure}}

At high energies $\rho > \lambda$, there are two separate effects to consider:
First, there is the modification of the universe's expansion at high
energies and the $\mathcal{O}(\rho/\lambda)$ corrections to the
perturbative equations of motion.
 Second, there is the
effect of the bulk degrees of freedom encapsulated by the bulk
master variable $\Omega$ (or, equivalently, the KK fluid ${\cal E}_{\mu \nu} $).
To separate out the two
effects, it is useful to introduce the 4-dimensional effective
theory where all $\mathcal{O}(\rho/\lambda)$
corrections to GR are retained, but the bulk effects are removed by
artificially setting $\Omega = 0$. In the case of radiation
domination, we obtain equations for the effective theory density
contrast $\Delta_{\eff}$ and curvature perturbation $\zeta_\eff$ from
Equations~(\ref{eq:explicit zeta}) and (\ref{eq:brane wave equation}) with
$\Omega_\b = 0$:
\begin{subequations}
\begin{align}
    0 & = \frac{d^2\Delta_\eff}{d\eta^2} + \left( \frac{k^2}{3} -
    \frac{4\rho a^2}{\lambda\ell^2} - \frac{18\rho^2
    a^2}{\lambda^2\ell^2}\right) \Delta_\eff, \\
    \zeta_\eff & = \left( \frac{1}{4}
    + \frac{3 \rho a^2}{2 \lambda k^2 \ell^2} + \frac{9\rho^2
    a^2}{4\lambda^2k^2\ell^2}\right)\Delta_\eff  + \frac{3Ha}
    {4k}\frac{d\Delta_\eff}{d\eta}.
\end{align}
\end{subequations}
These give a closed set of ODEs on the brane
that describe all of the $\mathcal{O}(\rho/\lambda)$ corrections to
GR.

Since in any given model we expect the primordial value of the curvature
perturbation to be fixed by inflation, it makes physical sense to
normalize the waveforms from each theory such that
$ \zeta_\fiveD \approx \zeta_\eff \approx \zeta_\GR \approx 1$
for $a \ll a_*$.
We can define a set of
``enhancement factors'', which are functions of $k$ that describe
the relative amplitudes of $\Delta$ after horizon crossing in the
various theories. Let the final
amplitudes of the density perturbation with wavenumber $k$ be
$\mathcal{C}_\fiveD(k)$, $\mathcal{C}_\eff(k)$ and
$\mathcal{C}_\GR(k)$ for the 5-dimensional, effective and GR
theories, respectively, given that the normalization $ \zeta_\fiveD \approx \zeta_\eff \approx \zeta_\GR \approx 1$ holds.
Then, we define enhancement factors as
\begin{equation}
    \mathcal{Q}_\eff(k) =
    \frac{\mathcal{C}_\eff(k)}{\mathcal{C}_\GR(k)}, \quad \mathcal{Q}_\KK(k) =
    \frac{\mathcal{C}_\fiveD(k)}{\mathcal{C}_\eff(k)}, \quad \mathcal{Q}_\fiveD(k) =
    \frac{\mathcal{C}_\fiveD(k)}{\mathcal{C}_\GR(k)}.
\end{equation}
It follows that $\mathcal{Q}_\eff(k)$ represents the
$\mathcal{O}(\rho/\lambda)$ enhancement to the density perturbation,
$\mathcal{Q}_\KK(k)$ gives the magnification due to KK modes, while
$\mathcal{Q}_\fiveD(k)$ gives the total 5-dimensional amplification
over the GR case. They all increase
as the scale is decreased, and that they all approach unity for $k
\rightarrow 0$.  Since $\mathcal{Q} = 1$ implies no enhancement of
the density perturbations over the standard result, this means we
recover general relativity on large scales.  For all wavenumbers we
see $\mathcal{Q}_\eff
> \mathcal{Q}_\KK > 1$, which implies that the amplitude magnification due
to the $\mathcal{O}(\rho/\lambda)$ corrections is always larger than
that due to the KK modes.  Interestingly, the $\mathcal{Q}$-factors
appear to approach asymptotically constant values for large $k$:
\begin{equation}
    \mathcal{Q}_\eff(k) \approx 3.0,
    \quad \mathcal{Q}_\KK(k) \approx
    2.4,
    \quad \mathcal{Q}_\fiveD(k) \approx
    7.1, \quad k \gg k_\c,
\end{equation}
where $k_c$ is the comoving wavenumber of the mode that enters the
horizon when $H=\ell^{-1}$.

In cosmological perturbation theory, transfer functions are very
important quantities. They allow one to transform the primordial
spectrum of some quantity set during inflation into the spectrum
of another quantity at a later time.  In this sense, they are
essentially the Fourier transform of the retarded Green's function
for cosmological perturbations.  There are many different transfer
functions one can define, but for our case it is useful to
consider a function $T(k)$ that will tell us how the initial
spectrum of curvature perturbations $\mathcal{P}_\zeta^\prim$ maps
onto the spectrum of density perturbations $\mathcal{P}_\Delta$ at
some low energy epoch within the radiation era. It is customary to normalize
transfer functions such that $T(k;\eta) \to  1 (k \to 0)$,
which leads us to the following definition
\begin{equation}
    T(k;\eta) = \frac{9}{4} \left[ \frac{k}{H(\eta)a(\eta)}
    \right]^{-2} \frac{\Delta_k(\eta)}{\zeta^\prim_k}.
\end{equation}
Here, $\zeta^\prim_k$ is the primordial value of the curvature
perturbation and $\Delta_k(\eta)$ is the maximum amplitude of the
density perturbation in the epoch of interest.  As demonstrated in
Figure~\ref{figurev2_02}, we know that we recover the GR result
in the extreme small scale limit $(k \to 0)$, which gives the transfer
function the correct normalization.
In the righthand panel of Figure~\ref{figurev2_03}, we show the
transfer functions derived from GR, the effective theory and the
5-dimensional simulations.  As expected, the $T(k;\eta)$ for each
formulation match one another on subcritical scales $k < k_\c$.
However, on supercritical scales we have $T_\fiveD > T_\eff >
T_\GR$.

\epubtkImage{figurev203.png}{%
  \begin{figure}[htbp]
    \centerline{\includegraphics[width=\textwidth]{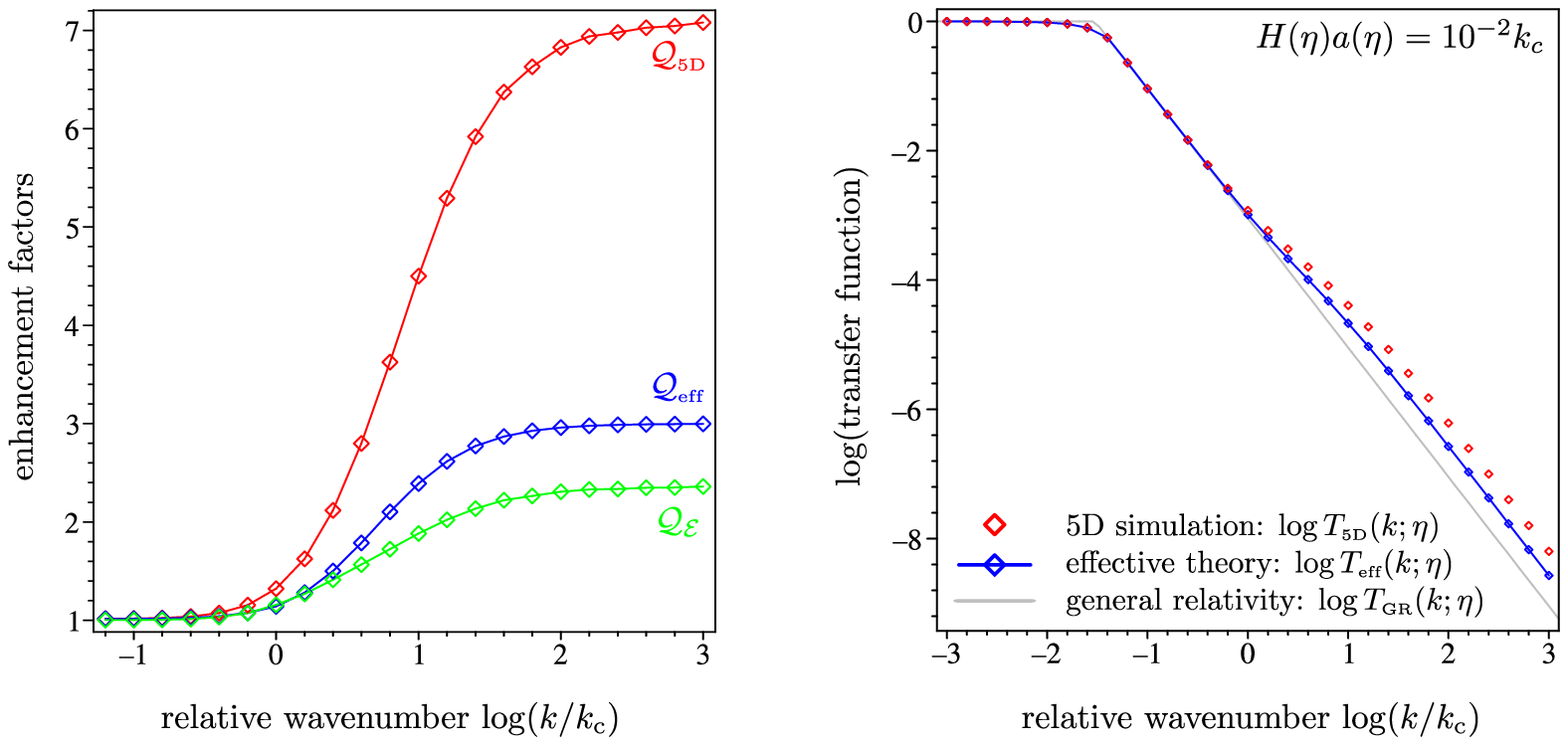}}
    \caption{Density perturbation enhancement factors (\emph{left})
    and transfer functions (\emph{right}) from simulations, effective
    theory, and general relativity. All of the $\mathcal{Q}$ factors
    monotonically increase with $k/k_\c$, and we see that the $\Delta$
    amplitude enhancement due to $\mathcal{O}(\rho/\lambda)$ effects
    $\mathcal{Q}_\eff$ is generally larger than the enhancement due to
    KK effects $\mathcal{Q}_\KK$. For asymptotically small scales $k
    \gg k_\c$, the enhancement seems to level off. The transfer
    functions in the right panel are evaluated at a given subcritical
    epoch in the radiation dominated era. The $T$ functions show how,
    for a fixed primordial spectrum of curvature perturbations
    $\mathcal{P}^\prim_\zeta$, the effective theory predicts excess
    power in the $\Delta$ spectrum $\mathcal{P}_\Delta \propto T^2
    \mathcal{P}^\prim_\zeta$ on supercritical/subhorizon scales
    compared to the GR result.  The excess small-scale power is even
    greater when KK modes are taken into account, as shown by
    $T_\fiveD(k;\eta)$. Figure taken from~\cite{Cardoso:2007zh}.}
    \label{figurev2_03}
\end{figure}}

Note that if we are interested in the transfer function at some
arbitrary epoch in the low-energy radiation regime $Ha \gg k_c$, it
is approximately given in terms of the enhancement factor as
follows:
\begin{equation}\label{eq:5D transfer function}
    T_\fiveD(k;\eta) \approx
    \begin{cases}
        1, & k < 3Ha, \\ (3Ha/k)^2 \mathcal{Q}_\fiveD(k), & k > 3Ha,
    \end{cases}
\end{equation}
Now, the spectrum of density fluctuations at any point in the
radiation era is given by
\begin{equation}
    \mathcal{P}_\Delta(k;\eta) = \frac{16}{81} T^2(k;\eta)
    \left( \frac{k}{Ha} \right)^4 \mathcal{P}^\prim_\zeta(k).
\end{equation}
Using Equation~(\ref{eq:5D transfer
function}), we see that the RS matter power spectrum (evaluated in
the low-energy regime) is $\sim 50$ times bigger than the GR
prediction on scales given by $k \sim 10^3 k_\c$.

The amplitude enhancement of perturbations is important on comoving
scales $\lesssim 10 \,\text{AU}$, which are far too small to be
relevant to present-day/cosmic microwave background measurements of
the matter power spectrum. However, it may have an important bearing
on the formation of compact objects such as primordial black holes
and boson stars at very high energies, i.e.\ the greater
gravitational force of attraction in the early universe will create
more of these objects than in GR (different aspects of primordial
black holes in RS cosmology in the context of various effective
theories have been considered in~\cite{Guedens:2002km, Guedens:2002sd,
  Clancy:2003zd, Sendouda:2003dc, Sendouda:2004hz, Sendouda:2006yc}).


\subsection{Vector perturbations}

The vorticity propagation equation on the brane is the same as in
general relativity,
\begin{equation}
  \dot{\omega}_\mu +2H\omega_\mu =-{1\over2}\curl A_\mu.
  \label{vor}
\end{equation}
Taking the $\curl$ of the conservation equation~(\ref{c2}) (for the
case of a perfect fluid, $q_\mu =0=\pi_{\mu\nu}$), and using the
identity in Equation~(\ref{vec}), one obtains
\begin{equation}
  \curl A_\mu =-6Hc_\mathrm{s}^2\omega_\mu
\end{equation}
as in general relativity, so that Equation~(\ref{vor}) becomes
\begin{equation}
  \dot{\omega}_\mu +\left(2-3c_\mathrm{s}^2\right) H\omega_\mu = 0,
\end{equation}
which expresses the conservation of angular momentum. In general
relativity, vector perturbations vanish when the vorticity is
zero. By contrast, in brane-world cosmology, bulk KK effects can
source vector perturbations even in the absence of
vorticity~\cite{m2}. This can be seen via the divergence equation
for the magnetic part $H_{\mu\nu}$ of the 4D Weyl tensor on the
brane,
\begin{equation}
  \D^2\bar{H}_{\mu} =
  2\kappa^2(\rho+p)\left[1+{\rho\over\lambda}\right] \omega_\mu +
  {4\over3} \kappa^2{\cu} \omega_\mu -
  {1\over2}\kappa^2 \curl{\bcq_\mu },
\end{equation}
where $H_{\mu\nu}=\D_{\langle\mu}\bar{H}_{\nu\rangle}$. Even when
$\omega_\mu =0$, there is a source for gravimagnetic terms on the
brane from the KK quantity $\curl{\bcq_\mu }$.

We define covariant dimensionless vector perturbation quantities
for the vorticity and the KK gravi-vector term:
\begin{equation}
  \bar{\alpha}_\mu =a\,\omega_\mu,
  \qquad
  \bar{\beta}_\mu ={a\over\rho}\curl{\bcq_\mu}.
\end{equation}
On large scales, we can find a closed system for these vector
perturbations on the brane~\cite{m2}:
\begin{eqnarray}
  \dot{\bar{\alpha}}_\mu +\left(1-3c_\mathrm{s}^2\right)
  H\bar{\alpha}_\mu &=&0,
  \\
  \dot{\bar{ \beta}}_\mu +(1-3w)H\bar{\beta}_\mu &=&
  {2\over 3}H\left[4\left( 3c_\mathrm{s}^2-1\right)
  {{\cu}\over\rho} - 9(1+w)^2{\rho\over\lambda}\right]
  \bar{\alpha}_\mu.
  \label{v2}
\end{eqnarray}%
Thus we can solve for $\bar{\alpha}_\mu $ and $\bar{\beta}_\mu $
on super-Hubble scales, as for density perturbations. Vorticity in
the brane matter is a source for the KK vector perturbation
$\bar{\beta}_\mu $ on large scales. Vorticity decays unless the
matter is ultra-relativistic or stiffer ($w\geq {1\over3}$), and
this source term typically provides a decaying mode. There is
another pure KK mode, independent of vorticity, but this mode
decays like vorticity. For $w\equiv p/\rho=\mathrm{const.}$, the
solutions are
\begin{eqnarray}
  \bar{\alpha}_\mu &=& b_\mu \left({a\over a_0}\right)^{3w-1}
  \!\!\!\!\!\!\!\!\!\!\!\!\!,
  \\
  \bar{\beta}_\mu  &=& c_\mu \left({a\over a_0}\right)^{3w-1}\!\!\!+
  b_\mu \left[ {8\rho_{{\cal E}\,0} \over 3 \rho_0}
  \left({a\over a_0}\right)^{2(3w-1)}\!\!\!+2(1+w){\rho_0\over \lambda}
  \left({a\over a_0}\right)^{-4}\right],
\end{eqnarray}%
where $\dot{b}_\mu =0=\dot{c}_\mu $.

Inflation will redshift away the vorticity and the KK mode.
Indeed, the massive KK vector modes are not excited during
slow-roll inflation~\cite{bmw_1, bmw_2}.


\subsection{Tensor perturbations}

The covariant description of tensor modes on the brane is via the
shear, which satisfies the wave equation~\cite{m2}
\begin{equation}
  \D^2{\bar{\sigma}}_{\mu\nu}-\ddot{\bar{\sigma}}_{\mu\nu}-
  5H\dot{\bar{\sigma}}_{\mu\nu}-
  \left[2\Lambda+{1\over2}\kappa^2
  \left(\rho-3p- (\rho+3p){\rho\over\lambda}\right)\right]
  {\bar{\sigma}}_{\mu\nu} -
  {\kappa^2}\left( { \dot{\bar{\pi}}^{\cal E}_{\mu\nu}}+
  2H {\bcp_{\mu\nu}} \right).
\end{equation}
Unlike the density and vector perturbations, there is no closed
system on the brane for large scales. The KK anisotropic stress
$\bcp_{\mu\nu}$ is an unavoidable source for tensor modes on the
brane. Thus it is necessary to use the 5D metric-based formalism.
This is the subject of the next Section~\ref{section_7}.

\newpage


\section{Gravitational Wave Perturbations in Brane-World Cosmology}
\label{section_7}

\subsection{Analytical approaches}

The tensor perturbations are given by Equation~(\ref{pertmetric}), i.e.,
(for a flat background brane),
\begin{equation}
  {}^{(5)\!}ds^2=-N^2(t,y)dt^2+
  A^2(t,y)\left[\delta_{ij}+f_{ij}\right]dx^idx^j+dy^2.
\end{equation}
The transverse traceless $f_{ij}$ satisfies Equation~(\ref{tensor}),
which implies, on splitting $f_{ij}$ into Fourier modes with
amplitude $f(t,y)$,
\begin{equation}
  {1\over N^2}\left[\ddot f+\left(3{\dot A \over A}-
  {\dot N \over N} \right)\dot f \right]+{k^2 \over A^2} f =
  f''+ \left(3{ A' \over A}+ { N' \over N} \right)f'.
  \label{tenwav}
\end{equation}
By the transverse traceless part of Equation~(\ref{metjun}), the
boundary condition is
\begin{equation}
  f'_{ij}\big|_\mathrm{brane} = \bar{\pi}_{ij},
\end{equation}
where $\bar{\pi}_{ij}$ is the tensor part of the anisotropic
stress of matter-radiation on the brane.

The wave equation~(\ref{tenwav}) cannot be solved analytically
except if the background metric functions are separable, and this
only happens for maximally symmetric branes, i.e., branes with
constant Hubble rate $H_0$. This includes the RS case $H_0=0$
already treated in Section~\ref{section_2}. The cosmologically
relevant case is the de Sitter brane, $H_0>0$. We can calculate the
spectrum of gravitational waves generated during brane
inflation~\cite{lmw, grs, fk, gwtan}, if we approximate slow-roll
inflation by a succession of de Sitter phases. The metric for a de
Sitter brane $\mathrm{dS}_4$ in $\mathrm{AdS}_5$ is given by
Equations~(\ref{gnm}, \ref{gnm1}, \ref{gnm2}) with
\begin{eqnarray}
  N(t,y)&=&n(y),
  \\
  A(t,y)&=&a(t) n(y),
  \\
  n(y)&=&\cosh\mu y-\left(1+{\rho_0\over\lambda}\right) \sinh\mu|y|,
  \\
  a(t)&=&a_0\exp{H_0(t-t_0)},
  \\
  H_0^2&=&{\kappa^2\over3}\rho_0\left(1+{\rho_0\over2\lambda}\right),
\end{eqnarray}%
where $\mu=\ell^{-1}$.

\epubtkImage{figure10.png}{%
 \begin{figure}[htbp]
   \def\epsfsize#1#2{1#1}
   \centerline{\epsfbox{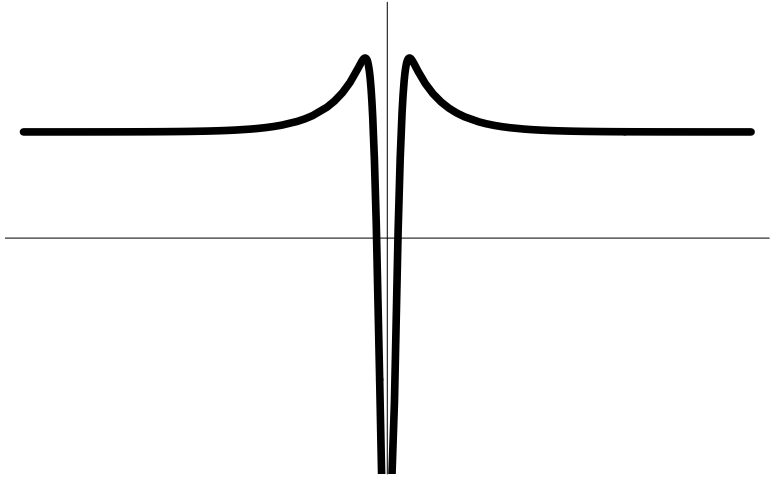}}
   \caption{Graviton ``volcano'' potential around the $\mathrm{dS}_4$
     brane, showing the mass gap. (Figure taken from~\cite{lan}.)}
   \label{figure_10}
 \end{figure}}

The linearized wave equation~(\ref{tenwav}) is separable. As
before, we separate the amplitude as $f=\sum \varphi_m(t) f_m(y)$
where $m$ is the 4D mass, and this leads to:
\begin{eqnarray}
  \ddot{\varphi}_m +3H_0\dot{\varphi}_m+
  \left[ m^2+{k^2\over a^2}\right] \varphi_m &=& 0,
  \label{varphieom} \\
  f_m''+4{n'\over n}f_m'+ {m^2\over n^2}f_m &=& 0.
  \label{Heom}
\end{eqnarray}%
The general solutions for $m>0$ are
\begin{eqnarray}
  \varphi_m(t) &=& \exp\left(-{3\over2}H_0t\right)
  B_\nu\left({k\over H_0}e^{-H_0t}\right),
  \label{vp2} \\
  f_m(y) &=& n(y)^{-3/2}L^\nu_{3/2}
  \left(\sqrt{1+{\mu^2\over H_0^2}n(y)^2}\right),
\end{eqnarray}%
where $B_\nu$ is a linear combination of Bessel functions,
$L^\nu_{3/2}$ is a linear combination of associated Legendre
functions, and
\begin{equation}
  \nu=i\sqrt{{m^2\over H_0^2}-{9\over4}}.
\end{equation}

It is more useful to reformulate Equation~(\ref{Heom}) as a
Schr\"odinger-type equation,
\begin{equation}
  {d^2\Psi_m\over dz^2} - V(z)\Psi_m =-m^2 \Psi_m,
  \label{SE}
\end{equation}
using the conformal coordinate
\begin{equation}
  z=z_\mathrm{b} +\int_0^y {d{\tilde y}\over n({\tilde y})},
  \qquad
  z_\mathrm{b} = {1\over H_0}\sinh^{-1}\left({H_0\over \mu}\right),
\end{equation}
and defining $\Psi_m\equiv n^{3/2}f_m$. The potential is given by
(see Figure~\ref{figure_10})
\begin{equation}
  V(z)= {15H_0^2 \over 4\sinh^2(H_0 z)} + {{9\over4}}H_0^2 -
  3\mu\left(1+{\rho_0\over\lambda}\right) \delta(z-z_\mathrm{b}),
\end{equation}
where the last term incorporates the boundary condition at the
brane. The ``volcano'' shape of the potential shows how the 5D
graviton is localized at the brane at low energies. (Note that
localization fails for an $\mathrm{AdS}_4$ brane~\cite{nonloc_1, nonloc_2}.)

The non-zero value of the Hubble parameter implies the existence
of a mass gap~\cite{gs},
\begin{equation}
  \Delta m={3\over2}H_0,
\end{equation}
between the zero mode and the continuum of massive KK modes. This
result has been generalized: For $\mathrm{dS}_4$ brane(s) with bulk
scalar field, a universal lower bound on the mass gap of the KK tower
is~\cite{fk}
\begin{equation}
  \Delta m \geq \sqrt{3\over2}H_0.
\end{equation}
The massive modes decay during inflation, according to
Equation~(\ref{vp2}), leaving only the zero mode, which is effectively
a 4D gravitational wave. The zero mode, satisfying the boundary
condition
\begin{equation}
  f'_0(x,0)=0,
\end{equation}
is given by
\begin{equation}
  f_0= \sqrt{\mu} \, F \left({H_0/\mu}\right),
\end{equation}
where the normalization condition
\begin{equation}
  2 \int_{z_\mathrm{b}}^\infty \!\!\! |\Psi_0^2| dz=1
\end{equation}
implies that the function $F$ is given by~\cite{lmw}
\begin{equation}
  F \left(x\right) =
  \left\{ \sqrt{1+x^2} - x^2 \ln
  \left[ {1\over x}+\sqrt{1+{1\over x^2}} \right] \right\}^{-1/2}
  \!\!\!\!\!\!\!\!\!\!\!.
  \label{dtot}
\end{equation}
At low energies ($H_0\ll \mu$) we recover the general relativity
amplitude: $F\to 1$. At high energies, the amplitude is
considerably enhanced:
\begin{equation}
  H_0 \gg \mu
  \quad \Rightarrow \quad
  F\approx \sqrt{3H_0\over 2\mu}.
\end{equation}
The factor $F$ determines the modification of the gravitational
wave amplitude relative to the standard 4D result:
\begin{equation}
  A_\mathrm{t}^2=
  \left[{8 \over M_\mathrm{p}^2}
  \left({H_0 \over 2\pi}\right)^2 \right]F^2(H_0/\mu).
\end{equation}
The modifying factor $F$ can also be interpreted as a change in the
effective Planck mass~\cite{fk}.

This enhanced zero mode produced by brane inflation remains frozen
outside the Hubble radius, as in general relativity, but when it
re-enters the Hubble radius during radiation or matter domination,
it will no longer be separated from the massive modes, since $H$
will not be constant. Instead, massive modes will be excited
during re-entry. In other words, energy will be lost from the zero
mode as 5D gravitons are emitted into the bulk, i.e., as massive
modes are produced on the brane. A phenomenological model of the
damping of the zero mode due to 5D graviton emission is given
in~\cite{l2}. Self-consistent low-energy approximations to compute
this effect are developed in~\cite{kkt, elmw}.

\epubtkUpdateA{Removed former Figure~11: Damping of brane-world gravity waves on horizon  re-entry due to massive mode generation.}

At zero order, the low-energy approximation is based on the
following~\cite{mbb_1, mbb_2, mbb_3}. In the radiation era, at low energy, the
background metric functions obey
\begin{equation}
  A(t,y)\to a(t)e^{-\mu y},
  \qquad
  N(t,y) \to e^{-\mu y}.
\end{equation}
To lowest order, the wave equation therefore separates, and the
mode functions can be found analytically~\cite{mbb_1, mbb_2, mbb_3}. The massive
modes in the bulk, $f_m(y)$, are the same as for a Minkowski
brane. On large scales, or at late times, the mode functions on
the brane are given in conformal time by
\begin{equation}
  \varphi^{(0)}_m(\eta)=\eta^{-1/2}B_{1/4}
  \left({ma_\mathrm{h}^2\,\mu\over \sqrt{2}}\eta^2 \right),
\end{equation}
where $a_\mathrm{h}$ marks the start of the low-energy regime
($\rho_\mathrm{h}=\lambda$), and $B_\nu$ denotes a linear combination
of Bessel functions. The massive modes decay on super-Hubble
scales, unlike the zero-mode. Expanding the wave equation in
$\rho_0/\lambda$, one arrives at the first order, where
mode-mixing arises. The massive modes $\varphi^{(1)}_m(\eta)$ on
sub-Hubble scales are sourced by the initial zero mode that is
re-entering the Hubble radius~\cite{elmw}:
\begin{equation}
  \left({\partial^2 \over \partial \eta^2}-{\partial^2_\eta a \over a}\right)
  a \varphi^{(1)}_m+ {k^2}a \varphi^{(1)}_m+ m^2 a^3 \varphi^{(1)}_m =
  -4{\rho_0 \over \lambda}I_{m0} {k^2}a\varphi^{(0)}_0,
\end{equation}
where $I_{m0}$ is a transfer matrix coefficient.
The numerical
integration of the equations~\cite{kkt} confirms the effect of
massive mode generation and consequent damping of the zero-mode.


\subsection{\new{Full numerical solutions}}
\label{section_7.2}

\epubtkUpdateA{Added Subsection~\ref{section_7.2} ``Full numerical
  solutions''.}

Full numerical solutions for the tensor perturbations have been
obtained by the two methods -- the pseudo-spectral (PS) method used
in~\cite{Hiramatsu:2006cv, Hiramatsu:2006bd} and the characteristic
integration (CI) algorithm developed in~\cite{Cardoso:2006nh}. It
was shown that both methods give identical results and the behaviour
of gravitational waves on a brane is quite insensitive to the initial
conditions in the bulk as for the scalar
perturbations~\cite{Seahra:2006tm}. Here we summarize the results
obtained in~\cite{Hiramatsu:2006bd}. It is convenient to use the
static bulk metric and consider a moving brane. The simplest initial
condition is $f(z, \tau)$~=~const when the mode is outside the hubble
horizon on the brane. If the brane is static, this would give a
zero-mode solution $f = \cos (k (\tau -\tau_0))$ where $\tau_0$ is an
initial time. However, due to the motion of the brane, which causes
the expansion of the brane universe, KK modes are excited and this
solution is modified. Figure~\ref{figurev2_04} demonstrate this
effect. Once the perturbation enters the horizon, non-trivial waves
are excited in the bulk and the amplitude of the tensor perturbation
is damped. Figure~\ref{figurev2_05} shows the behaviour of gravitational
waves for two different wave numbers. Here $\epsilon_* = \rho/\lambda$
at the time when the mode re-enters the horizon. As for scalar
perturbations, we can define the effective 4D solutions by ignoring
the bulk as a reference
\begin{equation}
\ddot{h}_{\mathrm{ref}} + 3 H \dot{h}_{\mathrm{ref}} + \frac{k^2}{a(t)^2} h_{\mathrm{ref}} =0.
\label{eq:reference}
\end{equation}
$h_{\mathrm{ref}}$ only takes into account the effect of the high-energy
modification of the Friedmann equation. Figure~\ref{figurev2_05} shows
that the full solution has an additional suppression of the amplitude
compared with $h_{\mathrm{ref}}$. This suppression is caused by the
excitations of KK modes at the horizon crossing as is seen in
Figure~\ref{figurev2_04}. The suppression is stronger for modes that
enter the horizon earlier $\epsilon_* >1$ and it becomes negligible at
low energies $\epsilon_* <1$.

\epubtkImage{figurev204.png}{%
  \begin{figure}[htbp]
    \centerline{
      \includegraphics[scale=0.6]{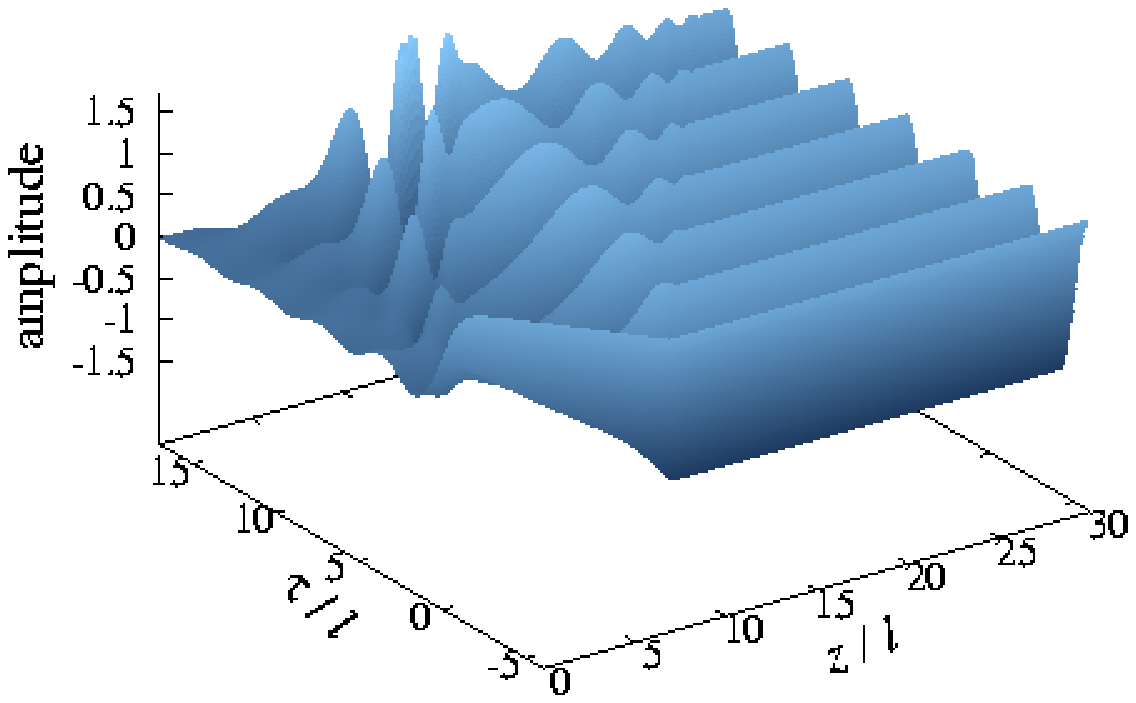}
      \includegraphics[scale=0.6]{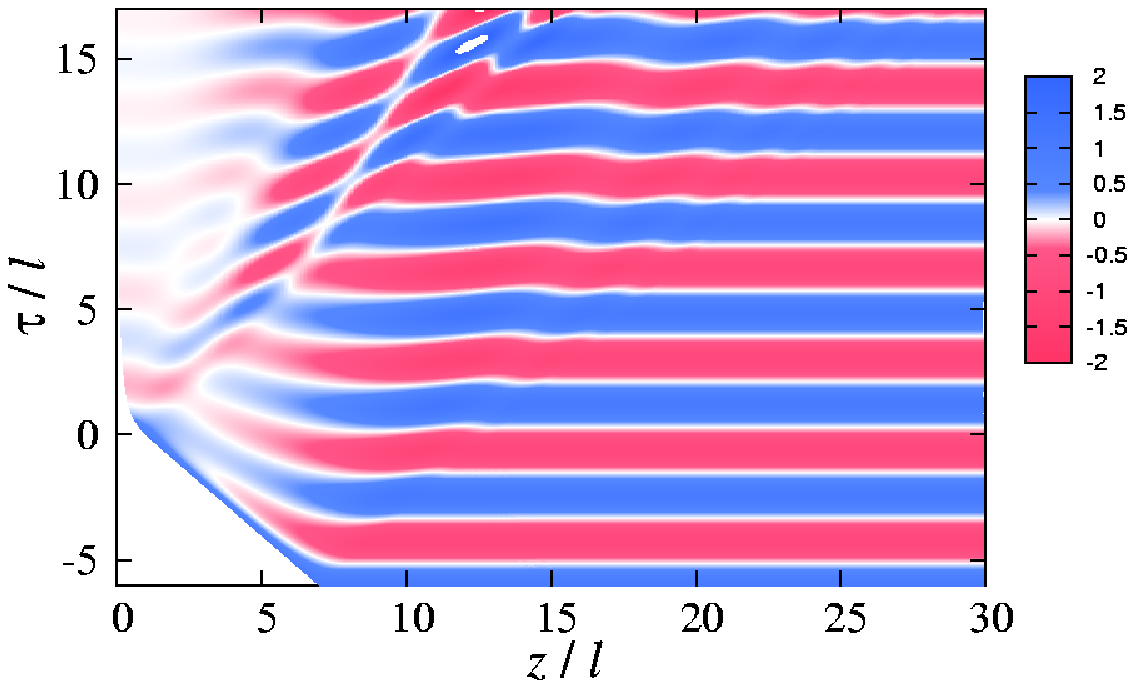}
    }
   \caption{The evolution of gravitational waves. We set the comoving
   wave number to $k=\sqrt{3}/\ell$ ($\epsilon_*=1.0$). The right
   panel depicts the projection of the three-dimensional waves of the
   left panel. Figure taken from~\cite{Hiramatsu:2006bd}.}
    \label{figurev2_04}
  \end{figure}}

The ratio $|h_{\mathrm{5D}}/h_{\mathrm{ref}}|$ evaluated at the low-energy
regime long after the horizon re-entry time monotonically decreases
with the frequency and the suppression of amplitude $h_{\mathrm{5D}}$ becomes
significant above the critical frequency $f_{\mathrm{crit}}$
given by
\begin{equation}
 \begin{aligned}
 f_{\mathrm{crit}} &= \frac{1}{2\pi\ell}\frac{a_{\mathrm{crit}}}{a_{\mathrm{eq}}}\frac{a_{\mathrm{eq}}}{a_{0}}\\
  &= 5.6\times 10^{-5}\mathrm{\ Hz}
  \left(\frac{\ell}{0.1\mathrm{\ mm}}\right)^{-1/2}
  \left(\frac{H_0}{72\mathrm{\ km/s}\cdot \mathrm{Mpc}}\right)^{1/2}
  \left(\frac{1+z_{\mathrm{eq}}}{3200}\right)^{-1/4}.
 \end{aligned}
\label{eq:critical_frequency}
\end{equation}
This corresponds to a frequency of the mode that enters the horizon when
$H_*=\ell^{-1}$ (cf.~\cite{Hogan:2000wz}). The ratio $|h_{\mathrm{5D}}/h_{\mathrm{ref}}|$
obtained from numerical solutions is fitted as
\begin{equation}
  \left|\frac{h_{\mathrm{5D}}}{h_{\mathrm{ref}}}\right| =
\alpha\left(\frac{f}{f_{\text{crit}}}\right)^{-\beta}
  \label{eq:fitting}
\end{equation}
with $\alpha=0.76\pm0.01$ and $\beta=0.67\pm0.01$.

\epubtkImage{figurev205.png}{%
  \begin{figure}[htbp]
    \def\epsfsize#1#2{0.8#1}
    \centerline{\epsfbox{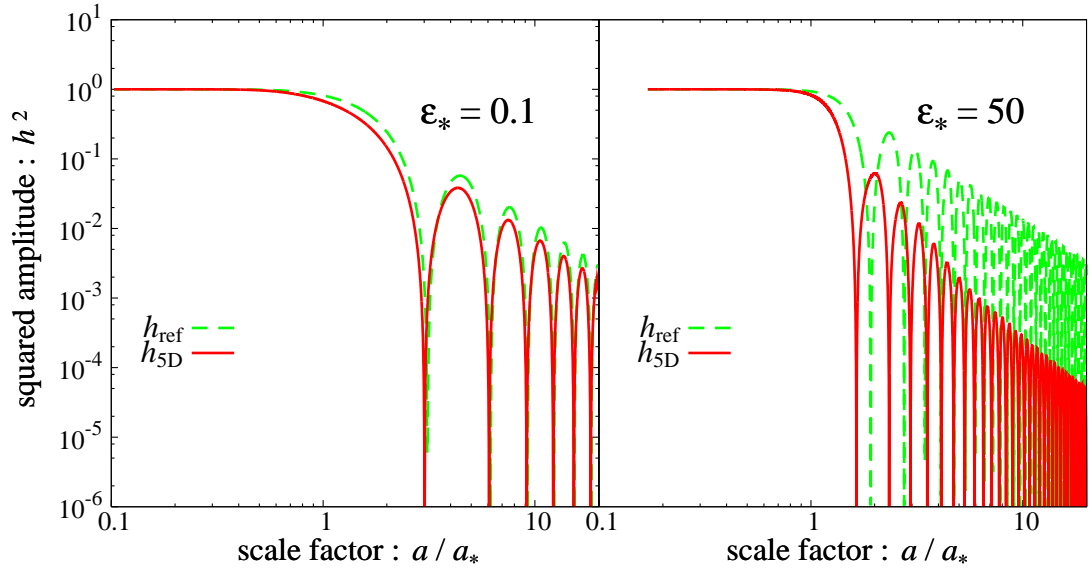}}
    \caption{Squared amplitude of gravitational waves on the brane in
    the low-energy (\emph{left}) and the high-energy (\emph{right})
    regimes. In both panels, solid lines represent the numerical
    solutions. The dashed lines are the amplitudes of reference
    gravitational waves $h_{\mathrm{ref}}$ obtained from
    Equation~(\ref{eq:reference}). Figure taken
    from~\cite{Hiramatsu:2006bd}.}
    \label{figurev2_05}
\end{figure}}

There are two important effects on the spectrum in the high-energy regime.
Let us first consider the non-standard cosmological expansion due to the
$\rho^2$-term. The spectrum of the stochastic gravitational waves is modified to
\begin{equation}
 \Omega_{\mathrm{ref}}=
 \begin{cases}
  \displaystyle f^{\frac{6w-2}{3w+1}} & \mathrm{for}\;\;f < f_{\mathrm{crit}},\\
  \displaystyle f^{\frac{6w+2}{3w+2}} & \mathrm{for}\;\;f > f_{\mathrm{crit}},
 \end{cases}
 \label{eq:Omega_w}
\end{equation}
where  $w$ is an equation of state. This is because gravitational waves re-enter the
horizon when the $\rho^2$-term dominates at high frequencies $f>f_{\mathrm{crit}}$.
In the high-energy radiation dominated phase, the spectrum of the stochastic gravitational
waves is modified to
\begin{equation}
 \Omega_{\mathrm{ref}} \propto
  f^{4/3} \qquad (f_{\mathrm{crit}} < f).
 \label{eq:O_propto_f_specific_5D}
\end{equation}
The other effect is the KK-mode excitations.
Taking account of the KK-mode excitations, the spectrum is calculated as
\begin{equation}
  \Omega_{\mathrm{GW}} =\left|\frac{h_{\mathrm{5D}}}{h_{\mathrm{ref}}}\right|^2\Omega_{\mathrm{ref}},
  \label{eq:Omega_GW_ref}
\end{equation}
where we used the fact $\Omega_{\mathrm{GW}}\propto h^2f^2$.
Combining it with the result~(\ref{eq:fitting}), the
spectrum becomes nearly flat above the critical frequency:
\begin{equation}
  \Omega_{\mathrm{GW}} \propto f^{0},
  \label{eq:fitting_Omega}
\end{equation}
which is shown in filled squares in Figure~\ref{figurev2_06}.
In this figure, the spectrum calculated from the
reference gravitational waves $\Omega_{\mathrm{ref}}$ is also shown in filled circles.
Note that the normalization factor of the spectrum is determined as
$\Omega_{\mathrm{GW}} = 10^{-14}$ from the CMB constraint.
The short-dashed line and the solid line represent
asymptotic behaviors in the high-frequency region. The spectrum taking
account of the two high-energy effects seems almost indistinguishable
from the standard four-dimensional prediction shown in long-dashed line
in the figure. In other words, while the effect due to the non-standard
cosmological expansion enhances the spectrum, the KK-mode effect reduces
the GW amplitude, which results in the same spectrum as the one
predicted in the four-dimensional theory. Note that the
amplitude taking account of the two effects near $f\approx f_{\mathrm{crit}}$
is slightly suppressed, which agrees with the results in the previous
study for $\epsilon_*\leq0.3$ using the Gaussian-normal
coordinates~\cite{kkt} discussed in the previous subsection.

\epubtkImage{figurev206.png}{%
  \begin{figure}[htbp]
   \def\epsfsize#1#2{0.8#1}
   \centerline{\epsfbox{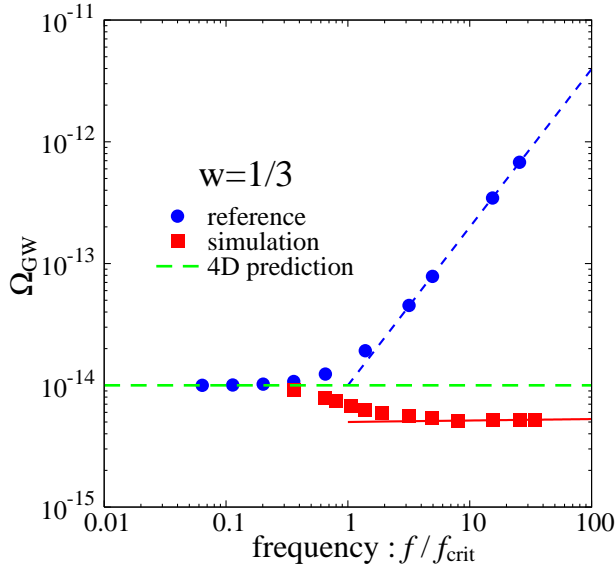}}
   \caption{The energy spectrum of the stochastic background of gravitational wave
   around the critical
 frequency $f_c$ in radiation dominated epoch. The filled circles represent the spectrum caused by
 the non-standard cosmological expansion of the
 universe. Taking account of the KK-mode excitations, the spectrum
 becomes the one plotted by filled squares. In the
 asymptotic region depicted in the solid line, the frequency dependence
 becomes almost the same as the one predicted in the four-dimensional theory
 (long-dashed line). Figure taken from~\cite{Hiramatsu:2006bd}.}
 \label{figurev2_06}
\end{figure}}
This cancelation of two high energy effects is valid only for $w=1/3$.
For other equations of state, the final spectrum at high frequencies are
different from 4D predictions. For example for $w=1$,
$\Omega_{\mathrm{GW}} \propto f^{2/5}$ for $f>f_c$ while
the 4D theory predicts $\Omega_{\mathrm{GW}} \propto f^{1}$.

\newpage


\section{CMB Anisotropies in Brane-World Cosmology}
\label{section_8}

For the CMB anisotropies, one needs to consider a multi-component
source. Linearizing the general nonlinear expressions for the
total effective energy-momentum tensor, we obtain
\begin{eqnarray}
  \rho_\mathrm{tot} &=&
  \rho\left(1 +\frac{\rho}{2\lambda} + \frac{\rho_{\cal E}}{\rho}\right),
  \\ \label{e:pressure1}
  p_\mathrm{tot}&=& p + \frac{\rho}{2\lambda}
  (2p+\rho)+\frac{\rho_{\cal E}}{3},
  \\
  q^\mathrm{tot}_\mu &=& q_\mu \left(1+ \frac{\rho}{\lambda}\right)+
  q^{\cal E}_\mu,
  \\ \label{ep2}
  \pi^\mathrm{tot}_{\mu\nu} &=& \pi_{\mu\nu}
  \left(1-\frac{\rho+3p}{2\lambda}\right)+\pi^{\cal E}_{\mu\nu},
\end{eqnarray}%
where
\begin{equation}
  \rho=\sum_i\rho_{(i)},
  \qquad
  p=\sum_ip_{(i)},
  \qquad
  q_\mu=\sum_iq^{(i)}_\mu
\end{equation}
are the total matter-radiation density, pressure, and momentum
density, respectively, and $\pi_{\mu\nu}$ is the photon anisotropic
stress (neglecting that of neutrinos, baryons, and CDM).

The perturbation equations in the previous Section~\ref{section_7}
form the basis for an analysis of scalar and tensor CMB anisotropies
in the brane-world. The full system of equations on the brane,
including the Boltzmann equation for photons, has been given for
scalar~\cite{l1} and tensor~\cite{l2} perturbations. But the
systems are not closed, as discussed above, because of the
presence of the KK anisotropic stress  $\cp_{\mu\nu}$, which acts
a source term.

In the tight-coupling radiation era, the scalar perturbation
equations may be decoupled to give an equation for the
gravitational potential $\Phi$, defined by the electric part of
the brane Weyl tensor (not to be confused with ${\cal
E}_{\mu\nu}$):
\begin{equation}
  E_{\mu\nu}=\D_{\langle\mu}\D_{\nu\rangle}\Phi.
\end{equation}
In general relativity, the equation in $\Phi$ has no source term,
but in the brane-world there is a source term made up of
$\cp_{\mu\nu}$ and its time-derivatives. At low energies
($\rho\ll\lambda$), and for a flat background ($K=0$), the
equation is~\cite{l1}
\begin{equation}
  3x\Phi_k''+12\Phi_k'+x\Phi_k={\mathrm{const.} \over \lambda}
  \left[ \pi_k^{{\cal E}\prime\prime}-
  {1\over x} {\pi_k^{{\cal E}\prime}} +
  \left({2\over x^3}- {3 \over x^2}+{1\over x} \right) \cp_k\right],
\end{equation}%
where $x=k/(aH)$, a prime denotes $d/dx$, and $\Phi_k$ and $\cp_k$
are the Fourier modes of $\Phi$ and $\cp_{\mu\nu}$, respectively. In
general relativity the right hand side is zero, so that the equation
may be solved for $\Phi_k$, and then for the remaining perturbative
variables, which gives the basis for initializing CMB numerical
integrations. At high energies, earlier in the radiation era, the
decoupled equation is fourth order~\cite{l1}:
\begin{eqnarray}
  && 729 x^2\Phi_k''''+3888x\Phi_k'''+(1782+54x^2) \Phi_k'' +
  144x \Phi_k'+(90+x^2)\Phi_k
  \nonumber \\
  && \qquad \qquad =\mathrm{const.} \times
  \left[243\left( {\cp_k\over\rho} \right)^{\prime\prime\prime\prime} \!\!\!-
  {810 \over x} \left({\cp_k\over\rho}\right)^{\prime\prime\prime} \!\!\!+
  {18(135+2x^2) \over x^2}\left({\cp_k\over\rho}\right)^{\prime\prime}\right.
  \nonumber \\
  && \qquad \qquad \qquad \qquad \quad \; \left. - {30(162+x^2)\over x^3}
  \left({\cp_k\over\rho}\right)^{\prime} + {x^4+30(162+x^2)\over x^4}
  \left({\cp_k\over\rho}\right)\right].
\end{eqnarray}%

The formalism and machinery are ready to compute the temperature
and polarization anisotro\-pies in brane-world cosmology, once a
solution, or at least an approximation, is given for
$\cp_{\mu\nu}$. The resulting power spectra will reveal the nature
of the brane-world imprint on CMB anisotropies, and would in
principle provide a means of constraining or possibly falsifying
the brane-world models. Once this is achieved, the implications
for the fundamental underlying theory, i.e., M~theory, would need
to be explored.

However, the first step required is the solution for
$\cp_{\mu\nu}$. This solution will be of the form given in
Equation~(\ref{e:soln}). Once ${\cal G}$ and $F_k$ are determined or
estimated, the numerical integration in Equation~(\ref{e:soln}) can in
principle be incorporated into a modified version of a CMB
numerical code. The full solution in this form represents a
formidable problem, and one is led to look for approximations.


\subsection{The low-energy approximation}

The basic idea of the low-energy approximation~\cite{sod_1, sod_2,
  sod_3, sod_4, sod_5} is to use a gradient expansion to exploit the
fact that, during most of the history of the universe, the curvature
scale on the observable brane is much greater than the curvature scale
of the bulk ($\ell < 1 \mathrm{\ mm}$):
\begin{equation}
  L\sim |R_{\mu\nu\alpha\beta}|^{-1/2} \gg
  \ell \sim | {}^{(5)\!}R_{ABCD}|^{-1/2}
  \quad \Rightarrow \quad
  |\nabla_\mu|\sim L^{-1} \ll |\partial_y|\sim \ell^{-1}.
  \label{grad}
\end{equation}
These conditions are equivalent to the low energy regime, since
$\ell^2\propto \lambda^{-1}$ and $|R_{\mu\nu\alpha\beta}|\sim
|T_{\mu\nu}|$:
\begin{equation}
  {\ell^2 \over L^2} \sim {\rho \over \lambda} \ll 1.
\end{equation}
Using Equation~(\ref{grad}) to neglect appropriate gradient terms in an
expansion in $\ell^2/L^2$, the low-energy equations can be solved.
However, two boundary conditions are needed to determine all
functions of integration. This is achieved by introducing a second
brane, as in the RS 2-brane scenario. This brane is to be thought
of either as a regulator brane, whose backreaction on the
observable brane is neglected (which will only be true for a
limited time), or as a shadow brane with physical fields, which
have a gravitational effect on the observable brane.

The background is given by low-energy FRW branes with tensions
$\pm\lambda$, proper times $t_\pm$, scale factors $a_\pm$, energy
densities $\rho_\pm$ and pressures $p_\pm$, and dark radiation
densities $\rho_{{\cal E}\,\pm}$. The physical distance between
the branes is $\ell \bar{d}(t)$, and
\begin{equation}
  {d\over dt_-}=e^{\bar{d}}\,{d \over dt_+},
  \qquad
  a_-=a_+e^{-{\bar{d}}},
  \qquad
  H_-=e^{\bar{d}}\left(H_+- \dot{\bar{d}}\right),
  \qquad
  \rho_{{\cal E}\,-}=e^{4{\bar{d}}}\rho_{{\cal E}\,+}.
\end{equation}
Then the background dynamics is given by
\begin{eqnarray}
  H_\pm^2 &=& \pm{\kappa^2 \over 3}
  \left(\rho_\pm \pm\rho_{{\cal E}\,\pm} \right),
  \\
  \ddot {\bar{d}}+3H_+\dot {\bar{d}}-\dot{{\bar{d}}}^2 &=&
  {\kappa^2 \over 6}
  \left[ \rho_+-3p_+ +e^{2{\bar{d}}}(\rho_--3p_-)\right].
  \label{dbar}
\end{eqnarray}%
(see~\cite{lang2_1, lang2_2} for the general background, including the
high-energy regime). The dark energy obeys
$\rho_{{\cal E}\,+} =C/a_+^4$, where $C$ is a constant. From now on,
we drop the +-subscripts which refer to the physical, observed
quantities.

The perturbed metric on the observable (positive tension) brane is
described, in longitudinal gauge, by the metric perturbations
$\psi$ and ${\cal R}$, and the perturbed radion is $d= {\bar{d}}+
N$. The approximation for the KK (Weyl) energy-momentum tensor on
the observable brane is
\begin{equation}
  {\cal E}^{\mu}{}_{\nu} = \frac{2}{e^{2d}-1}
  \left[-\frac{\kappa^2}{2}
  \left( T^{\mu}_{\nu}+e^{-2 d}T^{\mu}_{-~\nu}\right) -
  \nabla^{\mu}\nabla_{\nu} d +
  \delta^{\mu}_{\nu} \nabla^2 d -
  \left(\nabla^{\mu}d \nabla_{\nu} d +
  \frac{1}{2} \delta^{\mu}_{\nu} (\nabla d)^2 \right)\right]\!,
  \label{solE}
\end{equation}
and the field equations on the observable brane can be written in
scalar-tensor form as
\begin{equation}
  G^\mu{}_\nu = \frac{\kappa^2}{\chi} T^\mu_{\nu}+
  \frac{\kappa^2(1-\chi)^2}{\chi} T^\mu_{-~\nu} +
  \frac{1}{\chi}\left(\nabla^\mu \nabla_\nu \chi-
  \delta^\mu_\nu \nabla^2\chi \right) +
  \frac{\omega(\chi)}{\chi^2}\left(\nabla^\mu \chi \nabla_\nu \chi-
  \frac{1}{2}\delta^\mu_\nu (\nabla \chi)^2 \right)\!,
\end{equation}
where
\begin{equation}
  \chi=1-e^{-2d},
  \qquad
  \omega(\chi)=\frac{3}{2}\frac{\chi}{1-\chi}.
\end{equation}

The perturbation equations can then be derived as generalizations
of the standard equations. For example, the $\delta G^0{}_0$
equation is~\cite{koyn}
\begin{eqnarray}
  H^2 \psi - H \dot{\cal R}- \frac{1}{3}{ k^2 \over a^{2}} {\cal R}
  &=& - \frac{1}{6} \kappa^2 \frac{e^{2 \bar d}}{e^{2 \bar d}-1}
  \left(\delta \rho + e^{-4 \bar d} \delta \rho_-\right) +
  \frac{2}{3} \kappa^2 \frac{e^{2 \bar d}}{e^{2 \bar d}-1} \cu\, N
  \nonumber \\
  &&{} - \frac{1}{e^{2\bar d}-1} \left[ \left(\dot{\bar d}-H \right)
  \dot{N} + \left(\dot{\bar d}-H \right)^2 N -\dot{\bar d}^2 \psi +
  2 H \dot{\bar d} \psi - \dot{\bar d} \dot{\cal R}-
  \frac{1}{3}{k^2\over a^{2}} N \right].
  \nonumber \\
\end{eqnarray}%
The trace part of the perturbed field equation shows that the
radion perturbation determines the crucial quantity,
$\delta\pi_{\cal E}$:
\begin{equation}
  {\cal R}+\psi = -{2 \over e^{2\bar d}-1}N=
  -\kappa^2a^2\delta \pi_{\cal E},
  \label{rplusp}
\end{equation}
where the last equality follows from Equation~(\ref{metcurv}). The
radion perturbation itself satisfies the wave equation
\begin{eqnarray}
  \ddot{N} &+& \left(3H -2 \dot{\bar d}\right) \dot{N} -
  \left(2 \dot{H}+4 H^2+2 \dot{\bar d}^2-
  6 H \dot{\bar d}-2 \ddot{\bar d}\right)N + {k^2 \over a^{2}} N
  \nonumber \\
  &&{} -\dot{\bar d} \dot{\psi}+3 \dot{\bar d} \dot{\cal R}+
  \left(-2 \ddot{\bar d}-6 H \dot{\bar d}+2 \dot{\bar d}^2\right) \psi
  \nonumber \\
  &&{}= \frac{\kappa^2}{6} \left[ \delta \rho -3 \delta p +
  e^{-2 \bar d}(\delta \rho_--3 \delta p_-) \right].
\end{eqnarray}%

A new set of variables $\varphi_\pm, E$ turns out be very
useful~\cite{ksw, koy}:
\begin{eqnarray}
  {\cal R} &=&  -\varphi_+ - {a^{2}\over k^2} H \dot{E} + \frac{1}{3} E,
  \nonumber \\
  \psi &=& - \varphi_+ - {a^{2}\over k^2} (\ddot{E}+ 2H \dot{E}),
  \nonumber \\
  N &=& \varphi_- - \varphi_+ -  {a^{2}\over k^2}\dot{\bar d} \dot{E}.
\end{eqnarray}%
Equation~(\ref{rplusp}) gives
\begin{equation}
  \ddot{E}+ \left( 3 H +
  \frac{2 \dot{\bar d}}{e^{2 \bar d}-1}\right) \dot{E} -
  \frac{1}{3} {k^2 \over a^2} E =
  -\frac{2 e^{2 \bar d}}{e^{2 \bar d}-1} {k^2 \over a^2}
  \left(\varphi_+ - e^{-2\bar d} \varphi_-\right).
\end{equation}
The variable $E$ determines the metric shear in the bulk, whereas
$\varphi_\pm$ give the brane displacements in transverse
traceless gauge. The latter variables have a simple relation to
the curvature perturbations on large scales~\cite{ksw, koy}
(restoring the +-subscripts):
\begin{equation}
  \zeta_{\mathrm{tot}\,\pm} =-\varphi_\pm +{H_\pm^2 \over \dot H_\pm}
  \left({\dot{\varphi}_\pm \over H_\pm}+ \varphi_\pm \right),
\end{equation}
where $\dot{f}_\pm\equiv df_\pm/dt_\pm$.


\subsection{The simplest model}

The simplest model is the one in which
\begin{equation}
  \cu=0=\dot{\bar d}
\end{equation}
in the background, with $p_-/\rho_-=p/\rho$. The regulator brane
is assumed to be far enough away that its effects on the physical
brane can be neglected over the timescales of interest. By
Equation~(\ref{dbar}) it follows that
\begin{equation}
  \rho_-=-\rho e^{2\bar d},
\end{equation}
i.e., the matter on the regulator brane must have fine-tuned and
negative energy density to prevent the regulator brane from moving
in the background. With these assumptions, and further assuming
adiabatic perturbations for the matter, there is only one
independent brane-world parameter, i.e., the parameter measuring
dark radiation fluctuations:
\begin{equation}
  \delta C_{*}= {\delta\cu \over \rho_\mathrm{rad}}.
\end{equation}

\epubtkImage{figure12.png}{%
 \begin{figure}[htbp]
   \def\epsfsize#1#2{0.6#1}
   \centerline{\epsfbox{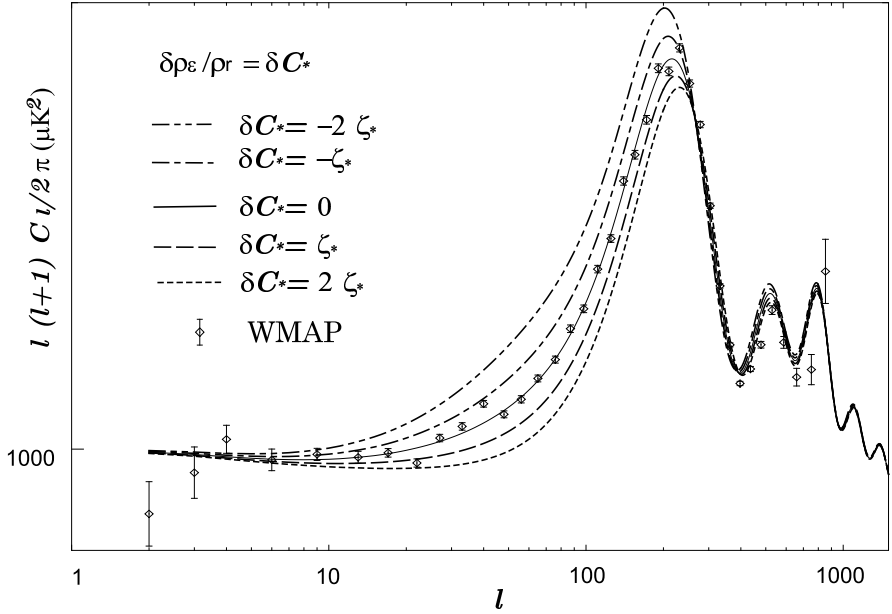}}
   \caption{The CMB power spectrum with brane-world effects,
     encoded in the dark radiation fluctuation parameter $\delta C_*$
     as a proportion of the large-scale curvature perturbation for
     matter (denoted $\zeta_*$ in the plot). (Figure taken
     from~\cite{koy}.)}
   \label{figure_12}
 \end{figure}}

This assumption has a remarkable consequence on large scales: The
Weyl anisotropic stress $\delta\pi_{\cal E}$ terms in the
Sachs--Wolfe formula~(\ref{sachsw}) cancel the entropy
perturbation from dark radiation fluctuations, so that there is no
difference on the largest scales from the standard general
relativity power spectrum. On small scales, beyond the first
acoustic peak, the brane-world corrections are negligible. On
scales up to the first acoustic peak, brane-world effects can be
significant, changing the height and the location of the first
peak. These features are apparent in Figure~\ref{figure_12}. However, it is not
clear to what extent these features are general brane-world
features (within the low-energy approximation), and to what extent
they are consequences of the simple assumptions imposed on the
background. Further work remains to be done.

A related low-energy approximation, using the moduli space
approximation, has been developed for certain 2-brane models with
bulk scalar field~\cite{rbbd_1, rbbd_2}. The effective gravitational action
on the physical brane, in the Einstein frame, is
\begin{equation}
  S_\mathrm{eff}={1\over 2\kappa^2}\int d^4x\sqrt{-g}
  \left[ R-{12\alpha^2 \over 1+2\alpha^2}(\partial \phi)^2-
  {6 \over 1+2\alpha^2} (\partial \chi)^2-V(\phi,\chi)\right],
  \label{mod}
\end{equation}
where $\alpha$ is a coupling constant, and $\phi$ and $\chi$ are
moduli fields (determined by the zero-mode of the bulk scalar
field and the radion). Figure~\ref{figure_13} shows how the CMB anisotropies
are affected by the $\chi$-field.

\epubtkImage{figure13.png}{%
 \begin{figure}[htbp]
   \def\epsfsize#1#2{0.85#1}
   \centerline{\epsfbox{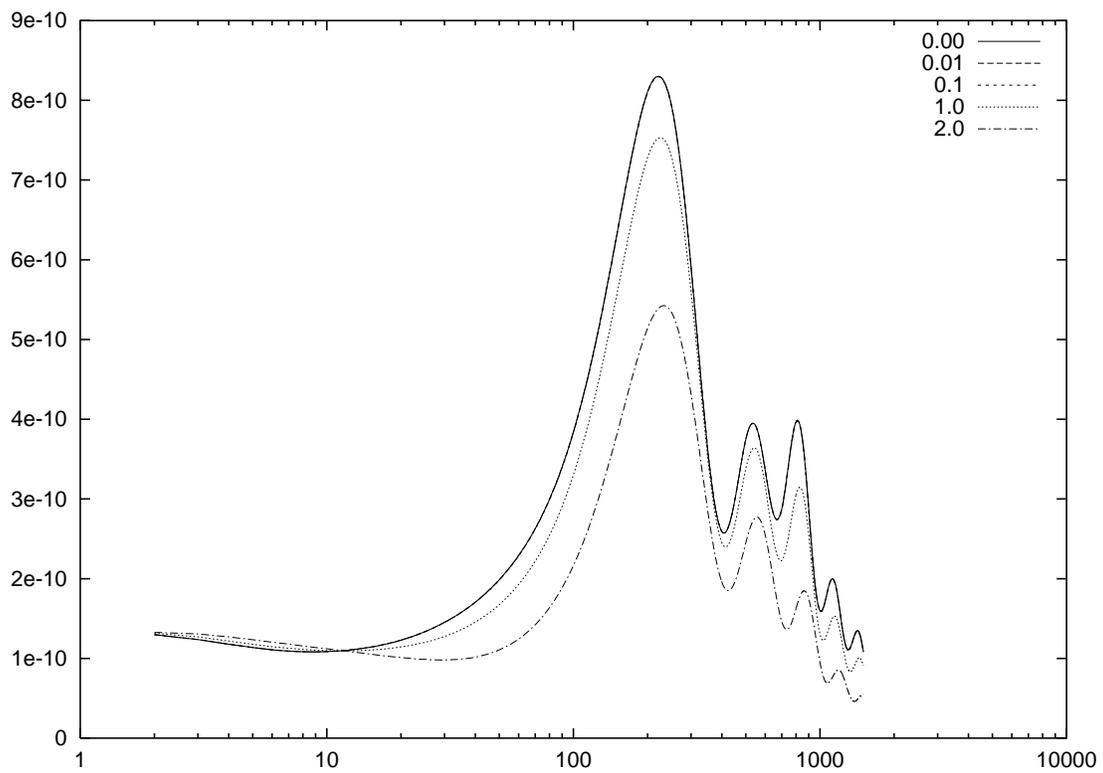}}
   \caption{The CMB power spectrum with brane-world moduli effects
     from the field $\chi$ in Equation~(\ref{mod}). The curves are labelled
     with the initial value of $\chi$. (Figure taken
     from~\cite{rbbd_1, rbbd_2}.)}
   \label{figure_13}
 \end{figure}}

\newpage


\section{\new{DGP Models: Modifying Gravity at Low Energies}}
\label{section_9}
\epubtkUpdateA{Added Section~\ref{section_9} ``DGP Models''.}

\subsection{`Self-accelerating' DGP}

Most brane-world models modify general relativity at high
energies. The Randall--Sundrum models discussed up to now are a
typical example. At low energies, $H\ell \ll 1$, the zero-mode of the
graviton dominates on the brane, and general relativity is recovered
to a good approximation. At high energies, $H\ell\gg 1$, the massive
modes of the graviton dominate over the zero-mode, and gravity on the
brane behaves increasingly 5-dimensional. On the unperturbed FRW
brane, the standard energy-conservation equation holds, but the
Friedmann equation is modified by an ultraviolet correction,
$(G\ell\rho)^2$. At high energies, gravity ``leaks'' off the brane and
$H^2\propto \rho^2$.

By contrast, the brane-world model of
Dvali--Gabadadze--Porrati~\cite{Dvali:2000hr} (DGP), which was
generalized to cosmology by Deffayet~\cite{Deffayet:2000uy},
modifies general relativity at \emph{low} energies. This model
produces `self-acceleration' of the late-time universe due to a
weakening of gravity at low energies. Like the RS model, the DGP
model is a 5D model with infinite extra dimensions.\epubtkFootnote{DGP
  brane-worlds without $Z_2$ symmetry have also been considered; see
  e.g.~\cite{varun2}.}

The action is given by
\begin{equation}
\label{DGPaction}
{-1\over 16\pi G}\left[ {1\over r_c}\int_{\mathrm{bulk}}
d^5x\,\sqrt{-g^{(5)}}\,R^{(5)}+\int_{\mathrm{brane}} d^4x\,\sqrt{-g}\,R
\right] \,.
\end{equation}
The bulk is assumed to be 5D Minkowski spacetime. Unlike the AdS
bulk of the RS model, the Minkowski bulk has infinite volume.
Consequently, there is no normalizable zero-mode of the 4D
graviton in the DGP brane-world. Gravity leaks off the 4D brane
into the bulk at large scales, $r\gg r_c$, where the first term in
the sum~(\ref{DGPaction}) dominates. On small scales, gravity is
effectively bound to the brane and 4D dynamics is recovered to a
good approximation, as the second term dominates. The transition
from 4D to 5D behaviour is governed by the crossover scale $r_c$.
For a Minkowski brane, the weak-field gravitational potential
behaves as
\begin{equation}
\psi \propto \left\{ \begin{array}{lll} r^{-1} & \mbox{for} & r\ll
r_c
\\ r^{-2} & \mbox{for} & r\gg r_c \end{array}\right. \,.
\end{equation}
On a Friedmann brane, gravity leakage at late times in the
cosmological evolution can initiate acceleration -- not due to any
negative pressure field, but due to the weakening of gravity on
the brane. 4D gravity is recovered at high energy via the lightest
massive modes of the 5D graviton, effectively via an ultra-light
meta-stable graviton.


\epubtkImage{figurev207.png}{%
  \begin{figure}[htbp]
    \centerline{
      \includegraphics[scale=0.7]{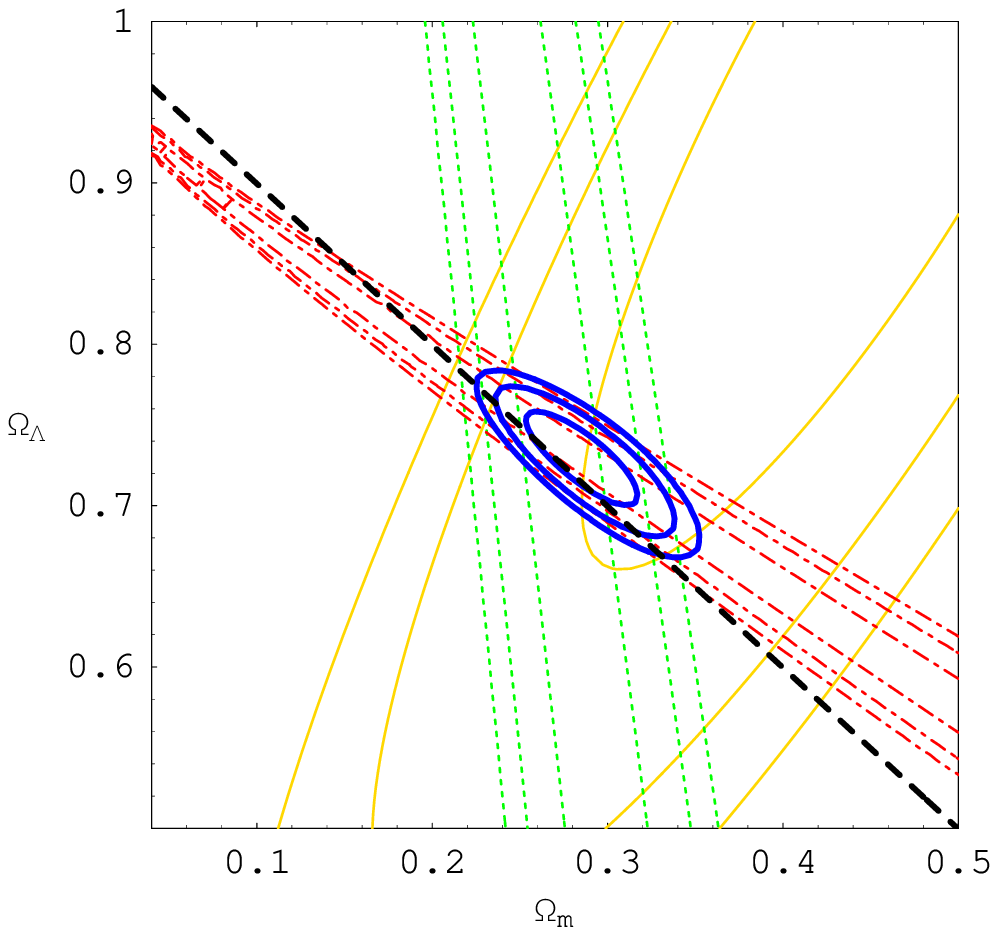}
      \includegraphics[scale=0.75]{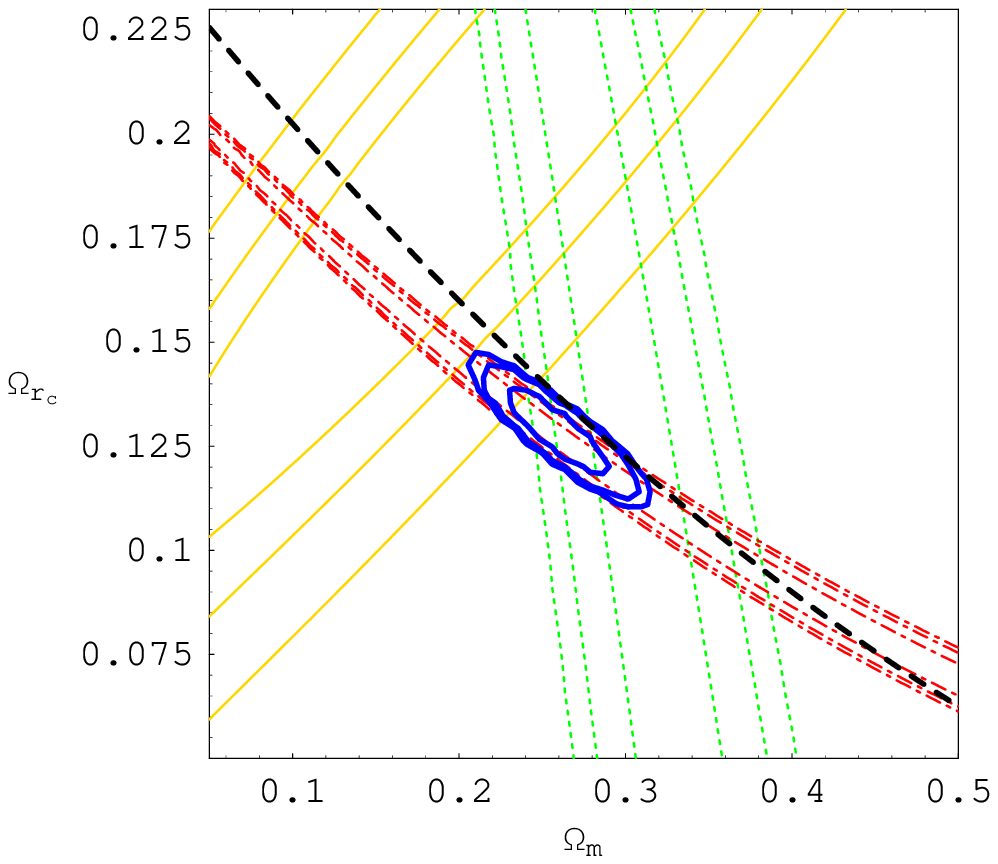}
    }
    \caption{Joint constraints [solid thick (blue)] from the SNLS data
      [solid thin (yellow)], the BAO peak at {\it z}~=~0.35 [dotted
      (green)] and the CMB shift parameter from WMAP3 [dot-dashed
      (red)]. The left plot shows LCDM models, the right plot shows
      DGP. The thick dashed (black) line represents the flat models,
      $\Omega_K=0$. (From~\cite{Maartens:2006yt}.)}
    \label{sac}
\end{figure}}

\epubtkImage{figurev208.png}{%
  \begin{figure}
    \centerline{\includegraphics[width=84mm]{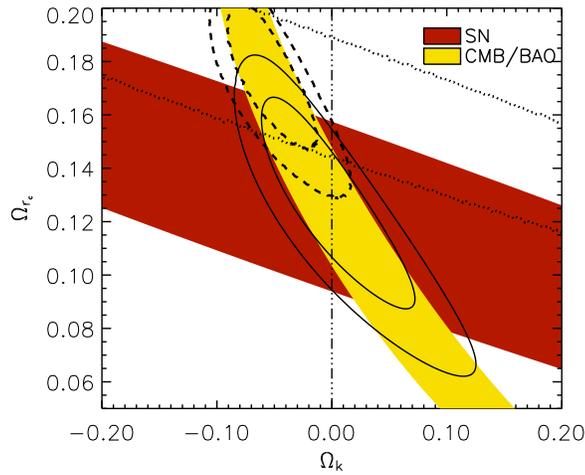}}
    \caption{The constraints from SNe and CMB/BAO on the parameters in
      the DGP model. The flat DGP model is indicated by the vertical
      dashed-dotted line; for the MLCS light curve fit, the flat model
      matches to the date very well. The SALT-II light curve fit to
      the SNe is again shown by the dotted contours. The combined
      constraints using the SALT-II SNe outlined by the dashed
      contours represent a poorer match to the CMB/BAO for the flat
      model. (From~\cite{Sollerman:2009yu}.)}
    \label{fig:dgpcomb}
\end{figure}}

The energy conservation equation remains the same as in general
relativity, but the Friedmann equation is modified~\cite{Deffayet:2000uy}:
\begin{eqnarray}
  \dot\rho+3H(\rho+p)&=&0\,,\label{ec} \\  H^2+{K\over a^2}-{1 \over
r_c}\sqrt{H^2+{K\over a^2}}&=& {8\pi G \over 3}\rho\,.
  \label{f}
\end{eqnarray}
The Friedmann equation can be derived from the junction condition
or the Gauss-Codazzi equation as in the RS model.
To arrive at Equation~(\ref{f}) we have to take a square root which
implies a choice of sign. As we shall see, the above choice has the
advantage of leading to acceleration -- but at the expense of a
`ghost' (negative kinetic energy) mode in the scalar graviton
sector. The `normal' (non-self-accelerating) DGP model, where the
opposite sign of the square root is chosen, has no ghost, and is
discussed below.

From Equation~(\ref{f}) we infer that at early times, i.e., $Hr_c \gg 1$,
the general relativistic Friedman equation is recovered. By contrast, at late
times in an expanding CDM universe, with $\rho\propto a^{-3}\to0$,
we have
\begin{equation}
H\to H_\infty= {1\over r_c}\,,
\end{equation}
so that expansion accelerates and is asymptotically de Sitter. The
above equations imply
\begin{equation}
\dot H - {K\over a^2}=-4\pi G\rho\left[1+ {1 \over \sqrt{1+32\pi
Gr_c^2\rho /3}} \right].
 \end{equation}
In order to achieve self-acceleration at late times, we require
 \begin{equation}
 r_c\gtrsim H_0^{-1}\,,
 \end{equation}
since $H_0\lesssim H_\infty$. This is confirmed by fitting
supernova observations, as shown in Figures~\ref{sac} and \ref{fig:dgpcomb}.
The dimensionless cross-over parameter is defined as
\begin{equation}
 \Omega_{r_c}={1\over 4(H_0r_c)^2}\,, \label{orc}
 \end{equation}
and the LCDM relation,
\begin{equation}
 \Omega_m+\Omega_\Lambda+\Omega_K=1\,,
\end{equation}
is modified to
\begin{equation}
 \Omega_m+ 2\sqrt{\Omega_{r_c}}\sqrt{1-\Omega_K}+\Omega_K=1\,.
\label{odgp}
\end{equation}

LCDM and DGP can both account for the supernova observations, with
the fine-tuned values $\Lambda\sim H_0^2$ and $r_c\sim H_0^{-1}$,
respectively. When we add further constraints to the expansion
history from the baryon acoustic oscillation peak at $z=0.35 $ and
the CMB shift parameter, the DGP flat models are in strong tension
with the data, whereas LCDM models provide a consistent fit. This is
evident in Figures~\ref{sac} and \ref{fig:dgpcomb} though this conclusion
depends on a choice of light curve fitters in the SNe observations.
The open DGP models provide a somewhat
better fit to the geometric data -- essentially because the lower
value of $\Omega_m$ favoured by supernovae reduces the distance to
last scattering and an open geometry is able to extend that
distance. For a combination of SNe, CMB shift and Hubble Key
Project data, the best-fit open DGP also performs better than the
flat DGP~\cite{Song:2006jk}, as shown in Figure~\ref{ssh}.

Observations based on structure formation provide further
evidence of the difference between DGP and LCDM, since the two
models suppress the growth of density perturbations in different
ways~\cite{Lue:2002sw, Lue:2004rj}. The distance-based observations draw only
upon the background 4D Friedman equation~(\ref{f}) in DGP models
-- and therefore there are quintessence models in general
relativity that can produce precisely the same supernova distances
as DGP. By contrast, structure formation observations require the
5D perturbations in DGP, and one cannot find equivalent
quintessence models~\cite{Koyama:2005kd}. One can find 4D general
relativity models, with dark energy that has anisotropic stress and
variable sound speed, which can in principle mimic
DGP~\cite{Kunz:2006ca}. However, these models are highly
unphysical and can probably be discounted on grounds of theoretical
consistency.

\epubtkImage{figurev209.png}{%
  \begin{figure}[htbp]
    \def\epsfsize#1#2{0.5#1}
    \centerline{\epsfbox{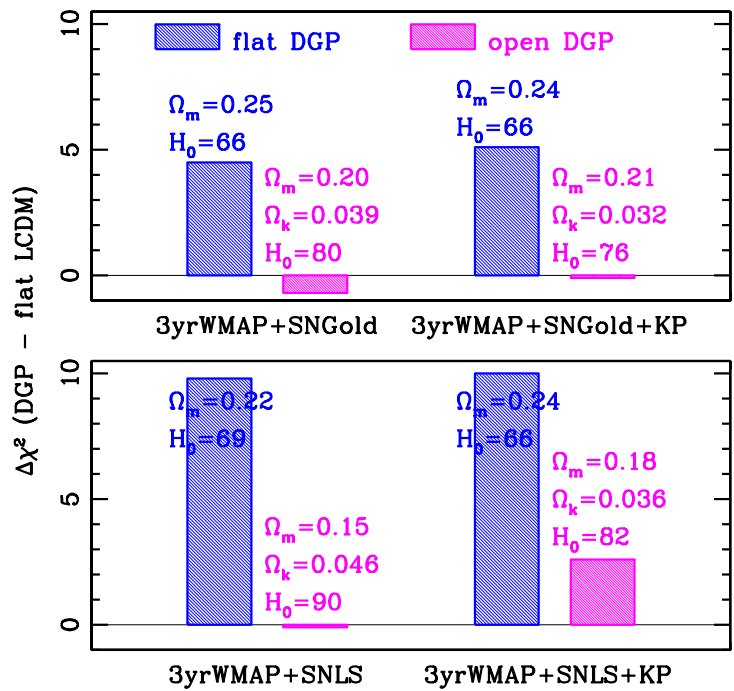}}
    \caption{The difference in $\chi^2$ between best-fit DGP (flat and
      open) and best-fit (flat) LCDM, using SNe, CMB shift and $H_0$
      Key Project data. (From~\cite{Song:2006jk}.)}
    \label{ssh}
\end{figure}}

For LCDM, the analysis of density perturbations is well
understood. For DGP the perturbations are much more subtle and
complicated~\cite{Koyama:2005kd}. Although matter is confined to the 4D
brane, gravity is fundamentally 5D, and the 5D bulk gravitational
field responds to and back-reacts on 4D density perturbations. The
evolution of density perturbations requires an analysis based on
the 5D nature of gravity. In particular, the 5D gravitational
field produces an effective ``dark'' anisotropic stress on the 4D
universe, as discussed in Section~\ref{coneq}. If one neglects this stress and
other 5D effects, and
simply treats the perturbations as 4D perturbations with a
modified background Hubble rate -- then as a consequence, the 4D
Bianchi identity on the brane is violated, i.e., $\nabla^\nu
G_{\mu\nu} \neq 0$, and the results are inconsistent. When the 5D
effects are incorporated~\cite{Koyama:2005kd,Cardoso:2007xc}, the
4D Bianchi identity is automatically satisfied. (See
Figure~\ref{fig:fig1}.)

There are three regimes governing structure formation in DGP
models:

\begin{itemize}
\item
On small scales, below the \emph{Vainshtein radius} (which for
cosmological purposes is roughly the scale of clusters), the
spin-0 scalar degree of freedom becomes strongly coupled, so that
the general relativistic limit is recovered~\cite{Koyama:2007ih}.

\item
On scales relevant for structure formation, i.e., between cluster
scales and the Hubble radius, the spin-0 scalar degree of freedom
produces a scalar-tensor behaviour. A quasi-static approximation
(as in the Newtonian approximation in standard 4D cosmology)
to the 5D perturbations shows that DGP gravity is like a
Brans--Dicke theory with parameter~\cite{Koyama:2005kd}
\begin{eqnarray}
 \omega_{BD} &=&{3 \over 2}(\beta -1),\label{obd}\\
\label{beta}
\beta &=& 1+2H^2r_c\left(H^2+{K \over a^2} \right)^{-1/2}\left[
1+{\dot H \over 3H^2}+{2K \over 3a^2H^2} \right].
\end{eqnarray}
At late times in an expanding universe, when $Hr_c\gtrsim 1$, it
follows that $\beta<1$, so that $\omega_{BD}<0$. (This is a signal of the ghost
pathology in DGP, which is discussed below.)

\item
Although the quasi-static approximation allows us to analytically
solve the 5D wave equation for the bulk degree of freedom, this
approximation breaks down near and beyond the Hubble radius. On
super-horizon scales, 5D gravity effects are dominant, and we need
to solve numerically the partial differential equation governing
the 5D bulk variable~\cite{Cardoso:2007xc}.

\end{itemize}

\epubtkImage{figurev210.png}{%
  \begin{figure}[htbp]
    \def\epsfsize#1#2{0.6#1}
    \centerline{\epsfbox{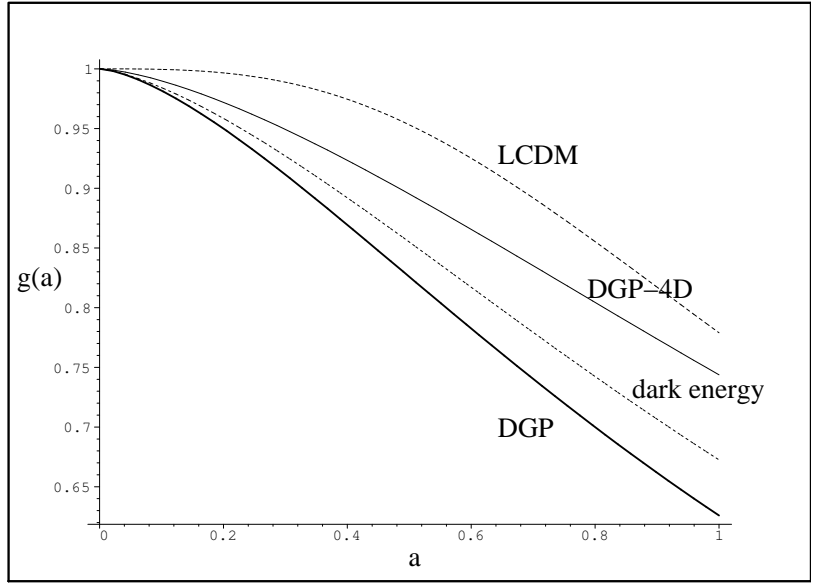}}
    \caption{The growth factor $g(a)=\Delta(a)/a$ for LCDM (long
      dashed) and DGP (solid, thick), as well as for a dark energy
      model with the same expansion history as DGP (short
      dashed). DGP-4D (solid, thin) shows the incorrect result in
      which the 5D effects are set to
      zero. (From~\cite{Koyama:2005kd}.)}
    \label{fig:fig1}
\end{figure}}

\epubtkImage{figurev211.png}{%
  \begin{figure}[htbp]
    \centerline{
      \includegraphics[scale=0.8]{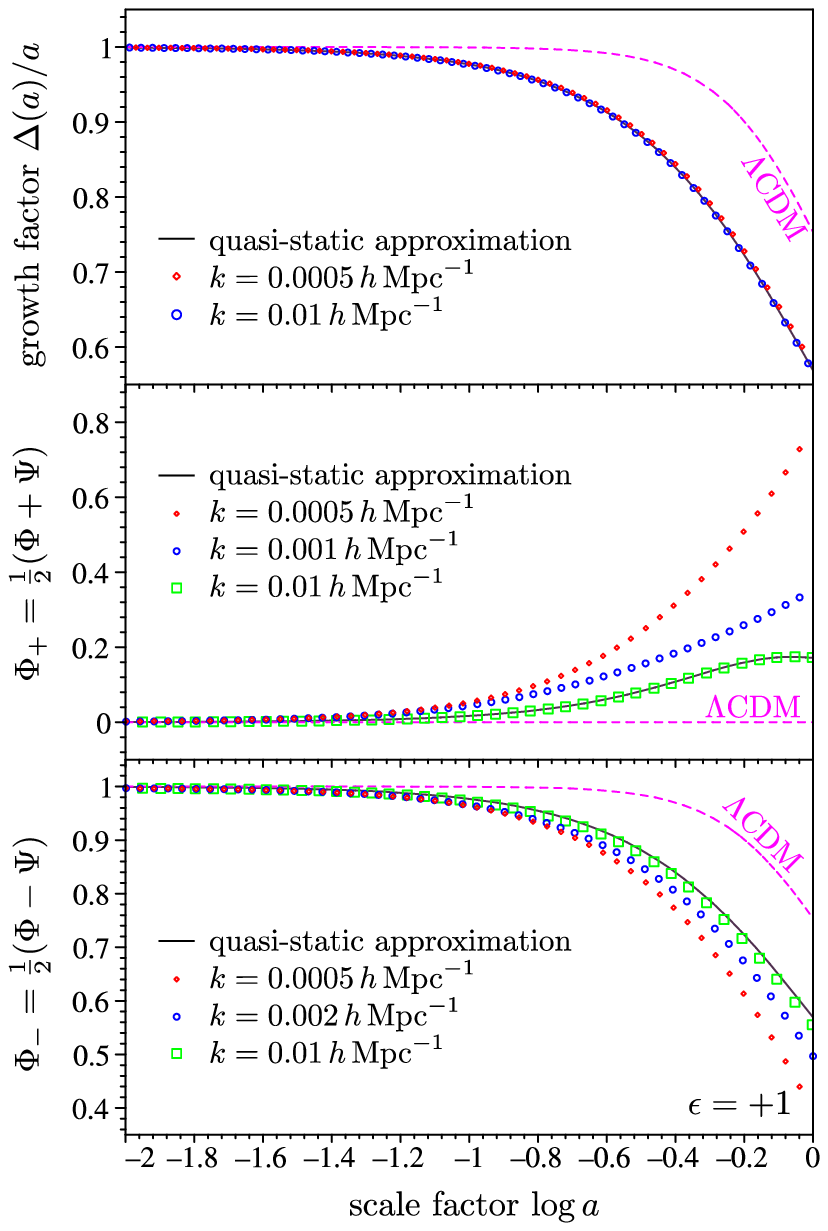}
      \includegraphics[scale=0.8]{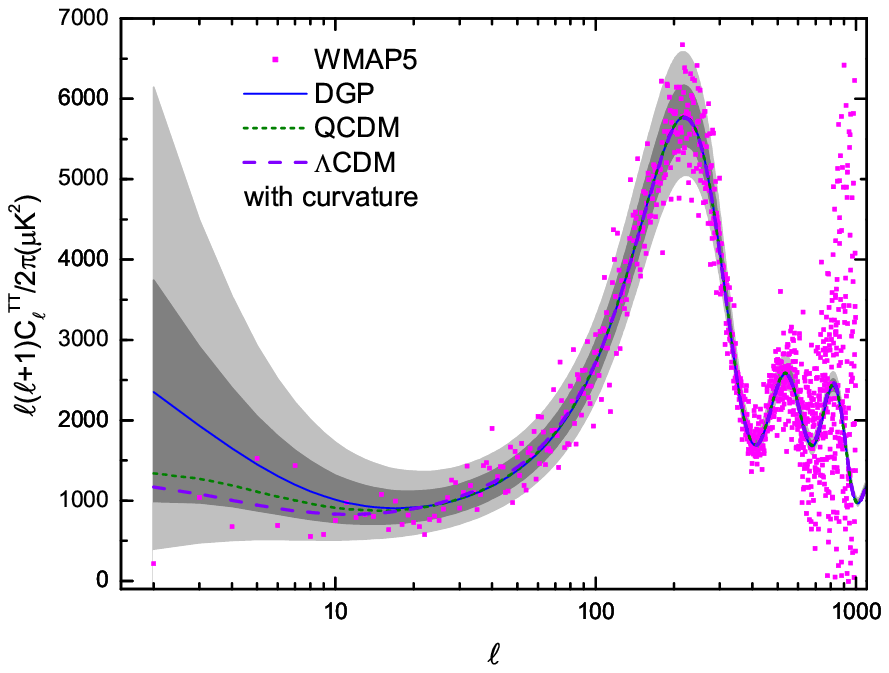}
    }
    \caption{\emph{Left:} Numerical solutions for DGP density and
      metric perturbations, showing also the quasi-static solution,
      which is an increasingly poor approximation as the scale is
      increased. (From~\cite{Cardoso:2007xc}.) \emph{Right:}
      Constraints on DGP (the open model in Figure~\ref{ssh} that
      provides a best fit to geometric data) from CMB anisotropies
      (WMAP5). The DGP model is the solid curve, QCDM (short-dashed
      curve) is the GR quintessence model with the same background
      expansion history as the DGP model, and LCDM is the dashed curve
      (a slightly closed model that gives the best fit to WMAP5, HST
      and SNLS data). (From~\cite{Fang:2008kc}.)}
    \label{card}
\end{figure}}

On subhorizon scales relevant for linear structure formation, 5D
effects produce a difference between $\phi$ and
$-\psi$~\cite{Koyama:2005kd}:
\begin{eqnarray}
 k^2\phi&=& 4\pi G a^2\left(1-{1\over 3\beta} \right)\rho\Delta\,,
\\ k^2\psi&=& -4\pi G a^2\left(1+{1\over 3\beta}
\right)\rho\Delta\,,
\end{eqnarray}
so that there is an effective dark anisotropic stress on the
brane:
\begin{equation}
 k^2(\phi+\psi) = -{8\pi Ga^2 \over 3\beta^2} \rho\Delta\,.
\end{equation}
The density perturbations evolve as
\begin{equation}
\ddot\Delta+2H\dot\Delta -4\pi G\left(1-{1\over 3\beta}
\right)\rho\Delta=0\,.
\end{equation}
The linear growth factor, $g(a)=\Delta(a)/a$ (i.e., normalized to
the flat CDM case, $\Delta \propto a$), is shown in
Figure~\ref{fig:fig1}. This illustrates the dramatic suppression of growth
in DGP relative to LCDM -- from both the background expansion and
the metric perturbations. If we parametrize the growth factor in
the usual way, we can quantify the deviation from general
relativity with smooth dark energy~\cite{Linder:2005in}:
\begin{equation}
f:={d\ln\Delta \over d\ln a}=\Omega_m(a)^\gamma\,,~~~~
\gamma\approx \left\{
\begin{array}{lll} 0.55+0.05[1+w(z=1)] & & \text{GR, smooth DE}\\ 0.68 & &
\text{DGP}\end{array} \right.
\end{equation}

Observational data on the growth factor~\cite{Guzzo:2008ac} are
not yet precise enough to provide meaningful constraints on the
DGP model. Instead, we can look at the large-angle anisotropies of
the CMB, i.e., the ISW effect. This requires a treatment of
perturbations near and beyond the horizon scale. The full
numerical solution has been given by~\cite{Cardoso:2007xc}, and is
illustrated in Figure~\ref{card}. The CMB anisotropies are also
shown in  Figure~\ref{card}, as computed in~\cite{Fang:2008kc} using
a scaling approximation to the super-Hubble
modes~\cite{Sawicki:2006jj} (the accuracy of the scaling ansatz is
verified by the numerical results~\cite{Cardoso:2007xc}).

It is evident from Figure~\ref{card} that the DGP model, which
provides a best fit to the geometric data (see Figure~\ref{ssh}), is
in serious tension with the WMAP5 data on large scales. The
problem arises because there is a large deviation of $\phi_-=(\phi-\psi)/2$
in the DGP model from the LCDM model. This deviation, i.e., a
stronger decay of $\phi_-$, leads to an over-strong ISW effect (which
is determined by $\dot{\phi}_-$), in tension with WMAP5 observations.

As a result of the combined observations of background expansion
history and large-angle CMB anisotropies, the DGP model provides a
worse fit to the data than LCDM at about the 5$\sigma$
level~\cite{Fang:2008kc}. Effectively, the DGP model is ruled out
by observations in comparison with the LCDM model.

In addition to the severe problems posed by cosmological
observations, a problem of theoretical consistency arises from the
fact that the late-time asymptotic de Sitter solution in DGP
cosmological models has a ghost. The ghost is signaled by the
negative Brans--Dicke parameter in the effective theory that
approximates the DGP on cosmological subhorizon scales:
The existence of the ghost is confirmed by detailed analysis of
the 5D perturbations in the de~Sitter
limit~\cite{Koyama:2007za, Gorbunov:2005zk,Charmousis:2006pn, Koyama:2007za}.
The DGP ghost is a
ghost mode in the scalar sector of the gravitational field --
which is more serious than the ghost in a phantom scalar field. It
effectively rules out the DGP, since it is hard to see how an
ultraviolet completion of the DGP can cure the \emph{infrared} ghost
problem. However, the DGP remains a valuable toy model for
illustrating the kinds of behaviour that can occur from a modification
to Einstein's equations -- and for developing cosmological tools to
test modified gravity and Einstein's theory itself.


\subsection{`Normal' DGP}

The self-accelerating DGP is effectively ruled out as a viable
cosmological model by observations and by the problem of the ghost
in the gravitational sector. Indeed, it may be the case that
self-acceleration generically comes with the price of ghost states. The
`normal' (i.e., non-self-accelerating and ghost-free) branch of the
DGP~\cite{Sahni:2002dx}, arises from a different embedding
of the DGP brane in the Minkowski bulk (see Figure~\ref{ndgp}). In
the background dynamics, this amounts to a replacement $r_c \to
-r_c $ in Equation~(\ref{f}) -- and there is no longer
late-time self-acceleration. Therefore, it is necessary to include
a $\Lambda$ term in order to accelerate the late universe:
\begin{eqnarray}
H^2+{K\over a^2}+{1 \over r_c}\sqrt{H^2+{K\over a^2}}= {8\pi G
\over 3}\rho +{\Lambda \over3}\,. \label{fn}
\end{eqnarray}
(Normal DGP models with a quintessence field have also been
investigated~\cite{Chimento:2006ac}.) Using the dimensionless
crossover parameter defined in Equation~(\ref{orc}), the densities are
related at the present time by
\begin{equation}
 \sqrt{1-\Omega_K}=-\sqrt{\Omega_{r_c}} +\sqrt{\Omega_{r_c} +
\Omega_m + \Omega_\Lambda }\,, \label{ondgp}
\end{equation}
which can be compared with the self-accelerating DGP
relation~(\ref{odgp}).

\epubtkImage{figurev212.png}{%
  \begin{figure}[htbp]
    \centerline{
      \includegraphics[scale=0.35]{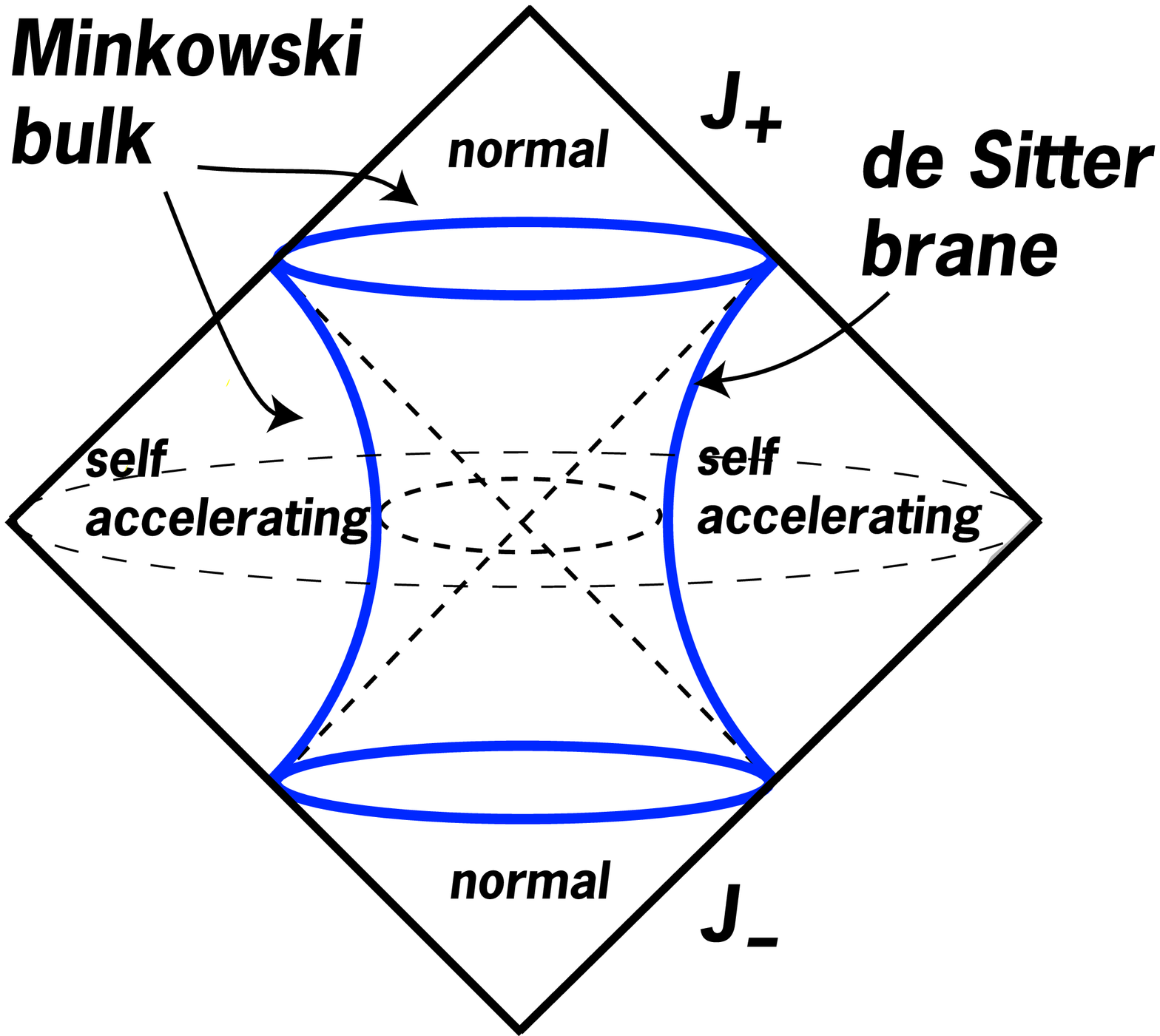}
      \includegraphics[scale=0.7]{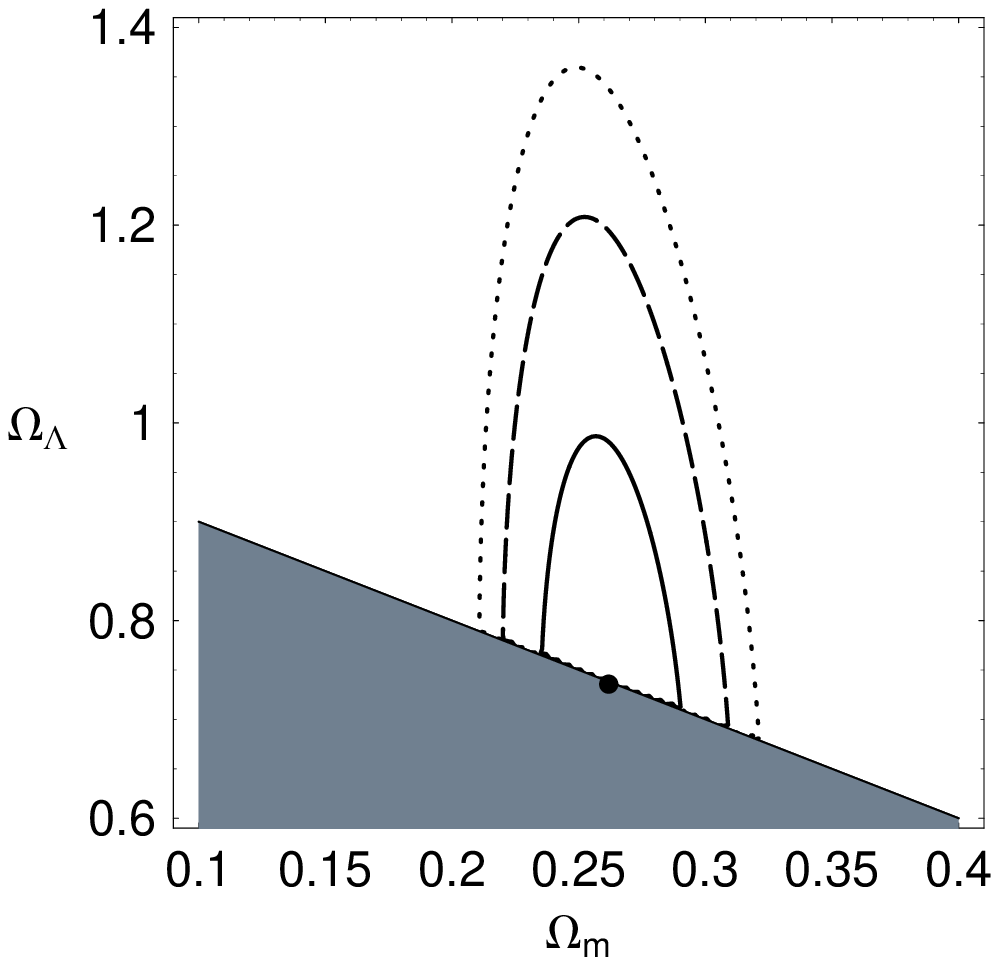}
    }
    \caption{\emph{Left:} The embedding of the self-accelerating and
    normal branches of the DGP brane in a Minkowski
    bulk. (From~\cite{Charmousis:2006pn}.) \emph{Right:} Joint
    constraints on normal DGP (flat, {\it K}~=~0) from SNLS, CMB shift
    (WMAP3) and BAO ({\it z}~=~0.35) data. The best-fit is the solid point,
    and is indistinguishable from the LCDM limit. The shaded region is
    unphysical and its upper boundary represents flat LCDM
    models. (From~\cite{Lazkoz:2006gp}.)}
    \label{ndgp}
\end{figure}}

An interesting feature of the normal branch is the `degravitation'
property, i.e., that $\Lambda$ is effectively screened by 5D gravity
effects. This follows from rewriting the modified Friedmann
equation~(\ref{fn}) in standard general relativistic form, with
\begin{equation}
 \Lambda_{\mathrm{eff}}=\Lambda -{3 \over r_c} \sqrt{H^2+{K\over a^2}} <
\Lambda \,.
\end{equation}
Thus, 5D gravity in normal DGP can in principle reduce the bare
vacuum energy significantly~\cite{Sahni:2002dx, Lue:2004za}.
However, Figure~\ref{ndgp} shows that best-fit
flat models, using geometric data, only admit insignificant
screening~\cite{Lazkoz:2006gp}. The closed models provide
a better fit to the data~\cite{Giannantonio:2008qr}, and can allow
a bare vacuum energy term with $\Omega_\Lambda >1$, as shown in
Figure~\ref{gsk}. This does not address the fundamental problem of
the smallness of $\Omega_\Lambda$, but is nevertheless an
interesting feature. We can define an effective equation-of-state
parameter via
\begin{equation}
 \dot{\Lambda}_{\mathrm{eff}}+3H(1+w_{\mathrm{eff}}) \Lambda_{\mathrm{eff}}=0\,.
\end{equation}
At the present time (setting $K=0$ for simplicity),
\begin{equation}
 w_{\mathrm{eff,0}}=-1 -{(\Omega_m+\Omega_\Lambda-1)\Omega_m \over
(1-\Omega_m) (\Omega_m+\Omega_\Lambda+1)}<-1\,,
\end{equation}
where the inequality holds because $\Omega_m<1$. This reveals
another important property of the normal DGP model: effective
phantom behaviour of the recent expansion history. This is
achieved without any pathological phantom field (similar to what
can be done in scalar-tensor theories~\cite{Boisseau:2000pr}).
Furthermore, there is no ``big rip'' singularity in the future
associated with this phantom acceleration, unlike the situation
that typically arises with phantom fields. The phantom behaviour
in the normal DGP model is also not associated with any ghost
problem -- indeed, the normal DGP branch is free of the ghost that
plagues the self-accelerating DGP~\cite{Charmousis:2006pn}.

\vspace{0.5cm}

\epubtkImage{figurev213.png}{%
  \begin{figure}[htbp]
    \centerline{
      \includegraphics[scale=0.4]{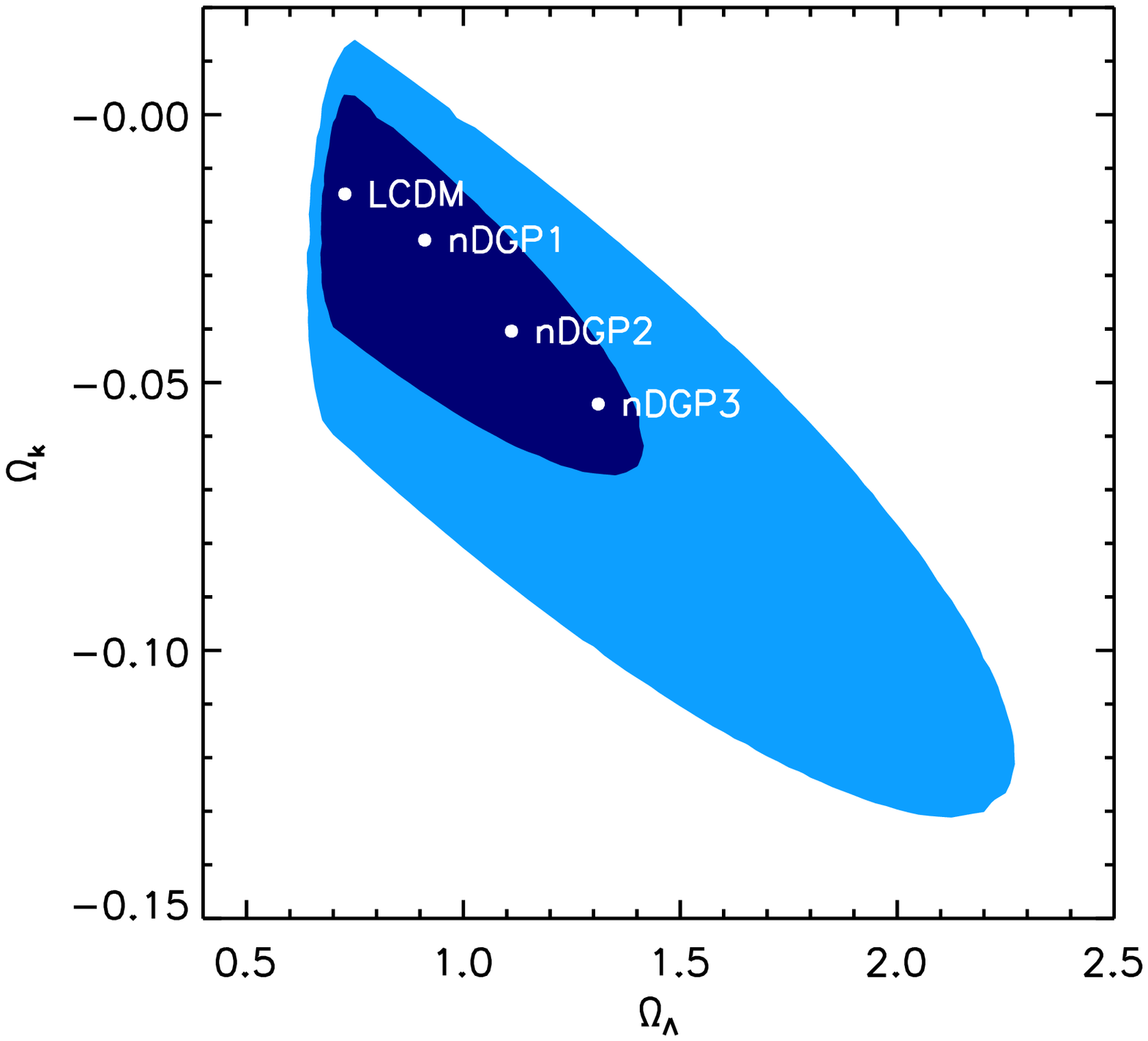} \hspace{0.5cm}
      \includegraphics[scale=0.8]{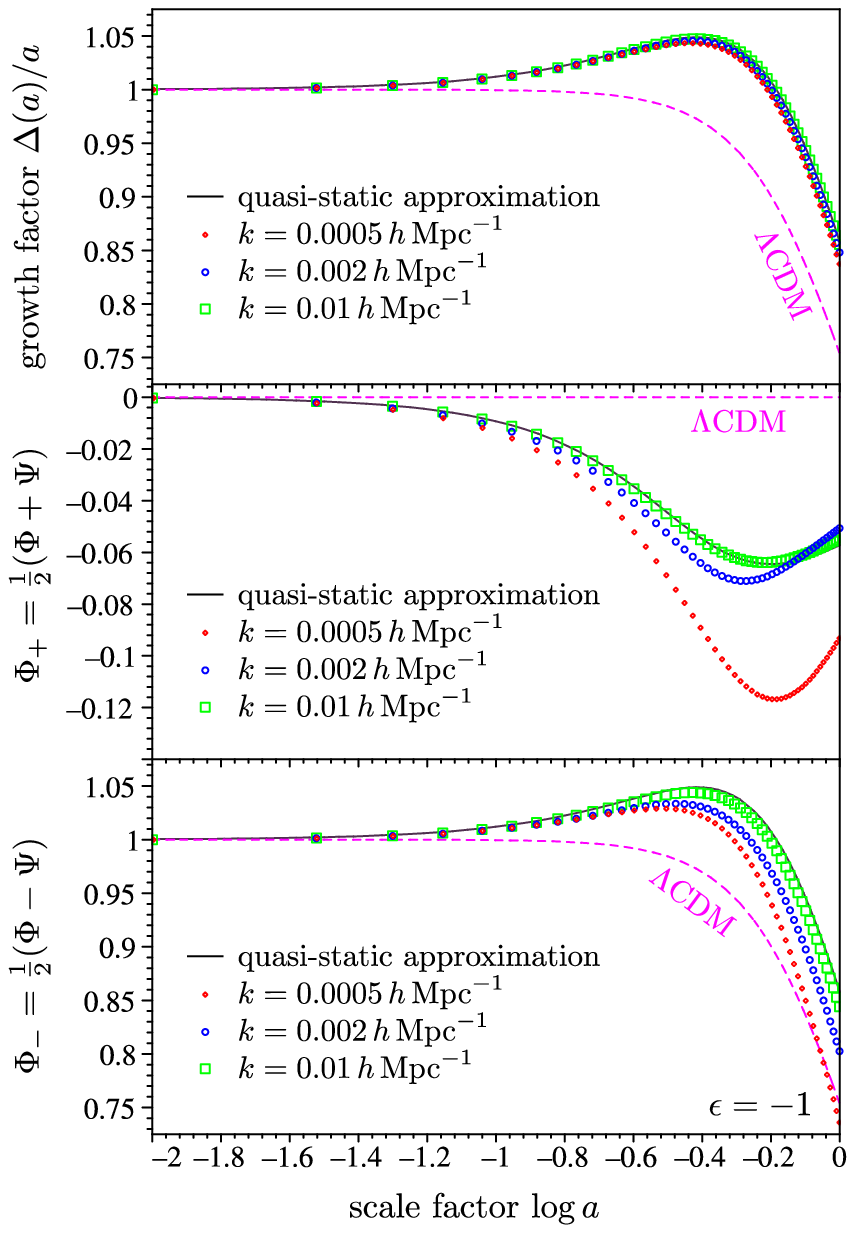}
    }
    \caption{\emph{Left:} Joint constraints on normal DGP from SNe
      Gold, CMB shift (WMAP3) and $H_0$ data in the projected
      curvature-$\Lambda$ plane, after marginalizing over other
      parameters. The best-fits are the solid points, corresponding to
      different values of
      $\Omega_m$. (From~\cite{Giannantonio:2008qr}.) \emph{Right:}
      Numerical solutions for the normal DGP density and metric
      perturbations, showing also the quasistatic solution, which is
      an increasingly poor approximation as the scale is
      increased. Compare with the self-accelerating DGP case in
      Figure~\ref{card}. (From~\cite{Cardoso:2007xc}.)}
    \label{gsk}
\end{figure}}

Perturbations in the normal branch have the same structure as
those in the self-accelerating branch, with the same regimes --
i.e., below the Vainshtein radius (recovering a GR limit), up to the
Hubble radius (Brans--Dicke behaviour), and beyond the Hubble
radius (strongly 5D behaviour). The quasistatic approximation and
the numerical integrations can be simply repeated with the
replacement $r_c \to -r_c$ (and the addition of $\Lambda$ to the
background). In the sub-Hubble regime, the effective Brans--Dicke
parameter is still given by Equations~(\ref{obd}) and (\ref{beta}), but
now we have $\omega_{BD}>0$ -- and this is consistent with the
absence of a ghost. Furthermore, a positive Brans--Dicke parameter
signals an extra positive contribution to structure formation from
the scalar degree of freedom, so that there is \emph{less}
suppression of structure formation than in LCDM -- the reverse of
what happens in the self-accelerating DGP. This is confirmed by
computations, as illustrated in Figure~\ref{gsk}.

The closed normal DGP models fit the background expansion data
reasonably well, as shown in Figure~\ref{gsk}. The key remaining
question is how well do these models fit the large-angle CMB
anisotropies, which is yet to be computed at the time of writing.
The derivative of the ISW potential $\dot{\phi}_-$ can be seen in
Figure~\ref{gsk}, and it is evident that the ISW contribution is
negative relative to LCDM at high redshifts, and goes through zero
at some redshift before becoming positive. This distinctive
behaviour may be contrasted with the behaviour in $f(R)$ models
(see Figure~\ref{sph}): both types of model lead to less suppression
of structure than LCDM, but they produce different ISW
effects. However, in the limit $r_r\to \infty$, normal DGP tends to ordinary
LCDM, hence observations which fit LCDM will always just provide a
lower limit for $r_c$.

\epubtkImage{figurev214.png}{%
  \begin{figure}[htbp]
    \centerline{\includegraphics[width=12cm]{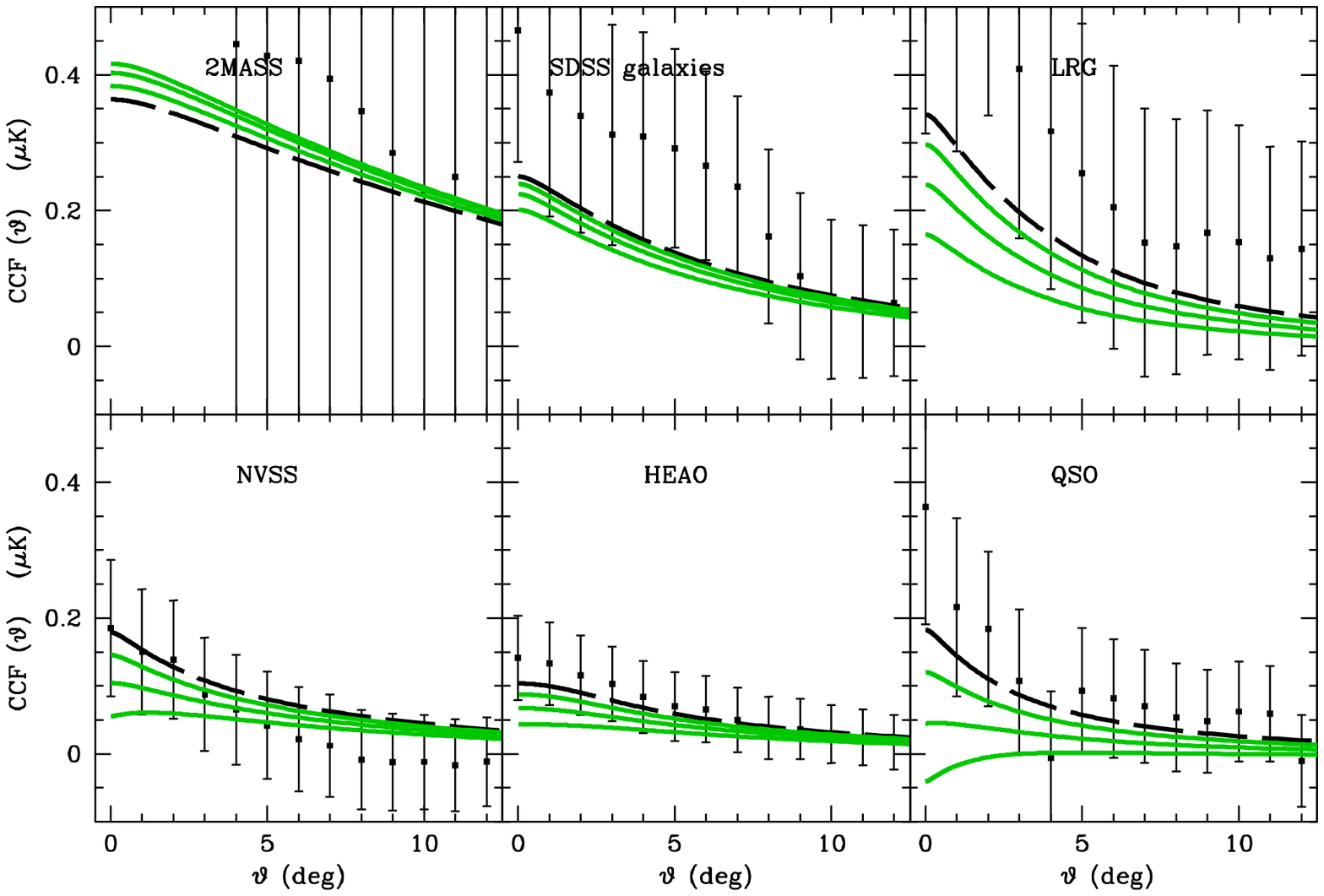}}
    \centerline{\includegraphics[width=12cm]{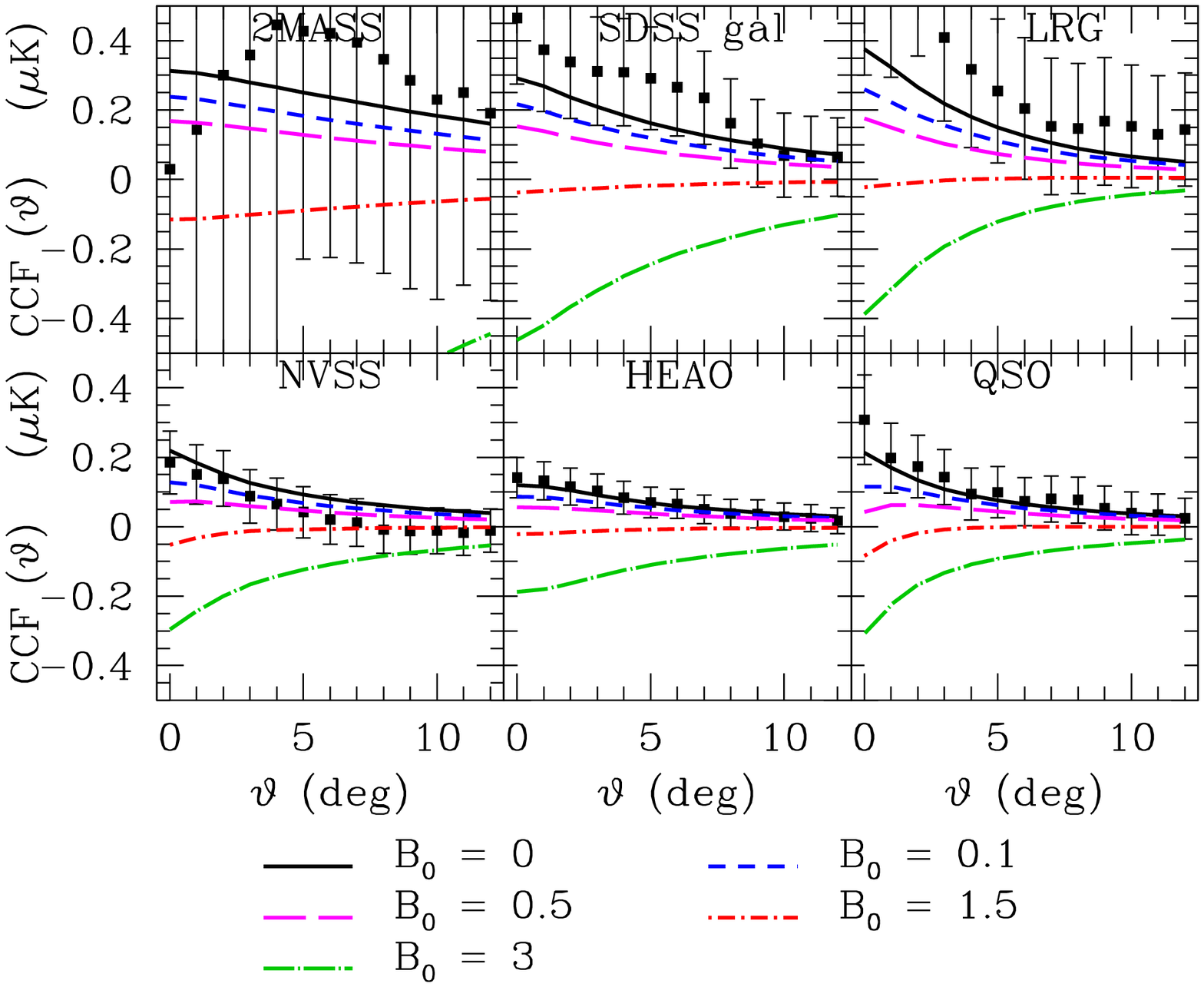}}
    \caption{\emph{Top:} Measurement of the cross-correlation
    functions between six different galaxy data sets and the CMB,
    reproduced from~\cite{Giannantonio:2008zi}. The curves show the
    theoretical predictions for the ISW-galaxy correlations at each
    redshift for the LCDM model (black, dashed) and the three nDGP
    models, which describe the $1-\sigma $ region of the geometry test
    from Figure~\ref{gsk}. (From~\cite{Giannantonio:2008qr}.)
    \emph{Bottom:} Theoretical predictions for a family of $f(R)$
    theories compared with the ISW data~\cite{Giannantonio:2008zi}
    measuring the angular CCF between the CMB and six galaxy
    catalogues. The model with $B_0 = 0 $ is equivalent to LCDM, while
    increasing departures from GR produce negative
    cross-correlations. (From~\cite{Giannantonio:2009gi}.)}
    \label{sph}
\end{figure}}

\newpage

\section{\new{6-Dimensional Models}}
\label{section_10}

\epubtkUpdateA{Added Section~\ref{section_10} ``6-Dimensional Models''.}

For brane-world models in 6-dimensional spacetime, the codimension of
a brane is two and the behaviour of gravity is qualitatively very
different from the codimension one brane-world models. Here we briefly
discuss some important examples and features of 6-dimensional models.


\subsection{Supersymmetric Large Extra Dimensions (SLED) Model}

This is a supersymmetric version of the Einstein--Maxwell
model~\cite{Carroll:2003db} (see~\cite{Burgess:2005wu, Burgess:2004ib,
  Burgess:2007ui, Koyama:2007rx} for reviews). The bosonic part of the
action is given by~\cite{Aghababaie:2002be, Aghababaie:2003wz,
  Aghababaie:2003ar}
\begin{equation}
S =\int d^6 x \sqrt{-g}
\left[ \frac{1}{2 \kappa_6^2}
\left(R -\partial_M \phi \partial^M \phi \right)
-\frac{1}{4} e^{-\phi} F_{MN}F^{MN}- e^{\phi} \Lambda_6
\right].
\end{equation}
There exists a solution where the dilaton $\phi$
is constant, $\phi=\phi_0$. The solution for the 6D spacetime is given by
\begin{equation}
ds^2  = \eta_{\mu \nu} dx^{\mu}dx^{\nu} + \gamma_{ij}
dx^{i} dx^{j}.
\end{equation}
The gauge field is taken to consist of magnetic flux threading
the extra dimensional space so that the field strength
takes the form
\begin{equation}
F_{ij} = \sqrt{\gamma} B_0 \epsilon_{ij},
\end{equation}
where $B_0$ is a constant, $\gamma$ is the determinant of $\gamma_{ij}$
and $\epsilon_{ij}$ is the antisymmetric tensor normalized as
$\epsilon_{12} =1$. All other components of $F_{ab}$ vanish.
A static and stable solution is obtained by choosing the
extra-dimensional space to be a two-sphere
\begin{equation}
\gamma_{ij} dx^{i} dx^{j} =a_0^2 (d \theta^2 + \sin^2 \theta d \varphi^2).
\end{equation}
The magnetic field strength $B_0$ and the radius $a_0$ are fixed by
the cosmological constant
\begin{equation}
B_0^2 = 2 \Lambda_6 e^{2 \phi_0}, \quad a_0^2 = \frac{M_6^4}{2 \Lambda_6 e^{\phi_0}. }.
\label{tuning}
\end{equation}
The constant value $\phi_0$ is determined by the condition
that the potential for $\phi$ has minimum~\cite{Garriga:2004tq}
\begin{equation}
V'(\phi_0) = -\frac{1}{2} B_0^2 e^{-\phi_0}
+\Lambda_6 e^{\phi_0} =0.
\end{equation}
This is exactly the condition to have a flat geometry
on the brane (see Equation~(\ref{tuning}))
\begin{equation}
B_0^2 e^{-\phi_0}= 2 \Lambda_6 e^{\phi_0}.
\label{finetune}
\end{equation}
Thus without tuning the 6D cosmological constant, it is possible to
obtain a flat 4D spacetime.

Now we add branes to this solution~\cite{Carroll:2003db}. The brane action is
given by
\begin{equation}
S_4 = - \sigma \int d^4 x \sqrt{-\gamma}.
\end{equation}
The solution for the extra dimensions is now given by
\begin{equation}
\gamma_{ij} dx^i dx^j = a_0^2 (d \theta^2 + \alpha^2
\sin^2 \theta d \varphi^2),
\end{equation}
where
\begin{equation}
\alpha = 1 - \frac{\sigma}{2 \pi M_6^2}, \quad
a_0^2 =\frac{M_6^4}{2 \Lambda_6 e^{\phi_0}}.
\end{equation}
The coordinate $\varphi$ ranges from $0$ to $2 \pi$. Thus the
effect of the brane makes a deficit angle
$\delta = 2 \pi(1- \alpha)$ in the bulk (see Figure~\ref{6d}).
This is a 6D realization of the ADD model
including the self-gravity of branes.
An interesting property of this model is that regardless of the tension of
the brane, the 4D specetime on the brane is flat. Thus this could solve the
cosmological constant problem - any vacuum energy on the brane only changes
a geometry of the extra-dimensions and does not curve the 4D spacetime.
This idea of solving the cosmological constant problem is known as
self-tuning.

\epubtkImage{figurev215.png}{%
  \begin{figure}[htbp]
    \def\epsfsize#1#2{0.8#1}
    \centerline{\epsfbox{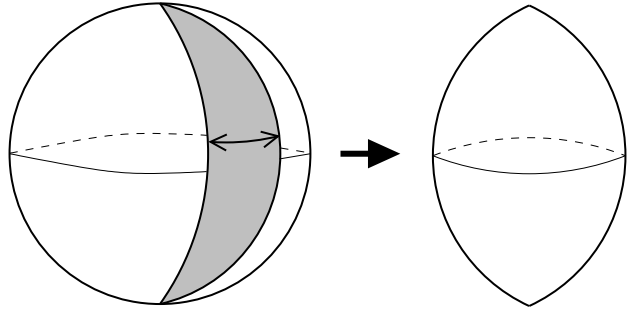}}
    \caption{Removing a wedge from a sphere and identifying opposite
	sides to obtain a rugby ball geometry. Two equal-tension
	branes with conical deficit angles are located at either pole;
	outside the branes there is constant spherical
	curvature. From~\cite{Carroll:2003db}.}
    \label{6d}
\end{figure}}

We should note that
there have been several objections to the idea of
self-tuning~\cite{Garriga:2004tq, Vinet:2004bk, Vinet:2005dg}.
Consider that a phase transition occurs and the tension
of the brane changes from $\sigma_1$ to $\sigma_2$. Accordingly,
$\alpha$ changes from $\alpha_1 =1-\sigma_1/(2 \pi M_6^4)$ to
$\alpha_2 = 1-\sigma_2/(2 \pi M_6^4)$. The magnetic flux is
conserved as the gauge field strength is a closed form,
$d F=0$. Then the magnetic flux which is obtained
by integrating the field strength over the extra dimensions
should be conserved
\begin{equation}
  \Phi_B= 4 \pi \alpha_1 B_{0,1} = 4 \pi \alpha_2 B_{0,2}.
\end{equation}
The relation between $\Lambda_6$ and $B_0$, Equation~(\ref{tuning}), that ensures
the existence of Minkowski branes cannot be imposed both
for $B_0 = B_{0,1}$ and $B_{0}=B_{0,2}$ when $\alpha_1 \neq
\alpha_2$ unless $\phi_0$ changes.  Moreover,
the quantization condition must be imposed on
the flux $\Phi_B$. What happens is that a modulus, which is a combination of
$\phi$ and the radion describing the size of extra-dimension,
acquires a runaway potential and the 4D spacetime
becomes non-static.

An unambiguous way to investigate this
problem is to study the dynamical solutions directly in the
6D spacetime. However,
once we consider the case where the tension becomes time
dependent, we encounter a difficulty to deal with the
branes~\cite{Vinet:2004bk}. This is because for co-dimension 2 branes, we
encounter a divergence of metric near the brane if
we put matter other than tension on a brane.
Hence, without specifying how we regularize the branes,
we cannot address the question what will happen if we change
the tension. Is the self-tuning mechanism at work and does it lead
to another static solution?
Or do we get a dynamical solution driven by the runaway
behaviour of the modulus field?

There was a negative conclusion on the self-tuning in this
supersymmetric model for a particular kind of
regularization~\cite{Vinet:2004bk, Vinet:2005dg}. However, the answer
could depend on the regularization of branes and the jury remains
out. It is important to study the time-dependent dynamics in the 6D
spacetime and the regularization of the branes in
detail~\cite{Tolley:2006ht, Burgess:2006ds, Burgess:2007vi,
  Tolley:2007et, Burgess:2008yx, Bayntun:2009im, deRham:2007pz,
  Navarro:2004di, Papantonopoulos:2006dv, Copeland:2007ur}.


\subsection{Cosmology in 6D brane-world models}

It is much harder to obtain cosmological solutions in 6D models than
in 5-dimensional models. In 5D brane-world models, cosmological
solutions can be obtained by considering a moving brane in a static
bulk spacetime. This is because the motion of the homogeneous and
isotropic brane does not change the bulk spacetime thanks to Birkoff's
theorem. However, this is no longer the case in 6D spacetime. Thus we
need to find time dependent bulk solutions that are coupled to a
motion of a brane, which requires us to solve non-linear partial
differential equations numerically. Moreover, as is mentioned above,
there appears a curvature singularity if one considers an infinitely
thin brane and puts matter other than tension on the
brane~\cite{Vinet:2004bk, Vinet:2005dg}. Thus we need some
regularization scheme to find cosmological solutions.

One of the popular ways to regularize a brane is to promote a brane
that is a point-like object in two extra-dimensions to a 5D
ring~\cite{Peloso:2006cq}. The ring is a codimension one object and it
is possible to consider a motion of this ring. However, a problem is
that this ring brane is not homogeneous as one of dimensions on the
brane is compact and this breaks Birkhoff's theorem. It has been shown
that cosmology obtained by the motion of a brane in a static bulk
shows pathological behaviour~\cite{Papantonopoulos:2007fk,
  Minamitsuji:2007fx}. This indicates that we need to find fully
time-dependent bulk solutions.

At the moment, the only accessible way is to solve the 6D bulk
spacetime using the gradient expansion methods assuming physical
scales on a brane are much lower than mass scales in the
bulk~\cite{Fujii:2007fi,Arroja:2007ss, Kobayashi:2008bm}. It is still
an open question what are high energy effects in 6D models.

As in 5D models, there are many generalizations such as the inclusion of
the induced gravity term on a brane and the Gauss--Bonnet term in the
bulk~\cite{Bostock:2003cv, Kanno:2004nr, Corradini:2004qr,
  Charmousis:2005ez, Papantonopoulos:2005nw, Kofinas:2005py,
  Charmousis:2008bt, Charmousis:2009uk}.


\subsection{Cascading brane-world model}

This is a 6D extension of the DGP model~\cite{deRham:2007xp,
  deRham:2007rw}. The action is given by
\begin{eqnarray}
S &=& \frac{1}{2 \kappa_6^2} \int d^6 x \sqrt{-{}^{(6)}g} {}^{(6)} R
+ \frac{1}{2 \kappa_5^2} \int_{\mathrm{4-brane}} d^5 x \sqrt{-{}^{(5)}g} {}^{(5)} R
\nonumber\\
&&+\frac{1}{2 \kappa_5^2} \int_{\mathrm{3-brane}} d^4 x \sqrt{-{}^{(4)}g}
\Big[{}^{(4)} R +{ L}_{\mathrm{matter}} \Big].
\end{eqnarray}

As in the 5D DGP model, we can define the cross over scale. In this
model, there are two cross over scales:
\begin{equation}
{}^{(5)}r_c = \frac{M_5^3}{M_4^2}, \quad {}^{(6)}r_c = \frac{M_6^3}{M_5^2}.
\end{equation}
Assuming that ${}^{(6)}r_c  \gg {}^{(5)}r_c$, the gravitational
potential on the 3-brane cascades from a $1/r$ (4D gravity) regime at
short scales, to a $1/r^2$ (5D gravity) regime at intermediate
distance and finally a $1/r^3$ (6D gravity) regime at large distances.

This model addresses several fundamental issues in induced gravity
models in 6D spacetime. Without the induced gravity term on the
4-brane, the 6D graviton propagator diverges logarithmically near the
position of the 3-brane~\cite{Geroch:1987qn}. On the other hand, the
graviton propagator on the 3-brane in this model behaves like $G(p)
\to \log(p {}^{(5)} r_c)$ in the $M_5 \to  0$ limit where $p$ is 4D
momentum. Thus the cross-over scale $ {}^{(5)} r_c $ acts as a cut-off
for the propagator and it remains finite even at the position of the
3-brane.

A more serious issue in the induced gravity model is that most
constructions seem to be plagued by ghost
instabilities~\cite{Dubovsky:2002jm, Gabadadze:2003ck}. Usually, the
regularization of the brane is necessary and it depends on the
regularization scheme whether there appears a ghost or
not~\cite{Kolanovic:2003am}. In the cascading model, there is still a
ghost if the 3-brane has no tension. However, it was shown that by
adding a tension to the 3-brane, this ghost
disappears~\cite{deRham:2007xp, deRham:2007rw, deRham:2010rw}. More
precisely, there is a critical tension
\begin{equation}
  \lambda_{\mathrm{crit}} = \frac{2}{3} {}^{(6)}r_c^{-2} M_4^2
\end{equation}
and above the critical tension $\lambda > \lambda_{\mathrm{crit}}$,
the model is free of ghosts. It is still unclear what is the meaning
of having the critical tension for the existence of the ghost. For
example, what happens if there is a phase transition on the 3-brane
and the tension changes from $\lambda> \lambda_{\mathrm{crit}}$ to
$\lambda < \lambda_{\mathrm{crit}}$? This remains a very interesting
open question. This model also provides an interesting insight into
the ``degravitation'' mechanism by which the cosmological constant
does not gravitate on the 3-brane~\cite{Dvali:2007kt, deRham:2007rw}.

Cosmological solutions in the cascading brane model are again
notoriously difficult to find because it is necessary to find
6-dimensional solutions that depend on time and two
extra-coordinates~\cite{Agarwal:2009gy}. The simplest de Sitter
solutions have been obtained~\cite{Minamitsuji:2008fz}. Interestingly,
there exists a self-accelerating solution for the 3-brane even when
the solution for the 4-brane is in the normal branch. It is still not
clear whether this self-accelerating solution is stable or not and it
is crucial to check the stability of this new self-accelerating
solution.

A similar class of models includes intersecting
branes~\cite{Kaloper:2004cy, Corradini:2007cz, Corradini:2008tu}. In
this model, we have two 4-branes that intersect and a 3-brane sits at
the intersection. Again there are self-accelerating de Sitter
solutions and cosmology has been studied by considering a motion of
one of the 4-branes. A model without a 4-brane has been studied by
regularizing a 3-brane by promoting it to a 5D ring
brane~\cite{Kaloper:2007ap, Kaloper:2007qh}.

\newpage


\section{Conclusion}
\label{conclusion}

Simple brane-world models provide a rich phenomenology
for exploring some of the ideas that are emerging from M~theory.
The higher-dimensional degrees of freedom for the gravitational
field, and the confinement of standard model fields to the visible
brane, lead to a complex but fascinating interplay between
gravity, particle physics, and geometry, that enlarges and enriches
general relativity in the direction of a quantum gravity theory.

This review has attempted to show some of the key features of
brane-world gravity from the perspective of astrophysics and
cosmology, emphasizing a geometric approach to dynamics and
perturbations. It has focused mainly on 1-brane RS-type brane-worlds,
but also considered the DGP brane-world models. The RS-type models
have some attractive features:
\begin{itemize}
\item They provide a simple 5D phenomenological realization of the
  Horava--Witten supergravity solutions in the limit where the hidden
  brane is removed to infinity, and the moduli effects from the 6
  further compact extra dimensions may be neglected.
\item They develop a new geometrical form of dimensional reduction
  based on a strongly curved (rather than flat) extra dimension.
\item They provide a realization to lowest order of the AdS/CFT
  correspondence.
\item They incorporate the self-gravity of the brane (via the brane
  tension).
\item They lead to cosmological models whose background dynamics are
  completely understood and reproduce general relativity results with
  suitable restrictions on parameters.
\end{itemize}

The review has highlighted both the successes and some remaining
open problems of the RS models and their generalizations. The open
problems stem from a common basic difficulty, i.e., understanding
and solving for the gravitational interaction between the bulk and
the brane (which is nonlocal from the brane viewpoint). The key
open problems of relevance to astrophysics and cosmology are
\begin{itemize}
\item to find the simplest realistic solution (or approximation to it)
  for an astrophysical black hole on the brane, and settle the
  questions about its staticity, Hawking radiation, and horizon; and
\item to develop realistic approximation schemes (building on recent
  work~\cite{sod_1, sod_2, sod_3, sod_4, sod_5, koy, rbbd_1, rbbd_2,
    kkt, elmw}) and manageable numerical codes (building on~\cite{koy,
    rbbd_1, rbbd_2, kkt, elmw}) to solve for the cosmological
  perturbations on all scales, to compute the CMB anisotropies and
  large-scale structure, and to impose observational constraints from
  high-precision data.
\end{itemize}

The RS-type models are the simplest brane-worlds with curved extra
dimension that allow for a meaningful approach to astrophysics and
cosmology. One also needs to consider generalizations that attempt
to make these models more realistic, or that explore other aspects
of higher-dimensional gravity which are not probed by these simple
models. Two important types of generalization are the following:
\begin{itemize}
\item \emph{The inclusion of dynamical interaction between the
    brane(s) and a bulk scalar field,} so that the action is
  \begin{eqnarray}
    S = {1 \over 2\kappa_5^2} \int \!
    d^5x\sqrt{-{}^{(5)\!}g}\left[{}^{(5)\!}R -
    \kappa_5^2\partial_A\Phi\partial^A\Phi-2\Lambda_5(\Phi) \right] +
    \int_\mathrm{brane(s)} \!\!\!\!\!\!\!\!\!\!\!\! d^4x\sqrt{-g}
    \left[\!-\lambda(\Phi)+{K\over \kappa_5^2}+L_\mathrm{matter}\!\right]
    \nonumber \\
  \end{eqnarray}%
  (see~\cite{mw, sca_1, sca_2, sca_3, sca_4, sca_5, sca_6, hs_1, hs_2,
    hs_3, hs_4, hs_5, hs_6, hs_7, hs_8, hs_9, hs_10, hs_11, hs_12,
    hs_13, hs_14, hs_15, hs_16, hs_17}).
  The scalar field could represent a bulk dilaton of the gravitational
  sector, or a modulus field encoding the dynamical influence on the
  effective 5D theory of an extra dimension other than the large fifth
  dimension~\cite{ek5_1, ek5_2, ek5_3, rbbd_1, rbbd_2, mod_1, mod_2,
    mod_3, mod_4, mod_5}.

  For two-brane models, the brane separation
  introduces a new scalar degree of freedom, the radion. For general
  potentials of the scalar field which provide radion stabilization,
  4D Einstein gravity is recovered at low energies on either
  brane~\cite{2b_1,2b_2,2b_3}. (By contrast, in the absence of a bulk
  scalar, low energy gravity is of Brans--Dicke type~\cite{gt}.)
  In particular, such models will allow some fundamental problems to
  be addressed:
  \begin{itemize}
  \item The hierarchy problem of particle physics.
  \item An extra-dimensional mechanism for initiating inflation (or the
    hot radiation era with super-Hubble correlations) via brane
    interaction (building on the initial work in~\cite{kss_1, kss_2,
      ek_1, ek_2, ek_3, ek_4, ek_5, ek_6, ek_7, ek5_1, ek5_2, ek5_3,
      bub_1, bub_2, bub_3}).
  \item An extra-dimensional explanation for the dark energy (and
    possibly also dark matter) puzzles: Could dark energy or late-time
    acceleration of the universe be a result of gravitational effects on
    the visible brane of the shadow brane, mediated by the bulk scalar
    field?
  \end{itemize}
\item \emph{The addition of stringy and quantum corrections to the
    Einstein--Hilbert action,} including the following:
  \begin{itemize}
  \item Higher-order curvature invariants, which arise in the AdS/CFT
    correspondence as next-to-leading order corrections in the CFT. The
    \emph{Gauss--Bonnet} combination in particular has unique properties
    in 5D, giving field equations which are second-order in the bulk
    metric (and linear in the second derivatives), and being
    ghost-free. The action is
    \begin{eqnarray}
      S&=&{1 \over 2\kappa_5^2} \!\int \!\!
      d^5x\sqrt{-{}^{(5)\!}g}\left[{}^{(5)\!}R-2\Lambda_5+
      \alpha\!\left( {}^{(5)\!}R^2\!-4 \:{}^{(5)\!}R_{AB} \:{}^{(5)\!}
      R^{AB}\!+{}^{(5)\!}R_{ABCD} \;{}^{(5)\!}R^{ABCD} \right)\!\right]
      \nonumber \\
      && +\int_\mathrm{brane} \!\!\!\!\!\!\!\!\!\! d^4x\sqrt{-g}
      \left[-\lambda+{K\over \kappa_5^2}+L_\mathrm{matter}\right]\!,
    \end{eqnarray}%
    where $\alpha$ is the Gauss--Bonnet coupling constant, related to
    the string scale. The cosmological dynamics of these brane-worlds is
    investigated in~\cite{gbon_1, gbon_2, gbon_3, gbon_4, gbon_5,
      gbon_6, gbon_7, gbon_8, gbon_9, gbon_10, gbon_11, gbon_12,
      gbon_13, gbon_14, gbon_15, gbon_16}. In~\cite{bmsv} it is shown
    that the black string solution of the form of Equation~(\ref{bs1})
    is ruled out by the Gauss--Bonnet term. In this sense, the
    Gauss--Bonnet correction removes an unstable and singular solution.

    In the early universe, the Gauss--Bonnet corrections to the
    Friedmann equation have the dominant form
    \begin{equation}
      H^2 \propto \rho^{2/3}
    \end{equation}
    at the highest energies. If the Gauss--Bonnet term is a small
    correction to the Einstein--Hilbert term, as may be expected if it
    is the first of a series of higher-order corrections, then there
    will be a regime of RS-dominance as the energy drops, when $H^2
    \propto \rho^2$. Finally at energies well below the brane tension,
    the general relativity behaviour is recovered.
  \item Quantum field theory corrections arising from the coupling
    between brane matter and bulk gravitons, leading to an induced
    4D Ricci term in the brane action. The original induced gravity
    brane-world is the DGP model~\cite{dgp_1, dgp_2, dgp_3, dgp_4},
    which we investigated in this review as an alternative to the
    RS-type models. Another viewpoint is to see the induced-gravity
    term in the action as a correction to the RS action:
    \begin{equation}
      S = {1 \over 2\kappa_5^2} \int d^5x
      \sqrt{-{}^{(5)\!}g}\left[{}^{(5)\!}R-2\Lambda_5 \right]+
      \int_\mathrm{brane} \!\!\!\!\!\!\!\!\!\! d^4x \sqrt{-g}
      \left[\beta{R}-\lambda+{K\over \kappa_5^2}+L_\mathrm{matter}\right],
    \end{equation}
    where $\beta$ is a positive coupling constant.

%
    The cosmological models have been analyzed in~\cite{ind_1, ind_2, ind_3, ind_4,
      ind_5, ind_6, ind_7, ind_8, ind_9, ind_10, ind_11, ind_12,
      ind_13, ind_14, kmp}. (Brane-world black holes with induced gravity
    are investigated in~\cite{kpz}.)
    Unlike RS-type models, DGP models lead to 5D behaviour on
    \emph{large scales} rather than small scales. Then on an FRW
    brane, the late-universe 5D behaviour of gravity can naturally
    produce a late-time acceleration, even \emph{without dark energy},
    although the self-accelerating models suffer from a
    ghost. Nevertheless, the DGP model is a critical example of
    modified gravity models in cosmology that act as alternatives to
    dark energy.


    %
  \end{itemize}
\end{itemize}

The RS and DGP models are 5-dimensional phenomenological models, and
so a key issue is how to realize such models in 10-dimensional string
theory. Some progress has been made. 6-dimensional cascading
brane-worlds are extensions of the DGP model. 10-dimensional type IIB
supergravity solutions have been found with the warped geometry that
generalizes the RS geometry. These models have also been important for
building inflationary models in string theory, based on the motion of
D3 branes in the warped throat \cite{Burgess:2001vr, Dvali:2001fw}
(see the reviews~\cite{Linde:2005dd, Baumann:2009ni} and references therein).
The action for D3 branes is described by the Dirac-Born-Infeld action and this
gives the possibility of generating a large non-Gaussianity in the
Cosmic Microwave Background temperature anisotropies, which can be tested
in the future experiments \cite{Silverstein:2003hf, Kachru:2003sx}
(see the reviews~\cite{Chen:2010xk, Koyama:2010xj}).

These models reply on the
effective 4-dimensional approach to deal with extra dimensions. For
example, the stabilization mechanism, which is necessary to fix moduli
fields in string theory, exploits non-perturbative effects and they
are often added in the 4-dimensional effective theory. Then it is not
clear whether the resultant 4-dimensional effective theory is
consistent with the 10-dimensional equations of
motion~\cite{deAlwis:2003sn, deAlwis:2004qh, Kodama:2005cz,
  Koyama:2006ni}. Recently there has been a new development and it has
become possible to calculate all significant contributions to the D3
brane potential in the single coherent framework of 10-dimensional
supergravity~\cite{Baumann:2007np, Baumann:2007ah, Baumann:2009qx, Baumann:2010sx}.
This will provide
us with a very interesting bridge between phenomenological brane-world
models, where dynamics of higher-dimensional gravity is studied in
detail, and string theory approaches, where 4D effective theory is
intensively used. It is crucial to identify the higher-dimensional
signature of the models in order to test a fundamental theory like
string theory.

In summary, brane-world gravity opens up exciting prospects for
subjecting M~theory ideas to the increasingly stringent tests
provided by high-precision astronomical observations. At the same
time, brane-world models provide a rich arena for probing the
geometry and dynamics of the gravitational field and its
interaction with matter.

\newpage


\section{Acknowledgments}

We thank our many collaborators and friends for discussions and
sharing of ideas. R.M.\ is supported by the UK's Science \& Technology
Facilities Council (STFC). K.K.\ is supported by the European Research
Council, Research Councils UK and STFC.

\newpage


\bibliography{refs}

\begin{thebibliography}{100}

\bibitem{Agarwal:2009gy}
Agarwal, N., Bean, R., Khoury, J., and Trodden, M., ``Cascading Cosmology'',
  arXiv e-print, (2009).
  {\small[\href{http://arxiv.org/abs/0912.3798}{{arXiv:0912.3798
  {\small[hep-th]}}}]}.

\bibitem{Aghababaie:2003ar}
Aghababaie, Y., Burgess, C.P., Cline, J.M., Firouzjahi, H., Parameswaran, S.L.,
  Quevedo, F., Tasinato, G., and Zavala~C, I., ``Warped brane worlds in six
  dimensional supergravity'', {\em J. High Energy Phys.}, {\bf 2003}(09), 037,
  (2003).
  {\small[\href{http://arxiv.org/abs/hep-th/0308064}{{hep-th/0308064}}]}.

\bibitem{Aghababaie:2002be}
Aghababaie, Y., Burgess, C.P., Parameswaran, S.L., and Quevedo, F., ``SUSY
  breaking and moduli stabilization from fluxes in gauged 6D supergravity'',
  {\em J. High Energy Phys.}, {\bf 2003}(03), 032, (2003).
  {\small[\href{http://arxiv.org/abs/hep-th/0212091}{{hep-th/0212091}}]}.

\bibitem{Aghababaie:2003wz}
Aghababaie, Y., Burgess, C.P., Parameswaran, S.L., and Quevedo, F., ``Towards a
  naturally small cosmological constant from branes in 6D supergravity'', {\em
  Nucl. Phys. B}, {\bf 680}, 389--414, (2004).
  {\small[\href{http://dx.doi.org/10.1016/j.nuclphysb.2003.12.015}{DOI}]},
  {\small[\href{http://arxiv.org/abs/hep-th/0304256}{{hep-th/0304256}}]}.

\bibitem{ruth_1}
Aguirregabiria, J.M., Chimento, L.P., and Lazkoz, R., ``Anisotropy and
  inflation in Bianchi I braneworlds'', {\em Class. Quantum Grav.}, {\bf 21},
  823--829, (2003).
  {\small[\href{http://arxiv.org/abs/gr-qc/0303096}{{gr-qc/0303096}}]}.

\bibitem{add_4}
Akama, K., ``An Early Proposal of `Brane World'\,'', in Kikkawa, K., Nakanishi,
  N., and Nariai, H., eds., {\em Gauge Theory and Gravitation}, Proceedings of
  the International Symposium on Gauge Theory and Gravitation, Nara, Japan,
  August 20\,--\,24, 1982, Lecture Notes in Physics, vol. 176, pp. 267--271,
  (Springer, Berlin; New York, 1982).
  {\small[\href{http://arxiv.org/abs/hep-th/0001113}{{hep-th/0001113}}]}.

\bibitem{ind_10}
Alam, U., and Sahni, V., ``Supernova Constraints on Braneworld Dark Energy'',
  arXiv e-print, (2002).
  {\small[\href{http://arxiv.org/abs/astro-ph/0209443}{{astro-ph/0209443}}]}.

\bibitem{dq_1}
Albrecht, A., Burgess, C.P., Ravndal, F., and Skordis, C., ``Natural
  quintessence and large extra dimensions'', {\em Phys. Rev. D}, {\bf 65},
  123507, 1--10, (2002).
  {\small[\href{http://arxiv.org/abs/astro-ph/0107573}{{astro-ph/0107573}}]}.

\bibitem{preheat_3}
Allahverdi, R., Mazumdar, A., and P{\'{e}}rez-Lorenzana, A., ``Final reheating
  temperature on a single brane'', {\em Phys. Lett. B}, {\bf 516}, 431--438,
  (2001).
  {\small[\href{http://arxiv.org/abs/hep-ph/0105125}{{hep-ph/0105125}}]}.

\bibitem{ksb_2}
Anchordoqui, L., Nu{\~{n}}ez, C., and Olsen, K., ``Quantum Cosmology and
  AdS/CFT'', {\em J. High Energy Phys.}, {\bf 2000}(10), 050, (2000).
  {\small[\href{http://arxiv.org/abs/hep-th/0007064}{{hep-th/0007064}}]}.

\bibitem{anlid}
Anderson, E., and Lidsey, J.E., ``Embeddings in non-vacuum spacetimes'', {\em
  Class. Quantum Grav.}, {\bf 18}, 4831--4843, (2001).
  {\small[\href{http://arxiv.org/abs/gr-qc/0106090}{{gr-qc/0106090}}]}.

\bibitem{antav}
Anderson, E., and Tavakol, R., ``PDE System Approach to Large Extra
  Dimensions'', arXiv e-print, (2003).
  {\small[\href{http://arxiv.org/abs/gr-qc/0309063}{{gr-qc/0309063}}]}.

\bibitem{add_3}
Antoniadis, I., ``A possible new dimension at a few TeV'', {\em Phys. Lett. B},
  {\bf 246}, 377--384, (1990).

\bibitem{add_2}
Antoniadis, I., Arkani-Hamed, N., Dimopoulos, S., and Dvali, G., ``New
  dimensions at a millimeter to a fermi and superstrings at a TeV'', {\em Phys.
  Lett. B}, {\bf 436}, 257--263, (1998).
  {\small[\href{http://arxiv.org/abs/hep-ph/9804398}{{hep-ph/9804398}}]}.

\bibitem{add_1}
Arkani-Hamed, N., Dimopoulos, S., and Dvali, G., ``The hierarchy problem and
  new dimensions at a millimeter'', {\em Phys. Lett. B}, {\bf 429}, 263--272,
  (1998).
  {\small[\href{http://arxiv.org/abs/hep-ph/9803315}{{hep-ph/9803315}}]}.

\bibitem{Arroja:2007ss}
Arroja, F., Kobayashi, T., Koyama, K., and Shiromizu, T., ``Low energy
  effective theory on a regularized brane in 6D gauged chiral supergravity'',
  {\em J. Cosmol. Astropart. Phys.}, {\bf 2007}(12), 006, (2007).
  {\small[\href{http://dx.doi.org/10.1088/1475-7516/2007/12/006}{DOI}]},
  {\small[\href{http://arxiv.org/abs/0710.2539}{{arXiv:0710.2539
  {\small[hep-th]}}}]}.

\bibitem{abd}
Ashcroft, P.R., van~de Bruck, C., and Davis, A.-C., ``Suppression of entropy
  perturbations in multifield inflation on the brane'', {\em Phys. Rev. D},
  {\bf 66}, 121302, 1--5, (2002).
  {\small[\href{http://arxiv.org/abs/astro-ph/0208411}{{astro-ph/0208411}}]}.

\bibitem{hs_17}
Ashcroft, P.R., van~de Bruck, C., and Davis, A.-C., ``Boundary inflation in the
  moduli space approximation'', {\em Phys. Rev. D}, {\bf 69}, 063519, 1--12,
  (2004).
  {\small[\href{http://arxiv.org/abs/astro-ph/0310643}{{astro-ph/0310643}}]}.

\bibitem{gbon_13}
Barcel{\'{o}}, C., Germani, C., and Sopuerta, C.F., ``Thin-shell limit of
  branes in the presence of Gauss--Bonnet interactions'', {\em Phys. Rev. D},
  {\bf 68}, 104007, 1--12, (2003).
  {\small[\href{http://arxiv.org/abs/gr-qc/0306072}{{gr-qc/0306072}}]}.

\bibitem{bmsv}
Barcel{\'{o}}, C., Maartens, R., Sopuerta, C.F., and Viniegra, F., ``Stacking a
  4D geometry into an Einstein--Gauss--Bonnet bulk'', {\em Phys. Rev. D}, {\bf
  67}, 064023, 1--6, (2003).
  {\small[\href{http://arxiv.org/abs/hep-th/0211013}{{hep-th/0211013}}]}.

\bibitem{sca_1}
Barcel{\'{o}}, C., and Visser, M., ``Braneworld gravity: Influence of the
  moduli fields'', {\em J. High Energy Phys.}, {\bf 2000}(10), 019, (2000).
  {\small[\href{http://arxiv.org/abs/hep-th/0009032}{{hep-th/0009032}}]}.

\bibitem{bv}
Barcel{\'{o}}, C., and Visser, M., ``Living on the edge: cosmology on the
  boundary of anti-de Sitter space'', {\em Phys. Lett. B}, {\bf 482}, 183--194,
  (2000).
  {\small[\href{http://arxiv.org/abs/hep-th/0004056}{{hep-th/0004056}}]}.

\bibitem{b1_3}
Barrow, J.D., and Hervik, S., ``Magnetic brane-worlds'', {\em Class. Quantum
  Grav.}, {\bf 19}, 155--172, (2002).
  {\small[\href{http://arxiv.org/abs/gr-qc/0109084}{{gr-qc/0109084}}]}.

\bibitem{bm}
Barrow, J.D., and Maartens, R., ``Kaluza--Klein anisotropy in the CMB'', {\em
  Phys. Lett. B}, {\bf 532}, 153--158, (2002).
  {\small[\href{http://arxiv.org/abs/gr-qc/0108073}{{gr-qc/0108073}}]}.

\bibitem{tsab}
Barrow, J.D., and Tsagas, C.G., ``G\"odel brane'', {\em Phys. Rev. D}, {\bf
  69}, 064007, 1--6, (2004).
  {\small[\href{http://arxiv.org/abs/gr-qc/0309030}{{gr-qc/0309030}}]}.

\bibitem{ek5_1}
Bastero-Gil, M., Copeland, E.J., Gray, J., Lukas, A., and Plumacher, M.,
  ``Baryogenesis by brane collision'', {\em Phys. Rev. D}, {\bf 66}, 066005,
  1--11, (2002).
  {\small[\href{http://arxiv.org/abs/hep-th/0201040}{{hep-th/0201040}}]}.

\bibitem{mbb_3}
Battye, R.A., van~de Bruck, C., and Mennim, A., ``Cosmological tensor
  perturbations in the Randall--Sundrum model: Evolution in the near-brane
  limit'', {\em Phys. Rev. D}, {\bf 69}, 064040, 1--14, (2004).
  {\small[\href{http://arxiv.org/abs/hep-th/0308134}{{hep-th/0308134}}]}.

\bibitem{Baumann:2009qx}
Baumann, D., Dymarsky, A., Kachru, S., Klebanov, I.R., and McAllister, L.,
  ``Compactification Effects in D-brane Inflation'', arXiv e-print, (2009).
  {\small[\href{http://arxiv.org/abs/0912.4268}{{arXiv:0912.4268
  {\small[hep-th]}}}]}.

\bibitem{Baumann:2010sx}
Baumann, D., Dymarsky, A., Kachru, S., Klebanov, I.R., and McAllister, L.,
  ``D3-brane Potentials from Fluxes in AdS/CFT'', arXiv e-print, (2010).
  {\small[\href{http://arxiv.org/abs/1001.5028}{{arXiv:1001.5028
  {\small[hep-th]}}}]}.

\bibitem{Baumann:2007ah}
Baumann, D., Dymarsky, A., Klebanov, I.R., and McAllister, L., ``Towards an
  Explicit Model of D-brane Inflation'', {\em J. Cosmol. Astropart. Phys.},
  {\bf 2008}(01), 024, (2008).
  {\small[\href{http://dx.doi.org/10.1088/1475-7516/2008/01/024}{DOI}]},
  {\small[\href{http://arxiv.org/abs/0706.0360}{{arXiv:0706.0360
  {\small[hep-th]}}}]}.

\bibitem{Baumann:2007np}
Baumann, D., Dymarsky, A., Klebanov, I.R., McAllister, L., and Steinhardt,
  P.J., ``A Delicate Universe'', {\em Phys. Rev. Lett.}, {\bf 99}, 141601,
  (2007).
  {\small[\href{http://dx.doi.org/10.1103/PhysRevLett.99.141601}{DOI}]},
  {\small[\href{http://arxiv.org/abs/0705.3837}{{arXiv:0705.3837
  {\small[hep-th]}}}]}.

\bibitem{Baumann:2009ni}
Baumann, D., and McAllister, L., ``Advances in Inflation in String Theory'',
  {\em Ann. Rev. Nucl. Part. Sci.}, {\bf 59}, 67--94, (2009).
  {\small[\href{http://dx.doi.org/10.1146/annurev.nucl.010909.083524}{DOI}]},
  {\small[\href{http://arxiv.org/abs/0901.0265}{{arXiv:0901.0265
  {\small[hep-th]}}}]}.

\bibitem{Bayntun:2009im}
Bayntun, A., Burgess, C.P., and van Nierop, L., ``Codimension-2 Brane-Bulk
  Matching: Examples from Six and Ten Dimensions'', (2009).
  {\small[\href{http://arxiv.org/abs/0912.3039}{{arXiv:0912.3039
  {\small[hep-th]}}}]}.

\bibitem{inf_8}
Bento, M.C., and Bertolami, O., ``$N = 1$ supergravity chaotic inflation in the
  braneworld scenario'', {\em Phys. Rev. D}, {\bf 65}, 063513, 1--5, (2002).
  {\small[\href{http://arxiv.org/abs/astro-ph/0111273}{{astro-ph/0111273}}]}.

\bibitem{inf_9}
Bento, M.C., Bertolami, O., and Sen, A.A., ``Supergravity Inflation on the
  Brane'', {\em Phys. Rev. D}, {\bf 67}, 023504, 1--5, (2003).
  {\small[\href{http://arxiv.org/abs/gr-qc/0204046}{{gr-qc/0204046}}]}.

\bibitem{inf_10}
Bento, M.C., Bertolami, O., and Sen, A.A., ``Tachyonic inflation in the
  braneworld scenario'', {\em Phys. Rev. D}, {\bf 67}, 063511, 1--4, (2003).
  {\small[\href{http://arxiv.org/abs/hep-th/0208124}{{hep-th/0208124}}]}.

\bibitem{gbon_8}
Bin{\'{e}}truy, P., Charmousis, C., Davis, S.C., and Dufaux, J.-F., ``Avoidance
  of naked singularities in dilatonic brane world scenarios with a
  Gauss--Bonnet term'', {\em Phys. Lett. B}, {\bf 544}, 183--191, (2002).
  {\small[\href{http://arxiv.org/abs/hep-th/0206089}{{hep-th/0206089}}]}.

\bibitem{bdel}
Bin{\'{e}}truy, P., Deffayet, C., Ellwanger, U., and Langlois, D., ``Brane
  cosmological evolution in a bulk with cosmological constant'', {\em Phys.
  Lett. B}, {\bf 477}, 285--291, (2000).
  {\small[\href{http://arxiv.org/abs/hep-th/9910219}{{hep-th/9910219}}]}.

\bibitem{lang2_1}
Bin{\'{e}}truy, P., Deffayet, C., and Langlois, D., ``The radion in brane
  cosmology'', {\em Nucl. Phys. B}, {\bf 615}, 219--236, (2001).
  {\small[\href{http://arxiv.org/abs/hep-th/0101234}{{hep-th/0101234}}]}.

\bibitem{bub_1}
Blanco-Pillado, J.J., and Bucher, M., ``Cosmological perturbations generated in
  the colliding bubble braneworld universe'', {\em Phys. Rev. D}, {\bf 65},
  083517, 1--15, (2002).
  {\small[\href{http://arxiv.org/abs/hep-th/0210189}{{hep-th/0210189}}]}.

\bibitem{pert_23}
Boehm, T., and Steer, D.A., ``Perturbations on a moving D3-brane and mirage
  cosmology'', {\em Phys. Rev. D}, {\bf 66}, 063510, 1--14, (2002).
  {\small[\href{http://arxiv.org/abs/hep-th/0206147}{{hep-th/0206147}}]}.

\bibitem{Boisseau:2000pr}
Boisseau, B., Esposito-Far\`ese, G., Polarski, D., and Starobinsky, A.A.,
  ``Reconstruction of a scalar-tensor theory of gravity in an accelerating
  universe'', {\em Phys. Rev. Lett.}, {\bf 85}, 2236, (2000).
  {\small[\href{http://dx.doi.org/10.1103/PhysRevLett.85.2236}{DOI}]},
  {\small[\href{http://arxiv.org/abs/gr-qc/0001066}{{gr-qc/0001066}}]}.

\bibitem{Bostock:2003cv}
Bostock, P., Gregory, R., Navarro, I., and Santiago, J., ``Einstein gravity on
  the codimension 2 brane?'', {\em Phys. Rev. Lett.}, {\bf 92}, 221601, (2004).
  {\small[\href{http://dx.doi.org/10.1103/PhysRevLett.92.221601}{DOI}]},
  {\small[\href{http://arxiv.org/abs/hep-th/0311074}{{hep-th/0311074}}]}.

\bibitem{ksb_3}
Bouhmadi-L{\'{o}}pez, M., Gonz{\'{a}}lez-Diaz, P.F., and Zhuk, A., ``On new
  gravitational instantons describing creation of brane-worlds'', {\em Class.
  Quantum Grav.}, {\bf 19}, 4863--4876, (2002).
  {\small[\href{http://arxiv.org/abs/hep-th/0208226}{{hep-th/0208226}}]}.

\bibitem{birk_2}
Bowcock, P., Charmousis, C., and Gregory, R., ``General brane cosmologies and
  their global spacetime structure'', {\em Class. Quantum Grav.}, {\bf 17},
  4745--4763, (2000).
  {\small[\href{http://arxiv.org/abs/hep-th/0007177}{{hep-th/0007177}}]}.

\bibitem{hs_12}
Brandenberger, R.H., Geshnizjani, G., and Watson, S., ``Initial conditions for
  brane inflation'', {\em Phys. Rev. D}, {\bf 67}, 123510, 1--10, (2003).
  {\small[\href{http://arxiv.org/abs/hep-th/0302222}{{hep-th/0302222}}]}.

\bibitem{dr_2}
Bratt, J.D., Gault, A.C., Scherrer, R.J., and Walker, T.P., ``Big Bang
  nucleosynthesis constraints on brane cosmologies'', {\em Phys. Lett. B}, {\bf
  546}, 19--22, (2002).
  {\small[\href{http://arxiv.org/abs/astro-ph/0208133}{{astro-ph/0208133}}]}.

\bibitem{sca_6}
Brax, P., Langlois, D., and Rodr{\'{\i}}guez-Mart{\'{\i}}nez, M., ``Fluctuating
  brane in a dilatonic bulk'', {\em Phys. Rev. D}, {\bf 67}, 104022, 1--11,
  (2003).
  {\small[\href{http://arxiv.org/abs/hep-th/0212067}{{hep-th/0212067}}]}.

\bibitem{rev_8}
Brax, P., and van~de Bruck, C., ``Cosmology and brane worlds: a review'', {\em
  Class. Quantum Grav.}, {\bf 20}, R201--R232, (2003).
  {\small[\href{http://arxiv.org/abs/hep-th/0303095}{{hep-th/0303095}}]}.

\bibitem{pert_14}
Brax, P., van~de Bruck, C., and Davis, A.-C., ``Brane-world cosmology, bulk
  scalars and perturbations'', {\em J. High Energy Phys.}, {\bf 2001}(10), 026,
  (2001).
  {\small[\href{http://arxiv.org/abs/hep-th/0108215}{{hep-th/0108215}}]}.

\bibitem{rbbd_2}
Brax, P., van~de Bruck, C., Davis, A.-C., and Rhodes, C.S., ``Brane World
  Moduli and the CMB'', arXiv e-print, (2003).
  {\small[\href{http://arxiv.org/abs/hep-ph/0309181}{{hep-ph/0309181}}]}.

\bibitem{bmwv}
Bridgman, H.A., Malik, K.A., and Wands, D., ``Cosmic vorticity on the brane'',
  {\em Phys. Rev. D}, {\bf 63}, 084012, 1--8, (2001).
  {\small[\href{http://arxiv.org/abs/hep-th/0010133}{{hep-th/0010133}}]}.

\bibitem{bmw_1}
Bridgman, H.A., Malik, K.A., and Wands, D., ``Cosmological perturbations in the
  bulk and on the brane'', {\em Phys. Rev. D}, {\bf 65}, 043502, 1--20, (2002).
  {\small[\href{http://arxiv.org/abs/astro-ph/0107245}{{astro-ph/0107245}}]}.

\bibitem{pert_24}
Bruni, M., and Dunsby, P.K.S., ``Singularities on the brane are not
  isotropic'', {\em Phys. Rev. D}, {\bf 66}, 101301, 1--5, (2002).
  {\small[\href{http://arxiv.org/abs/hep-th/0207189}{{hep-th/0207189}}]}.

\bibitem{mod_1}
Brustein, R., de~Alwis, S.P., and Novak, E.G., ``M-theory moduli space and
  cosmology'', {\em Phys. Rev. D}, {\bf 68}, 043507, 1--18, (2003).
  {\small[\href{http://arxiv.org/abs/hep-th/0212344}{{hep-th/0212344}}]}.

\bibitem{dq_4}
Burgess, C.P., ``Natural Quintessence and the Brane World'', arXiv e-print,
  (2002).
  {\small[\href{http://arxiv.org/abs/astro-ph/0207174}{{astro-ph/0207174}}]}.

\bibitem{Burgess:2005wu}
Burgess, C.P., ``Supersymmetric large extra dimensions and the cosmological
  constant problem'', arXiv e-print, (2005).
  {\small[\href{http://arxiv.org/abs/hep-th/0510123}{{hep-th/0510123}}]}.

\bibitem{Burgess:2004ib}
Burgess, C.P., ``Towards a natural theory of dark energy: Supersymmetric large
  extra dimensions'', in {\em The New Cosmology}, Conference on Strings and
  Cosmology and The Mitchell Symposium on Observational Cosmology, College
  Station, Texas, 14\,--\,17 March 2004, AIP Conf. Proc., vol. 743, pp.
  417--449, (2005). {\small[\href{http://dx.doi.org/10.1063/1.1848343}{DOI}]},
  {\small[\href{http://arxiv.org/abs/hep-th/0411140}{{hep-th/0411140}}]}.

\bibitem{Burgess:2007ui}
Burgess, C.P., ``Extra Dimensions and the Cosmological Constant Problem'',
  arXiv e-print, (2007).
  {\small[\href{http://arxiv.org/abs/0708.0911}{{arXiv:0708.0911
  {\small[hep-ph]}}}]}.

\bibitem{Burgess:2006ds}
Burgess, C.P., de~Rham, C., Hoover, D., Mason, D., and Tolley, A.J., ``Kicking
  the rugby ball: Perturbations of 6D gauged chiral supergravity'', {\em J.
  Cosmol. Astropart. Phys.}, {\bf 2007}(02), 009, (2007).
  {\small[\href{http://arxiv.org/abs/hep-th/0610078}{{hep-th/0610078}}]}.

\bibitem{Burgess:2008yx}
Burgess, C.P., Hoover, D., de~Rham, C., and Tasinato, G., ``Effective Field
  Theories and Matching for Codimension-2 Branes'', {\em J. High Energy Phys.},
  {\bf 2009}(03), 124, (2009).
  {\small[\href{http://dx.doi.org/10.1088/1126-6708/2009/03/124}{DOI}]},
  {\small[\href{http://arxiv.org/abs/0812.3820}{{arXiv:0812.3820
  {\small[hep-th]}}}]}.

\bibitem{Burgess:2007vi}
Burgess, C.P., Hoover, D., and Tasinato, G., ``UV Caps and Modulus
  Stabilization for 6D Gauged Chiral Supergravity'', {\em J. High Energy
  Phys.}, {\bf 2007}(09), 124, (2007).
  {\small[\href{http://dx.doi.org/10.1088/1126-6708/2007/09/124}{DOI}]},
  {\small[\href{http://arxiv.org/abs/0705.3212}{{arXiv:0705.3212
  {\small[hep-th]}}}]}.

\bibitem{Burgess:2001vr}
Burgess, C.P., Martineau, P., Quevedo, F., Rajesh, G., and Zhang, R.J., ``Brane
  antibrane inflation in orbifold and orientifold models'', {\em J. High Energy
  Phys.}, {\bf 2002}(03), 052, (2002).
  {\small[\href{http://arxiv.org/abs/hep-th/0111025}{{hep-th/0111025}}]}.

\bibitem{cal}
Calcagni, G., ``Consistency equations in Randall--Sundrum cosmology: a test for
  braneworld inflation'', {\em J. Cosmol. Astropart. Phys.}, {\bf 2003}(11),
  009, (2003).
  {\small[\href{http://arxiv.org/abs/hep-ph/0310304}{{hep-ph/0310304}}]}.

\bibitem{speed_3}
Caldwell, R.R., and Langlois, D., ``Shortcuts in the fifth dimension'', {\em
  Phys. Lett. B}, {\bf 511}, 129--135, (2001).
  {\small[\href{http://arxiv.org/abs/gr-qc/0103070}{{gr-qc/0103070}}]}.

\bibitem{ruth_3}
Campos, A., Maartens, R., Matravers, D.R., and Sopuerta, C.F., ``Braneworld
  cosmological models with anisotropy'', {\em Phys. Rev. D}, {\bf 68}, 103520,
  1--9, (2003).
  {\small[\href{http://arxiv.org/abs/hep-th/0308158}{{hep-th/0308158}}]}.

\bibitem{cs_2}
Campos, A., and Sopuerta, C.F., ``Bulk effects in the cosmological dynamics of
  brane-world scenarios'', {\em Phys. Rev. D}, {\bf 64}, 104011, 1--13, (2001).
  {\small[\href{http://arxiv.org/abs/hep-th/0105100}{{hep-th/0105100}}]}.

\bibitem{cs_1}
Campos, A., and Sopuerta, C.F., ``Evolution of cosmological models in the
  brane-world scenario'', {\em Phys. Rev. D}, {\bf 63}, 104012, 1--14, (2001).
  {\small[\href{http://arxiv.org/abs/hep-th/0101060}{{hep-th/0101060}}]}.

\bibitem{Cardoso:2007zh}
Cardoso, A., Hiramatsu, T., Koyama, K., and Seahra, S.S., ``Scalar
  perturbations in braneworld cosmology'', {\em J. Cosmol. Astropart. Phys.},
  {\bf 2007}(07), 008, (2007).
  {\small[\href{http://dx.doi.org/10.1088/1475-7516/2007/07/008}{DOI}]},
  {\small[\href{http://arxiv.org/abs/0705.1685}{{arXiv:0705.1685
  {\small[astro-ph]}}}]}.

\bibitem{Cardoso:2006nh}
Cardoso, A., Koyama, K., Mennim, A., Seahra, S.S., and Wands, D., ``Coupled
  bulk and brane fields about a de Sitter brane'', {\em Phys. Rev. D}, {\bf
  75}, 084002, (2007).
  {\small[\href{http://dx.doi.org/10.1103/PhysRevD.75.084002}{DOI}]},
  {\small[\href{http://arxiv.org/abs/hep-th/0612202}{{hep-th/0612202}}]}.

\bibitem{Cardoso:2007xc}
Cardoso, A., Koyama, K., Seahra, S.S., and Silva, F.P., ``Cosmological
  perturbations in the DGP braneworld: Numeric solution'', {\em Phys. Rev. D},
  {\bf 77}, 083512, (2008).
  {\small[\href{http://dx.doi.org/10.1103/PhysRevD.77.083512}{DOI}]},
  {\small[\href{http://arxiv.org/abs/0711.2563}{{arXiv:0711.2563
  {\small[astro-ph]}}}]}.

\bibitem{Carroll:2003db}
Carroll, S.M., and Guica, M.M., ``Sidestepping the cosmological constant with
  football-shaped extra dimensions'', arXiv e-print, (2003).
  {\small[\href{http://arxiv.org/abs/hep-th/0302067}{{hep-th/0302067}}]}.

\bibitem{bhsol_2}
Casadio, R., Fabbri, A., and Mazzacurati, L., ``New black holes in the brane
  world?'', {\em Phys. Rev. D}, {\bf 65}, 084040, 1--5, (2002).
  {\small[\href{http://arxiv.org/abs/gr-qc/0111072}{{gr-qc/0111072}}]}.

\bibitem{cas}
Casadio, R., and Mazzacurati, L., ``Bulk Shape of Brane-World Black Holes'',
  {\em Mod. Phys. Lett. A}, {\bf 18}, 651--60, (2003).
  {\small[\href{http://arxiv.org/abs/gr-qc/0205129}{{gr-qc/0205129}}]}.

\bibitem{cav}
Cavagli{\`{a}}, M., ``Black hole and brane production in TeV gravity: A
  review'', {\em Int. J. Mod. Phys. A}, {\bf 18}, 1843--1882, (2003).
  {\small[\href{http://dx.doi.org/10.1142/S0217751X03013569}{DOI}]},
  {\small[\href{http://arxiv.org/abs/hep-ph/0210296}{{hep-ph/0210296}}]}.

\bibitem{chr}
Chamblin, A., Hawking, S.W., and Reall, H.S., ``Brane-world black holes'', {\em
  Phys. Rev. D}, {\bf 61}, 065007, 1--6, (2000).
  {\small[\href{http://arxiv.org/abs/hep-th/9909205}{{hep-th/9909205}}]}.

\bibitem{ckn_1}
Chamblin, A., Karch, A., and Nayeri, A., ``Thermal equilibration of
  brane-worlds'', {\em Phys. Lett. B}, {\bf 509}, 163--167, (2001).
  {\small[\href{http://arxiv.org/abs/hep-th/0007060}{{hep-th/0007060}}]}.

\bibitem{num_2}
Chamblin, A., Reall, H.S., Shinkai, H.A., and Shiromizu, T., ``Charged
  brane-world black holes'', {\em Phys. Rev. D}, {\bf 63}, 064015, 1--11,
  (2001).
  {\small[\href{http://arxiv.org/abs/hep-th/0008177}{{hep-th/0008177}}]}.

\bibitem{gbon_15}
Charmousis, C., Davis, S.C., and Dufaux, J.-F., ``Scalar brane backgrounds in
  higher order curvature gravity'', {\em J. High Energy Phys.}, {\bf 2000}(12),
  029, (2003).
  {\small[\href{http://arxiv.org/abs/hep-th/0309083}{{hep-th/0309083}}]}.

\bibitem{gbon_6}
Charmousis, C., and Dufaux, J.-F., ``General Gauss--Bonnet brane cosmology'',
  {\em Class. Quantum Grav.}, {\bf 19}, 4671--4682, (2002).
  {\small[\href{http://arxiv.org/abs/hep-th/0202107}{{hep-th/0202107}}]}.

\bibitem{Charmousis:2006pn}
Charmousis, C., Gregory, R., Kaloper, N., and Padilla, A., ``DGP
  specteroscopy'', {\em J. High Energy Phys.}, {\bf 2006}(10), 066, (2006).
  {\small[\href{http://dx.doi.org/10.1088/1126-6708/2006/10/066}{DOI}]},
  {\small[\href{http://arxiv.org/abs/hep-th/0604086}{{hep-th/0604086}}]}.

\bibitem{Charmousis:2009uk}
Charmousis, C., Kofinas, G., and Papazoglou, A., ``The consistency of
  codimension-2 braneworlds and their cosmology'', {\em J. Cosmol. Astropart.
  Phys.}, {\bf 2010}(01), 022, (2010).
  {\small[\href{http://dx.doi.org/10.1088/1475-7516/2010/01/022}{DOI}]},
  {\small[\href{http://arxiv.org/abs/0907.1640}{{arXiv:0907.1640
  {\small[hep-th]}}}]}.

\bibitem{Charmousis:2008bt}
Charmousis, C., and Papazoglou, A., ``Self-properties of codimension-2
  braneworlds'', {\em J. High Energy Phys.}, {\bf 2008}(07), 062, (2008).
  {\small[\href{http://dx.doi.org/10.1088/1126-6708/2008/07/062}{DOI}]},
  {\small[\href{http://arxiv.org/abs/0804.2121}{{arXiv:0804.2121
  {\small[hep-th]}}}]}.

\bibitem{Charmousis:2005ez}
Charmousis, C., and Zegers, R., ``Einstein gravity on an even codimension
  brane'', {\em Phys. Rev. D}, {\bf 72}, 064005, (2005).
  {\small[\href{http://dx.doi.org/10.1103/PhysRevD.72.064005}{DOI}]},
  {\small[\href{http://arxiv.org/abs/hep-th/0502171}{{hep-th/0502171}}]}.

\bibitem{pert_20}
Chen, C.-M., Harko, T., Kao, W.F., and Mak, M.K., ``Rotational perturbations of
  Friedmann--Robertson--Walker type brane-world cosmological models'', {\em
  Nucl. Phys. B}, {\bf 636}, 159--178, (2002).
  {\small[\href{http://arxiv.org/abs/hep-th/0201012}{{hep-th/0201012}}]}.

\bibitem{Chen:2010xk}
Chen, X., ``Primordial Non-Gaussianities from Inflation Models'', arXiv
  e-print, (2010).
  {\small[\href{http://arxiv.org/abs/1002.1416}{{arXiv:1002.1416
  {\small[astro-ph.CO]}}}]}.

\bibitem{cheung}
Cheung, K., ``Collider Phenomenology for models of extra dimensions'', arXiv
  e-print, (2003).
  {\small[\href{http://arxiv.org/abs/hep-ph/0305003}{{hep-ph/0305003}}]}.

\bibitem{Chimento:2006ac}
Chimento, L.P., Lazkoz, R., Maartens, R., and Quiros, I., ``Crossing the
  phantom divide without phantom matter'', {\em J. Cosmol. Astropart. Phys.},
  {\bf 2006}(09), 004, (2006).
  {\small[\href{http://dx.doi.org/10.1088/1475-7516/2006/09/004}{DOI}]},
  {\small[\href{http://arxiv.org/abs/astro-ph/0605450}{{astro-ph/0605450}}]}.

\bibitem{speed_1}
Chung, D.J.H., and Freese, K., ``Can geodesics in extra dimensions solve the
  cosmological horizon problem?'', {\em Phys. Rev. D}, {\bf 62}, 063513, 1--7,
  (2000).
  {\small[\href{http://arxiv.org/abs/hep-ph/9910235}{{hep-ph/9910235}}]}.

\bibitem{pert_21}
Chung, D.J.H., and Freese, K., ``Lensed density perturbations in braneworlds:
  Towards an alternative to perturbations from inflation'', {\em Phys. Rev. D},
  {\bf 67}, 103505, 1--14, (2003).
  {\small[\href{http://arxiv.org/abs/astro-ph/0202066}{{astro-ph/0202066}}]}.

\bibitem{Clancy:2003zd}
Clancy, D., Guedens, R., and Liddle, A.R., ``Primordial black holes in
  braneworld cosmologies: Astrophysical constraints'', {\em Phys. Rev. D}, {\bf
  68}, 023507, 1--17, (2003).
  {\small[\href{http://dx.doi.org/10.1103/PhysRevD.68.023507}{DOI}]},
  {\small[\href{http://arxiv.org/abs/astro-ph/0301568}{{astro-ph/0301568}}]}.

\bibitem{inf_2}
Cline, J.M., Grojean, C., and Servant, G., ``Cosmological Expansion in the
  Presence of an Extra Dimension'', {\em Phys. Rev. Lett.}, {\bf 83},
  4245--4248, (1999).
  {\small[\href{http://arxiv.org/abs/hep-ph/9906523}{{hep-ph/9906523}}]}.

\bibitem{b1_4}
Coley, A.A., ``Dynamics of brane-world cosmological models'', {\em Phys. Rev.
  D}, {\bf 66}, 023512, 1--5, (2002).
  {\small[\href{http://arxiv.org/abs/hep-th/0110049}{{hep-th/0110049}}]}.

\bibitem{ruth_2}
Coley, A.A., and Hervik, S., ``Braneworld singularities'', {\em Class. Quantum
  Grav.}, {\bf 20}, 3061--3070, (2003).
  {\small[\href{http://arxiv.org/abs/gr-qc/0303003}{{gr-qc/0303003}}]}.

\bibitem{dgp_2}
Collins, H., and Holdom, B., ``Brane cosmologies without orbifolds'', {\em
  Phys. Rev. D}, {\bf 62}, 105009, 1--7, (2000).
  {\small[\href{http://arxiv.org/abs/hep-ph/0003173}{{hep-ph/0003173}}]}.

\bibitem{preheat_5}
Collins, H., Holman, R., and Martin, M.R., ``Radion-Induced Brane Preheating'',
  {\em Phys. Rev. Lett.}, {\bf 90}, 231301, 1--4, (2003).
  {\small[\href{http://arxiv.org/abs/hep-ph/0205240}{{hep-ph/0205240}}]}.

\bibitem{que_3}
Copeland, E.J., ``String Cosmology'', arXiv e-print, (2002).
  {\small[\href{http://arxiv.org/abs/hep-th/0202028}{{hep-th/0202028}}]}.

\bibitem{ek5_2}
Copeland, E.J., Gray, J., Lukas, A., and Skinner, D., ``Five-dimensional moving
  brane solutions with four-dimensional limiting behavior'', {\em Phys. Rev.
  D}, {\bf 66}, 124007, 1--11, (2002).
  {\small[\href{http://arxiv.org/abs/hep-th/0207281}{{hep-th/0207281}}]}.

\bibitem{steep_1}
Copeland, E.J., Liddle, A.R., and Lidsey, J.E., ``Steep inflation: Ending
  braneworld inflation by gravitational particle production'', {\em Phys. Rev.
  D}, {\bf 64}, 023509, 1--5, (2001).
  {\small[\href{http://arxiv.org/abs/astro-ph/0006421}{{astro-ph/0006421}}]}.

\bibitem{Copeland:2007ur}
Copeland, E.J., and Seto, O., ``Dynamical solutions of warped six dimensional
  supergravity'', {\em J. High Energy Phys.}, {\bf 2007}(08), 001, (2007).
  {\small[\href{http://dx.doi.org/10.1088/1126-6708/2007/08/001}{DOI}]},
  {\small[\href{http://arxiv.org/abs/0705.4169}{{arXiv:0705.4169
  {\small[hep-th]}}}]}.

\bibitem{Corradini:2004qr}
Corradini, O., ``4D gravity on a brane from bulk higher-curvature terms'', {\em
  Mod. Phys. Lett. A}, {\bf 20}, 2775--2784, (2005).
  {\small[\href{http://dx.doi.org/10.1142/S021773230501889X}{DOI}]},
  {\small[\href{http://arxiv.org/abs/hep-th/0405038}{{hep-th/0405038}}]}.

\bibitem{Corradini:2007cz}
Corradini, O., Koyama, K., and Tasinato, G., ``Induced gravity on intersecting
  brane-worlds Part I: Maximally symmetric solutions'', {\em Phys. Rev. D},
  {\bf 77}, 084006, (2008).
  {\small[\href{http://dx.doi.org/10.1103/PhysRevD.77.084006}{DOI}]},
  {\small[\href{http://arxiv.org/abs/0712.0385}{{arXiv:0712.0385
  {\small[hep-th]}}}]}.

\bibitem{Corradini:2008tu}
Corradini, O., Koyama, K., and Tasinato, G., ``Induced gravity on intersecting
  brane-worlds Part II: Cosmology'', {\em Phys. Rev. D}, {\bf 78}, 124002,
  (2008). {\small[\href{http://dx.doi.org/10.1103/PhysRevD.78.124002}{DOI}]},
  {\small[\href{http://arxiv.org/abs/0803.1850}{{arXiv:0803.1850
  {\small[hep-th]}}}]}.

\bibitem{dad}
Dadhich, N., ``Negative Energy Condition and Black Holes on the Brane'', {\em
  Phys. Lett. B}, {\bf 492}, 357--360, (2000).
  {\small[\href{http://arxiv.org/abs/hep-th/0009178}{{hep-th/0009178}}]}.

\bibitem{dmpr}
Dadhich, N., Maartens, R., Papadopoulos, P., and Rezania, V., ``Black holes on
  the brane'', {\em Phys. Lett. B}, {\bf 487}, 1--6, (2000).
  {\small[\href{http://arxiv.org/abs/hep-th/0003061}{{hep-th/0003061}}]}.

\bibitem{dms}
Dadhich, N., Maartens, R., Shiromizu, T., and Singh, P., in preparation.

\bibitem{inf_6}
Davis, S.C., Perkins, W.B., Davis, A.-C., and Vernon, I.R., ``Cosmological
  phase transitions in a brane world'', {\em Phys. Rev. D}, {\bf 63}, 083518,
  1--6, (2001).
  {\small[\href{http://arxiv.org/abs/hep-ph/0012223}{{hep-ph/0012223}}]}.

\bibitem{deAlwis:2003sn}
de~Alwis, S.P., ``On potentials from fluxes'', {\em Phys. Rev. D}, {\bf 68},
  126001, (2003).
  {\small[\href{http://dx.doi.org/10.1103/PhysRevD.68.126001}{DOI}]},
  {\small[\href{http://arxiv.org/abs/hep-th/0307084}{{hep-th/0307084}}]}.

\bibitem{deAlwis:2004qh}
de~Alwis, S.P., ``Brane worlds in 5D and warped compactifications in IIB'',
  {\em Phys. Lett. B}, {\bf 603}, 230--238, (2004).
  {\small[\href{http://dx.doi.org/10.1016/j.physletb.2004.10.035}{DOI}]},
  {\small[\href{http://arxiv.org/abs/hep-th/0407126}{{hep-th/0407126}}]}.

\bibitem{deRham:2004yt}
de~Rham, C., ``Beyond the low energy approximation in braneworld cosmology'',
  {\em Phys. Rev. D}, {\bf 71}, 024015, (2005).
  {\small[\href{http://dx.doi.org/10.1103/PhysRevD.71.024015}{DOI}]},
  {\small[\href{http://arxiv.org/abs/hep-th/0411021}{{hep-th/0411021}}]}.

\bibitem{deRham:2007pz}
de~Rham, C., ``The Effective Field Theory of Codimension-two Branes'', {\em J.
  High Energy Phys.}, {\bf 2008}(01), 060, (2008).
  {\small[\href{http://dx.doi.org/10.1088/1126-6708/2008/01/060}{DOI}]},
  {\small[\href{http://arxiv.org/abs/0707.0884}{{arXiv:0707.0884
  {\small[hep-th]}}}]}.

\bibitem{deRham:2007rw}
de~Rham, C., Hofmann, S., Khoury, J., and Tolley, A.J., ``Cascading gravity and
  degravitation'', {\em J. Cosmol. Astropart. Phys.}, {\bf 2008}(02), 011,
  (2008).
  {\small[\href{http://dx.doi.org/10.1088/1475-7516/2008/02/011}{DOI}]},
  {\small[\href{http://arxiv.org/abs/0712.2821}{{arXiv:0712.2821
  {\small[hep-th]}}}]}.

\bibitem{deRham:2010rw}
de~Rham, C., Khoury, J., and Tolley, A.J., ``Cascading Gravity is Ghost Free'',
  arXiv e-print, (2010).
  {\small[\href{http://arxiv.org/abs/1002.1075}{{arXiv:1002.1075
  {\small[hep-th]}}}]}.

\bibitem{deRham:2007db}
de~Rham, C., and Watson, S., ``Living on a dS brane: Effects of KK modes on
  inflation'', {\em Class. Quant. Grav.}, {\bf 24}, 4219--4234, (2007).
  {\small[\href{http://dx.doi.org/10.1088/0264-9381/24/16/015}{DOI}]},
  {\small[\href{http://arxiv.org/abs/hep-th/0702048}{{hep-th/0702048}}]}.

\bibitem{deRham:2007xp}
de~Rham, C. et~al., ``Cascading gravity: Extending the Dvali-Gabadadze-Porrati
  model to higher dimension'', {\em Phys. Rev. Lett.}, {\bf 100}, 251603,
  (2008).
  {\small[\href{http://dx.doi.org/10.1103/PhysRevLett.100.251603}{DOI}]},
  {\small[\href{http://arxiv.org/abs/0711.2072}{{arXiv:0711.2072
  {\small[hep-th]}}}]}.

\bibitem{Deffayet:2000uy}
Deffayet, C., ``Cosmology on a brane in Minkowski bulk'', {\em Phys. Lett. B},
  {\bf 502}, 199--208, (2001).
  {\small[\href{http://dx.doi.org/10.1016/S0370-2693(01)00160-5}{DOI}]},
  {\small[\href{http://arxiv.org/abs/hep-th/0010186}{{hep-th/0010186}}]}.

\bibitem{pert_22}
Deffayet, C., ``On brane world cosmological perturbations'', {\em Phys. Rev.
  D}, {\bf 66}, 103504, 1--22, (2002).
  {\small[\href{http://arxiv.org/abs/hep-th/0205084}{{hep-th/0205084}}]}.

\bibitem{ind_1}
Deffayet, C., Dvali, G., and Gabadadze, G., ``Accelerated universe from gravity
  leaking to extra dimensions'', {\em Phys. Rev. D}, {\bf 65}, 044023, 1--9,
  (2002).
  {\small[\href{http://arxiv.org/abs/astro-ph/0105068}{{astro-ph/0105068}}]}.

\bibitem{ind_5}
Deffayet, C., Landau, S.J., Raux, J., Zaldarriaga, M., and Astier, P.,
  ``Supernovae, CMB, and gravitational leakage into extra dimensions'', {\em
  Phys. Rev. D}, {\bf 66}, 024019, 1--10, (2002).
  {\small[\href{http://arxiv.org/abs/astro-ph/0201164}{{astro-ph/0201164}}]}.

\bibitem{germ_2}
Deruelle, N., ``Stars on branes: the view from the brane'', arXiv e-print,
  (2001). {\small[\href{http://arxiv.org/abs/gr-qc/0111065}{{gr-qc/0111065}}]}.

\bibitem{rev_7}
Deruelle, N., ``Cosmological perturbations of an expanding brane in an anti-de
  Sitter bulk: a short review'', {\em Astrophys. Space Sci.}, {\bf 283},
  619--626, (2003).
  {\small[\href{http://arxiv.org/abs/gr-qc/0301035}{{gr-qc/0301035}}]}.

\bibitem{gbon_1}
Deruelle, N., and Dolezel, T., ``Brane versus shell cosmologies in Einstein and
  Einstein--Gauss--Bonnet theories'', {\em Phys. Rev. D}, {\bf 62}, 103502,
  1--8, (2000).
  {\small[\href{http://arxiv.org/abs/gr-qc/0004021}{{gr-qc/0004021}}]}.

\bibitem{morepert_2}
Deruelle, N., and Dolezel, T., ``Linearized gravity in the Randall--Sundrum
  scenario'', {\em Phys. Rev. D}, {\bf 64}, 103506, 1--7, (2001).
  {\small[\href{http://arxiv.org/abs/gr-qc/0105118}{{gr-qc/0105118}}]}.

\bibitem{pert_11}
Deruelle, N., Dolezel, T., and Katz, J., ``Perturbations of brane worlds'',
  {\em Phys. Rev. D}, {\bf 63}, 083513, 1--10, (2001).
  {\small[\href{http://arxiv.org/abs/hep-th/0010215}{{hep-th/0010215}}]}.

\bibitem{gbon_14}
Deruelle, N., and Germani, C., ``Smooth branes and junction conditions in
  Einstein Gauss--Bonnet gravity'', {\em Nuovo Cimento B}, {\bf 118}, 977--988,
  (2003). {\small[\href{http://arxiv.org/abs/gr-qc/0306116}{{gr-qc/0306116}}]}.

\bibitem{gbon_12}
Deruelle, N., and Madore, J., ``On the quasi-linearity of the Einstein
  `Gauss--Bonnet' gravity field equations'', arXiv e-print, (2003).
  {\small[\href{http://arxiv.org/abs/gr-qc/0305004}{{gr-qc/0305004}}]}.

\bibitem{ind_3}
Diamandis, G.A., Georgalas, B.C., Mavromatos, N.E., Papantonopoulos, E., and
  Pappa, I., ``Cosmological Evolution in a Type-0 String Theory'', {\em Int. J.
  Mod. Phys. A}, {\bf 17}, 2241--2265, (2002).
  {\small[\href{http://dx.doi.org/10.1142/S0217751X02010534}{DOI}]},
  {\small[\href{http://arxiv.org/abs/hep-th/0107124}{{hep-th/0107124}}]}.

\bibitem{pert_15}
Dorca, M., and van~de Bruck, C., ``Cosmological perturbations in brane worlds:
  brane bending and anisotropic stresses'', {\em Nucl. Phys. B}, {\bf 605},
  215--233, (2001).
  {\small[\href{http://arxiv.org/abs/hep-th/0012116}{{hep-th/0012116}}]}.

\bibitem{Dubovsky:2002jm}
Dubovsky, S.L., and Rubakov, V.A., ``Brane-induced gravity in more than one
  extra dimensions: Violation of equivalence principle and ghost'', {\em Phys.
  Rev. D}, {\bf 67}, 104014, (2003).
  {\small[\href{http://dx.doi.org/10.1103/PhysRevD.67.104014}{DOI}]},
  {\small[\href{http://arxiv.org/abs/hep-th/0212222}{{hep-th/0212222}}]}.

\bibitem{acft}
Duff, M.J., and Liu, J.T., ``Complementarity of the Maldacena and
  Randall--Sundrum Pictures'', {\em Phys. Rev. Lett.}, {\bf 85}, 2052--2055,
  (2000).
  {\small[\href{http://arxiv.org/abs/hep-th/0003237}{{hep-th/0003237}}]}.

\bibitem{durkoc}
Durrer, R., and Kocian, P., ``Testing extra dimensions with the binary
  pulsar'', {\em Class. Quantum Grav.}, {\bf 21}, 2127--2137, (2004).
  {\small[\href{http://arxiv.org/abs/hep-th/0305181}{{hep-th/0305181}}]}.

\bibitem{dgp_1}
Dvali, G., Gabadadze, G., and Porrati, M., ``4D gravity on a brane in 5D
  Minkowski space'', {\em Phys. Lett. B}, {\bf 485}, 208--214, (2000).
  {\small[\href{http://adsabs.harvard.edu/abs/2000PhLB..485..208D}{ADS}]},
  {\small[\href{http://arxiv.org/abs/hep-th/0005016}{{hep-th/0005016}}]}.

\bibitem{Dvali:2007kt}
Dvali, G., Hofmann, S., and Khoury, J., ``Degravitation of the cosmological
  constant and graviton width'', {\em Phys. Rev. D}, {\bf 76}, 084006, (2007).
  {\small[\href{http://dx.doi.org/10.1103/PhysRevD.76.084006}{DOI}]},
  {\small[\href{http://arxiv.org/abs/hep-th/0703027}{{hep-th/0703027}}]}.

\bibitem{kss_1}
Dvali, G., and Tye, S.-H.H., ``Brane inflation'', {\em Phys. Lett. B}, {\bf
  450}, 72--82, (1999).
  {\small[\href{http://arxiv.org/abs/hep-ph/9812483}{{hep-ph/9812483}}]}.

\bibitem{Dvali:2000hr}
Dvali, G.~R., Gabadadze, G., and Porrati, M., ``4D gravity on a brane in 5D
  Minkowski space'', {\em Phys. Lett. B}, {\bf 485}, 208--214, (2000).
  {\small[\href{http://dx.doi.org/10.1016/S0370-2693(00)00669-9}{DOI}]},
  {\small[\href{http://arxiv.org/abs/hep-th/0005016}{{hep-th/0005016}}]}.

\bibitem{Dvali:2001fw}
Dvali, G.R., Shafi, Q., and Solganik, S., ``D-brane inflation'', arXiv e-print,
  (2001).
  {\small[\href{http://arxiv.org/abs/hep-th/0105203}{{hep-th/0105203}}]}.

\bibitem{elmw}
Easther, R., Langlois, D., Maartens, R., and Wands, D., ``Evolution of
  gravitational waves in Randall--Sundrum cosmology'', {\em J. Cosmol.
  Astropart. Phys.}, {\bf 2003}(10), 014, (2003).
  {\small[\href{http://arxiv.org/abs/hep-th/0308078}{{hep-th/0308078}}]}.

\bibitem{covp}
Ellis, G.F.R., and van Elst, H., ``Cosmological Models. Carg\`ese lectures
  1998'', in Lachi{\`{e}}ze-Rey, M., ed., {\em Theoretical and Observational
  Cosmology}, Proceedings of the NATO Advanced Study Institute, Carg\`ese,
  France, August 17\,--\,29, 1998, NATO Science Series C, vol. 451, (Kluwer,
  Dordrecht; Boston, 1999).
  {\small[\href{http://arxiv.org/abs/gr-qc/9812046}{{gr-qc/9812046}}]}.

\bibitem{gid_2}
Emparan, R., ``Black Hole Production at a TeV'', arXiv e-print, (2003).
  {\small[\href{http://arxiv.org/abs/hep-ph/0302226}{{hep-ph/0302226}}]}.

\bibitem{efk}
Emparan, R., Fabbri, A., and Kaloper, N., ``Quantum Black Holes as Holograms in
  AdS Braneworlds'', {\em J. High Energy Phys.}, {\bf 2002}(08), 043, (2002).
  {\small[\href{http://arxiv.org/abs/hep-th/0206155}{{hep-th/0206155}}]}.

\bibitem{egk}
Emparan, R., Garc{\'{\i}}a-Bellido, J., and Kaloper, N., ``Black Hole
  Astrophysics in AdS Braneworlds'', {\em J. High Energy Phys.}, {\bf
  2003}(01), 079, (2003).
  {\small[\href{http://arxiv.org/abs/hep-th/0212132}{{hep-th/0212132}}]}.

\bibitem{ehm}
Emparan, R., Horowitz, G.T., and Meyers, R.C., ``Exact Description of Black
  Holes on Branes'', {\em J. High Energy Phys.}, {\bf 2000}(01), 007, (2000).
  {\small[\href{http://arxiv.org/abs/hep-th/9911043}{{hep-th/9911043}}]}.

\bibitem{Fang:2008kc}
Fang, W., Wang, S., Hu, W., Haiman, Z., Hui, L., and May, M., ``Challenges to
  the DGP model from horizon-scale growth and geometry'', {\em Phys. Rev. D},
  {\bf 78}, 103509, (2008).
  {\small[\href{http://dx.doi.org/10.1103/PhysRevD.78.103509}{DOI}]},
  {\small[\href{http://arxiv.org/abs/0808.2208}{{arXiv:0808.2208
  {\small[astro-ph]}}}]}.

\bibitem{sca_3}
Feinstein, A., Kunze, K.E., and Vazquez-Mozo, M.A., ``Curved dilatonic brane
  worlds'', {\em Phys. Rev. D}, {\bf 64}, 084015, 1--9, (2001).
  {\small[\href{http://arxiv.org/abs/hep-th/0105182}{{hep-th/0105182}}]}.

\bibitem{sca_5}
Felder, G.N., Frolov, A., and Kofman, L., ``Warped geometry of brane worlds'',
  {\em Class. Quantum Grav.}, {\bf 19}, 2983--3002, (2002).
  {\small[\href{http://arxiv.org/abs/hep-th/0112165}{{hep-th/0112165}}]}.

\bibitem{morers2_7}
Flanagan, {\'{E}}.{\'{E}}., Henry~Tye, S.-H., and Wasserman, I., ``Cosmological
  expansion in the Randall--Sundrum brane world scenario'', {\em Phys. Rev. D},
  {\bf 62}, 044039, 1--6, (2000).
  {\small[\href{http://arxiv.org/abs/hep-ph/9910498}{{hep-ph/9910498}}]}.

\bibitem{hs_3}
Flanagan, {\'{E}}.{\'{E}}., Henry~Tye, S.-H., and Wasserman, I., ``Brane world
  models with bulk scalar fields'', {\em Phys. Lett. B}, {\bf 522}, 155--165,
  (2001).
  {\small[\href{http://arxiv.org/abs/hep-th/0110070}{{hep-th/0110070}}]}.

\bibitem{que_1}
F{\"{o}}rste, S., ``Strings, Branes and Extra Dimensions'', {\em Fortschr.
  Phys.}, {\bf 50}, 221--403, (2002).
  {\small[\href{http://arxiv.org/abs/hep-th/0110055}{{hep-th/0110055}}]}.

\bibitem{fk}
Frolov, A.V., and Kofman, L.A., ``Gravitational Waves from Braneworld
  Inflation'', arXiv e-print, (2002).
  {\small[\href{http://arxiv.org/abs/hep-th/0209133}{{hep-th/0209133}}]}.

\bibitem{hs_16}
Frolov, A.V., and Kofman, L.A., ``Can inflating braneworlds be stabilized?'',
  {\em Phys. Rev. D}, {\bf 69}, 044021, 1--7, (2004).
  {\small[\href{http://arxiv.org/abs/hep-th/0309002}{{hep-th/0309002}}]}.

\bibitem{Fujii:2007fi}
Fujii, S., Kobayashi, T., and Shiromizu, T., ``Low energy effective theory on a
  regularized brane in six-dimensional flux compactifications'', {\em Phys.
  Rev. D}, {\bf 76}, 104052, (2007).
  {\small[\href{http://dx.doi.org/10.1103/PhysRevD.76.104052}{DOI}]},
  {\small[\href{http://arxiv.org/abs/0708.2534}{{arXiv:0708.2534
  {\small[hep-th]}}}]}.

\bibitem{r_4}
Gabadadze, G., ``ICTP Lectures on Large Extra Dimensions'', arXiv e-print,
  (2003).
  {\small[\href{http://arxiv.org/abs/hep-ph/0308112}{{hep-ph/0308112}}]}.

\bibitem{Gabadadze:2003ck}
Gabadadze, G., and Shifman, M., ``Softly massive gravity'', {\em Phys. Rev. D},
  {\bf 69}, 124032, (2004).
  {\small[\href{http://dx.doi.org/10.1103/PhysRevD.69.124032}{DOI}]},
  {\small[\href{http://arxiv.org/abs/hep-th/0312289}{{hep-th/0312289}}]}.

\bibitem{Garriga:2004tq}
Garriga, J., and Porrati, M., ``Football shaped extra dimensions and the
  absence of self-tuning'', {\em J. High Energy Phys.}, {\bf 2004}(08), 028,
  (2004).
  {\small[\href{http://arxiv.org/abs/hep-th/0406158}{{hep-th/0406158}}]}.

\bibitem{gs}
Garriga, J., and Sasaki, M., ``Brane-world creation and black holes'', {\em
  Phys. Rev. D}, {\bf 62}, 043523, 1--8, (2000).
  {\small[\href{http://arxiv.org/abs/hep-th/9912118}{{hep-th/9912118}}]}.

\bibitem{gt}
Garriga, J., and Tanaka, T., ``Gravity in the Randall--Sundrum Brane World'',
  {\em Phys. Rev. Lett.}, {\bf 84}, 2778--2781, (2000).
  {\small[\href{http://arxiv.org/abs/hep-th/9911055}{{hep-th/9911055}}]}.

\bibitem{bub_2}
Garriga, J., and Tanaka, T., ``Cosmological perturbations in the 5D big bang'',
  {\em Phys. Rev. D}, {\bf 65}, 103506, 1--7, (2002).
  {\small[\href{http://arxiv.org/abs/hep-th/0112028}{{hep-th/0112028}}]}.

\bibitem{bub_3}
Gen, U., Ishibashi, A., and Tanaka, T., ``Brane collisions and braneworld
  cosmology'', {\em Prog. Theor. Phys. Suppl.}, {\bf 148}, 267--275, (2002).
  {\small[\href{http://arxiv.org/abs/hep-th/0207140}{{hep-th/0207140}}]}.

\bibitem{pert_12}
Gen, U., and Sasaki, M., ``Radion on the de Sitter Brane'', {\em Prog. Theor.
  Phys.}, {\bf 105}, 591--606, (2001).
  {\small[\href{http://arxiv.org/abs/gr-qc/0011078}{{gr-qc/0011078}}]}.

\bibitem{bgm}
Germani, C., Bruni, M., and Maartens, R., ``Gravitational Collapse on the
  Brane: A No-Go Theorem'', {\em Phys. Rev. Lett.}, {\bf 87}, 231302, 1--4,
  (2001). {\small[\href{http://arxiv.org/abs/gr-qc/0108013}{{gr-qc/0108013}}]}.

\bibitem{germ_1}
Germani, C., and Maartens, R., ``Stars in the braneworld'', {\em Phys. Rev. D},
  {\bf 64}, 124010, 1--6, (2001).
  {\small[\href{http://arxiv.org/abs/hep-th/0107011}{{hep-th/0107011}}]}.

\bibitem{gbon_5}
Germani, C., and Sopuerta, C.F., ``String Inspired Brane World Cosmology'',
  {\em Phys. Rev. Lett.}, {\bf 88}, 231101, 1--4, (2002).
  {\small[\href{http://arxiv.org/abs/hep-th/0202060}{{hep-th/0202060}}]}.

\bibitem{Geroch:1987qn}
Geroch, R.P., and Traschen, J.H., ``Strings and Other Distributional Sources in
  General Relativity'', {\em Phys. Rev. D}, {\bf 36}, 1017, (1987).
  {\small[\href{http://dx.doi.org/10.1103/PhysRevD.36.1017}{DOI}]}.

\bibitem{giaren}
Giannakis, I., and Ren, H., ``Possible extensions of the 4D Schwarzschild
  horizon in the brane world'', {\em Phys. Rev. D}, {\bf 63}, 125017, 1--6,
  (2001).
  {\small[\href{http://arxiv.org/abs/hep-th/0010183}{{hep-th/0010183}}]}.

\bibitem{gr1}
Giannakis, I., and Ren, H., ``Recovery of the Schwarzschild metric in theories
  with localized gravity beyond linear order'', {\em Phys. Rev. D}, {\bf 63},
  024001, 1--9, (2001).
  {\small[\href{http://arxiv.org/abs/hep-th/0007053}{{hep-th/0007053}}]}.

\bibitem{Giannantonio:2009gi}
Giannantonio, T., Martinelli, M., Silvestri, A., and Melchiorri, A., ``New
  constraints on parametrised modified gravity from correlations of the CMB
  with large scale structure'', (2009).
  {\small[\href{http://arxiv.org/abs/0909.2045}{{arXiv:0909.2045
  {\small[astro-ph.CO]}}}]}.

\bibitem{Giannantonio:2008zi}
Giannantonio, T., Scranton, R., Crittenden, R.G., Nichol, R.C., Boughn, S.P.,
  Myers, A.D., and Richards, G.T., ``Combined analysis of the integrated
  Sachs-Wolfe effect and cosmological implications'', {\em Phys. Rev. D}, {\bf
  77}, 123520, (2008).
  {\small[\href{http://dx.doi.org/10.1103/PhysRevD.77.123520}{DOI}]},
  {\small[\href{http://arxiv.org/abs/0801.4380}{{arXiv:0801.4380
  {\small[astro-ph]}}}]}.

\bibitem{Giannantonio:2008qr}
Giannantonio, T., Song, Y.-S., and Koyama, K., ``Detectability of a
  phantom-like braneworld model with the integrated Sachs-Wolfe effect'', {\em
  Phys. Rev. D}, {\bf 78}, 044017, (2008).
  {\small[\href{http://dx.doi.org/10.1103/PhysRevD.78.044017}{DOI}]},
  {\small[\href{http://arxiv.org/abs/0803.2238}{{arXiv:0803.2238
  {\small[astro-ph]}}}]}.

\bibitem{add_7}
Gibbons, G.W., and Wiltshire, D.L., ``Spacetime as a membrane in higher
  dimensions'', {\em Nucl. Phys. B}, {\bf 287}, 717--742, (1987).
  {\small[\href{http://arxiv.org/abs/hep-th/0109093}{{hep-th/0109093}}]}.

\bibitem{gid_1}
Giddings, S.B., ``Black holes in the lab?'', {\em Gen. Relativ. Gravit.}, {\bf
  34}, 1775--1779, (2002).
  {\small[\href{http://arxiv.org/abs/hep-th/0205205}{{hep-th/0205205}}]}.

\bibitem{Giddings:2001yu}
Giddings, S.B., Kachru, S., and Polchinski, J., ``Hierarchies from fluxes in
  string compactifications'', {\em Phys. Rev. D}, {\bf 66}, 106006, (2002).
  {\small[\href{http://dx.doi.org/10.1103/PhysRevD.66.106006}{DOI}]},
  {\small[\href{http://arxiv.org/abs/hep-th/0105097}{{hep-th/0105097}}]}.

\bibitem{morepert_1}
Giddings, S.B., Katz, E., and Randall, L., ``Linearized Gravity in Brane
  Backgrounds'', {\em J. High Energy Phys.}, {\bf 2000}(03), 023, (2000).
  {\small[\href{http://arxiv.org/abs/hep-th/0002091}{{hep-th/0002091}}]}.

\bibitem{gklr}
Giudice, G.F., Kolb, E.W., Lesgourgues, J., and Riotto, A., ``Transdimensional
  physics and inflation'', {\em Phys. Rev. D}, {\bf 66}, 083512, 1--15, (2002).
  {\small[\href{http://arxiv.org/abs/hep-ph/0207145}{{hep-ph/0207145}}]}.

\bibitem{add_8}
Gogberashvili, M., ``Our world as an expanding shell'', {\em Europhys. Lett.},
  {\bf 49}, 396, (2000).
  {\small[\href{http://arxiv.org/abs/hep-ph/9812365}{{hep-ph/9812365}}]}.

\bibitem{goldwise}
Goldberger, W.D., and Wise, M.B., ``Modulus Stabilization with Bulk Fields'',
  {\em Phys. Rev. Lett.}, {\bf 83}, 4922--4925, (1999).
  {\small[\href{http://arxiv.org/abs/hep-ph/9907447}{{hep-ph/9907447}}]}.

\bibitem{Gorbunov:2005zk}
Gorbunov, D., Koyama, K., and Sibiryakov, S., ``More on ghosts in the
  Dvali-Gabadaze-Porrati model'', {\em Phys. Rev. D}, {\bf 73}, 044016, (2006).
  {\small[\href{http://dx.doi.org/10.1103/PhysRevD.73.044016}{DOI}]},
  {\small[\href{http://arxiv.org/abs/hep-th/0512097}{{hep-th/0512097}}]}.

\bibitem{grs}
Gorbunov, D.S., Rubakov, V.A., and Sibiryakov, S.M., ``Gravity waves from
  inflating brane or Mirrors moving in AdS$_5$'', {\em J. High Energy Phys.},
  {\bf 2001}(10), 015, (2001).
  {\small[\href{http://arxiv.org/abs/hep-th/0108017}{{hep-th/0108017}}]}.

\bibitem{gm}
Gordon, C., and Maartens, R., ``Density perturbations in the brane-world'',
  {\em Phys. Rev. D}, {\bf 63}, 044022, 1--7, (2001).
  {\small[\href{http://arxiv.org/abs/hep-th/0009010}{{hep-th/0009010}}]}.

\bibitem{govdad}
Govender, M., and Dadhich, N., ``Collapsing sphere on the brane radiates'',
  {\em Phys. Lett. B}, {\bf 538}, 233--238, (2002).
  {\small[\href{http://arxiv.org/abs/hep-th/0109086}{{hep-th/0109086}}]}.

\bibitem{mod_5}
Gray, J., and Lukas, A., ``Gauge five-brane moduli in four-dimensional
  heterotic models'', {\em Phys. Rev. D}, {\bf 70}, 086003, 1--15, (2004).
  {\small[\href{http://arxiv.org/abs/hep-th/0309096}{{hep-th/0309096}}]}.

\bibitem{acftcosmo_7}
Gregory, J.P., and Padilla, A., ``Exact braneworld cosmology induced from bulk
  black holes'', {\em Class. Quantum Grav.}, {\bf 19}, 4071--4083, (2002).
  {\small[\href{http://arxiv.org/abs/hep-th/0204218}{{hep-th/0204218}}]}.

\bibitem{gbon_11}
Gregory, J.P., and Padilla, A., ``Braneworld holography in Gauss--Bonnet
  gravity'', {\em Class. Quantum Grav.}, {\bf 20}, 4221--4238, (2003).
  {\small[\href{http://arxiv.org/abs/hep-th/0304250}{{hep-th/0304250}}]}.

\bibitem{g}
Gregory, R., ``Black string instabilities in anti-de Sitter space'', {\em
  Class. Quantum Grav.}, {\bf 17}, L125--L131, (2000).
  {\small[\href{http://arxiv.org/abs/hep-th/0004101}{{hep-th/0004101}}]}.

\bibitem{morers2_4}
Gubser, S.S., ``AdS/CFT and gravity'', {\em Phys. Rev. D}, {\bf 63}, 084017,
  1--13, (2001).
  {\small[\href{http://arxiv.org/abs/hep-th/9912001}{{hep-th/9912001}}]}.

\bibitem{Guedens:2002sd}
Guedens, R., Clancy, D., and Liddle, A.R., ``Primordial black holes in
  braneworld cosmologies: Accretion after formation'', {\em Phys. Rev. D}, {\bf
  66}, 083509, (2002).
  {\small[\href{http://dx.doi.org/10.1103/PhysRevD.66.083509}{DOI}]},
  {\small[\href{http://arxiv.org/abs/astro-ph/0208299}{{astro-ph/0208299}}]}.

\bibitem{Guedens:2002km}
Guedens, R., Clancy, D., and Liddle, A.R, ``Primordial black holes in
  braneworld cosmologies: Formation, cosmological evolution and evaporation'',
  {\em Phys. Rev. D}, {\bf 66}, 043513, (2002).
  {\small[\href{http://dx.doi.org/10.1103/PhysRevD.66.043513}{DOI}]},
  {\small[\href{http://arxiv.org/abs/astro-ph/0205149}{{astro-ph/0205149}}]}.

\bibitem{ind_14}
Gumjudpai, B., ``Brane-Cosmology Dynamics with Induced Gravity'', {\em Gen.
  Relativ. Gravit.}, {\bf 36}, 747--766, (2004).
  {\small[\href{http://arxiv.org/abs/gr-qc/0308046}{{gr-qc/0308046}}]}.

\bibitem{ggm}
Gumjudpai, B., Maartens, R., and Gordon, C., ``Density perturbations in a
  braneworld universe with dark radiation'', {\em Class. Quantum Grav.}, {\bf
  20}, 3295--3306, (2003).
  {\small[\href{http://arxiv.org/abs/gr-qc/0304067}{{gr-qc/0304067}}]}.

\bibitem{Guzzo:2008ac}
Guzzo, L., Pierleoni, M., Meneux, B., Branchini, E., Le~F\`evre, O., Marinoni,
  C., Garilli, B., Blaizot, J., De~Lucia, G., Pollo, A., McCracken, H.~J.,
  Bottini, D., Le~Brun, V., Maccagni, D., Picat, J.~P., Scaramella, R.,
  Scodeggio, M., Tresse, L., Vettolani, G., Zanichelli, A., Adami, C., Arnouts,
  S., Bardelli, S., Bolzonella, M., Bongiorno, A., Cappi, A., Charlot, S.,
  Ciliegi, P., Contini, T., Cucciati, O., de~la Torre, S., Dolag, K., Foucaud,
  S., Franzetti, P., Gavignaud, I., Ilbert, O., Iovino, A., Lamareille, F.,
  Marano, B., Mazure, A., Memeo, P., Merighi, R., Moscardini, L., Paltani, S.,
  Pell\`o, R., Perez-Montero, E., Pozzetti, L., Radovich, M., Vergani, D.,
  Zamorani, G., and Zucca, E., ``A test of the nature of cosmic acceleration
  using galaxy redshift distortions'', {\em Nature}, {\bf 451}, 541--544,
  (2008). {\small[\href{http://dx.doi.org/10.1038/nature06555}{DOI}]},
  {\small[\href{http://arxiv.org/abs/0802.1944}{{arXiv:0802.1944
  {\small[astro-ph]}}}]}.

\bibitem{hanraf_1}
Hannestad, S., and Raffelt, G.G., ``Stringent Neutron-Star Limits on Large
  Extra Dimensions'', {\em Phys. Rev. Lett.}, {\bf 88}, 071301, 1--4, (2002).
  {\small[\href{http://dx.doi.org/10.1103/PhysRevLett.88.071301}{DOI}]},
  {\small[\href{http://arxiv.org/abs/hep-ph/0110067}{{hep-ph/0110067}}]}.

\bibitem{pert_1}
Hawking, S.W., Hertog, T., and Reall, H.S., ``Brane new world'', {\em Phys.
  Rev. D}, {\bf 62}, 043501, 1--16, (2000).
  {\small[\href{http://arxiv.org/abs/hep-th/0003052}{{hep-th/0003052}}]}.

\bibitem{pert_10}
Hawking, S.W., Hertog, T., and Reall, H.S., ``Trace anomaly driven inflation'',
  {\em Phys. Rev. D}, {\bf 63}, 083504, 1--23, (2001).
  {\small[\href{http://arxiv.org/abs/hep-th/0010232}{{hep-th/0010232}}]}.

\bibitem{inf_12}
Hawkins, R., and Lidsey, J.E., ``Inflationary energy scale in braneworld
  cosmology'', {\em Phys. Rev. D}, {\bf 68}, 083505, 1--10, (2003).
  {\small[\href{http://arxiv.org/abs/astro-ph/0306311}{{astro-ph/0306311}}]}.

\bibitem{acftcosmo_8}
Hebecker, A., and March-Russell, J., ``Randall--Sundrum II cosmology, AdS/CFT,
  and the bulk black hole'', {\em Nucl. Phys. B}, {\bf 608}, 375--393, (2001).
  {\small[\href{http://arxiv.org/abs/hep-ph/0103214}{{hep-ph/0103214}}]}.

\bibitem{hanraf_2}
Hewett, J., and Spiropulu, M., ``Particle Physics Probes of Extra Spacetime
  Dimensions'', {\em Annu. Rev. Nucl. Part. Sci.}, {\bf 52}, 397--424, (2002).
  {\small[\href{http://dx.doi.org/10.1146/annurev.nucl.52.050102.090706}{DOI}]%
}, {\small[\href{http://arxiv.org/abs/hep-ph/0205106}{{hep-ph/0205106}}]}.

\bibitem{hs_2}
Himemoto, Y., and Sasaki, M., ``Brane-world inflation without inflaton on the
  brane'', {\em Phys. Rev. D}, {\bf 63}, 044015, 1--8, (2001).
  {\small[\href{http://arxiv.org/abs/gr-qc/0010035}{{gr-qc/0010035}}]}.

\bibitem{hs_11}
Himemoto, Y., and Sasaki, M., ``Braneworld Inflation Driven by Dynamics of a
  Bulk Scalar Field'', {\em Prog. Theor. Phys. Suppl.}, {\bf 148}, 235--244,
  (2002). {\small[\href{http://arxiv.org/abs/gr-qc/0302054}{{gr-qc/0302054}}]}.

\bibitem{hs_6}
Himemoto, Y., and Tanaka, T., ``Braneworld reheating in the bulk inflaton
  model'', {\em Phys. Rev. D}, {\bf 67}, 084014, 1--5, (2003).
  {\small[\href{http://arxiv.org/abs/gr-qc/0212114}{{gr-qc/0212114}}]}.

\bibitem{hs_5}
Himemoto, Y., Tanaka, T., and Sasaki, M., ``Bulk scalar field in the braneworld
  can mimic the 4D inflaton dynamics'', {\em Phys. Rev. D}, {\bf 65}, 104020,
  1--9, (2002).
  {\small[\href{http://arxiv.org/abs/gr-qc/0112027}{{gr-qc/0112027}}]}.

\bibitem{Hiramatsu:2006bd}
Hiramatsu, T., ``High-energy effects on the spectrum of inflationary
  gravitational wave background in braneworld cosmology'', {\em Phys. Rev. D},
  {\bf 73}, 084008, (2006).
  {\small[\href{http://dx.doi.org/10.1103/PhysRevD.73.084008}{DOI}]},
  {\small[\href{http://arxiv.org/abs/hep-th/0601105}{{hep-th/0601105}}]}.

\bibitem{Hiramatsu:2006cv}
Hiramatsu, T., and Koyama, K., ``Numerical study of curvature perturbations in
  a brane-world inflation at high-energies'', {\em J. Cosmol. Astropart.
  Phys.}, {\bf 2006}(12), 009, (2006).
  {\small[\href{http://dx.doi.org/10.1088/1475-7516/2006/12/009}{DOI}]},
  {\small[\href{http://arxiv.org/abs/hep-th/0607068}{{hep-th/0607068}}]}.

\bibitem{kkt}
Hiramatsu, T., Koyama, K., and Taruya, A., ``Evolution of gravitational waves
  from inflationary brane-world: numerical study of high-energy effects'', {\em
  Phys. Lett. B}, {\bf 578}, 269--275, (2004).
  {\small[\href{http://arxiv.org/abs/hep-th/0308072}{{hep-th/0308072}}]}.

\bibitem{Hogan:2000wz}
Hogan, C.J., ``Brane-World Astronomy'', in Wheeler, J.C., and Martle, H., eds.,
  {\em Relativistic Astrophysics: 20th Texas Symposium}, Austin, Texas (USA),
  10\,--\,15 December 2000, AIP Conference Proceedings, vol. 586, pp. 11--21,
  (American Institute of Physics, Melville, NY, 2000).
  {\small[\href{http://dx.doi.org/10.1063/1.1419526}{DOI}]},
  {\small[\href{http://arxiv.org/abs/astro-ph/0104105}{{astro-ph/0104105}}]}.

\bibitem{hv}
Horava, P., and Witten, E., ``Heterotic and Type I string dynamics from eleven
  dimensions'', {\em Nucl. Phys. B}, {\bf 460}, 506--524, (1996).
  {\small[\href{http://dx.doi.org/10.1016/0550-3213(95)00621-4}{DOI}]},
  {\small[\href{http://arxiv.org/abs/hep-th/9510209}{{hep-th/9510209}}]}.

\bibitem{hovmy}
Hovdebo, J.L., and Myers, R.C., ``Bouncing braneworlds go crunch!'', {\em J.
  Cosmol. Astropart. Phys.}, {\bf 2003}(11), 012, (2003).
  {\small[\href{http://arxiv.org/abs/hep-th/0308088}{{hep-th/0308088}}]}.

\bibitem{hulid1}
Huey, G., and Lidsey, J.E., ``Inflation, braneworlds and quintessence'', {\em
  Phys. Lett. B}, {\bf 514}, 217--225, (2001).
  {\small[\href{http://arxiv.org/abs/astro-ph/0104006}{{astro-ph/0104006}}]}.

\bibitem{hulid2}
Huey, G., and Lidsey, J.E., ``Inflation and braneworlds: Degeneracies and
  consistencies'', {\em Phys. Rev. D}, {\bf 66}, 043514, 1--8, (2002).
  {\small[\href{http://arxiv.org/abs/astro-ph/0205236}{{astro-ph/0205236}}]}.

\bibitem{ichnak_2}
Ichiki, K., and Nakamura, K., ``Causal structure and gravitational waves in
  brane world cosmology'', {\em Phys. Rev. D}, {\bf 70}, 064017, 1--9, (2004).
  {\small[\href{http://arxiv.org/abs/hep-th/0310282}{{hep-th/0310282}}]}.

\bibitem{dr_1}
Ichiki, K., Yahiro, M., Kajino, T., Orito, M., and Mathews, G.J.,
  ``Observational constraints on dark radiation in brane cosmology'', {\em
  Phys. Rev. D}, {\bf 66}, 043521, 1--5, (2002).
  {\small[\href{http://arxiv.org/abs/astro-ph/0203272}{{astro-ph/0203272}}]}.

\bibitem{morers2_6}
Ida, D., ``Brane-world cosmology'', {\em J. High Energy Phys.}, {\bf 2000}(09),
  014, (2000).
  {\small[\href{http://arxiv.org/abs/gr-qc/9912002}{{gr-qc/9912002}}]}.

\bibitem{inta}
Inoue, K.T., and Tanaka, T., ``Gravitational Waves from Sub-Lunar-Mass
  Primordial Black-Hole Binaries: A New Probe of Extradimensions'', {\em Phys.
  Rev. Lett.}, {\bf 91}, 021101, 1--4, (2003).
  {\small[\href{http://arxiv.org/abs/gr-qc/0303058}{{gr-qc/0303058}}]}.

\bibitem{speed_2}
Ishihara, H., ``Causality of the Brane Universe'', {\em Phys. Rev. Lett.}, {\bf
  86}, 381--384, (2001).
  {\small[\href{http://arxiv.org/abs/gr-qc/0007070}{{gr-qc/0007070}}]}.

\bibitem{Kachru:2003sx}
Kachru, S., Kallosh, R., Linde, A., Maldacena, J.M., McAllister, L., and
  Trivedi, S.P., ``Towards inflation in string theory'', {\em J. Cosmol.
  Astropart. Phys.}, {\bf 2003}(10), 013, (2003).
  {\small[\href{http://dx.doi.org/10.1088/1475-7516/2003/10/013}{DOI}]},
  {\small[\href{http://arxiv.org/abs/hep-th/0308055}{{hep-th/0308055}}]}.

\bibitem{mod_3}
Kachru, S., Kallosh, R.E., Linde, A.D., Maldacena, J.M., McAllister, L., and
  Trivedi, S.P., ``Towards inflation in string theory'', {\em J. Cosmol.
  Astropart. Phys.}, {\bf 2003}(10), 013, (2003).
  {\small[\href{http://arxiv.org/abs/hep-th/0308055}{{hep-th/0308055}}]}.

\bibitem{mtheory_1}
Kallosh, R.E., ``Supergravity, M theory and Cosmology'', arXiv e-print, (2002).
  {\small[\href{http://arxiv.org/abs/hep-th/0205315}{{hep-th/0205315}}]}.

\bibitem{ek_2}
Kallosh, R.E., Kofman, L.A., and Linde, A., ``Pyrotechnic universe'', {\em
  Phys. Rev. D}, {\bf 64}, 123523, 1--18, (2001).
  {\small[\href{http://arxiv.org/abs/hep-th/0104073}{{hep-th/0104073}}]}.

\bibitem{morers2_2}
Kaloper, N., ``Bent domain walls as braneworlds'', {\em Phys. Rev. D}, {\bf
  60}, 123506, 1--14, (1999).
  {\small[\href{http://arxiv.org/abs/hep-th/9905210}{{hep-th/9905210}}]}.

\bibitem{Kaloper:2004cy}
Kaloper, N., ``Origami world'', {\em J. High Energy Phys.}, {\bf 2004}(05),
  061, (2004).
  {\small[\href{http://arxiv.org/abs/hep-th/0403208}{{hep-th/0403208}}]}.

\bibitem{Kaloper:2007qh}
Kaloper, N., ``Brane Induced Gravity: Codimension-2'', {\em Mod. Phys. Lett.
  A}, {\bf 23}, 781--796, (2008).
  {\small[\href{http://dx.doi.org/10.1142/S0217732308026819}{DOI}]},
  {\small[\href{http://arxiv.org/abs/0711.3210}{{arXiv:0711.3210
  {\small[hep-th]}}}]}.

\bibitem{Kaloper:2007ap}
Kaloper, N., and Kiley, D., ``Charting the Landscape of Modified Gravity'',
  {\em J. High Energy Phys.}, {\bf 2007}(05), 045, (2007).
  {\small[\href{http://arxiv.org/abs/hep-th/0703190}{{hep-th/0703190}}]}.

\bibitem{kss_2}
Kanno, S., Sasaki, M., and Soda, J., ``Born-Again Braneworld'', {\em Prog.
  Theor. Phys.}, {\bf 109}, 357--369, (2003).
  {\small[\href{http://arxiv.org/abs/hep-th/0210250}{{hep-th/0210250}}]}.

\bibitem{hs_14}
Kanno, S., and Soda, J., ``Low Energy Effective Action for Dilatonic Braneworld
  --A Formalism for Inflationary Braneworld--'', {\em Gen. Relativ. Gravit.},
  {\bf 36}, 689--712, (2004).
  {\small[\href{http://arxiv.org/abs/hep-th/0303203}{{hep-th/0303203}}]}.

\bibitem{Kanno:2004nr}
Kanno, S., and Soda, J., ``Quasi-thick codimension 2 braneworld'', {\em J.
  Cosmol. Astropart. Phys.}, {\bf 2004}(07), 002, (2004).
  {\small[\href{http://dx.doi.org/10.1088/1475-7516/2004/07/002}{DOI}]},
  {\small[\href{http://arxiv.org/abs/hep-th/0404207}{{hep-th/0404207}}]}.

\bibitem{bhsol_3}
Kanti, P., Olasagasti, I., and Tamvakis, K., ``Quest for localized 4D black
  holes in brane worlds. II. Removing the bulk singularities'', {\em Phys. Rev.
  D}, {\bf 68}, 124001, 1--10, (2003).
  {\small[\href{http://arxiv.org/abs/hep-th/0307201}{{hep-th/0307201}}]}.

\bibitem{bhsol_1}
Kanti, P., and Tamvakis, K., ``Quest for localized 4D black holes in brane
  worlds'', {\em Phys. Rev. D}, {\bf 65}, 084010, 1--12, (2002).
  {\small[\href{http://arxiv.org/abs/hep-th/0110298}{{hep-th/0110298}}]}.

\bibitem{nonloc_1}
Karch, A., and Randall, L., ``Locally localized gravity'', {\em J. High Energy
  Phys.}, {\bf 2001}(05), 008, (2001).
  {\small[\href{http://arxiv.org/abs/hep-th/0011156}{{hep-th/0011156}}]}.

\bibitem{morers2_3}
Kehagias, A., and Kiritsis, E., ``Mirage Cosmology'', {\em J. High Energy
  Phys.}, {\bf 1999}(11), 022, (1999).
  {\small[\href{http://arxiv.org/abs/hep-th/9910174}{{hep-th/9910174}}]}.

\bibitem{gergasym}
Keresztes, Z., and Gergely, L.{\'{A}}., ``3+1+1 dimensional covariant
  gravitational dynamics on an asymmetrically embedded brane'', {\em Annalen
  Phys.}, {\bf 19}, 249--253, (2010).
  {\small[\href{http://dx.doi.org/10.1002/andp.201010421}{DOI}]},
  {\small[\href{http://arxiv.org/abs/0911.2495}{{arXiv:0911.2495
  {\small[gr-qc]}}}]}.

\bibitem{gergnew}
Keresztes, Z., and Gergely, L.{\'{A}}., ``Covariant gravitational dynamics in
  3+1+1 dimensions'', {\em Class. Quant. Grav.}, {\bf 27}, 105009, (2010).
  {\small[\href{http://dx.doi.org/10.1088/0264-9381/27/10/105009}{DOI}]},
  {\small[\href{http://arxiv.org/abs/0909.0490}{{arXiv:0909.0490
  {\small[gr-qc]}}}]}.

\bibitem{ek_1}
Khoury, J., Ovrut, B.A., Steinhardt, P.J., and Turok, N., ``The Ekpyrotic
  Universe: Colliding Branes and the Origin of the Hot Big Bang'', {\em Phys.
  Rev. D}, {\bf 64}, 123522, (2001).
  {\small[\href{http://arxiv.org/abs/hep-th/0103239}{{hep-th/0103239}}]}.

\bibitem{ind_4}
Kiritsis, E., Tetradis, N., and Tomaras, T.N., ``Induced Gravity on RS
  Branes'', {\em J. High Energy Phys.}, {\bf 2002}(03), 019, (2002).
  {\small[\href{http://arxiv.org/abs/hep-th/0202037}{{hep-th/0202037}}]}.

\bibitem{hs_1}
Kobayashi, S., Koyama, K., and Soda, J., ``Quantum fluctuations of bulk
  inflaton in inflationary brane world'', {\em Phys. Lett. B}, {\bf 501},
  157--164, (2001).
  {\small[\href{http://arxiv.org/abs/hep-th/0009160}{{hep-th/0009160}}]}.

\bibitem{gwtan}
Kobayashi, T., Kudoh, H., and Tanaka, T., ``Primordial gravitational waves in
  an inflationary braneworld'', {\em Phys. Rev. D}, {\bf 68}, 044025, 1--12,
  (2003). {\small[\href{http://arxiv.org/abs/gr-qc/0305006}{{gr-qc/0305006}}]}.

\bibitem{Kobayashi:2008bm}
Kobayashi, T., Shiromizu, T., and de~Rham, C., ``Curvature corrections to the
  low energy effective theory in 6D regularized braneworlds'', {\em Phys. Rev.
  D}, {\bf 77}, 124012, (2008).
  {\small[\href{http://dx.doi.org/10.1103/PhysRevD.77.124012}{DOI}]},
  {\small[\href{http://arxiv.org/abs/0802.0103}{{arXiv:0802.0103
  {\small[hep-th]}}}]}.

\bibitem{pert_13}
Kodama, H., ``Behavior of Cosmological Perturbations in the Brane-World Mode'',
  arXiv e-print, (2000).
  {\small[\href{http://arxiv.org/abs/hep-th/0012132}{{hep-th/0012132}}]}.

\bibitem{pert_2}
Kodama, H., Ishibashi, A., and Seto, O., ``Brane world cosmology:
  Gauge-invariant formalism for perturbation'', {\em Phys. Rev. D}, {\bf 62},
  064022, 1--19, (2000).
  {\small[\href{http://arxiv.org/abs/hep-th/0004160}{{hep-th/0004160}}]}.

\bibitem{metp_1}
Kodama, H., and Sasaki, M., ``Evolution of Isocurvature Perturbations I:
  Photon-Baryon Universe'', {\em Int. J. Mod. Phys. A}, {\bf 1}, 265--301,
  (1986).

\bibitem{Kodama:2005cz}
Kodama, H., and Uzawa, K., ``Comments on the four-dimensional effective theory
  for warped compactification'', {\em J. High Energy Phys.}, {\bf 2006}(03),
  053, (2006).
  {\small[\href{http://arxiv.org/abs/hep-th/0512104}{{hep-th/0512104}}]}.

\bibitem{ind_2}
Kofinas, G., ``General brane cosmology with $^{(4)}R$ term in $(A)dS_{5}$ or
  Minkowski bulk'', {\em J. High Energy Phys.}, {\bf 2001}(08), 034, (2001).
  {\small[\href{http://dx.doi.org/10.1088/1126-6708/2001/08/034}{DOI}]},
  {\small[\href{http://arxiv.org/abs/hep-th/0108013}{{hep-th/0108013}}]}.

\bibitem{Kofinas:2005py}
Kofinas, G., ``On braneworld cosmologies from six dimensions, and absence
  thereof'', {\em Phys. Lett. B}, {\bf 633}, 141--148, (2006).
  {\small[\href{http://dx.doi.org/10.1016/j.physletb.2005.11.064}{DOI}]},
  {\small[\href{http://arxiv.org/abs/hep-th/0506035}{{hep-th/0506035}}]}.

\bibitem{kmp}
Kofinas, G., Maartens, R., and Papantonopoulos, E., ``Brane cosmology with
  curvature corrections'', {\em J. High Energy Phys.}, {\bf 2003}(10), 066,
  (2003).
  {\small[\href{http://arxiv.org/abs/hep-th/0307138}{{hep-th/0307138}}]}.

\bibitem{kpz}
Kofinas, G., Papantonopoulos, E., and Zamarias, V., ``Black hole solutions in
  braneworlds with induced gravity'', {\em Phys. Rev. D}, {\bf 66}, 104028,
  1--8, (2002).
  {\small[\href{http://arxiv.org/abs/hep-th/0208207}{{hep-th/0208207}}]}.

\bibitem{mod_2}
Kofman, L.A., ``Probing String Theory with Modulated Cosmological
  Fluctuations'', arXiv e-print, (2003).
  {\small[\href{http://arxiv.org/abs/astro-ph/0303614}{{astro-ph/0303614}}]}.

\bibitem{Kolanovic:2003am}
Kolanovic, M., Porrati, M., and Rombouts, J.-W., ``Regularization of brane
  induced gravity'', {\em Phys. Rev. D}, {\bf 68}, 064018, (2003).
  {\small[\href{http://dx.doi.org/10.1103/PhysRevD.68.064018}{DOI}]},
  {\small[\href{http://arxiv.org/abs/hep-th/0304148}{{hep-th/0304148}}]}.

\bibitem{koyn}
Koyama, K., unpublished manuscript.

\bibitem{ksw}
Koyama, K., ``Radion and large scale anisotropy on the brane'', {\em Phys. Rev.
  D}, {\bf 66}, 084003, 1--18, (2002).
  {\small[\href{http://arxiv.org/abs/gr-qc/0204047}{{gr-qc/0204047}}]}.

\bibitem{koy}
Koyama, K., ``Cosmic Microwave Background Radiation Anisotropies in Brane
  Worlds'', {\em Phys. Rev. Lett.}, {\bf 91}, 221301, 1--4, (2003).
  {\small[\href{http://arxiv.org/abs/astro-ph/0303108}{{astro-ph/0303108}}]}.

\bibitem{Koyama:2007za}
Koyama, K., ``Ghosts in the self-accelerating universe'', {\em Class. Quant.
  Grav.}, {\bf 24}, R231--R253, (2007).
  {\small[\href{http://dx.doi.org/10.1088/0264-9381/24/24/R01}{DOI}]},
  {\small[\href{http://arxiv.org/abs/0709.2399}{{arXiv:0709.2399
  {\small[hep-th]}}}]}.

\bibitem{Koyama:2007rx}
Koyama, K., ``The cosmological constant and dark energy in braneworlds'', {\em
  Gen. Rel. Grav.}, {\bf 40}, 421--450, (2008).
  {\small[\href{http://dx.doi.org/10.1007/s10714-007-0552-x}{DOI}]},
  {\small[\href{http://arxiv.org/abs/0706.1557}{{arXiv:0706.1557
  {\small[astro-ph]}}}]}.

\bibitem{Koyama:2010xj}
Koyama, K., ``Non-Gaussianity of quantum fields during inflation'', arXiv
  e-print, (2010).
  {\small[\href{http://arxiv.org/abs/1002.0600}{{arXiv:1002.0600
  {\small[hep-th]}}}]}.

\bibitem{Koyama:2006ni}
Koyama, K., Koyama, K., and Arroja, F., ``On the 4D effective theory in warped
  compactifications with fluxes and branes'', {\em Phys. Lett. B}, {\bf 641},
  81--87, (2006).
  {\small[\href{http://dx.doi.org/10.1016/j.physletb.2006.08.029}{DOI}]},
  {\small[\href{http://arxiv.org/abs/hep-th/0607145}{{hep-th/0607145}}]}.

\bibitem{Koyama:2004ap}
Koyama, K., Langlois, D., Maartens, R., and Wands, D., ``Scalar perturbations
  from brane-world inflation'', {\em J. Cosmol. Astropart. Phys.}, {\bf
  2004}(11), 002, (2004).
  {\small[\href{http://dx.doi.org/10.1088/1475-7516/2004/11/002}{DOI}]},
  {\small[\href{http://arxiv.org/abs/hep-th/0408222}{{hep-th/0408222}}]}.

\bibitem{Koyama:2005kd}
Koyama, K., and Maartens, R., ``Structure formation in the
  Dvali--Gabadadze--Porrati cosmological model'', {\em J. Cosmol. Astropart.
  Phys.}, {\bf 2006}(01), 016, (2006).
  {\small[\href{http://arxiv.org/abs/astro-ph/0511634}{{astro-ph/0511634}}]}.

\bibitem{Koyama:2007as}
Koyama, K., Mennim, A., Rubakov, V.A., Wands, D., and Hiramatsu, T.,
  ``Primordial perturbations from slow-roll inflation on a brane'', {\em J.
  Cosmol. Astropart. Phys.}, {\bf 2007}(04), 001, (2007).
  {\small[\href{http://dx.doi.org/10.1088/1475-7516/2007/04/001}{DOI}]},
  {\small[\href{http://arxiv.org/abs/hep-th/0701241}{{hep-th/0701241}}]}.

\bibitem{Koyama:2005gh}
Koyama, K., Mennim, A., and Wands, D., ``Coupled boundary and bulk fields in
  anti-de Sitter spacetime'', {\em Phys. Rev. D}, {\bf 72}, 064001, (2005).
  {\small[\href{http://dx.doi.org/10.1103/PhysRevD.72.064001}{DOI}]},
  {\small[\href{http://arxiv.org/abs/hep-th/0504201}{{hep-th/0504201}}]}.

\bibitem{Koyama:2007su}
Koyama, K., Mennim, A., and Wands, D., ``Brane-world inflation: slow-roll
  corrections to the spectral index'', {\em Phys. Rev. D}, {\bf 77}, 021501,
  (2008). {\small[\href{http://dx.doi.org/10.1103/PhysRevD.77.021501}{DOI}]},
  {\small[\href{http://arxiv.org/abs/0709.0294}{{arXiv:0709.0294
  {\small[astro-ph]}}}]}.

\bibitem{Koyama:2005ek}
Koyama, K., Mizuno, S., and Wands, D., ``Slow-roll corrections to inflaton
  fluctuations on a brane'', {\em J. Cosmol. Astropart. Phys.}, {\bf 2005}(08),
  009, (2005).
  {\small[\href{http://dx.doi.org/10.1088/1475-7516/2005/08/009}{DOI}]},
  {\small[\href{http://arxiv.org/abs/hep-th/0506102}{{hep-th/0506102}}]}.

\bibitem{Koyama:2007ih}
Koyama, K., and Silva, F.P., ``Nonlinear interactions in a cosmological
  background in the Dvali-Gabadadze-Porrati braneworld'', {\em Phys. Rev. D},
  {\bf 75}, 084040, (2007).
  {\small[\href{http://dx.doi.org/10.1103/PhysRevD.75.084040}{DOI}]},
  {\small[\href{http://arxiv.org/abs/hep-th/0702169}{{hep-th/0702169}}]}.

\bibitem{ksb_1}
Koyama, K., and Soda, J., ``Birth of the brane world'', {\em Phys. Lett. B},
  {\bf 483}, 432--442, (2000).
  {\small[\href{http://arxiv.org/abs/gr-qc/0001033}{{gr-qc/0001033}}]}.

\bibitem{pert_5}
Koyama, K., and Soda, J., ``Evolution of cosmological perturbations in the
  brane world'', {\em Phys. Rev. D}, {\bf 62}, 123502, 1--14, (2000).
  {\small[\href{http://arxiv.org/abs/hep-th/0005239}{{hep-th/0005239}}]}.

\bibitem{hs_9}
Koyama, K., and Takahashi, K., ``Primordial fluctuations in bulk inflaton
  model'', {\em Phys. Rev. D}, {\bf 67}, 103503, 1--8, (2003).
  {\small[\href{http://arxiv.org/abs/hep-th/0301165}{{hep-th/0301165}}]}.

\bibitem{morers2_1}
Kraus, P., ``Dynamics of Anti-de Sitter Domain Walls'', {\em J. High Energy
  Phys.}, {\bf 1999}(12), 011, (1999).
  {\small[\href{http://arxiv.org/abs/hep-th/9910149}{{hep-th/9910149}}]}.

\bibitem{r_3}
Kubyshin, Y.A., ``Models with Extra Dimensions and Their Phenomenology'', arXiv
  e-print, (2001).
  {\small[\href{http://arxiv.org/abs/hep-ph/0111027}{{hep-ph/0111027}}]}.

\bibitem{ktn}
Kudoh, H., Tanaka, T., and Nakamura, T., ``Small localized black holes in a
  braneworld: Formulation and numerical method'', {\em Phys. Rev. D}, {\bf 68},
  024035, 1--11, (2003).
  {\small[\href{http://arxiv.org/abs/gr-qc/0301089}{{gr-qc/0301089}}]}.

\bibitem{Kunz:2006ca}
Kunz, M., and Sapone, D., ``Dark Energy versus Modified Gravity'', {\em Phys.
  Rev. Lett.}, {\bf 98}, 121301, (2007).
  {\small[\href{http://dx.doi.org/10.1103/PhysRevLett.98.121301}{DOI}]},
  {\small[\href{http://arxiv.org/abs/astro-ph/0612452}{{astro-ph/0612452}}]}.

\bibitem{inf_13}
Kunze, K.E., ``Stochastic inflation on the brane'', {\em Phys. Lett. B}, {\bf
  587}, 1--6, (2004).
  {\small[\href{http://arxiv.org/abs/hep-th/0310200}{{hep-th/0310200}}]}.

\bibitem{pert_3}
Langlois, D., ``Brane cosmological perturbations'', {\em Phys. Rev. D}, {\bf
  62}, 126012, 1--8, (2000).
  {\small[\href{http://arxiv.org/abs/hep-th/0005025}{{hep-th/0005025}}]}.

\bibitem{pert_9}
Langlois, D., ``Evolution of Cosmological Perturbations in a Brane-Universe'',
  {\em Phys. Rev. Lett.}, {\bf 86}, 2212--2215, (2001).
  {\small[\href{http://arxiv.org/abs/hep-th/0010063}{{hep-th/0010063}}]}.

\bibitem{rev_9}
Langlois, D., ``Brane Cosmology'', {\em Prog. Theor. Phys. Suppl.}, {\bf 148},
  181--212, (2002).
  {\small[\href{http://arxiv.org/abs/hep-th/0209261}{{hep-th/0209261}}]}.

\bibitem{rev_5}
Langlois, D., ``Gravitation and cosmology in a brane-universe'', arXiv e-print,
  (2002). {\small[\href{http://arxiv.org/abs/gr-qc/0207047}{{gr-qc/0207047}}]}.

\bibitem{rev_2}
Langlois, D., ``Gravitational and Cosmological Properties of a
  Brane-Universe'', {\em Int. J. Mod. Phys. A}, {\bf 17}, 2701--2705, (2002).
  {\small[\href{http://arxiv.org/abs/gr-qc/0205004}{{gr-qc/0205004}}]}.

\bibitem{lan}
Langlois, D., ``Cosmology with an extra-dimension'', arXiv e-print, (2003).
  {\small[\href{http://arxiv.org/abs/astro-ph/0301021}{{astro-ph/0301021}}]}.

\bibitem{lmsw}
Langlois, D., Maartens, R., Sasaki, M., and Wands, D., ``Large-scale
  cosmological perturbations on the brane'', {\em Phys. Rev. D}, {\bf 63},
  084009, 1--10, (2001).
  {\small[\href{http://arxiv.org/abs/hep-th/0012044}{{hep-th/0012044}}]}.

\bibitem{lmw}
Langlois, D., Maartens, R., and Wands, D., ``Gravitational waves from inflation
  on the brane'', {\em Phys. Lett. B}, {\bf 489}, 259--267, (2000).
  {\small[\href{http://arxiv.org/abs/hep-th/0006007}{{hep-th/0006007}}]}.

\bibitem{ek_5}
Langlois, D., Maeda, K., and Wands, D., ``Conservation Laws for Collisions of
  Branes and Shells in General Relativity'', {\em Phys. Rev. Lett.}, {\bf 88},
  181301, 1--4, (2002).
  {\small[\href{http://arxiv.org/abs/gr-qc/0111013}{{gr-qc/0111013}}]}.

\bibitem{sca_4}
Langlois, D., and Rodr{\'{\i}}guez-Mart{\'{\i}}nez, M., ``Brane cosmology with
  a bulk scalar field'', {\em Phys. Rev. D}, {\bf 64}, 123507, 1--9, (2001).
  {\small[\href{http://arxiv.org/abs/hep-th/0106245}{{hep-th/0106245}}]}.

\bibitem{hs_10}
Langlois, D., and Sasaki, M., ``Massive scalar states localized on a de Sitter
  brane'', {\em Phys. Rev. D}, {\bf 68}, 064012, 1--11, (2003).
  {\small[\href{http://arxiv.org/abs/hep-th/0302069}{{hep-th/0302069}}]}.

\bibitem{lang2_2}
Langlois, D., and Sorbo, L., ``Effective action for the homogeneous radion in
  brane cosmology'', {\em Phys. Lett. B}, {\bf 543}, 155--162, (2002).
  {\small[\href{http://arxiv.org/abs/hep-th/0203036}{{hep-th/0203036}}]}.

\bibitem{ckn_3}
Langlois, D., and Sorbo, L., ``Bulk gravitons from a cosmological brane'', {\em
  Phys. Rev. D}, {\bf 68}, 084006, 1--11, (2003).
  {\small[\href{http://arxiv.org/abs/hep-th/0306281}{{hep-th/0306281}}]}.

\bibitem{ckn_2}
Langlois, D., Sorbo, L., and Rodr{\'{\i}}guez-Mart{\'{\i}}nez, M., ``Cosmology
  of a Brane Radiating Gravitons into the Extra Dimension'', {\em Phys. Rev.
  Lett.}, {\bf 89}, 171301, 1--4, (2002).
  {\small[\href{http://arxiv.org/abs/hep-th/0206146}{{hep-th/0206146}}]}.

\bibitem{Lazkoz:2006gp}
Lazkoz, R., Maartens, R., and Majerotto, E., ``Observational constraints on
  phantom-like braneworld cosmologies'', {\em Phys. Rev. D}, {\bf 74}, 083510,
  (2006). {\small[\href{http://dx.doi.org/10.1103/PhysRevD.74.083510}{DOI}]},
  {\small[\href{http://arxiv.org/abs/astro-ph/0605701}{{astro-ph/0605701}}]}.

\bibitem{ckn_4}
Leeper, E., Maartens, R., and Sopuerta, C.F., ``Dynamics of radiating
  braneworlds'', {\em Class. Quantum Grav.}, {\bf 21}, 1125--1133, (2004).
  {\small[\href{http://arxiv.org/abs/gr-qc/0309080}{{gr-qc/0309080}}]}.

\bibitem{l2}
Leong, B., Challinor, A.D., Maartens, R., and Lasenby, A.N., ``Braneworld
  tensor anisotropies in the CMB'', {\em Phys. Rev. D}, {\bf 66}, 104010, 1--6,
  (2002).
  {\small[\href{http://arxiv.org/abs/astro-ph/0208015}{{astro-ph/0208015}}]}.

\bibitem{l1}
Leong, B., Dunsby, P.K.S., Challinor, A.D., and Lasenby, A.N., ``1 + 3
  covariant dynamics of scalar perturbations in braneworlds'', {\em Phys. Rev.
  D}, {\bf 65}, 104012, 1--17, (2002).
  {\small[\href{http://arxiv.org/abs/gr-qc/0111033}{{gr-qc/0111033}}]}.

\bibitem{2b_3}
Lesgourgues, J., and Sorbo, L., ``Goldberger--Wise variations: Stabilizing
  brane models with a bulk scalar'', {\em Phys. Rev. D}, {\bf 69}, 084010,
  1--8, (2000).
  {\small[\href{http://arxiv.org/abs/hep-th/0310007}{{hep-th/0310007}}]}.

\bibitem{lidsmi}
Liddle, A.R., and Smith, A.J., ``Observational constraints on braneworld
  chaotic inflation'', {\em Phys. Rev. D}, {\bf 68}, 061301, 1--5, (2003).
  {\small[\href{http://arxiv.org/abs/astro-ph/0307017}{{astro-ph/0307017}}]}.

\bibitem{inf_7}
Liddle, A.R., and Taylor, A.N., ``Inflaton potential reconstruction in the
  braneworld scenario'', {\em Phys. Rev. D}, {\bf 65}, 041301, 1--5, (2002).
  {\small[\href{http://arxiv.org/abs/astro-ph/0109412}{{astro-ph/0109412}}]}.

\bibitem{steep_5}
Liddle, A.R., and Ure{\~{n}}a-L{\'{o}}pez, L.A., ``Curvaton reheating: An
  application to braneworld inflation'', {\em Phys. Rev. D}, {\bf 68}, 043517,
  1--8, (2003).
  {\small[\href{http://arxiv.org/abs/astro-ph/0302054}{{astro-ph/0302054}}]}.

\bibitem{lidrev}
Lidsey, J.E., ``Inflation and Braneworlds'', in Bret{\'{o}}n, N.,
  Cervantes-Cota, J.L., and Salgado, M., eds., {\em The Early Universe and
  Observational Cosmology}, Fifth Mexican School on Gravitation and
  Mathematical Physics, November 2002, Lecture Notes in Physics, vol. 646,
  (Springer, Berlin; New York, 2004).
  {\small[\href{http://arxiv.org/abs/astro-ph/0305528}{{astro-ph/0305528}}]}.

\bibitem{dq_3}
Lidsey, J.E., Matos, T., and Ure{\~{n}}a-L{\'{o}}pez, L.A., ``Inflaton field as
  self-interacting dark matter in the braneworld scenario'', {\em Phys. Rev.
  D}, {\bf 66}, 023514, 1--5, (2002).
  {\small[\href{http://arxiv.org/abs/astro-ph/0111292}{{astro-ph/0111292}}]}.

\bibitem{gbon_7}
Lidsey, J.E., Nojiri, S., and Odintsov, S.D., ``Braneworld cosmology in
  (anti)-de Sitter Einstein--Gauss--Bonnet--Maxwell gravity'', {\em J. High
  Energy Phys.}, {\bf 2002}(06), 026, (2002).
  {\small[\href{http://arxiv.org/abs/hep-th/0202198}{{hep-th/0202198}}]}.

\bibitem{acftcosmo_5}
Lidsey, J.E., Nojiri, S., Odintsov, S.D., and Ogushi, S., ``The AdS/CFT
  correspondence and logarithmic corrections to braneworld cosmology and the
  Cardy--Verlinde formula'', {\em Phys. Lett. B}, {\bf 544}, 337--345, (2002).
  {\small[\href{http://arxiv.org/abs/hep-th/0207009}{{hep-th/0207009}}]}.

\bibitem{gbon_10}
Lidsey, J.E., and Nunes, N.J., ``Inflation in Gauss--Bonnet brane cosmology'',
  {\em Phys. Rev. D}, {\bf 67}, 103510, 1--9, (2003).
  {\small[\href{http://arxiv.org/abs/astro-ph/0303168}{{astro-ph/0303168}}]}.

\bibitem{Linde:2005dd}
Linde, A.D., ``Towards inflation in string theory'', {\em J. Phys. Conf. Ser.},
  {\bf 24}, 151--160, (2004).
  {\small[\href{http://dx.doi.org/10.1088/1742-6596/24/1/018}{DOI}]},
  {\small[\href{http://arxiv.org/abs/hep-th/0503195}{{hep-th/0503195}}]}.

\bibitem{Linder:2005in}
Linder, E.V., ``Cosmic growth history and expansion history'', {\em Phys. Rev.
  D}, {\bf 72}, 043529, (2005).
  {\small[\href{http://dx.doi.org/10.1103/PhysRevD.72.043529}{DOI}]},
  {\small[\href{http://arxiv.org/abs/astro-ph/0507263}{{astro-ph/0507263}}]}.

\bibitem{exp}
Long, J.C., Chan, H.W., Churnside, A.B., Gulbis, E.A., Varney, M.C.M., and
  Price, J.C., ``Upper limits to submillimetre-range forces from extra
  space-time dimensions'', {\em Nature}, {\bf 421}, 922--925, (2003).
  {\small[\href{http://arxiv.org/abs/hep-ph/0210004}{{hep-ph/0210004}}]}.

\bibitem{Lue:2004rj}
Lue, A., Scoccimarro, R., and Starkman, G.D., ``Probing Newton's constant on
  vast scales: Dvali-Gabadadze-Porrati gravity, cosmic acceleration, and large
  scale structure'', {\em Phys. Rev. D}, {\bf 69}, 124015, (2004).
  {\small[\href{http://dx.doi.org/10.1103/PhysRevD.69.124015}{DOI}]},
  {\small[\href{http://arxiv.org/abs/astro-ph/0401515}{{astro-ph/0401515}}]}.

\bibitem{Lue:2002sw}
Lue, A., and Starkman, G., ``Gravitational leakage into extra dimensions:
  Probing dark energy using local gravity'', {\em Phys. Rev. D}, {\bf 67},
  064002, (2003).
  {\small[\href{http://dx.doi.org/10.1103/PhysRevD.67.064002}{DOI}]},
  {\small[\href{http://arxiv.org/abs/astro-ph/0212083}{{astro-ph/0212083}}]}.

\bibitem{ind_12}
Lue, A., and Starkman, G., ``Gravitational leakage into extra dimensions:
  Probing dark energy using local gravity'', {\em Phys. Rev. D}, {\bf 67},
  064002, 1--9, (2003).
  {\small[\href{http://arxiv.org/abs/astro-ph/0212083}{{astro-ph/0212083}}]}.

\bibitem{Lue:2004za}
Lue, A., and Starkman, G.D., ``How a brane cosmological constant can trick us
  into thinking that $w < -1$'', {\em Phys. Rev. D}, {\bf 70}, 101501, (2004).
  {\small[\href{http://dx.doi.org/10.1103/PhysRevD.70.101501}{DOI}]},
  {\small[\href{http://arxiv.org/abs/astro-ph/0408246}{{astro-ph/0408246}}]}.

\bibitem{ek5_3}
Lukas, A., ``Moving Five-Branes and Cosmology'', arXiv e-print, (2002).
  {\small[\href{http://arxiv.org/abs/hep-th/0210026}{{hep-th/0210026}}]}.

\bibitem{low_1}
Lukas, A., Ovrut, B.A., Stelle, K.S., and Waldram, D., ``Universe as a domain
  wall'', {\em Phys. Rev. D}, {\bf 59}, 086001, 1--9, (1999).
  {\small[\href{http://arxiv.org/abs/hep-th/9803235}{{hep-th/9803235}}]}.

\bibitem{low_2}
Lukas, A., Ovrut, B.A., and Waldram, D., ``Cosmological solutions of
  Horava--Witten theory'', {\em Phys. Rev. D}, {\bf 60}, 086001, 1--11, (1999).
  {\small[\href{http://arxiv.org/abs/hep-th/9806022}{{hep-th/9806022}}]}.

\bibitem{low_3}
Lukas, A., Ovrut, B.A., and Waldram, D., ``Boundary inflation'', {\em Phys.
  Rev. D}, {\bf 61}, 023506, 1--18, (2000).
  {\small[\href{http://arxiv.org/abs/hep-th/9902071}{{hep-th/9902071}}]}.

\bibitem{m1}
Maartens, R., ``Cosmological dynamics on the brane'', {\em Phys. Rev. D}, {\bf
  62}, 084023, 1--14, (2000).
  {\small[\href{http://arxiv.org/abs/hep-th/0004166}{{hep-th/0004166}}]}.

\bibitem{m2}
Maartens, R., ``Geometry and dynamics of the brane-world'', in
  Pascual-S{\'{a}}nchez, J.-F., Flor{\'{\i}}a, L., San~Miguel, A., and Vicente,
  F., eds., {\em Reference Frames and Gravitomagnetism}, Proceedings of the
  XXIII Spanish Relativity Meeting (Eres2000), pp. 93--120, (World Scientific,
  Singapore, 2001).
  {\small[\href{http://arxiv.org/abs/gr-qc/0101059}{{gr-qc/0101059}}]}.

\bibitem{mjap}
Maartens, R., ``Brane-World Cosmological Perturbations --A Covariant
  Approach--'', {\em Prog. Theor. Phys. Suppl.}, {\bf 148}, 213--234, (2002).
  {\small[\href{http://arxiv.org/abs/gr-qc/0304089}{{gr-qc/0304089}}]}.

\bibitem{Maartens:2006yt}
Maartens, R., and Majerotto, E., ``Observational constraints on
  self-accelerating cosmology'', {\em Phys. Rev. D}, {\bf 74}, 023004, (2006).
  {\small[\href{http://dx.doi.org/10.1103/PhysRevD.74.023004}{DOI}]},
  {\small[\href{http://arxiv.org/abs/astro-ph/0603353}{{astro-ph/0603353}}]}.

\bibitem{mss}
Maartens, R., Sahni, V., and Saini, T.D., ``Anisotropy dissipation in
  brane-world inflation'', {\em Phys. Rev. D}, {\bf 63}, 063509, 1--6, (2001).
  {\small[\href{http://arxiv.org/abs/gr-qc/0011105}{{gr-qc/0011105}}]}.

\bibitem{mwbh}
Maartens, R., Wands, D., Bassett, B.A., and Heard, I.P.C., ``Chaotic inflation
  on the brane'', {\em Phys. Rev. D}, {\bf 62}, 041301, 1--5, (2000).
  {\small[\href{http://arxiv.org/abs/hep-ph/9912464}{{hep-ph/9912464}}]}.

\bibitem{ind_11}
Maeda, K., Mizuno, S., and Torii, T., ``Effective gravitational equations on a
  brane world with induced gravity'', {\em Phys. Rev. D}, {\bf 68}, 024033,
  1--8, (2003).
  {\small[\href{http://arxiv.org/abs/gr-qc/0303039}{{gr-qc/0303039}}]}.

\bibitem{gbon_16}
Maeda, K., and Torii, T., ``Covariant gravitational equations on a brane world
  with a Gauss--Bonnet term'', {\em Phys. Rev. D}, {\bf 69}, 024002, 1--11,
  (2004).
  {\small[\href{http://arxiv.org/abs/hep-th/0309152}{{hep-th/0309152}}]}.

\bibitem{mw}
Maeda, K., and Wands, D., ``Dilaton-gravity on the brane'', {\em Phys. Rev. D},
  {\bf 62}, 124009, 1--9, (2000).
  {\small[\href{http://arxiv.org/abs/hep-th/0008188}{{hep-th/0008188}}]}.

\bibitem{steep_2}
Majumdar, A.S., ``From brane assisted inflation to quintessence through a
  single scalar field'', {\em Phys. Rev. D}, {\bf 64}, 083503, 1--6, (2001).
  {\small[\href{http://arxiv.org/abs/astro-ph/0105518}{{astro-ph/0105518}}]}.

\bibitem{clan_3}
Majumdar, A.S., ``Domination of Black Hole Accretion in Brane Cosmology'', {\em
  Phys. Rev. Lett.}, {\bf 90}, 031303, 1--4, (2003).
  {\small[\href{http://arxiv.org/abs/astro-ph/0208048}{{astro-ph/0208048}}]}.

\bibitem{rev_1}
March-Russell, J., ``Classical and Quantum Brane Cosmology'', arXiv e-print,
  (2000).
  {\small[\href{http://arxiv.org/abs/hep-ph/0012151}{{hep-ph/0012151}}]}.

\bibitem{hs_15}
Martin, J., Felder, G.N., Frolov, A.V., Peloso, M., and Kofman, L.A.,
  ``Braneworld dynamics with the BRANECODE'', {\em Phys. Rev. D}, {\bf 69},
  084017, 1--18, (2004).
  {\small[\href{http://arxiv.org/abs/hep-th/0309001}{{hep-th/0309001}}]}.

\bibitem{que_2}
Mavromatos, N.E., ``String Cosmology'', in Cotsakis, S., and Papantonopoulos,
  E., eds., {\em Cosmological Crossroads: An Advanced Course in Mathematical,
  Physical and String Cosmology}, Lecture Notes in Phyics, vol. 592, pp.
  392--457, (Springer, Berlin, New York, 2002).
  {\small[\href{http://arxiv.org/abs/hep-th/0111275}{{hep-th/0111275}}]}.

\bibitem{ek_6}
Mavromatos, N.E., ``Vacuum Energy, Cosmological Supersymmetry Breaking and
  Inflation from Colliding Brane Worlds'', arXiv e-print, (2002).
  {\small[\href{http://arxiv.org/abs/hep-th/0210008}{{hep-th/0210008}}]}.

\bibitem{gbon_9}
Mavromatos, N.E., and Rizos, J., ``Exact Solutions and the Cosmological
  Constant Problem in Dilatonic-Domain-Wall Higher-Curvature String Gravity'',
  {\em Int. J. Mod. Phys. A}, {\bf 18}, 57--84, (2003).
  {\small[\href{http://arxiv.org/abs/hep-th/0205299}{{hep-th/0205299}}]}.

\bibitem{inf_5}
Mazumdar, A., ``Interesting consequences of brane cosmology'', {\em Phys. Rev.
  D}, {\bf 64}, 027304, 1--4, (2001).
  {\small[\href{http://arxiv.org/abs/hep-ph/0007269}{{hep-ph/0007269}}]}.

\bibitem{inf_4}
Mendes, L., and Liddle, A.R., ``Initial conditions for hybrid inflation'', {\em
  Phys. Rev. D}, {\bf 62}, 103511, 1--6, (2000).
  {\small[\href{http://arxiv.org/abs/astro-ph/0006020}{{astro-ph/0006020}}]}.

\bibitem{mbb_1}
Mennim, A., {\em Gravitation and cosmology on a brane-world}, Ph.D. Thesis,
  (University of Cambridge, Cambridge, 2003).

\bibitem{sca_2}
Mennim, A., and Battye, R.A., ``Cosmological expansion on a dilatonic
  brane-world'', {\em Class. Quantum Grav.}, {\bf 18}, 2171--2194, (2001).
  {\small[\href{http://arxiv.org/abs/hep-th/0008192}{{hep-th/0008192}}]}.

\bibitem{mbb_2}
Mennim, A., Battye, R.A., and van~de Bruck, C., ``Cosmological tensor
  perturbations in brane world models'', {\em Astrophys. Space Sci.}, {\bf
  283}, 633--638, (2003).

\bibitem{Minamitsuji:2008fz}
Minamitsuji, M., ``Self-accelerating solutions in cascading DGP braneworld'',
  {\em Phys. Lett. B}, {\bf 684}, 92--95, (2010).
  {\small[\href{http://dx.doi.org/10.1016/j.physletb.2010.01.010}{DOI}]},
  {\small[\href{http://arxiv.org/abs/0806.2390}{{arXiv:0806.2390
  {\small[gr-qc]}}}]}.

\bibitem{hs_13}
Minamitsuji, M., Himemoto, Y., and Sasaki, M., ``Geometry and cosmological
  perturbations in the bulk inflaton model'', {\em Phys. Rev. D}, {\bf 68},
  024016, 1--17, (2003).
  {\small[\href{http://arxiv.org/abs/gr-qc/0303108}{{gr-qc/0303108}}]}.

\bibitem{Minamitsuji:2007fx}
Minamitsuji, M., and Langlois, D., ``Cosmological evolution of regularized
  branes in 6D warped flux compactifications'', {\em Phys. Rev. D}, {\bf 76},
  084031, (2007).
  {\small[\href{http://dx.doi.org/10.1103/PhysRevD.76.084031}{DOI}]},
  {\small[\href{http://arxiv.org/abs/0707.1426}{{arXiv:0707.1426
  {\small[hep-th]}}}]}.

\bibitem{dq_2}
Mizuno, S., and Maeda, K., ``Quintessence in a brane world'', {\em Phys. Rev.
  D}, {\bf 64}, 123521, 1--16, (2001).
  {\small[\href{http://arxiv.org/abs/hep-ph/0108012}{{hep-ph/0108012}}]}.

\bibitem{inf_11}
Mizuno, S., Maeda, K., and Yamamoto, K., ``Dynamics of a scalar field in a
  brane world'', {\em Phys. Rev. D}, {\bf 67}, 023516, 1--15, (2003).
  {\small[\href{http://arxiv.org/abs/hep-ph/0205292}{{hep-ph/0205292}}]}.

\bibitem{metp_2}
Mukhanov, V.F., Feldman, H.A., and Brandenberger, R.H., ``Theory of
  cosmological perturbations'', {\em Phys. Rep.}, {\bf 215}, 203--333, (1992).
  {\small[\href{http://dx.doi.org/10.1016/0370-1573(92)90044-Z}{DOI}]}.

\bibitem{ichnak_1}
Mukohyama, S., unpublished manuscript.

\bibitem{morers2_5}
Mukohyama, S., ``Brane-world solutions, standard cosmology, and dark
  radiation'', {\em Phys. Lett. B}, {\bf 473}, 241--245, (2000).
  {\small[\href{http://arxiv.org/abs/hep-th/9911165}{{hep-th/9911165}}]}.

\bibitem{mu_1}
Mukohyama, S., ``Gauge-invariant gravitational perturbations of maximally
  symmetric spacetimes'', {\em Phys. Rev. D}, {\bf 62}, 084015, 1--14, (2000).
  {\small[\href{http://arxiv.org/abs/hep-th/0004067}{{hep-th/0004067}}]}.

\bibitem{pert_6}
Mukohyama, S., ``Perturbation of the junction condition and doubly
  gauge-invariant variables'', {\em Class. Quantum Grav.}, {\bf 17},
  4777--4797, (2000).
  {\small[\href{http://arxiv.org/abs/hep-th/0006146}{{hep-th/0006146}}]}.

\bibitem{mu_2}
Mukohyama, S., ``Integro-differential equation for brane-world cosmological
  perturbations'', {\em Phys. Rev. D}, {\bf 64}, 064006, 1--10, (2001).
  {\small[\href{http://arxiv.org/abs/hep-th/0104185}{{hep-th/0104185}}]}.

\bibitem{mukcol}
Mukohyama, S., and Coley, A.A., ``Scaling solution, radion stabilization, and
  initial condition for brane-world cosmology'', {\em Phys. Rev. D}, {\bf 69},
  064029, 1--13, (2004).
  {\small[\href{http://arxiv.org/abs/hep-th/0310140}{{hep-th/0310140}}]}.

\bibitem{2b_2}
Mukohyama, S., and Kofman, L., ``Brane gravity at low energy'', {\em Phys. Rev.
  D}, {\bf 65}, 124025, 1--16, (2002).
  {\small[\href{http://arxiv.org/abs/hep-th/0112115}{{hep-th/0112115}}]}.

\bibitem{birk_1}
Mukohyama, S., Shiromizu, T., and Maeda, K., ``Global structure of exact
  cosmological solutions in the brane world'', {\em Phys. Rev. D}, {\bf 62},
  024028, 1--6, (2000).
  {\small[\href{http://arxiv.org/abs/hep-th/9912287}{{hep-th/9912287}}]}.

\bibitem{ind_13}
Multam{\"{a}}ki, T., Gazta{\~{n}}aga, E., and Manera, M., ``Large-scale
  structure in non-standard cosmologies'', {\em Mon. Not. R. Astron. Soc.},
  {\bf 344}, 761--775, (2003).
  {\small[\href{http://arxiv.org/abs/astro-ph/0303526}{{astro-ph/0303526}}]}.

\bibitem{Navarro:2004di}
Navarro, I., and Santiago, J., ``Gravity on codimension 2 brane worlds'', {\em
  J. High Energy Phys.}, {\bf 2005}(02), 007, (2005).
  {\small[\href{http://dx.doi.org/10.1088/1126-6708/2005/02/007}{DOI}]},
  {\small[\href{http://arxiv.org/abs/hep-th/0411250}{{hep-th/0411250}}]}.

\bibitem{ek_3}
Neronov, A., ``Brane collisions in anti-de Sitter space'', {\em J. High Energy
  Phys.}, {\bf 2001}(11), 007, (2001).
  {\small[\href{http://arxiv.org/abs/hep-th/0109090}{{hep-th/0109090}}]}.

\bibitem{pert_16}
Neronov, A., and Sachs, I., ``Metric perturbations in brane-world scenarios'',
  {\em Phys. Lett. B}, {\bf 513}, 173--178, (2001).
  {\small[\href{http://arxiv.org/abs/hep-th/0011254}{{hep-th/0011254}}]}.

\bibitem{gbon_4}
Neupane, I.P., ``Consistency of higher derivative gravity in the brane
  background'', {\em J. High Energy Phys.}, {\bf 2000}(09), 040, (2000).
  {\small[\href{http://arxiv.org/abs/hep-th/0008190}{{hep-th/0008190}}]}.

\bibitem{acftcosmo_1}
Nojiri, S., and Odintsov, S.D., ``AdS/CFT correspondence in cosmology'', {\em
  Phys. Lett. B}, {\bf 494}, 135--140, (2000).
  {\small[\href{http://dx.doi.org/10.1016/S0370-2693(00)01176-X}{DOI}]},
  {\small[\href{http://arxiv.org/abs/hep-th/0008160}{{hep-th/0008160}}]}.

\bibitem{gbon_2}
Nojiri, S., and Odintsov, S.D., ``Brane-world cosmology in higher derivative
  gravity or warped compactification in the next-to-leading order of AdS/CFT
  correspondence'', {\em J. High Energy Phys.}, {\bf 2000}(07), 049, (2000).
  {\small[\href{http://arxiv.org/abs/hep-th/0006232}{{hep-th/0006232}}]}.

\bibitem{dgp_3}
Nojiri, S., and Odintsov, S.D., ``Brane world inflation induced by quantum
  effects'', {\em Phys. Lett. B}, {\bf 484}, 119--123, (2000).
  {\small[\href{http://arxiv.org/abs/hep-th/0004097}{{hep-th/0004097}}]}.

\bibitem{gbon_3}
Nojiri, S., and Odintsov, S.D., ``Can we live on the brane in
  Schwarzschild--anti de Sitter black hole?'', {\em Phys. Lett. B}, {\bf 493},
  153--161, (2000).
  {\small[\href{http://dx.doi.org/10.1016/S0370-2693(00)01116-3}{DOI}]},
  {\small[\href{http://arxiv.org/abs/hep-th/0007205}{{hep-th/0007205}}]}.

\bibitem{steep_4}
Nunes, N.J., and Copeland, E.J., ``Tracking quintessential inflation from brane
  worlds'', {\em Phys. Rev. D}, {\bf 66}, 043524, 1--7, (2002).
  {\small[\href{http://arxiv.org/abs/astro-ph/0204115}{{astro-ph/0204115}}]}.

\bibitem{acftcosmo_6}
Padilla, A., {\em Braneworld Cosmology and Holography}, Ph.D. Thesis,
  (University of Durham, Durham, 2002).
  {\small[\href{http://arxiv.org/abs/hep-th/0210217}{{hep-th/0210217}}]}.

\bibitem{rev_4}
Papantonopoulos, E., ``Brane Cosmology'', in Cotsakis, S., and Papantonopoulos,
  E., eds., {\em Cosmological Crossroads: An Advanced Course in Mathematical,
  Physical and String Cosmology}, Lecture Notes in Physics, vol. 592, pp.
  458--477, (Springer, Berlin, New York, 2002).
  {\small[\href{http://arxiv.org/abs/hep-th/0202044}{{hep-th/0202044}}]}.

\bibitem{Papantonopoulos:2005nw}
Papantonopoulos, E., and Papazoglou, A., ``Cosmological evolution of a purely
  conical codimension-2 brane world'', {\em J. High Energy Phys.}, {\bf
  2005}(09), 012, (2005).
  {\small[\href{http://arxiv.org/abs/hep-th/0507278}{{hep-th/0507278}}]}.

\bibitem{Papantonopoulos:2006dv}
Papantonopoulos, E., Papazoglou, A., and Zamarias, V., ``Regularization of
  conical singularities in warped six-dimensional compactifications'', {\em J.
  High Energy Phys.}, {\bf 2007}(03), 002, (2007).
  {\small[\href{http://arxiv.org/abs/hep-th/0611311}{{hep-th/0611311}}]}.

\bibitem{Papantonopoulos:2007fk}
Papantonopoulos, E., Papazoglou, A., and Zamarias, V., ``Induced cosmology on a
  regularized brane in six-dimensional flux compactification'', {\em Nucl.
  Phys. B}, {\bf 797}, 520--536, (2008).
  {\small[\href{http://dx.doi.org/10.1016/j.nuclphysb.2007.12.031}{DOI}]},
  {\small[\href{http://arxiv.org/abs/0707.1396}{{arXiv:0707.1396
  {\small[hep-th]}}}]}.

\bibitem{mod_4}
Peloso, M., and Poppitz, E., ``Quintessence from shape moduli'', {\em Phys.
  Rev. D}, {\bf 68}, 125009, 1--7, (2003).
  {\small[\href{http://arxiv.org/abs/hep-ph/0307379}{{hep-ph/0307379}}]}.

\bibitem{Peloso:2006cq}
Peloso, M., Sorbo, L., and Tasinato, G., ``Standard 4d gravity on a brane in
  six dimensional flux compactifications'', {\em Phys. Rev. D}, {\bf 73},
  104025, (2006).
  {\small[\href{http://dx.doi.org/10.1103/PhysRevD.73.104025}{DOI}]},
  {\small[\href{http://arxiv.org/abs/hep-th/0603026}{{hep-th/0603026}}]}.

\bibitem{r_1}
P{\'{e}}rez-Lorenzana, A., ``Theories in more than four dimensions'', in
  Herrera~Corrall, G., and Nellen, L., eds., {\em Particles and Fields: Ninth
  Mexican School}, Metepec, Puebla, Mexico, 9\,--\,19 August 2000, AIP
  Conference Proceedings, vol. 562, pp. 53--85, (AIP, Melville, NY, 2001).
  {\small[\href{http://arxiv.org/abs/hep-ph/0008333}{{hep-ph/0008333}}]}.

\bibitem{Polchinski:1998rr}
Polchinski, J., {\em String Theory. Vol. 2: Superstring theory and beyond},
  Cambridge Monographs on Mathematical Physics, (Cambridge University Press,
  Cambridge; New York, 1998).

\bibitem{mtheory_2}
Polchinski, J., ``M Theory: Uncertainty and Unification'', arXiv e-print,
  (2002).
  {\small[\href{http://arxiv.org/abs/hep-th/0209105}{{hep-th/0209105}}]}.

\bibitem{que_4}
Quevedo, F., ``Lectures on string/brane cosmology'', {\em Class. Quantum
  Grav.}, {\bf 19}, 5721--5779, (2002).
  {\small[\href{http://arxiv.org/abs/hep-th/0210292}{{hep-th/0210292}}]}.

\bibitem{rs2}
Randall, L., and Sundrum, R., ``An Alternative to Compactification'', {\em
  Phys. Rev. Lett.}, {\bf 83}, 4690--4693, (1999).
  {\small[\href{http://arxiv.org/abs/hep-th/9906064}{{hep-th/9906064}}]}.

\bibitem{rs1}
Randall, L., and Sundrum, R., ``Large Mass Hierarchy from a Small Extra
  Dimension'', {\em Phys. Rev. Lett.}, {\bf 83}, 3370--3373, (1999).
  {\small[\href{http://arxiv.org/abs/hep-ph/9905221}{{hep-ph/9905221}}]}.

\bibitem{rev_6}
R{\"{a}}s{\"{a}}nen, S., ``A primer on the ekpyrotic scenario'', arXiv e-print,
  (2002).
  {\small[\href{http://arxiv.org/abs/astro-ph/0208282}{{astro-ph/0208282}}]}.

\bibitem{rbbd_1}
Rhodes, C.S., van~de Bruck, C., Brax, P., and Davis, A.-C., ``CMB anisotropies
  in the presence of extra dimensions'', {\em Phys. Rev. D}, {\bf 68}, 083511,
  1--13, (2003).
  {\small[\href{http://arxiv.org/abs/astro-ph/0306343}{{astro-ph/0306343}}]}.

\bibitem{bmw_2}
Riazuelo, A., Vernizzi, F., Steer, D.A., and Durrer, R., ``Gauge invariant
  cosmological perturbation theory for braneworlds'', arXiv e-print, (2002).
  {\small[\href{http://arxiv.org/abs/hep-th/0205220}{{hep-th/0205220}}]}.

\bibitem{inf_14}
Rinaldi, M., ``Brane-worlds in T-dual bulks'', {\em Phys. Lett. B}, {\bf 582},
  249--256, (2004).
  {\small[\href{http://arxiv.org/abs/hep-th/0311147}{{hep-th/0311147}}]}.

\bibitem{pert_25}
Ringeval, C., Boehm, T., and Durrer, R., ``CMB anisotropies from vector
  perturbations in the bulk'', arXiv e-print, (2003).
  {\small[\href{http://arxiv.org/abs/hep-th/0307100}{{hep-th/0307100}}]}.

\bibitem{loop_1}
Rovelli, C., ``Loop Quantum Gravity'', {\em Living Rev. Relativity}, {\bf 11},
  lrr-2008-5, (2008). URL (cited on 19 January 2010):
  \newline\url{http://www.livingreviews.org/lrr-2008-5}.

\bibitem{r_2}
Rubakov, V.A., ``Large and infinite extra dimensions'', {\em Phys. Usp.}, {\bf
  44}, 871--893, (2001).
  {\small[\href{http://arxiv.org/abs/hep-ph/0104152}{{hep-ph/0104152}}]}.

\bibitem{add_5}
Rubakov, V.A., and Shaposhnikov, M.E., ``Do we live inside a domain wall?'',
  {\em Phys. Lett. B}, {\bf 125}, 136--138, (1983).

\bibitem{hs_4}
Sago, N., Himemoto, Y., and Sasaki, M., ``Quantum fluctuations in brane-world
  inflation without inflaton on the brane'', {\em Phys. Rev. D}, {\bf 65},
  024014, 1--9, (2002).
  {\small[\href{http://arxiv.org/abs/gr-qc/0104033}{{gr-qc/0104033}}]}.

\bibitem{steep_3}
Sahni, V., Sami, M., and Souradeep, T., ``Relic gravity waves from braneworld
  inflation'', {\em Phys. Rev. D}, {\bf 65}, 023518, 1--16, (2002).
  {\small[\href{http://arxiv.org/abs/gr-qc/0105121}{{gr-qc/0105121}}]}.

\bibitem{ind_8}
Sahni, V., and Shtanov, Y., ``New Vistas in Braneworld Cosmology'', {\em Int.
  J. Mod. Phys. D}, {\bf 11}, 1515--1521, (2002).
  {\small[\href{http://arxiv.org/abs/gr-qc/0205111}{{gr-qc/0205111}}]}.

\bibitem{Sahni:2002dx}
Sahni, V., and Shtanov, Y., ``Braneworld models of dark energy'', {\em J.
  Cosmol. Astropart. Phys.}, {\bf 2003}(11), 014, (2003).
  {\small[\href{http://dx.doi.org/10.1088/1475-7516/2003/11/014}{DOI}]},
  {\small[\href{http://arxiv.org/abs/astro-ph/0202346}{{astro-ph/0202346}}]}.

\bibitem{ind_6}
Sahni, V., and Shtanov, Y., ``Braneworld models of dark energy'', {\em J.
  Cosmol. Astropart. Phys.}, {\bf 2003}(11), 014, (2003).
  {\small[\href{http://arxiv.org/abs/astro-ph/0202346}{{astro-ph/0202346}}]}.

\bibitem{b1_2}
Santos, M.G., Vernizzi, F., and Ferreira, P.G., ``Isotropy and stability of the
  brane'', {\em Phys. Rev. D}, {\bf 64}, 063506, 1--8, (2001).
  {\small[\href{http://arxiv.org/abs/hep-ph/0103112}{{hep-ph/0103112}}]}.

\bibitem{ssm}
Sasaki, M., Shiromizu, T., and Maeda, K., ``Gravity, stability, and energy
  conservation on the Randall--Sundrum brane world'', {\em Phys. Rev. D}, {\bf
  62}, 024008, 1--8, (2000).
  {\small[\href{http://arxiv.org/abs/hep-th/9912233}{{hep-th/9912233}}]}.

\bibitem{acftcosmo_2}
Savonije, I., and Verlinde, E., ``CFT and entropy on the brane'', {\em Phys.
  Lett. B}, {\bf 507}, 305--311, (2001).
  {\small[\href{http://arxiv.org/abs/hep-th/0102042}{{hep-th/0102042}}]}.

\bibitem{Sawicki:2006jj}
Sawicki, I., Song, Y.-S., and Hu, W., ``Near-horizon solution for
  Dvali-Gabadadze-Porrati perturbations'', {\em Phys. Rev. D}, {\bf 75},
  064002, (2007).
  {\small[\href{http://dx.doi.org/10.1103/PhysRevD.75.064002}{DOI}]},
  {\small[\href{http://arxiv.org/abs/astro-ph/0606285}{{astro-ph/0606285}}]}.

\bibitem{mtheory_3}
Schwarz, J.H., ``Update on String Theory'', arXiv e-print, (2003).
  {\small[\href{http://arxiv.org/abs/astro-ph/0304507}{{astro-ph/0304507}}]}.

\bibitem{Seahra:2006tm}
Seahra, S.S., ``Gravitational waves and cosmological braneworlds: A
  characteristic evolution scheme'', {\em Phys. Rev. D}, {\bf 74}, 044010,
  (2006). {\small[\href{http://dx.doi.org/10.1103/PhysRevD.74.044010}{DOI}]},
  {\small[\href{http://arxiv.org/abs/hep-th/0602194}{{hep-th/0602194}}]}.

\bibitem{ksb_4}
Seahra, S.S., Sepangi, H.R., and Ponce~de Leon, J., ``Brane classical and
  quantum cosmology from an effective action'', {\em Phys. Rev. D}, {\bf 68},
  066009, 1--23, (2003).
  {\small[\href{http://arxiv.org/abs/gr-qc/0303115}{{gr-qc/0303115}}]}.

\bibitem{dq_5}
Seery, D., and Bassett, B.A., ``Radiative constraints on brane quintessence'',
  {\em J. Cosmol. Astropart. Phys.}, {\bf 2004}(02), 010, (2004).
  {\small[\href{http://arxiv.org/abs/astro-ph/0310208}{{astro-ph/0310208}}]}.

\bibitem{seetay}
Seery, D., and Taylor, A.N., ``Consistency relation in braneworld inflation'',
  {\em Phys. Rev. D}, {\bf 71}, 063508, 1--16, (2005).
  {\small[\href{http://arxiv.org/abs/astro-ph/0309512}{{astro-ph/0309512}}]}.

\bibitem{Sendouda:2006yc}
Sendouda, Y., ``Cosmic rays from primordial black holes in the Randall-Sundrum
  braneworld'', vol. 861, pp. 1023--1030, (2006).
  {\small[\href{http://dx.doi.org/10.1063/1.2399694}{DOI}]}.

\bibitem{Sendouda:2004hz}
Sendouda, Y., Kohri, K., Nagataki, S., and Sato, K., ``Sub-GeV galactic
  cosmic-ray antiprotons from PBHs in the Randall-Sundrum braneworld'', {\em
  Phys. Rev. D}, {\bf 71}, 063512, (2005).
  {\small[\href{http://dx.doi.org/10.1103/PhysRevD.71.063512}{DOI}]},
  {\small[\href{http://arxiv.org/abs/astro-ph/0408369}{{astro-ph/0408369}}]}.

\bibitem{Sendouda:2003dc}
Sendouda, Y., Nagataki, S., and Sato, K., ``Constraints on the mass spectrum of
  primordial black holes and braneworld parameters from the high-energy diffuse
  photon background'', {\em Phys. Rev. D}, {\bf 68}, 103510, (2003).
  {\small[\href{http://dx.doi.org/10.1103/PhysRevD.68.103510}{DOI}]},
  {\small[\href{http://arxiv.org/abs/astro-ph/0309170}{{astro-ph/0309170}}]}.

\bibitem{pert_17}
Seto, O., and Kodama, H., ``Gravitational waves in cosmological models of
  Horava--Witten theory'', {\em Phys. Rev. D}, {\bf 63}, 123506, 1--8, (2001).
  {\small[\href{http://arxiv.org/abs/hep-th/0012102}{{hep-th/0012102}}]}.

\bibitem{acftcosmo_3}
Shiromizu, T., and Ida, D., ``Anti-de Sitter no-hair, AdS/CFT and the
  brane-world'', {\em Phys. Rev. D}, {\bf 64}, 044015, 1--6, (2001).
  {\small[\href{http://arxiv.org/abs/hep-th/0102035}{{hep-th/0102035}}]}.

\bibitem{sod_3}
Shiromizu, T., and Koyama, K., ``Low energy effective theory for a two branes
  system: Covariant curvature formulation'', {\em Phys. Rev. D}, {\bf 67},
  084022, 1--4, (2003).
  {\small[\href{http://arxiv.org/abs/hep-th/0210066}{{hep-th/0210066}}]}.

\bibitem{sms}
Shiromizu, T., Maeda, K., and Sasaki, M., ``The Einstein equations on the
  3-brane world'', {\em Phys. Rev. D}, {\bf 62}, 024012, 1--6, (2000).
  {\small[\href{http://arxiv.org/abs/gr-qc/9910076}{{gr-qc/9910076}}]}.

\bibitem{num_1}
Shiromizu, T., and Shibata, M., ``Black holes in the brane world: Time
  symmetric initial data'', {\em Phys. Rev. D}, {\bf 62}, 127502, 1--4, (2000).
  {\small[\href{http://dx.doi.org/10.1103/PhysRevD.62.127502}{DOI}]},
  {\small[\href{http://adsabs.harvard.edu/abs/2000PhRvD..62l7502S}{ADS}]},
  {\small[\href{http://arxiv.org/abs/hep-th/0007203}{{hep-th/0007203}}]}.

\bibitem{acftcosmo_4}
Shiromizu, T., Torii, T., and Ida, D., ``Brane-World and Holography'', {\em J.
  High Energy Phys.}, {\bf 2002}(03), 007, (2002).
  {\small[\href{http://arxiv.org/abs/hep-th/0105256}{{hep-th/0105256}}]}.

\bibitem{ind_7}
Shtanov, Y., and Sahni, V., ``New cosmological singularities in braneworld
  models'', {\em Class. Quantum Grav.}, {\bf 19}, L101--L107, (2002).
  {\small[\href{http://arxiv.org/abs/gr-qc/0204040}{{gr-qc/0204040}}]}.

\bibitem{varun1}
Shtanov, Y., and Sahni, V., ``Bouncing braneworlds'', {\em Phys. Lett. B}, {\bf
  557}, 1--6, (2003).
  {\small[\href{http://dx.doi.org/10.1016/S0370-2693(03)00179-5}{DOI}]},
  {\small[\href{http://arxiv.org/abs/gr-qc/0208047}{{arXiv:gr-qc/0208047}}]}.

\bibitem{varun2}
Shtanov, Y., Sahni, V., Shafieloo, A., and Toporensky, A., ``Induced
  cosmological constant and other features of asymmetric brane embedding'',
  {\em J. Cosmol. Astropart. Phys.}, {\bf 2009}(04), 023, (2009).
  {\small[\href{http://dx.doi.org/10.1088/1475-7516/2009/04/023}{DOI}]},
  {\small[\href{http://arxiv.org/abs/0901.3074}{{arXiv:0901.3074
  {\small[gr-qc]}}}]}.

\bibitem{dgp_4}
Shtanov, Y.V., ``On Brane-World Cosmology'', arXiv e-print, (2000).
  {\small[\href{http://arxiv.org/abs/hep-th/0005193}{{hep-th/0005193}}]}.

\bibitem{Silverstein:2003hf}
Silverstein, E., and Tong, D., ``Scalar Speed Limits and Cosmology:
  Acceleration from D-cceleration'', {\em Phys. Rev. D}, {\bf 70}, 103505,
  (2004). {\small[\href{http://dx.doi.org/10.1103/PhysRevD.70.103505}{DOI}]},
  {\small[\href{http://arxiv.org/abs/hep-th/0310221}{{hep-th/0310221}}]}.

\bibitem{nonloc_2}
Singh, P., and Dadhich, N., ``Localized gravity on FRW branes'', arXiv e-print,
  (2002).
  {\small[\href{http://arxiv.org/abs/hep-th/0208080}{{hep-th/0208080}}]}.

\bibitem{ind_9}
Singh, P., Vishwakarma, R.G., and Dadhich, N., ``Brane curvature and supernovae
  Ia observations'', arXiv e-print, (2002).
  {\small[\href{http://arxiv.org/abs/hep-th/0206193}{{hep-th/0206193}}]}.

\bibitem{sod_1}
Soda, J., and Kanno, S., ``Radion and holographic brane gravity'', {\em Phys.
  Rev. D}, {\bf 66}, 083506, 1--15, (2002).
  {\small[\href{http://arxiv.org/abs/hep-th/0207029}{{hep-th/0207029}}]}.

\bibitem{sod_4}
Soda, J., and Kanno, S., ``Holographic View of Cosmological Perturbations'',
  {\em Astrophys. Space Sci.}, {\bf 283}, 639--644, (2003).
  {\small[\href{http://arxiv.org/abs/gr-qc/0209086}{{gr-qc/0209086}}]}.

\bibitem{sod_5}
Soda, J., and Kanno, S., ``Low Energy Effective Action for Dilatonic Braneworld
  --A Formalism for Inflationary Braneworld--'', {\em Gen. Relativ. Gravit.},
  {\bf 36}, 689--712, (2004).
  {\small[\href{http://arxiv.org/abs/hep-th/0303203}{{hep-th/0303203}}]}.

\bibitem{Sollerman:2009yu}
Sollerman, J., M\"ortsell, E., Davis, T.M., Blomqvist, M., Bassett, B., Becker,
  A.C., Cinabro, D., Filippenko, A.V., Foley, R.J., Frieman, J., Garnavich, P.,
  Lampeitl, H., Marriner, J., Miquel, R., Nichol, R.C., Richmond, M.W., Sako,
  M., Schneider, D.P., Smith, M., Vanderplas, J.T., and Wheeler, J.C.,
  ``First-Year Sloan Digital Sky Survey-II (SDSS-II) Supernova Results:
  Constraints on Non-Standard Cosmological Models'', {\em Astrophys. J.}, {\bf
  703}, 1374--1385, (2009).
  {\small[\href{http://dx.doi.org/10.1088/0004-637X/703/2/1374}{DOI}]},
  {\small[\href{http://arxiv.org/abs/0908.4276}{{arXiv:0908.4276
  {\small[astro-ph.CO]}}}]}.

\bibitem{Song:2006jk}
Song, Y.-S., Sawicki, I., and Hu, W., ``Large-scale tests of the
  Dvali-Gabadadze-Porrati model'', {\em Phys. Rev. D}, {\bf 75}, 064003,
  (2007). {\small[\href{http://dx.doi.org/10.1103/PhysRevD.75.064003}{DOI}]},
  {\small[\href{http://arxiv.org/abs/astro-ph/0606286}{{astro-ph/0606286}}]}.

\bibitem{ek_4}
Steinhardt, P.J., and Turok, N., ``Cosmic evolution in a cyclic universe'',
  {\em Phys. Rev. D}, {\bf 65}, 126003, 1--20, (2002).
  {\small[\href{http://arxiv.org/abs/hep-th/0111098}{{hep-th/0111098}}]}.

\bibitem{Stewart:1993bc}
Stewart, E.D., and Lyth, D.H., ``A more accurate analytic calculation of the
  spectrum of cosmological perturbations produced during inflation'', {\em
  Phys. Lett. B}, {\bf 302}, 171--175, (1993).
  {\small[\href{http://dx.doi.org/10.1016/0370-2693(93)90379-V}{DOI}]},
  {\small[\href{http://arxiv.org/abs/gr-qc/9302019}{{gr-qc/9302019}}]}.

\bibitem{inf_3}
Stoica, H., Tye, S.H.H., and Wasserman, I., ``Cosmology in the Randall--Sundrum
  brane world scenario'', {\em Phys. Lett. B}, {\bf 482}, 205--212, (2000).
  {\small[\href{http://arxiv.org/abs/hep-th/0004126}{{hep-th/0004126}}]}.

\bibitem{t}
Tanaka, T., ``Classical Black Hole Evaporation in Randall--Sundrum Infinite
  Braneworld'', {\em Prog. Theor. Phys. Suppl.}, {\bf 148}, 307--316, (2002).
  {\small[\href{http://arxiv.org/abs/gr-qc/0203082}{{gr-qc/0203082}}]}.

\bibitem{hs_7}
Tanaka, T., and Himemoto, Y., ``Generation of dark radiation in the bulk
  inflaton model'', {\em Phys. Rev. D}, {\bf 67}, 104007, 1--6, (2003).
  {\small[\href{http://arxiv.org/abs/gr-qc/0301010}{{gr-qc/0301010}}]}.

\bibitem{2b_1}
Tanaka, T., and Montes, X., ``Gravity in the brane-world for two-branes model
  with stabilized modulus'', {\em Nucl. Phys. B}, {\bf 582}, 259--276, (2000).
  {\small[\href{http://arxiv.org/abs/hep-th/0001092}{{hep-th/0001092}}]}.

\bibitem{loop_2}
Thiemann, T., ``Lectures on Loop Quantum Gravity'', in Giulini, D., Kiefer, C.,
  and L{\"{a}}mmerzahl, C., eds., {\em Quantum Gravity: From Theory to
  Experimental Search}, 271th WE-Heraeus Seminar `Aspects of Quantum Gravity',
  Bad Honnef, Germany, 24 February\,--\,1 March 2002, Lecture Notes in Physics,
  vol. 631, pp. 41--135, (Springer, Berlin; New York, 2003).
  {\small[\href{http://arxiv.org/abs/gr-qc/0210094}{{gr-qc/0210094}}]}.

\bibitem{Tolley:2006ht}
Tolley, A.J., Burgess, C.P., de~Rham, C., and Hoover, D., ``Scaling solutions
  to 6D gauged chiral supergravity'', {\em New J. Phys.}, {\bf 8}, 324, (2006).
  {\small[\href{http://dx.doi.org/10.1088/1367-2630/8/12/324}{DOI}]},
  {\small[\href{http://arxiv.org/abs/hep-th/0608083}{{hep-th/0608083}}]}.  URL
  (cited on 21 April 2010):
  \newline\url{http://stacks.iop.org/1367-2630/8/12/324}.

\bibitem{Tolley:2007et}
Tolley, A.J., Burgess, C.P., de~Rham, C., and Hoover, D., ``Exact Wave
  Solutions to 6D Gauged Chiral Supergravity'', {\em J. High Energy Phys.},
  {\bf 2008}(07), 075, (2008).
  {\small[\href{http://dx.doi.org/10.1088/1126-6708/2008/07/075}{DOI}]},
  {\small[\href{http://arxiv.org/abs/0710.3769}{{arXiv:0710.3769
  {\small[hep-th]}}}]}.

\bibitem{ek_7}
Tolley, A.J., Turok, N., and Steinhardt, P.J., ``Cosmological perturbations in
  a big-crunch--bang space-time'', {\em Phys. Rev. D}, {\bf 69}, 106005, 1--26,
  (2004).
  {\small[\href{http://arxiv.org/abs/hep-th/0306109}{{hep-th/0306109}}]}.

\bibitem{b1_1}
Toporensky, A.V., ``The shear dynamics in Bianchi I cosmological model on the
  brane'', {\em Class. Quantum Grav.}, {\bf 18}, 2311--2316, (2001).
  {\small[\href{http://arxiv.org/abs/gr-qc/0103093}{{gr-qc/0103093}}]}.

\bibitem{preheat_1}
Tsujikawa, S., ``Preheating with extra dimensions'', {\em J. High Energy
  Phys.}, {\bf 2001}(07), 024, (2001).
  {\small[\href{http://arxiv.org/abs/hep-ph/0005105}{{hep-ph/0005105}}]}.

\bibitem{preheat_4}
Tsujikawa, S., Maeda, K., and Mizuno, S., ``Brane preheating'', {\em Phys. Rev.
  D}, {\bf 63}, 123511, 1--5, (2001).
  {\small[\href{http://arxiv.org/abs/hep-ph/0012141}{{hep-ph/0012141}}]}.

\bibitem{pert_4}
van~de Bruck, C., Dorca, M., Brandenberger, R.H., and Lukas, A., ``Cosmological
  perturbations in brane-world theories: Formalism'', {\em Phys. Rev. D}, {\bf
  62}, 123515, 1--13, (2000).
  {\small[\href{http://arxiv.org/abs/hep-th/0005032}{{hep-th/0005032}}]}.

\bibitem{pert_7}
van~de Bruck, C., Dorca, M., Martins, C.J., and Parry, M., ``Cosmological
  consequences of the brane/bulk interaction'', {\em Phys. Lett. B}, {\bf 495},
  183--192, (2000).
  {\small[\href{http://arxiv.org/abs/hep-th/0009056}{{hep-th/0009056}}]}.

\bibitem{Verlinde:1999fy}
Verlinde, H.L., ``Holography and compactification'', {\em Nucl. Phys. B}, {\bf
  580}, 264--274, (2000).
  {\small[\href{http://dx.doi.org/10.1016/S0550-3213(00)00224-8}{DOI}]},
  {\small[\href{http://arxiv.org/abs/hep-th/9906182}{{hep-th/9906182}}]}.

\bibitem{Vinet:2004bk}
Vinet, J., and Cline, J.M., ``Can codimension-two branes solve the cosmological
  constant problem?'', {\em Phys. Rev. D}, {\bf 70}, 083514, (2004).
  {\small[\href{http://dx.doi.org/10.1103/PhysRevD.70.083514}{DOI}]},
  {\small[\href{http://arxiv.org/abs/hep-th/0406141}{{hep-th/0406141}}]}.

\bibitem{Vinet:2005dg}
Vinet, J., and Cline, J.M., ``Codimension-two branes in six-dimensional
  supergravity and the cosmological constant problem'', {\em Phys. Rev. D},
  {\bf 71}, 064011, (2005).
  {\small[\href{http://dx.doi.org/10.1103/PhysRevD.71.064011}{DOI}]},
  {\small[\href{http://arxiv.org/abs/hep-th/0501098}{{hep-th/0501098}}]}.

\bibitem{add_6}
Visser, M., ``An exotic class of Kaluza-Klein models'', {\em Phys. Lett. B},
  {\bf 159}, 22--25, (1985).
  {\small[\href{http://arxiv.org/abs/hep-th/9910093}{{hep-th/9910093}}]}.

\bibitem{germ_3}
Visser, M., and Wiltshire, D.L., ``On-brane data for braneworld stars'', {\em
  Phys. Rev. D}, {\bf 67}, 104004, 1--6, (2003).
  {\small[\href{http://arxiv.org/abs/hep-th/0212333}{{hep-th/0212333}}]}.

\bibitem{wald}
Wald, R.M., {\em General Relativity}, (University of Chicago Press, Chicago,
  1984).

\bibitem{rev_3}
Wands, D., ``String-inspired cosmology'', {\em Class. Quantum Grav.}, {\bf 19},
  3403--3416, (2002).
  {\small[\href{http://arxiv.org/abs/hep-th/0203107}{{hep-th/0203107}}]}.

\bibitem{wmll}
Wands, D., Malik, K.A., Lyth, D.H., and Liddle, A.R., ``New approach to the
  evolution of cosmological perturbations on large scales'', {\em Phys. Rev.
  D}, {\bf 62}, 043527, 1--8, (2000).
  {\small[\href{http://arxiv.org/abs/astro-ph/0003278}{{astro-ph/0003278}}]}.

\bibitem{hs_8}
Wang, B., Xue, L.-H., Zhang, X., and Hwang, W.-Y.P., ``Modification to the
  power spectrum in the brane world inflation driven by the bulk inflaton'',
  {\em Phys. Rev. D}, {\bf 67}, 123519, 1--6, (2003).
  {\small[\href{http://arxiv.org/abs/hep-th/0301072}{{hep-th/0301072}}]}.

\bibitem{w}
Wiseman, T., ``Relativistic stars in Randall--Sundrum gravity'', {\em Phys.
  Rev. D}, {\bf 65}, 124007, 1--32, (2002).
  {\small[\href{http://arxiv.org/abs/hep-th/0111057}{{hep-th/0111057}}]}.

\bibitem{sod_2}
Wiseman, T., ``Strong brane gravity and the radion at low energies'', {\em
  Class. Quantum Grav.}, {\bf 19}, 3083--3105, (2002).
  {\small[\href{http://arxiv.org/abs/hep-th/0201127}{{hep-th/0201127}}]}.

\bibitem{preheat_2}
Yokoyama, J., and Himemoto, Y., ``After bulky brane inflation'', {\em Phys.
  Rev. D}, {\bf 64}, 083511, 1--5, (2001).
  {\small[\href{http://arxiv.org/abs/hep-ph/0103115}{{hep-ph/0103115}}]}.

\bibitem{Yoshino:2008rx}
Yoshino, H., ``On the existence of a static black hole on a brane'', {\em J.
  High Energy Phys.}, {\bf 2009}(01), 068, (2009).
  {\small[\href{http://dx.doi.org/10.1088/1126-6708/2009/01/068}{DOI}]},
  {\small[\href{http://arxiv.org/abs/0812.0465}{{arXiv:0812.0465
  {\small[gr-qc]}}}]}.

\end{thebibliography}

\end{document}